\documentclass[aps,prx,reprint,groupedaddress,superscriptaddress,floatfix,longbibliography]{revtex4-2}

\usepackage[T1]{fontenc}
\usepackage{newtxtext}
\usepackage{newtxmath}

\usepackage{titlesec}

\usepackage{amsmath}
\usepackage{mathtools}
\usepackage{amsfonts}
\usepackage{bbold}
\usepackage{physics}
\usepackage{color}
\usepackage{xcolor}
\usepackage[colorlinks,citecolor=blue,linkcolor=blue,urlcolor=blue]{hyperref}

\usepackage{graphicx}
\usepackage[all]{hypcap}

\usepackage{enumitem}

\usepackage{tabularx}
\usepackage{makecell}

\begin{document}

\title{Exact Solvability Of Entanglement For Arbitrary Initial State in an Infinite-Range Floquet System}

\author{Harshit Sharma}
\email{ds21phy007@students.vnit.ac.in}
\affiliation{Department of Physics, National Institute of Technology, Nagpur 440010, India}

\author{Udaysinh T. Bhosale}
\email{udaysinhbhosale@phy.vnit.ac.in}
\affiliation{Department of Physics, National Institute of Technology, Nagpur 440010, India}

\date{\today}
\begin{abstract}
Sharma and Bhosale [\href{https://journals.aps.org/prb/abstract/10.1103/PhysRevB.109.014412}{Phys. Rev. B \textbf{109}, 014412 (2024)}; \href{https://journals.aps.org/prb/abstract/10.1103/PhysRevB.110.064313}{Phys. Rev. B \textbf{110}, 064313,(2024)}] recently introduced an $N$-spin Floquet model with infinite-range Ising interactions. There, we have shown that the model exhibits the signatures of quantum integrability for specific parameter values $J=1,1/2$ and $\tau=\pi/4$. We have found analytically the eigensystem and the time evolution of the unitary operator for finite values of $N$ up to $12$ qubits. We have calculated the reduced density matrix, its eigensystem, time-evolved linear entropy, and the time-evolved concurrence for the initial states $\ket{0,0}$ and $\ket{\pi/2,-\pi/2}$. For the general case $N>12$, we have provided sufficient numerical evidences for the signatures of quantum integrability, such as the degenerate spectrum, the exact periodic nature of entanglement dynamics, and the time-evolved unitary operator. In this paper, we have extended these calculations to arbitrary initial state $\ket{\theta_0,\phi_0}$, such that $\theta_0 \in [0,\pi]$ and $\phi_0 \in [-\pi,\pi]$. Along with that, we have analytically calculated the expression for the average linear entropy for arbitrary initial states. We numerically find that the  average value of time-evolved concurrence for arbitrary initial states  decreases with
$N$, implying the multipartite nature of entanglement. We numerically show that the values $\langle S\rangle/S_{Max} \rightarrow 1$ for Ising strength ($J\neq1,1/2$), while for $J=1$ and $1/2$, it deviates from $1$ for arbitrary initial states even though the
thermodynamic limit does not exist in our model. This deviation is shown to be a signature of integrability in earlier studies where the thermodynamic limit exist. We also discuss possible experiments that could be conducted to verify our results.
\end{abstract}
\maketitle
\section{ Introduction}
Entanglement is a fundamental property of the quantum realm that has been studied extensively, which has no counterpart in classical mechanics \cite{schrodinger1935present,horodecki2009quantum,amico2008entanglement}. Presently, the entanglement theory has been used in several important discoveries like quantum cryptography  \cite{ekert1991quantum,zeng2023controlled}, quantum teleportation \cite{bennett1895experimental,hu2023progress}, quantum computing \cite{shor1995scheme,conlon2023approaching}, quantum phase transition \cite{osborne2002entanglement,osterloh2002scaling,vidal2003entanglement,larsson2006single,mueller2023quantum,amico2008entanglement}, and quantum dense coding \cite{bennett1992communication}. All these effects have been experimentally demonstrated in numerous studies \cite{mattle1996dense,bouwmeester1997experimental,boschi1998experimental,furusawa1998unconditional,pan1998experimental,jennewein2000quantum,naik2000entangled,tittel2000quantum}. Furthermore, entanglement plays a crucial role in understanding various physical phenomena, such as super-radiance \cite{lambert2004entanglement}, disordered systems \cite{dur2005entanglement}, the emergence of classicality \cite{zurek2003decoherence}, and superconductivity \cite{vedral2004high}. The concept of entanglement witness has been employed to various problems in statistical systems \cite{wiesniak2005magnetic,cavalcanti2006entanglement}, quantum optics \cite{stobinska2006witnessing}, bound entanglement \cite{hyllus2004generation}, experimental realization of cluster states \cite{vallone2007realization}, hidden nonlocality \cite{masanes2008all}, quantum information \cite{bouwmeester2000physics,nielsen2010quantum}, quantum gravity \cite{nishioka2009holographic}, condensed matter physics \cite{amico2008entanglement,laflorencie2016quantum}, quantum spin chains \cite{turkeshi2023entanglement,williamson2024many}, and long-range interaction \cite{lerose2020origin,defenu2023long}. In recent times, there has been growing interest in quantum long-range systems to explore the fundamental physics of nonlocal systems and examine the connection between local and long-distance properties \cite{defenu2023long}. Additionally, researchers are investigating how these systems differ from their classical counterparts \cite{defenu2023long}.

Long-range interaction plays a vital role in various domains of physics \cite{dauxois2002dynamics,sutherland2004beautiful,campa2014physics,wormer1977quantum,salam2009molecular,dolgov1999long,nusser2005structure,van2012dark,esteban2021long,chomaz2002phase,elskens2019microscopic,miller1990statistical,schuckert2020nonlocal,morningstar2023hydrodynamics,roca2024transformer}. They decay according to power law $1/r^{\alpha}$ as a function of distance $r$, where the exponent $\alpha$ is the main character for defining the strong and weak long-range interaction. For the case $\alpha<d$,  the energy is not extensive, where $d$ is the physical dimension of the system  \cite{campa2014physics,dauxois2002dynamics,defenu2023long}. Models that satisfy this criterion are characterized as long-range interactions. These interactions were implemented across various experimental platforms, including Rydberg atoms \cite{saffman2010quantum}, dipolar quantum gases \cite{lahaye2009physics,chomaz2022dipolar,smith2023supersolidity}, polar molecules  \cite{yan2013observation}, cold atoms
in cavities \cite{baumann2010dicke}, quantum gases coupled to optical cavities, magnetic atoms \cite{griesmaier2005bose,beaufils2008all,baier2016extended}, nonlinear optical media \cite{firstenberg2013attractive}, solid-state defects \cite{childress2006coherent}, and trapped ions \cite{blatt2012quantum,monroe2021programmable}. On tuning the exponent $\alpha$, long-range interactions fall into several categories like van der Waals interaction ($\alpha=6$), Coulomb interaction ($\alpha=1$), dipole-dipole interaction ($\alpha=2$), and infinite range interaction ($\alpha=0$) \cite{britton2012engineered,schauss2012observation,peter2012anomalous,yan2013observation,
jurcevic2014quasiparticle,hazzard2014many,richerme2014non,douglas2015quantum,gil2016nonequilibrium,Marino2019,sauerwein2023engineering,liu2024signature,vzunkovivc2024mean,offei2020quantum,delmonte2024measurement,lerose2020origin}, etc. Such interactions are helpful in multiple quantum technology applications like quantum computing \cite{inoue2015infinite,lewis2021optimal,lewis2023ion}, quantum heat engine \cite{solfanelli2023quantum}, ion trap \cite{gambetta2020long}, quantum metrology \cite{pezze2018quantum}, entanglement spreading \cite{pappalardi2018scrambling}, and generation of faster entanglement \cite{hauke2013spread,eldredge2017fast,colmenarez2020lieb}. Quantum spin chains have been extensively studied in two main families of models: nearest-neighbor interactions and long-range interactions \cite{ExactDogra2019,pal2018entangling,mishra2015protocol,sharma2024exactly,sharma2024signatures,delmonte2024measurement}. There are various integrable models corresponding to these interactions \cite{bargheer2008boosting,mishra2015protocol,pal2018entangling,kumari2022eigenstate,sharma2024exactly,sharma2024signatures,wierzchucka2024integrability}. However, our main focus in this work is on quantum integrability in models with infinite range interaction. Notably, some of these models, such as the well-known Lipkin-Meshkov-Glick (LMG)  models \cite{kumari2022eigenstate} and the model with Ising interaction in a transverse field, exhibit quantum integrability \cite{sharma2024exactly,sharma2024signatures}

Integrable models have significantly contributed to our understanding of physical systems \cite{negro2023deterministic,retore2022introduction,alcaraz2020free}. In classical mechanics, the integrability of a system requires that the number of degrees of freedom should be equal to the number of constant of motion in involution \cite{babelon2003introduction,retore2022introduction,torrielli2016classical}. On the other hand, quantum integrability generally associated with the exact solution of the models \cite{thacker1981exact,owusu2008link,doikou2010introduction,gubin2012quantum,yuzbashyan2013quantum,
gombor2021integrable,tang2023integrability,vernier2024strong,claeys2018integrable}. This is often achieved using the Yang-Baxter equation \cite{wadati1993quantum,doikou2010introduction,lambe2013introduction,baxter2016exactly,gaudin2014bethe,zheng2024exact} and techniques like Bethe ansatz \cite{bethe1931theorie,faddeev1995algebraic,pan1999analytical,bargheer2008boosting,wierzchucka2024integrability}. Additionally, quantum integrability can be identified with other features like sufficient number of independent conserved quantities, and/or Poissonian level statistics of the Hamiltonian \cite{berry1977level,yuzbashyan2013quantum}. Recent studies have shown that the quantum integrability in a system can also be identified through additional signatures, such as the exact periodicity of entanglement dynamics, the time evolution of the Floquet operator, and degenerate spectra \cite{yuzbashyan2013quantum,mishra2015protocol,doikou2010introduction,pal2018entangling,naik2019controlled,sharma2024exactly,sharma2024signatures}. Furthermore, numerous studies have demonstrated that the integrability and nonintegrability in systems can be distinguished by their average entanglement entropy \cite{vidmar2017entanglement,hackl2019average,lydzba2020eigenstate,kumari2022eigenstate,page1993average}. In integrable cases, the average entanglement entropy significantly diverges from their maximum value. In contrast, in non-integrable cases, it is close to the maximum possible value i.e., $\langle S\rangle/S_{Max}\rightarrow 1$ in the thermodynamic limit. For an integrable system at half bipartitaion, it has been shown analytically that the ratio $ \langle S\rangle/S_{Max}$ for free fermions \cite{vidmar2017entanglement} and $XY$ chain \cite{hackl2019average} falls within the interval $[0.52,0.59]$; for random quadratic model around $0.557$  \cite{lydzba2020eigenstate}; for the Dicke basis and LMG model, it is around $0.7213$ and $0.5$ respectively \cite{kumari2022eigenstate}.

In recent studies, we introduced a many-body Floquet spin model with an infinite-range Ising interaction and showed that it possesses the signatures of quantum integrability for specific values of the parameters \cite{sharma2024exactly,sharma2024signatures}. There we have analytically evaluated the eigensystem,  time evolution of the unitary operator, reduced density matrix, and the entanglement dynamics for the system with $5\le N\le 12$ qubits. However, for the general case of $N>12$ qubits, we employed numerical methods due to the complexity and fairly large calculations, as finding analytical solutions presents substantial mathematical challenges. We have used linear entropy, entanglement entropy, and concurrence to quantify the entanglement \cite{buscemi2007linear,nielsenbook,benenti2004principles,wootters1998entanglement,wootters2001entanglement}. The quantum integrability in the system was identified through signatures like the periodicity of the entanglement dynamics \cite{ArulLakshminarayan2005,naik2019controlled,mishra2015protocol,
pal2018entangling} and that of the Floquet operator dynamics \cite{pal2018entangling}, and highly degenerated spectra \cite{yuzbashyan2013quantum,naik2019controlled}. We observed these signatures in our system for the parameter value of  Ising strength $J=1,1/2$ and $\tau=\pi/4$. In Refs. \cite{sharma2024exactly,sharma2024signatures}, the entanglement dynamics were calculated analytically $N<12$  and numerically $N>12$ for the initial coherent states $\ket{0,0}$ and $\ket{\pi/2,-\pi/2}$ with the parameter values mentioned above.

In our earlier studies  \cite{sharma2024exactly,sharma2024signatures}, we have shown that our model has a close connection to the integrable LMG \cite{kumari2022eigenstate}, and Quantum kicked Top (QKT) \cite{haake1987classical,Haakebook, UdaysinhPeriodicity2018,anand2024quantum} model for the special values of the parameter. Through mapping with QKT, the integrability of the system was limited to only up to four qubits  \cite{UdaysinhPeriodicity2018, ExactDogra2019}. In a very recent work \cite{sharma2024exactly}, we have shown that for $J=1,\tau=\pi/4$, the system is integrable, and its pairwise concurrence vanishes for any $N$-qubits for the aforementioned coherent states. In contrast, in Ref. \cite{sharma2024signatures}, for the parameter $J=1/2$ and same $\tau$, integrability in the system was observed only for even $N$ qubits, while no such signatures were observed for odd $N$. Additionally, the concurrence decays with increasing $N$, indicating the multipartite nature of the system in both the cases. In our previous works \cite{sharma2024exactly,sharma2024signatures}, we dealt with only two initial states $\ket{0,0}$ and $\ket{\pi/2,-\pi/2}$, but the calculation for arbitrary initial states were left open for future. The details of the model will be explained in the subsequent part of this paper.

In this work, our main focus is on deriving the exact results for any arbitrary initial coherent state (To be defined in Sec. \ref{sec:example-section2}; see Eq. (\ref{Eq:oddBasis11})). This is motivated by the fact that as the initial state for time evolution changes distinct outcomes emerge, such as variations in the nature of entanglement dynamics, critical exponents, critical disorder strength, and other essential properties of the system \cite{madhok2015comment,ghose2008chaos,lakshminarayan2024chaos}. From this perspective, we analytically calculate the entanglement dynamics for a given arbitrary initial coherent state. We derive expressions for the linear entropy, entanglement entropy, and time average linear entropy for qubits ranging from $4$ to $10$ with parameters $J=1$ and $\tau=\pi/4$. We derive similar expressions for the parameters $J=1/2$ and  $\tau=\pi/4$ for even number of qubits in the range $4$ to $10$. We observe that the entanglement dynamics are periodic in nature for arbitrary initial unentangled states with the same parameters values.

For the general case $N>10$, we numerically observe a similar periodic nature for any arbitrary initial states across all $N$  when $J=1$, and for even number of qubits when $J=1/2$. We numerically show that the time average concurrence approaches zero with $N$, indicating the multipartite nature of entanglement for both the cases. We identify the initial states where the average entanglement dynamics attain their maximum and minimum values. In Refs.\cite{sharma2024exactly,sharma2024signatures}, we have shown that the signatures of quantum integrability exhibits in our model for the initial states $\ket{0,0}$ and $\ket{\pi/2,-\pi/2}$. Here, we extend these signatures of quantum integrability for arbitrary initial states. It is obvious but worth noting that the spectral signatures and dynamics of the time-evolution operator are independent of the initial state.

Various studies have shown that, in integral systems, the average entanglement entropy diverges from its maximal. It serves as a good indicator in distinguishing the integral and non-integral systems in thermodynamic limits \cite{vidmar2017entanglement,hackl2019average,lydzba2020eigenstate,kumari2022eigenstate}.
In this work, we numerically investigate the normalized linear entropy ($\langle S\rangle/S_{Max}$) with $J$ for various initial states across different numbers of odd and even qubits. We also study the normalized linear entropy with $N$ for specific values of $J$. We observe that for certain values of $J$, the ratio $\langle S\rangle/S_{Max}$ diverges from $1$ for an arbitrary initial state even though the thermodynamic limit does not
exist in our model. It serves as an additional valuable signature, alongside previously known signatures, for identifying the quantum integrability in our system.

The rest of this paper is organized as follows. In Sec. \ref{sec:example-section2}, we provide a brief introduction to the model under investigation. In Sec. \ref{sec:example-section3}, we present an exact analytical solution for entanglement measures, such as linear entropy and entanglement entropy, along with the expression for the average linear entropy for the parameter $J=1$ and $\tau=\pi/4$, considering qubit systems ranging from $4$ to $10$  for arbitrary initial unentangled states. In Sec. \ref{sec:example-section4}, we derive similar expressions for $J=1/2$ and  $\tau=\pi/4$ for even  $N=4,6,8$ and $10$. In Sec. \ref{sec:example-section5}, we provide extensive numerical results analyzing the impact of Ising coupling strength on the average linear entropy for arbitrary initial states in systems with $N$ qubits. We also discuss specific parameter values where the system exhibits quantum integrability. In Sec. \ref{sec:example-section6}, we summarize the main results and conclusions of our work.
\section{The Spin Model} \label{sec:example-section2}
The generalized  Hamiltonian model from the  Ref. \cite{sharma2024exactly} is given as  follows:
\begin{equation}
\label{Eq:QKT}
H(t)= H_I(h)+\sum_{n = -\infty}^{ \infty} \delta(n-t/\tau)~ H_k,
\end{equation}
where $\delta(t)$ is the Dirac delta function and we define,
\begin{eqnarray}
H_I={J} \sum_{ l< l'}\sigma^z_{l} \sigma^z_{l'}~~ \mbox{and} ~~
H_k= \sum_{l=1}^{N}\sigma^y_l.
\end{eqnarray}
The nature of Ising interaction in our model is uniform and all-to-all. The field strength of the Ising interaction is $J$ (first term), and $\tau$ is the time period of the magnetic field, which is applied periodically along the y-axis (second term). The corresponding Floquet operator is given  as follows:
\begin{eqnarray}\nonumber \label{Eq:QKT1}
\mathcal{U} &=& \exp\left [-i~ \tau H_I(h) \right]\exp\left[-i~ \tau H_k\right]\\
  &=& \exp\left (-i~ J \tau \sum_{ l< l'} \sigma^z_{l} \sigma^z_{l'}   \right)  \exp\left(-i~ \tau \sum_{l=1}^{N}\sigma^y_l \right).
\end{eqnarray}
Symmetries in a given model simplifies its structure by imposing additional constraints. This reduces the degrees of freedom thus, easing analytical calculations and numerical computations. Particularly, in our model, the presence of permutation symmetry under the exchange of qubit reduces the Hilbert space dimension from $2^N$ to $N+1$ \cite{ExactDogra2019,sharma2024exactly,sharma2024signatures,seshadri2018tripartite}. This will be explained in further parts of the paper. One of the special case of our model is one with the nearest-neighbor interaction. This is an integrable model and has been studied extensively \cite{ArulLakshminarayan2005,mishra2015protocol,apollaro2016entanglement,bertini2019entanglement,naik2019controlled,shukla2022characteristic}. In the limit as $\tau\rightarrow 0$, where the kicking become increasingly frequent, the kicked model effectively transitions to a time-independent Hamiltonian \cite{pal2018entangling}. As shown in earlier works,  our model is quantum integrable for the special case $J=1,1/2$ and $\tau=\pi/4$.

To achieve an analytical solution for the system involving any number of $N$ qubits, we employ the general basis defined in Ref. \cite{sharma2024exactly}.
The  basis when $N$ is even is given as follows:
\begin{eqnarray}\label{Eq:evenBasis1}
\begin{split}
\ket{\phi_q^{\pm}}&=\frac{1}{\sqrt{2}}\left(\ket{w_q}\pm {(-1)^{\left(j-q\right)}} \ket{\overline{w_q}}\right), 0\leq q\leq j-1\\
\mbox{and} &\;\;
\ket{\phi_{\frac{N}{2}}^+}=\left({1}/{\sqrt{\binom{N}{\frac{N}{2}}}}\right)\sum_{\mathcal{P}}\left(\otimes^{\frac{N}{2}}\ket{0}\otimes^{\frac{N}{2}}\ket{1}\right)_\mathcal{P},
\end{split}
\end{eqnarray}
whereas for odd $N$ it is :
\begin{equation}\label{Eq:oddBasis1}
\ket{\phi_q^{\pm}}=\frac{1}{\sqrt{2}}\left(\ket{w_q}\pm {i^{\left(N-2q\right)}} \ket{\overline{w_q}}\right),
0\leq q\leq \dfrac{N-1}{2};
\end{equation}
where
$\ket{w_q}=\sum_\mathcal{P}\left(\otimes^q \ket{1} \otimes^{(N-q)}\ket{0}\right)_\mathcal{P}$ and
$\ket{\overline{w_q}}=\sum_\mathcal{P}\left(\otimes^{q}\ket{0}\otimes^{(N-q)}\ket{1}\right)_\mathcal{P}$,
both being definite particle states \cite{Vikram11}. The $\sum_\mathcal{P}$ denotes the sum over all possible permutations.
These basis states $\ket{\phi_j^{\pm}}$ are characterized as the eigenstate of parity operator having eigenvalues $\pm1$ i.e. $\otimes_{l=1}^{N}\sigma_l^y\ket{\phi_j^{\pm}}=\pm\ket{\phi_j^{\pm}}$. The advantage of these permutation symmetric basis is that $\mathcal U$ becomes block-diagonalized into two blocks, $\mathcal U^{+}$ and $\mathcal U^{-}$. This block-diagonalization simplifies the computation of its $n${th}  power and facilitates more efficient analysis, as shown in the further part of this paper. In this work, we have studied the time evolution of arbitrary initial states localized within spherical phase space, which lie on a unit sphere with spherical coordinates ($\theta_0,\phi_0$).
The coherent state in the computational basis is given as follows:
\begin{equation}\label{Eq:oddBasis11}
  \ket{\psi_0}=|\theta_0,\phi_0\rangle = \cos(\theta_0/2) |0\rangle + e^{-i \phi_0} \sin(\theta_0/2) |1\rangle.
\end{equation}
We initialized each of the $N$ qubits in the state $\ket{\psi_0}$, such that $N$-qubit unentangled arbitrary initial state is $\ket{\psi}=\otimes^{N}\ket{\psi_0}$, where $\theta_0 \in [0,\pi]$ and $\phi_0 \in [-\pi,\pi]$. This state when expanded fully will have $2^N$ coefficients but as discussed earlier due to permutation symmetry Hilbert space dimensions are just $N+1$. Thus, only $N+1$ coefficients are enough to express the state and its time evolution. To use this fact, we express the arbitrary initial state for any  odd  number of qubits $\ket{\psi}$ in $\ket{\phi}$ basis  as follows:
\begin{equation}
 \ket{\psi}= \sum_{q=1}^{(N+1)/2}\frac{1}{\sqrt{2}}\left( a_q \ket{\phi_{q-1}^+} +b_q \ket{\phi_{q-1}^-}\right),
\end{equation}
where the coefficients $a_q$ and $b_q$ are given as follows, where $q$ lies in the interval $[1,\frac{N+1}{2}]$:
\begin{eqnarray}\nonumber \label{Eq:arbitaray2}
a_q&=&\sqrt{\binom{N}{q-1}}\left(\cos^{N-(q-1)}\left(\theta_0/2\right) e^{-i (q-1)\phi_0} \right. \\   &&\left.\sin^{(q-1)}\left(\theta_0/2\right) -~i^{N-2(q-1)}\cos^{(q-1)}\left(\theta_0/2\right) \right. \\ \nonumber  &&\left.e^{-i (N-(q-1))\phi_0}\sin^{N-(q-1)}\left(\theta_0/2\right)\right),~~\mbox{and}\\ \nonumber \label{Eq:arbitaray3}
b_q&=&\sqrt{\binom{N}{(q-1)}}\left(\cos^{N-(q-1)}\left(\theta_0/2\right) e^{-i (q-1) \phi_0} \right. \\   &&\left.\sin^{q-1}\left(\theta_0/2\right) +~i^{N-2(q-1)}\cos^{(q-1)}\left(\theta_0/2\right)\right. \\ \nonumber   &&\left. e^{-i (N-(q-1))\phi_0}\sin^{N-(q-1)}\left(\theta_0/2\right)\right).
\end{eqnarray}
 The arbitrary initial state for any even number of qubits can be expressed as,
\begin{equation}
 \ket{\psi}= \sum_{q=1}^{N/2}\frac{1}{\sqrt{2}}\left( a_q \ket{\phi_{q-1}^+} +b_q \ket{\phi_{q-1}^-}\right)+a_{\frac{N+2}{2}} \ket{\phi_{\frac{N}{2}}^+},
\end{equation}
where the coefficients $a_q$, $b_q$ and $a_{\frac{N+2}{2}}$ are given as follows, with $q$ lying in the interval $[1,\frac{N}{2}]$:
\begin{eqnarray}\nonumber \label{Eq:arbitaray}
a_q&=&\sqrt{\binom{N}{q-1}}\left(\cos^{N-(q-1)}\left(\theta_0/2\right) e^{-i (q-1)\phi_0} \right. \\  &&\left.\sin^{(q-1)}\left(\theta_0/2\right) +~i^{N-2(q-1)}\cos^{(q-1)}\left(\theta_0/2\right) \right. \\ \nonumber  &&\left.e^{-i (N-(q-1))\phi_0}\sin^{N-(q-1)}\left(\theta_0/2\right)\right),\\ \nonumber
\end{eqnarray}
\begin{eqnarray}\label{Eq:arbitaray1} \nonumber
b_q&=&\sqrt{\binom{N}{(q-1)}}\left(\cos^{N-(q-1)}\left(\theta_0/2\right) e^{-i (q-1) \phi_0} \right. \\   &&\left.\sin^{q-1}\left(\theta_0/2\right) -~i^{N-2(q-1)}\cos^{(q-1)}\left(\theta_0/2\right) \right. \\ \nonumber  &&\left.e^{-i (N-(q-1))\phi_0}\sin^{N-(q-1)}\left(\theta_0/2\right)\right),~~\mbox{and}\\  \label{Eq:arbitaray4}
a_{\frac{N+2}{2}}&=&\sqrt{\binom{N}{\frac{N}{2}}}\left(e^{-i\frac{N}{2} \phi_0}\cos^{\frac{N}{2}}\left(\theta_0/2\right)  \sin^{\frac{N}{2}}\left(\theta_0/2\right)\right).
\end{eqnarray}
Now, in the subsequent part of this paper we use these compact coefficients (Eqs. (\ref{Eq:arbitaray2}), (\ref{Eq:arbitaray3}), (\ref{Eq:arbitaray}), (\ref{Eq:arbitaray1}) and (\ref{Eq:arbitaray4})) for obtaining various analytical expressions exactly. To be precise, we study the entanglement dynamics of $N$-qubit for unentangled arbitrary initial state $\ket{\psi}$ using the Floquet operator $\mathcal{U}$. This formalism has been used earlier in Ref. \cite{ExactDogra2019,seshadri2018tripartite}. To quantify the entanglement in the system, we employ measures such as linear entropy, entanglement entropy (analytically), and concurrence (numerically).
\section{The case for $J=1$}\label{sec:example-section3}
In this section, we analytically calculated the linear entropy and entanglement entropy for arbitrary initial states for qubits ranging from $4$ to $10$ with parameters $J=1$ and $\tau=\pi/4$. In Ref. \cite{sharma2024exactly}, we have shown that our model exhibits signatures of quantum integrability for the initial state $\ket{0,0}$ and $\ket{\pi/2,-\pi/2}$ for the said parameter. Here, we extend these  signatures of integrability for arbitrary initial states. We also calculated the expression of  time-average linear entropy analytically and time-average concurrence numerically for  arbitrary initial state.
\subsection{Exact solution for $4$ qubit}
\begin{figure}[t]\vspace{0.4cm}
\includegraphics[width=0.45\textwidth,height=0.25\textheight]{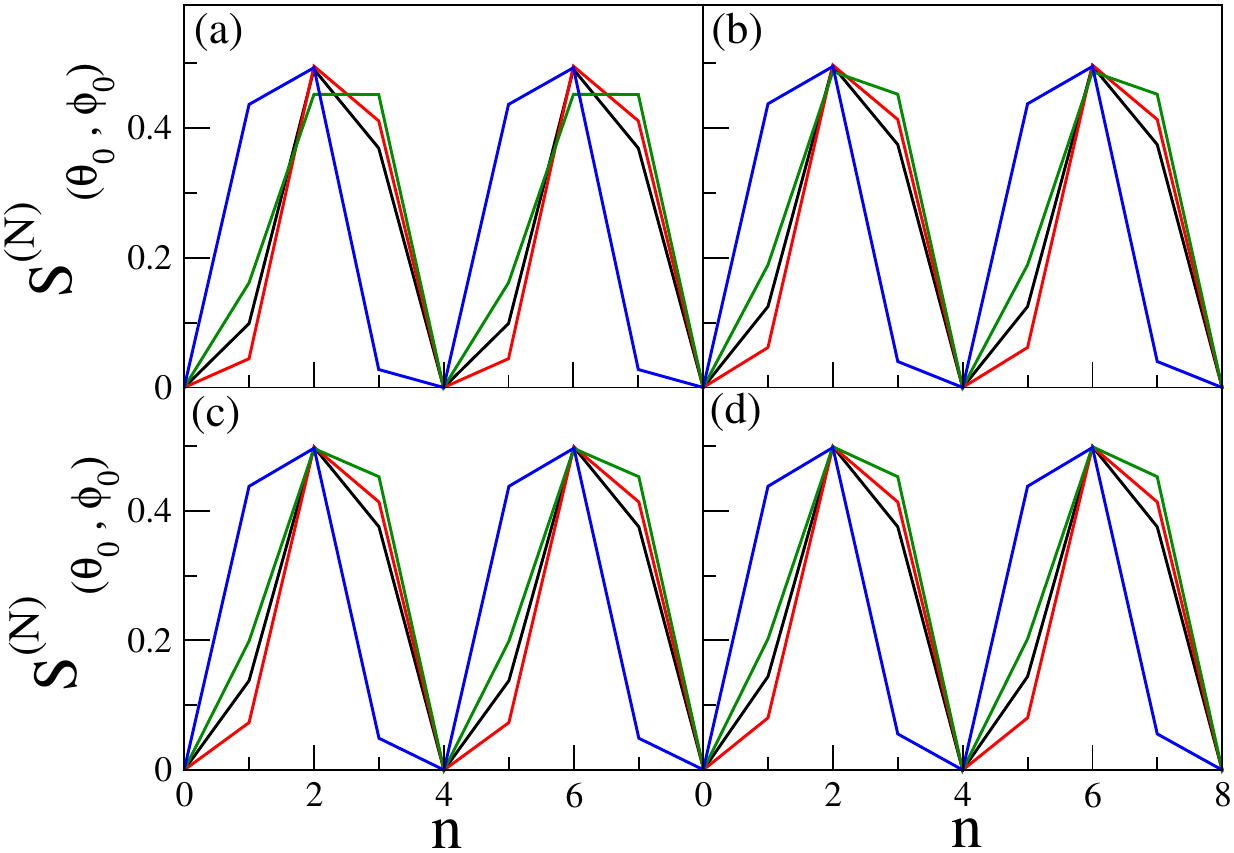}
\caption{The linear entropy for various initial states ($\ket{\theta_0,\phi_0}$ = $\ket{2\pi/3,-\pi/12}$ (black), $\ket{3,-2}$ (Red), $\ket{8\pi/5,-4\pi/5}$ (green) and $\ket{\pi/8,\pi/8}$ (blue)) are plotted for (a) $4$ qubits  (b) $6$ qubits (c) $8$ qubits and (d) $10$ qubits.}
\label{fig:4.795qubitavg}
\end{figure}

\begin{figure}[htbp!]\vspace{0.4cm}
\includegraphics[width=0.45\textwidth,height=0.25\textheight]{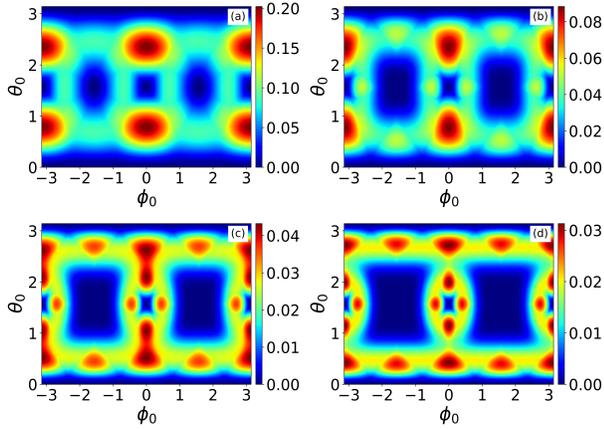}
\caption{The time average concurrence are plotted  for (a) 4 qubits (b) 6 qubits (c) 8 qubits and (d) 10 qubits  for the arbitrary  initial states $\ket{\theta_0,\phi_0}$.}
\label{fig:4.781qubitavg}
\end{figure}

Using Eq. (\ref{Eq:QKT1}), the unitary operator $\mathcal{U}$ for $4$ qubits in $\ket{\phi}$  basis can be written as follows:
\begin{equation}\label{Eq:4qubit}
\mathcal{U}= \left(
\begin{array}{ccccc}
  -1&  0 & 0 & 0&0 \\
 0 & i/2 & i\sqrt{3}/2 & 0 &0\\
 0 &  i\sqrt{3}/2 & -i/2 & 0&0 \\
 0 & 0 & 0  & 0 &1 \\
 0 & 0 & 0  & -i&0 \\
\end{array}
\right).
 \end{equation}
 From the Eq. (\ref{Eq:4qubit}), it can be seen that, the unitary operator $\mathcal{U}$
 can be written in terms of two blocks ($\mathcal{U}_{\pm}$), each consisting of matrices with different dimensions, and is expressed as,
 \begin{equation}
\mathcal{U} =\left(
\begin{array}{cc}
            \mathcal{U}_{+} & 0_A \\ 0_B & \mathcal{U}_{-}
     \end{array}
\right),
\end{equation}
where $\mathcal{U}_{+}~(\mathcal{U}_{-})$ are $3\times3~(2\times2)$ dimensional matrices and $0_A~(0_B)$ is null matrices with dimensions $3\times2~(2\times3)$. The eigenvalues of $\mathcal{U}$ for the case $J=1$ and $\tau=\pi/4$ are $\left\lbrace -1,\pm i,\pm \exp({\frac{3i\pi}{4}})\right\rbrace$, which implies that $\mathcal{U}^8=I$ \cite{sharma2024exactly}. The block dignolisation of $\mathcal{U}$ in two blocks makes it easier to calculate its $n${th} power, which helps in simplifying further analysis. Thus, the $n$th time evolution of the  blocks $\mathcal{U_{\pm}}$ can be expressed  as,
 \begin{equation}
\mathcal{U}_{+}^n= \left(
\begin{array}{ccc}
  (-1)^n&  0 & 0  \\
 0 & \left[(-i)^n+3(i)^n\right]/4 & \left[i\sqrt{3}\sin\left(\frac{n\pi}{2}\right)\right]/2 \\
 0 &   \left[i\sqrt{3}\sin\left(\frac{n\pi}{2}\right)\right]/2  & \left[3(-i)^n+(i)^n\right]/4 \\
\end{array}
\right)
 \end{equation}
 \begin{equation}
\mbox{and}~~\mathcal{U}_{-}^n= e^{\frac{-in\pi}{4}}\left(
\begin{array}{cc}
  \cos^2\left(\frac{n\pi}{2}\right) & e^{\frac{i\pi}{4}}\sin^2\left(\frac{n\pi}{2}\right)\\
 \sin^2\left(\frac{n\pi}{2}\right)  & \cos^2\left(\frac{n\pi}{2}\right)\\
\end{array}
\right).
 \end{equation}
The initial state $\ket{\psi}$ after the $n${th} implementations of the unitary operator $\mathcal{U}$ can be expressed as follows:
\begin{eqnarray}
\ket{\psi_n}&=&\mathcal{U}^n\ket{\psi}\\ \nonumber
&=& c_{1n}\ket{\phi_0^+}+c_{2n}\ket{\phi_1^+}+c_{3n}\ket{\phi_2^+}+c_{4n}\ket{\phi_0^-} +c_{5n}\ket{\phi_1^-},\nonumber
\end{eqnarray}
where the coefficients are presented in compact form as follows:
\begin{equation}\label{Eq:qkt4}
c_{jn}=\sum_{q=1}^{\frac{N+2}{2}}\mathcal{U}^n_{j,q}~ a_q+\sum_{q=\frac{N+4}{2}}^{N+1}\mathcal{U}^n_{j,q}~ b_{q-\frac{N+2}{2}}, 1\leq j\leq N+1.
\end{equation}
\begin{figure}[t!]\vspace{0.4cm}
\includegraphics[width=0.47\textwidth,height=0.15\textheight]{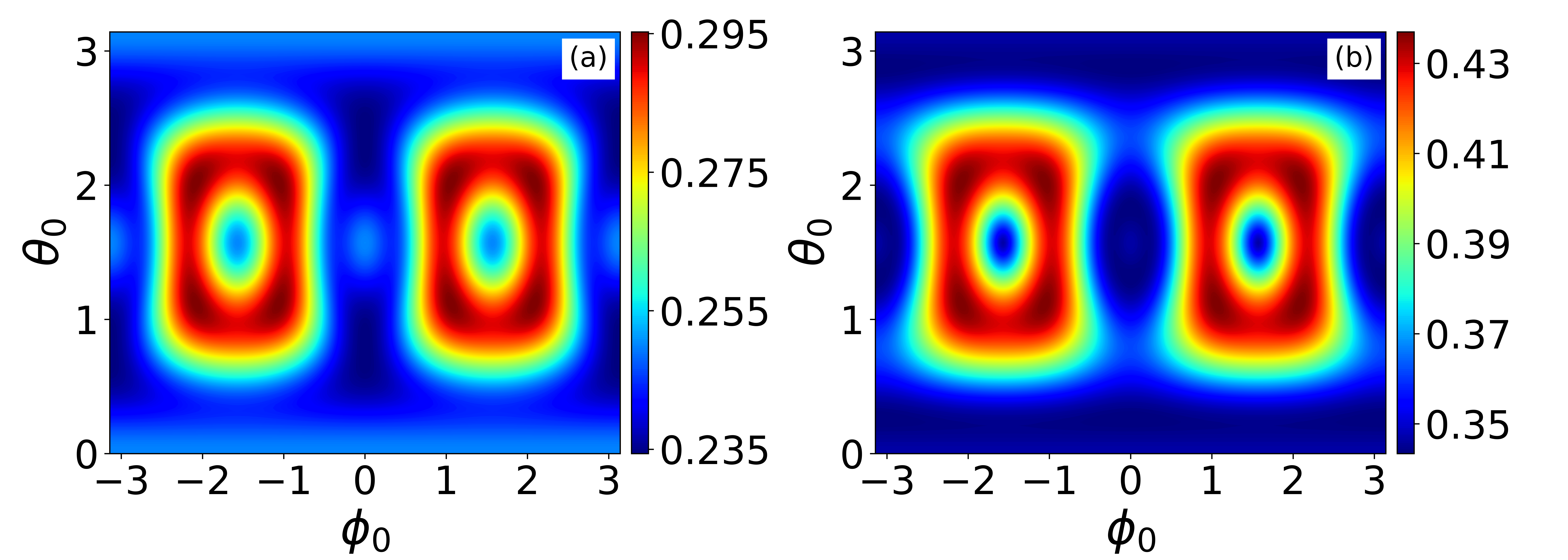}
\caption{Contour plot of time-averaged values of (a) linear entropy and (b) entanglement entropy for any  arbitrary initial states $\ket{\theta_{0},\phi{_0}}$ for $4$ qubits.}
\label{fig:4qubitavg}
\end{figure}
The expressions of the coefficients $c_{jn}$ can be calculated using  Eqs.  (\ref{Eq:arbitaray}), (\ref{Eq:arbitaray1}), (\ref{Eq:arbitaray4}) and   (\ref{Eq:qkt4}),  as follows:
\begin{widetext}
\begin{eqnarray}\nonumber
c_{1n}&=&e^{i n \pi } \left(\sqrt{2} e^{-i \phi_0 } \cos^3\left(\frac{\theta_0 }{2}\right) \sin\left(\frac{\theta_0 }{2}\right)-\sqrt{2}
e^{-3 i \phi_0 } \cos\left(\frac{\theta_0 }{2}\right) \sin^3\left(\frac{\theta_0 }{2}\right)\right),\\ \nonumber
c_{2n}&=& \left[12 i e^{-2 i \phi_0 } \cos^2\left(\frac{\theta_0 }{2}\right) \sin\left(\frac{n \pi }{2}\right) \sin^2\left(\frac{\theta_0
}{2}\right)+\frac{e^{\frac{i n \pi }{2}}}{4}  \left(3+e^{i n \pi }\right) \left(\cos^4\left(\frac{\theta_0 }{2}\right)+e^{-4
i \phi_0 } \sin^4\left(\frac{\theta_0 }{2}\right)\right)\right]\Big{/}4\sqrt{2},\\ \nonumber
c_{3n}&=& \left[\sqrt{3}~e^{-\frac{1}{2} i n \pi -2 i \phi_0 }  \left(3+e^{i n \pi }\right) \cos^2\left(\frac{\theta_0
}{2}\right) \sin^2\left(\frac{\theta_0 }{2}\right)+i \sqrt{3} \sin\left(\frac{n \pi }{2}\right) \left(\cos^4\left(\frac{\theta_0
}{2}\right)+e^{-4 i \phi_0 } \sin^4\left(\frac{\theta_0 }{2}\right)\right)\right]\Big{/}2\sqrt{2},\\ \nonumber
c_{4n}&=&e^{-\frac{3}{4} i n \pi } \left[\sqrt{2}\cos\left(\frac{n \pi }{2}\right) \left(e^{-i \phi_0 } \cos^3\left(\frac{\theta_0
}{2}\right) \sin\left(\frac{\theta_0 }{2}\right)+ e^{-3 i \phi_0 } \cos\left(\frac{\theta_0 }{2}\right) \sin^3\left(\frac{\theta_0
}{2}\right)\right)+\frac{e^{\frac{3 i \pi }{4}}}{\sqrt{2}} \sin\left(\frac{n \pi }{2}\right) \right.\\  \nonumber && \left.\left(\cos^4\left(\frac{\theta_0 }{2}\right)-e^{-4 i \phi_0 } \sin^4\left(\frac{\theta_0 }{2}\right)\right)\right]~~~~\mbox{and}\\ \nonumber
c_{5n}&=&-e^{-\frac{3}{4} i n \pi }\left[\sqrt{2}~e^{-\frac{3 i \pi }{4}} \sin\left(\frac{n \pi }{2}\right) \left( e^{-i \phi_0 } \cos\left(\frac{\theta_0
}{2}\right) \sin\left(\frac{\theta_0 }{2}\right)+ e^{-3 i \phi_0 } \cos^3\left(\frac{\theta_0 }{2}\right) \sin^3\left(\frac{\theta_0
}{2}\right)\right)- \frac{\cos\left(\frac{n \pi }{2}\right)}{\sqrt{2}}\right.\\  \nonumber && \left. \left(\cos^4\left(\frac{\theta_0 }{2}\right)-e^{-4 i \phi_0 } \sin^4\left(\frac{\theta_0 }{2}\right)\right)\right].
\end{eqnarray}
\end{widetext}
The single-qubit reduced density matrix (RDM) can be obtain by tracing out any of the $N-1$ qubits from the $N$ qubit density operator ($\rho_n=\ket{\psi_n}\bra{\psi_n}$), which can be expressed as follows:
\begin{equation}
\rho_1(n)=\frac{1}{2}\left(
\begin{array}{cc}
 \bar{t}_n & \bar{v}_n \\
\bar{ v}_n^* & 2-\bar{t}_n \\
\end{array}
\right),
\end{equation}
where the coefficient $\bar{t}_n$ and $\bar{v}_n$ can be expressed as,
\begin{eqnarray} \nonumber
 \bar{t}_n&=&\frac{1}{2}+{c}_{2n}{c}_{5n}^*+{c}_{5n}{c}_{2n}^*+\frac{1}{2}\left({c}_{1n}{c}_{4n}^*+{c}_{4n}{c}_{1n}^*\right)~~\mbox{and}\\ \nonumber
 \bar{v}_n&=&\left[\left(\left({c}_{2n}+{c}_{5n}\right)\left({c}_{1n}^*+{c}_{4n}^*\right)+\left({c}_{4n}-{c}_{1n}\right)\left({c}_{2n}^*-{c}_{5n}^*\right)\right)\right.\\ \nonumber &&\left.+{\sqrt{3}}(\left({c}_{1n}+{c}_{4n}\right){c}_{3n}^*+\left({c}_{4n}^*-{c}_{1n}^*\right)){c}_{3n}\right]\Big{/}{2}.
\end{eqnarray}
The eigenvalues of $\rho_1(n)$ are $\frac{1}{2} \left(1\pm\sqrt{1-\bar{t}_n\left(2-\bar{t}_n\right)-|{\bar{v}_n}|^2}\right)$. The linear entropy of the single qubit RDM as a function of initial states parameters  $\ket{\theta_{0},\phi{_0}}$ is given as follows:
\begin{eqnarray}\nonumber
 S_{(\theta_0,\phi_0)}^{(4)}(n)&=&1-\mbox{tr}\left[\rho_1^2(n)\right]\\
 &=&[\bar{t}_n(2-\bar{t}_n)-|{\bar{v}_n}|^2]/2.
 \end{eqnarray}
 The entanglement entropy can be evaluated using
 $-(\lambda_1\ln\lambda_1 +\lambda_2\ln\lambda_2)$, where $\lambda_1$ and $\lambda_2$ are the eigenvalues of $\rho_1(n)$. We find that  the entanglement dynamics are periodic in nature having period $4$  for any arbitrary initial  state, except for the  initial state $\ket{\pi/2,\pm\pi/2}$, where the period is $2$. Which is shown in Fig. \ref{fig:4.795qubitavg}~(a) for various initial states. Due to this periodic nature the infinite time averages  can be found easily by considering the values over only one period. Thus, the  expression of time-averaged linear entropy for an arbitrary initial  states, is given  as follows:
\begin{eqnarray}\nonumber
  \langle S_{(\theta_0,\phi_0)}^{(4)}\rangle &=&\left[16877-872\cos\left(2 \theta_0\right)-156\cos\left(4 \theta_0\right)+424\right.\\ \nonumber && \left.\cos\left(6 \theta_0\right)-96 (67+60\cos\left(2 \theta_0\right)+\cos\left(4 \theta_0\right))\right.\\ \nonumber && \left.\cos\left(2 \phi_0 \right)\sin^4\left(\theta_0\right)-1024 (2+\cos\left(2 \theta_0\right))\right.\\ \nonumber && \left.\cos\left(4 \phi_0
\right)\sin^6\left(\theta_0\right)+111\cos(8\theta_0)+128 \left(6\right.\right.\\  && \left.\left.\cos\left(6 \phi_0 \right)+\cos\left(8 \phi_0 \right)\right)\sin^8\left(\theta_0\right)\right]\Big{/}{65536}.
\end{eqnarray}
It takes values within the narrow interval $[0.2343, 0.2953]$, and is shown in Fig. \ref{fig:4qubitavg}~(a). The time-averaged entanglement entropy is also plotted in the same figure.  The advantage of a contour plot is that it  provides clear insights  to easily identify possible regions where the quantity exhibits minima or maxima. Based on these insights  from the contour plot, the maximum value corresponds to eight initial states as follows: $\ket{1.1169,\pm1.061155}$, $\ket{1.1169,\pm2.080439}$, $\ket{2.024689,\pm2.080439}$, $\ket{2.024689,\pm1.061155}$. Whereas the minimum for the states $\ket{\pi/4,0}$, $\ket{3\pi/4,\pm \pi}$, $\ket{3\pi/4,0}$, $\ket{\pi/4,\pm\pi}$. These extremes can also be seen in Fig. \ref{fig:4qubitavg}~(a). We have  numerically obtain the average values of concurrence for any arbitrary initial state and plotted in Fig. \ref{fig:4.781qubitavg}~(a). From the  same figure, we observe that for the states  $\ket{0,0}$, $\ket{\pi/2,\pm\pi/2}$ and $\ket{\pi/2,0}$  pairwise  concurrence  vanishes \cite{sharma2024exactly}, which indicates the multipartite nature of entanglement. We also observe that, for these states the linear and entanglement entropy attain their maximum upper bound   $0.5$ and $\ln2~ (\approx0.6932$) respectively \cite{sharma2024exactly}. The average value of concurrence is maximum of $0.2022542486$  for the states $\ket{\pi/4,0}$, $\ket{3\pi/4,\pm \pi}$, $\ket{3\pi/4,0}$, $\ket{\pi/4,\pm\pi}$. Thus, we conclude that for states where the average linear entropy is at its maximum, the concurrence takes low values in the order of $10^{-2}$.  On the other hand, for the initial states when the linear entropy is minimum, the concurrence is maximized, as shown in the Figs. \ref{fig:4qubitavg} and \ref{fig:4.781qubitavg}~(a). We follow the same procedure in the subsequent part of this paper.
\subsection{Exact solution for $5$ qubits}
\begin{figure}[t]\vspace{0.4cm}
\includegraphics[width=0.47\textwidth,height=0.25\textheight]{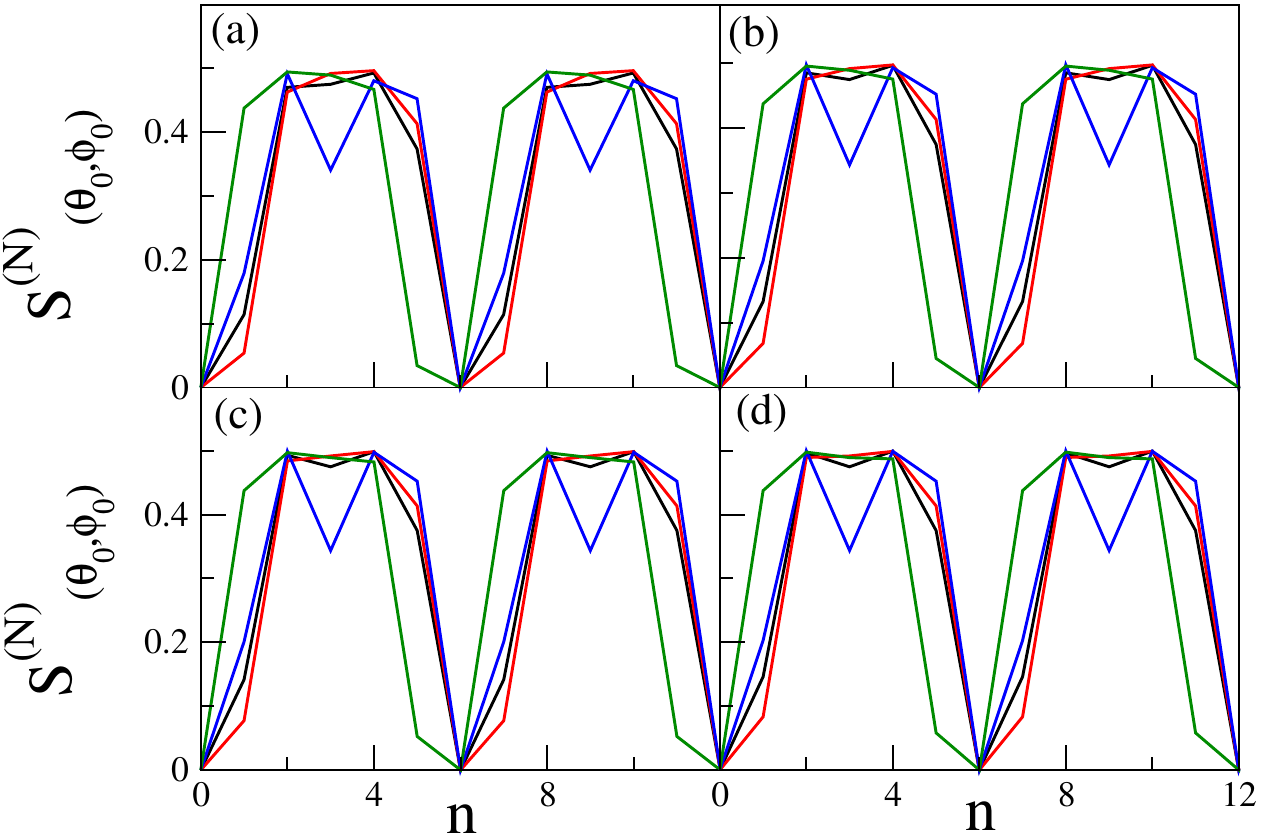}
\caption{The linear entropy for various initial states ($\ket{\theta_0,\phi_0}$ = $\ket{2\pi/3,-\pi/12}$ (black), $\ket{3,-2}$ (Red), $\ket{8\pi/5,-4\pi/5}$ (blue) and $\ket{\pi/8,\pi/8}$ (green)) are plotted for (a) $5$ qubits and (b) $7$ qubits (c) $9$ qubits and (d) $11$ qubits.}
\label{fig:4.79qubitavg}
\end{figure}

\begin{figure}[t!]\vspace{0.4cm}
\includegraphics[width=0.45\textwidth,height=0.25\textheight]{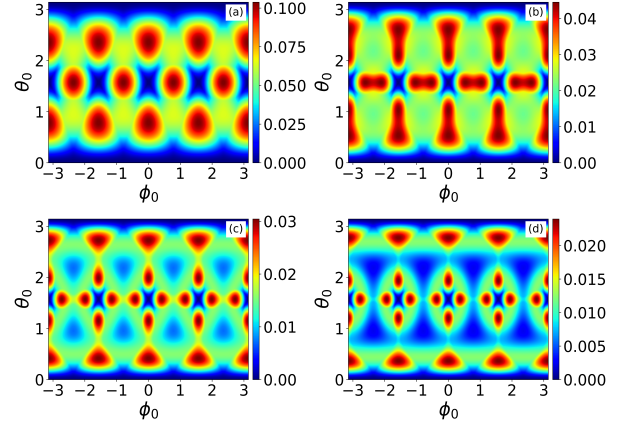}
\caption{ The time-average concurrence are plotted  for (a) 5 qubits (b) 7 qubits (c) 9 qubits and (d) 11 qubits for the arbitrary  initial states $\ket{\theta_0,\phi_0}$.}
\label{fig:4.88qubitavg}
\end{figure}

As we mention earlier, the unitary operator $\mathcal{U}$ is block diagonlised in  $\mathcal{U}_{+}~(\mathcal{U}_{-})$  having dimension $3\times3~(3\times 3)$ respectively. The block $\mathcal{U}_{+}$ and $\mathcal{U}_{-}$ are given as follows:
\begin{equation}
  \mathcal{U}_{\pm}= \frac{1}{4} e^{\pm{\frac{ i \pi }{4}}}
\begin{pmatrix}
 \mp{1} &  i\sqrt{5}  &  \mp{\sqrt{10}} \\
- i\sqrt{5} & \pm 3 & - i\sqrt{2} \\
  \pm \sqrt{10} & - i\sqrt{2} & \mp 2 \\
\end{pmatrix}.
\end{equation}
Applying the unitary operator $\mathcal{U}$ $n$ times on the state $\ket{\psi}$ we obtain,
\begin{eqnarray}
\ket{\psi_n}&=&\mathcal{U}^n\ket{\psi}\\ \nonumber
&=& a_{1n}\ket{\phi_0^+}+a_{2n}\ket{\phi_1^+}+a_{3n}\ket{\phi_2^+}+a_{4n}\ket{\phi_0^-}\\ \nonumber&& +a_{5n}\ket{\phi_1^-}+a_{6n}\ket{\phi_2^-},\nonumber
\end{eqnarray}
where the coefficients are given as follows:
\begin{equation}\label{Eq:qkt5}
 a_{jn}=\sum_{q=1}^{\frac{N+1}{2}}\mathcal{U}^n_{j,q}~ a_q+\sum_{q=\frac{N+3}{2}}^{N+1}\mathcal{U}^n_{j,q}~ b_{q-\frac{N+1}{2}},1\leq j\leq N+1.
\end{equation}
The expressions of the coefficients $a_{jn}$ can be calculated using  Eqs. (\ref{Eq:arbitaray2}), (\ref{Eq:arbitaray3}) and (\ref{Eq:qkt5}) for $5$ qubits. The detailed calculations  regarding the $n${th} time evolution of $\mathcal{U}$ and  the coefficient $a_{jn}$ are provided in the supplemental material \cite{supplementry2023}. We have moved these calculations and the coefficients in the supplemental material, as their sizes are very large. The single qubit RDM for $5$ qubit is given as,
\begin{equation}
\rho_1(n)=\frac{1}{2}\left(
\begin{array}{cc}
 r_n & w_n \\
 w_n^* & 2-r_n \\
\end{array}
\right),
\end{equation}
where the coefficients $r_n$ and $w_n$ are given as follows:
\begin{eqnarray} \nonumber
 r_n&=&\frac{1}{2}+a_{1n}a_{4n}^*+a_{4n}a_{1n}^*+\frac{3}{5}\left(a_{2n}a_{5n}^*+a_{5n}a_{2n}^*\right)+\\ \nonumber &&\frac{1}{5}\left(a_{3n}a_{6n}^*+a_{6n}a_{3n}^*\right) ~~\mbox{and} \\ \nonumber
 w_n&=&\left[\sqrt{5}\left(\left(a_{1n}+a_{4n}\right)\left(a_{5n}^*+a_{2n}^*\right)+\left(a_{5n}-a_{2n}\right)\left(a_{1n}^*-a_{4n}^*\right)\right)\right.\\ \nonumber &&\left.+{2\sqrt{2}}\left(\left(a_{2n}+a_{5n}\right)\left(a_{3n}^*+a_{6n}^*\right)+\left(a_{3n}-a_{6n}\right)\right.\right.\\ \nonumber &&\left.\left.\left(-a_{2n}^*+a_{5n}^*\right)\right)-{3i}\left(a_{3n}+a_{6n}\right)\left(a_{6n}^*-a_{3n}^*\right)\right]\big{/}{5}.
\end{eqnarray}
The eigenvalues of $\rho_1(n)$ are $\frac{1}{2} \left(1\pm\sqrt{1-{r}_n\left(2-{r}_n\right)-|{{w}_n}|^2}\right)$. The linear entropy of single qubit RDM is given as follows:
\begin{equation}
 S_{(\theta_0,\phi_0)}^{(5)}(n,1)=[r_n(2-r_n)-|{w_n}|^2]/2.
\end{equation}
The eigenvalues of $\mathcal{U}$ are $\exp({\frac{i\pi}{4}})\left\lbrace 1,\exp({\frac{i\pi}{2}}),\exp({\frac{\pm 2i\pi}{3}}),i~\exp({\frac{\pm 2i\pi}{3}})\right\rbrace$, which implies that $\mathcal{U}^{12}=I$. We find that  the entanglement dynamics are periodic in nature having period $6$  for any arbitrary initial coherent states, except for the initial state $\ket{\pi/2,\pm\pi/2}$ where the  period is $3$ and plotted in Fig. \ref{fig:4.79qubitavg}~(a) for various initial states. Thus, the time-averaged linear entropy is given  as follows:
\begin{eqnarray}\nonumber
  \langle S_{(\theta_0,\phi_0)}^{(5)}\rangle&=&\left\lbrace 127418+658\cos\left(2 \theta_0 \right)+1069\cos\left(6 \theta_0 \right)+\right.\\ \nonumber &&\left.744\cos\left(4 \theta_0 \right)+1118\cos\left(8 \theta_0 \right)+65\cos\left(10
\theta_0 \right)\right.\\ \nonumber &&\left.-256\left[(52+60\cos\left(2 \theta_0 \right))\cos\left(4 \phi_0 \right)-(1-5\right.\right.\\ &&\left.\left.\cos\left(2 \theta_0 \right))\cos\left(8 \phi_0 \right)\right] \sin^8\left(\theta_0 \right)\right\rbrace\Big{/}{393216}.
\end{eqnarray}
\begin{figure}[t!]\vspace{0.4cm}
\includegraphics[width=0.47\textwidth,height=0.15\textheight]{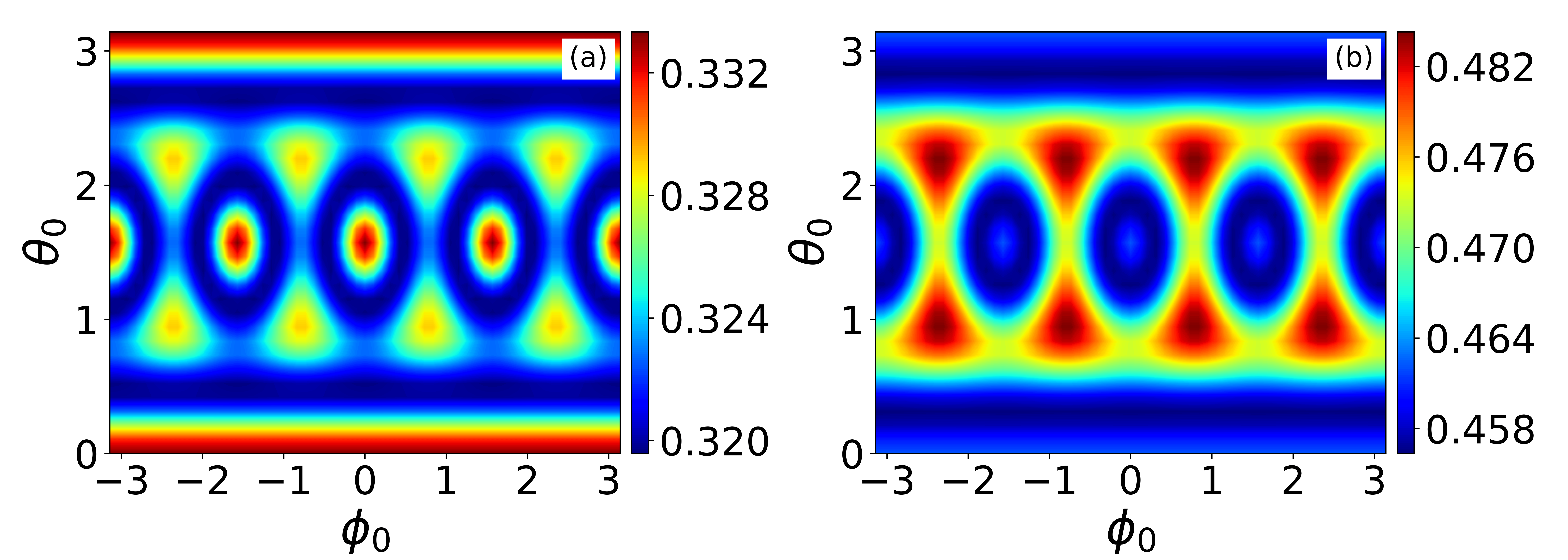}
\caption{ Contour plot of time-averaged values of (a) linear entropy and (b) entanglement entropy for any arbitrary initial  states $\ket{\theta_{0},\phi{_0}}$ for $5$ qubits.}
\label{fig:5qubitavg}
\end{figure}
It takes values from the  narrow interval $[0.3194, 1/3]$ and is shown in Fig. \ref{fig:5qubitavg}~(a). The maximum value corresponds to various initial states as follows: $\ket{0,\phi_0}$, $\ket{\pi/2,\pm \pi/2}$, $\ket{\pi,\phi_0}$, $\ket{\pi/2,0}$, $\ket{\pi/2,\pm \pi}$. In contrast the  average entanglement entropy attains lower value for these initial states. On the other hand, the minimum value of average linear entropy is observed for the states $\ket{0.477656,\pm \pi/2}$, $\ket{\pi/2,\pm 2.66393}$, $\ket{\pi/2,\pm 0.477656}$, $\ket{2.66393,\pi/2}$, $\ket{1.093138,\pi/2}$, $\ket{\pi/2,\pm 1.093138}$, $\ket{2.04845,\pm\pi/2}$, $\ket{\pi/2,\pm2.04845}$, $\ket{1.093138,0}$, $\ket{2.04845,0}$, $\ket{1.093138,\pm\pi}$ and $\ket{2.04845,\pm\pi}$, as can be seen in Fig. \ref{fig:5qubitavg}~(a). We numerically obtain the average values of concurrence for any arbitrary initial state and plotted in Fig.  \ref{fig:4.88qubitavg}. From  Fig. \ref{fig:4.88qubitavg}~(a), we observe that for the states such as $\ket{0,0}$, $\ket{\pi/2,\pm\pi/2}$ and $\ket{\pi/2,0}$, the linear entropy and entanglement entropy are maximum, while the concurrence is zero \cite{sharma2024exactly}, indicates  the presence of  multipartite nature of entanglement. For the states $\ket{\pi/4,\pm\pi}$, $\ket{3\pi/4,\pm\pi}$, $\ket{3\pi/4,\pm\pi/2}$, $\ket{\pi/4,0}$, $\ket{3\pi/4,0}$, $\ket{\pi/4,\pm\pi/2}$, $\ket{\pi/2,\pm3\pi/4}$, and $\ket{\pi/2,\pm\pi/4}$, the average value of concurrence is maximum $0.1037915448$.

\subsection{Exact solution for $6$ qubit}
In Ref. \cite{sharma2024exactly}, we have shown that in $\ket{\phi}$ basis, the  unitary operator $\mathcal{U}$ is block diagonalized in two blocks $\mathcal{U_+}$ and $\mathcal{U_-}$ having dimension $4\times4$ and $3\times3$. The blocks are given as follows:
 \begin{equation}
\mathcal{U_+}= \frac{e^{\frac{ i \pi}{4}}}{2\sqrt{2}}\left(
\begin{array}{cccc}
 0 & -\sqrt{3} & 0 & -\sqrt{5} \\
i\sqrt{3} & 0 & i\sqrt{5}& 0 \\
 0 & -\sqrt{5} & 0 & \sqrt{3} \\
 i\sqrt{5} & 0 & -i\sqrt{3} & 0 \\
\end{array}
\right)~~\mbox{and}
 \end{equation}
 \begin{equation}
\mathcal{U_-}= \frac{e^{\frac{ i \pi}{4}}}{4}{\left(
\begin{array}{ccc}
  1 & 0 &  \sqrt{15}  \\
 0 & -4i & 0 \\
 \sqrt{15}  & 0 & - 1 \\
\end{array}
\right)}.
 \end{equation}
The state $\ket{\psi_n}$ can be obtain after the $n$ implementations of $\mathcal{U}$ on the state $\ket{\psi}$. Thus,
\begin{eqnarray}\nonumber
\ket{\psi_n}&=&\mathcal{U}^n\ket{\psi}\\ \nonumber
&=& g_{1n}\ket{\phi_0^+}+g_{2n}\ket{\phi_1^+}+g_{3n}\ket{\phi_2^+}+g_{4n}\ket{\phi_3^+}+g_{5n}\ket{\phi_0^-}\\ && + ~g_{6n}\ket{\phi_1^-}+g_{7n}\ket{\phi_2^-},
\end{eqnarray}
where the coefficients are given as,
\begin{equation}\label{Eq:qkt50}
g_{jn}=\sum_{q=1}^{\frac{N+2}{2}}\mathcal{U}^n_{j,q}~ a_q+\sum_{q=\frac{N+4}{2}}^{N+1}\mathcal{U}^n_{j,q}~ b_{q-\frac{N+2}{2}},1\leq j\leq N+1.
\end{equation}
The expressions of the coefficients $g_{jn}$ can be calculated using  Eqs.  (\ref{Eq:arbitaray}), (\ref{Eq:arbitaray1}), (\ref{Eq:arbitaray4}) and (\ref{Eq:qkt50}) for $6$ qubits. The detailed calculations  regarding the $n${th} time evolution of $\mathcal{U}$ and  the coefficient $g_{jn}$ are provided in the supplemental material \cite{supplementry2023}.
The single qubit RDM is given as follows:
\begin{equation}
\rho_1(n)=\frac{1}{2}\left(
\begin{array}{cc}
 q_n & \bar{b}_n \\
\bar{b}_n^* & 2-q_n \\
\end{array}
\right),
\end{equation}
\begin{figure}[htbp!]\vspace{0.4cm}
\includegraphics[width=0.47\textwidth,height=0.15\textheight]{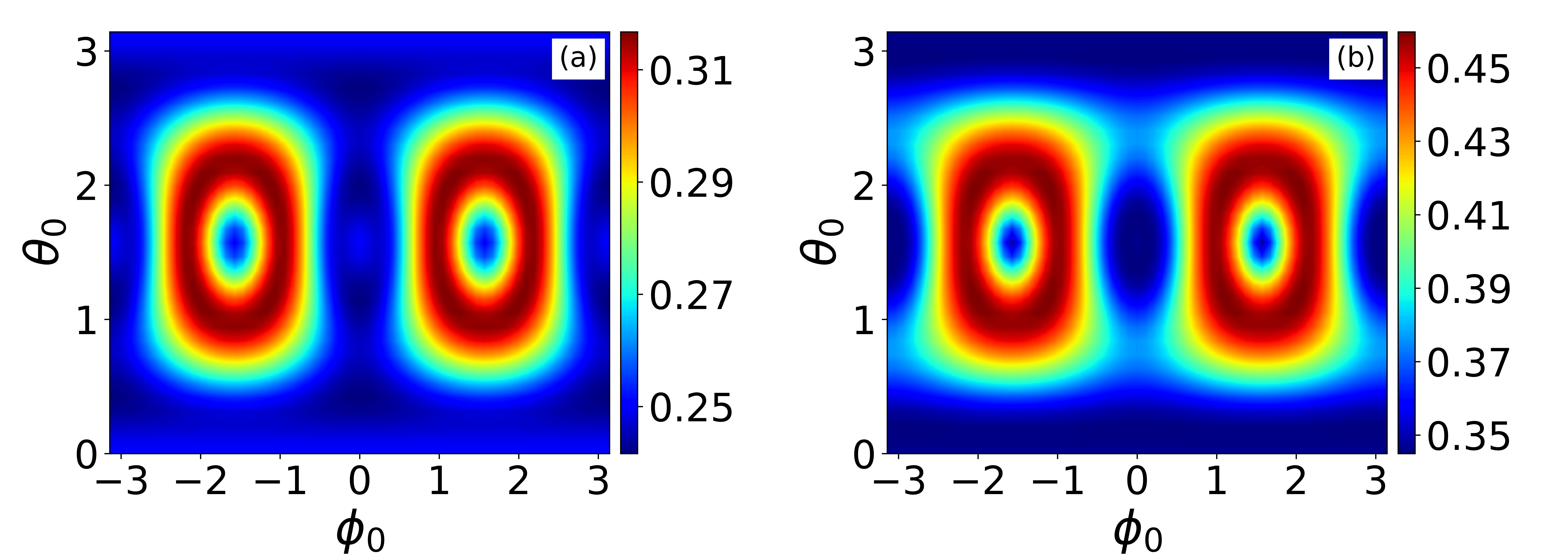}
\caption{Contour plot of time-averaged values of (a) linear entropy and (b) entanglement entropy for any arbitrary  initial states $\ket{\theta_{0},\phi{_0}}$ for $6$ qubits.}
\label{fig:6qubitavg}
\end{figure}
where the coefficient $q_n$ and $\bar{b}_n$ are given as follows:
\begin{eqnarray} \nonumber
 q_n&=&\frac{1}{2}+g_{1n}g_{5n}^*+g_{5n}g_{1n}^*+\frac{2}{3}\left(g_{6n}g_{2n}^*+g_{2n}g_{6n}^*\right)\\ \nonumber&&+\frac{1}{3}\left(g_{7n}g_{3n}^*+g_{3n}g_{7n}^*\right)~~ \mbox{and} \\ \nonumber
 \bar{b}_n&=&\dfrac{1}{\sqrt{6}}\left(\left(g_{1n}+g_{5n}\right)\left(g_{2n}^*+g_{6n}^*\right)+\left(g_{2n}-g_{6n}\right)\left(g_{5n}^*-g_{1n}^*\right)\right)\\ \nonumber &&+\dfrac{5}{3\sqrt{10}}\left(\left(g_{2n}+g_{6n}\right)\left(g_{3n}^*+g_{7n}^*\right)+\left(g_{7n}-g_{3n}\right) \right.\\ \nonumber &&\left.\left(g_{2n}^*-g_{6n}^*\right)\right)+\sqrt{\frac{2}{3}}\left(\left(g_{3n}+g_{7n}\right)g_{4n}^*+\left(g_{7n}^*-g_{3n}^*\right)g_{4n}\right).
\end{eqnarray}
The eigenvalues of $\rho_1(n)$ are $\frac{1}{2} \left(1\pm\sqrt{1-{q}_n\left(2-{q}_n\right)-|\bar{{b}}_n}|^2\right)$. The linear entropy of single qubit is given as follows:
\begin{equation}
 S_{(\theta_0,\phi_0)}^{(6)}(n,1)=[q_n(2-q_n)-|{\bar{b}_n}|^2]/2.
\end{equation}
The eigenvalues of $\mathcal{U}$ for the case $J=1$ and $\tau=\pi/4$ are $\left\lbrace \pm1,\pm1,\pm \exp({\frac{i\pi}{4}}),-\exp({\frac{3i\pi}{4}})\right\rbrace$, which implies that $\mathcal{U}^8=I$. We find that  the entanglement dynamics are periodic in nature having period $4$  for any arbitrary initial state, except for the initial state $\ket{\pi/2,\pm\pi/2}$ where the period $2$. Which is shown in Fig. \ref{fig:4.795qubitavg}~(b) for various initial states. Thus, the time-averaged linear entropy
for an arbitrary initial coherent states, is given  as follows:
\begin{eqnarray}\nonumber
 \langle S_{(\theta_0,\phi_0)}^{(6)}\rangle&=&\left[2248542-170488 \cos\left(2 \theta_0\right)+873\cos\left(12 \theta_0\right)\right.\\ \nonumber && \left. -57465 \cos\left(4 \theta_0\right)+16818 \cos\left(8
\theta_0\right)+5892 \right.\\ \nonumber && \left. \cos\left(10 \theta_0\right)+52980\cos\left(6 \theta_0\right)-160 (7413+\right.\\ \nonumber && \left.7288 \cos\left(2 \theta_0\right)+1220 \cos\left(4 \theta_0\right)+456 \cos\left(6
\theta_0\right)  \right.\\ \nonumber && \left.+7 \cos\left(8 \theta_0\right))\cos\left(2 \phi_0 \right) \sin^{4}\left(\theta_0\right)-512(15 (47  \right.\\ \nonumber &&\left.+17 \cos\left(2 \theta_0\right))\cos\left(4 \phi_0 \right)+2 (17+23 \cos\left(2
\theta_0\right)) \right.\\  \nonumber&&\left.\cos\left(8 \phi_0 \right))\sin^{10}\left(\theta_0\right)+1024 (130 \cos\left(6 \phi_0 \right) +10\right.\\ \nonumber  &&\left.  \cos\left(10 \phi_0 \right)+\cos\left(12 \phi_0 \right))\sin^{12}\left(\theta\right)\right]\Big{/}{8388608}.
\end{eqnarray}
It takes values from the  narrow interval $[0.2416, 0.317]$ and is shown in Fig. \ref{fig:6qubitavg}~(a). The maximum value corresponds to the initial states: $\ket{1.1522423,\pm 1.1098}$, $\ket{1.1522423,\pm 2.031791}$, $\ket{1.98973,\pm 1.1098}$, $\ket{1.98935,\pm 2.031791}$, while the minimum values are associated with states such as, $\ket{0.422007,0}$, $\ket{2.719585,\pm \pi}$, $\ket{0.422007,\pm\pi}$, $\ket{2.719585,0}$, which can be seen from same figure. We numerically computed the average values of concurrence for any arbitrary initial state and presented the results in  Fig. \ref{fig:4.781qubitavg}~(b). From the results, we observe that for the initial states $\ket{0,0}$,$\ket{\pi/2,\pm\pi/2}$ and $\ket{\pi/2,0}$, concurrence vanishes, while the linear and entanglement entropy are maximized \cite{sharma2024exactly}. The average value of concurrence is maximum of $8.8388\times10^{-2}$  for the states $\ket{\pi/4,\pm\pi}$,$\ket{3\pi/4,\pm\pi}$, $\ket{\pi/4,0}$ and $\ket{3\pi/4,0}$.
\subsection{Exact solution for $7$ qubit}
The unitary operator in $\ket{\phi}$ basis can be written in two blocks $\mathcal{U}_{+}(\mathcal{U}_{-})$  \cite{sharma2024exactly}, as follows:
\begin{equation}
\mathcal{U}_{+} ={\frac{1}{8}\left(
\begin{array}{cccc}
 -1 & -i \sqrt{7} &  -\sqrt{21}  &- i \sqrt{35}
\\
 -i \sqrt{7}  & - 5  & - 3i\sqrt{3}  & - \sqrt{5} \\
  \sqrt{21}  &  3 i \sqrt{3}  &  1 & -i\sqrt{15}
\\
 i \sqrt{35}   &  \sqrt{5}  & -i \sqrt{15}  & -3  \\
\end{array}
\right)} ~~~~\mbox{and}
\end{equation}
\begin{equation}
\mathcal{U}_{-} ={\frac{1}{8}\left(
\begin{array}{cccc}
 i &  \sqrt{7} &  i\sqrt{21}  & \sqrt{35}
\\
  \sqrt{7}  &  5i  & 3\sqrt{3}  & i\sqrt{5} \\
  -i\sqrt{21}  &  -3 \sqrt{3}  &  -i & \sqrt{15}
\\
 -\sqrt{35}   &  -i\sqrt{5}  & \sqrt{15}  & 3i  \\
\end{array}
\right)}.
\end{equation}
The initial state $\ket{\psi}$ after the $n${th} implementations of the unitary operator $\mathcal{U}$ can be expressed as follows:
\begin{eqnarray}
\ket{\psi_n}&=&\mathcal{U}^n\ket{\psi}\\ \nonumber
&=& b_{1n}\ket{\phi_1^+}+b_{2n}\ket{\phi_2^+}+b_{3n}\ket{\phi_3^+}+b_{4n}\ket{\phi_4^+}\\ \nonumber&& +b_{5n}\ket{\phi_1^-}+b_{6n}\ket{\phi_2^-}+b_{7n}\ket{\phi_3^-}+b_{8n}\ket{\phi_4^-},\nonumber
\end{eqnarray}
where the coefficients can be expressed as follows:
\begin{equation}\label{Eq:qkt52}
 b_{jn}=\sum_{q=1}^{\frac{N+1}{2}}\mathcal{U}^n_{j,q}~ a_q+\sum_{q=\frac{N+3}{2}}^{N+1}\mathcal{U}^n_{j,q}~ b_{q-\frac{N+1}{2}},1\leq j\leq N+1.
 \end{equation}
 The expressions of the coefficients $b_{jn}$ can be calculated using  Eqs. (\ref{Eq:arbitaray2}), (\ref{Eq:arbitaray3}) and (\ref{Eq:qkt52}) for $7$ qubits. The detailed calculations  regarding the $n${th} time evolution of $\mathcal{U}$ and  the coefficient $b_{jn}$ are provided in the supplemental material \cite{supplementry2023}.
The single qubit RDM, $\rho_1(n)$, is given as follows:
\begin{equation}
\rho_1(n)=\frac{1}{2}\left(
\begin{array}{cc}
 \bar{r}_n & \bar{w}_n \\
\bar{ w}_n^* & 2-\bar{r} \\
\end{array}
\right),
\end{equation}
\begin{figure}[htbp!]\vspace{0.4cm}
\includegraphics[width=0.47\textwidth,height=0.15\textheight]{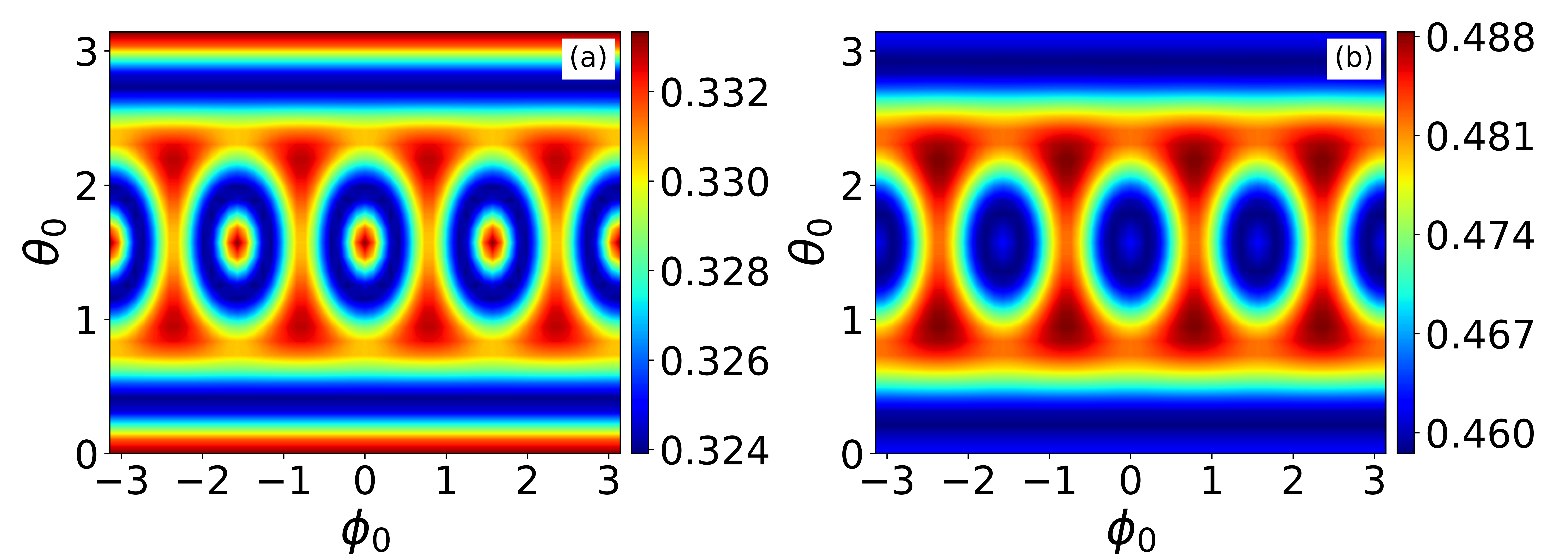}
\caption{ Contour plot of time-averaged values of (a) linear entropy and (b) entanglement entropy for any arbitrary  initial states $\ket{\theta_{0},\phi{_0}}$ for $7$ qubits.}
\label{fig:7qubitavg}
\end{figure}
where the coefficient $\bar{r}_n$ and $\bar{w}_n$ are expressed as,
\begin{eqnarray} \nonumber
 \bar{r}_n&=&\frac{1}{2}+b_{1n}b_{5n}^*+b_{5n}b_{1n}^*+\frac{5}{7}\left(b_{6n}b_{2n}^*+b_{2n}b_{6n}^*\right)+\\ \nonumber &&\frac{3}{7}\left(b_{7n}b_{3n}^*+b_{3n}b_{7n}^*\right)+\frac{1}{7}\left(b_{8n}b_{4n}^*+b_{4n}b_{8n}^*\right)~~ \mbox{and} \\ \nonumber
 \bar{w}_n&=&\left[{\sqrt{7}}\left(\left(b_{1n}+b_{5n}\right)\left(b_{2n}^*+b_{6n}^*\right)+\left(b_{2n}-b_{6n}\right)\right.\right.\\ \nonumber &&\left.\left.\left(b_{5n}^*-b_{1n}^*\right)\right)+{\sqrt{12}}\left(\left(b_{2n}+b_{6n}\right)\left(b_{3n}^*+b_{7n}^*\right)+\right.\right.\\ \nonumber &&\left.\left.\left(b_{7n}-b_{3n}\right)\left(b_{2n}^*-b_{6n}^*\right)\right)+{\sqrt{15}}\left(\left(b_{3n}+b_{7n}\right)\right.\right.\\ \nonumber &&\left.\left.\left(b_{4n}^*+b_{8n}^*\right)+\left(b_{4n}-b_{8n}\right)\left(b_{7n}^*-b_{3n}^*\right)\right)-\right.\\ \nonumber &&{4i}\left(\left(b_{4n}+b_{8n}\right)\left(b_{4n}^*-b_{8n}^*\right)\right]\Big{/}{7}.
\end{eqnarray}
The eigenvalues  of $\rho_1(n)$ are $\frac{1}{2} \left(1\pm\sqrt{1-\bar{r}_n\left(2-\bar{r}_n\right)-|\bar{w}_n}|^2\right)$. The linear entropy of single qubit RDM can be expressed as follows:
\begin{equation}
 S_{\theta_0,\phi_0}^{(7)}(n,1)=[\bar{r_n}(2-\bar{r_n})-|{\bar{w}_n}|^2]/2.
\end{equation}
The eigenvalues of $\mathcal{U}$  are $\left\lbrace -1,-1,i,i,\exp({\frac{\pm i\pi}{3}}),i~\exp({\frac{\pm 2i\pi}{3}})\right\rbrace$, which implies that $\mathcal{U}^{12}=I$. We find that  the entanglement dynamics are periodic in nature having period $6$  for any arbitrary initial coherent states, except for the initial state $\ket{\pi/2,\pm\pi/2}$ where the  period $3$ and plotted in Fig. \ref{fig:4.79qubitavg}~(b) for various initial states. Thus, the time-averaged linear entropy is given  as follows:
\begin{widetext}
\begin{eqnarray}\nonumber
 \langle S_{(\theta_0,\phi_0)}^{(7)}\rangle&=&\left[16524436+6369\cos\left(2 \theta_0 \right)-33374 \cos\left(4 \theta_0 \right)+101035 \cos\left(6 \theta_0 \right)+136588 \cos\left(8 \theta_0 \right)+27169 \cos\left(10 \theta_0 \right) \right. \\  \nonumber&& \left.+14398\cos\left(12 \theta_0 \right)+595 \cos\left(14 \theta_0 \right)-2048 (11 (89+91 \cos\left(2 \theta_0 \right)) \cos\left(4 \phi_0 \right)+2 (41+91 \cos\left(2 \theta_0 \right)) \cos\left(8 \phi_0 \right)\right. \\  && \left.-(3-7 \cos\left(2 \theta_0 \right)) \cos\left(12 \phi_0 \right)) \sin^{12}\left(\theta_0 \right)\right]\Big{/}{50331648}.
\end{eqnarray}
\end{widetext}
The time-averaged linear entropy  is confined to the narrow interval $[0.32388, 1/3]$, as illustrated in Fig. \ref{fig:7qubitavg}~(a). The maximum value corresponds to various initial states as follows: $\ket{0,\phi_0},\ket{\pi/2,\pm \pi/2},\ket{\pi,\phi_0}$, $\ket{\pi/2,0}$,$\ket{\pi/2,\pm\pi}$, while the minimum  values is associated with the states $\ket{0.3877284,\pm\pi/2}$, $\ket{2.75386,\pm\pi/2}$, $\ket{\pi/2,\pm2.7538}$,$\ket{\pi/2,\pm 0.3877284}$, $\ket{\pi/2,\pm 1.958524}$, $\ket{\pi/2,\pm1.183068}$ as shown in Fig. \ref{fig:7qubitavg}. We numerically computed the average values of concurrence for arbitrary initial states and found that for the states $\ket{0,0}$, $\ket{\pi/2,\pm\pi/2}$, and $\ket{\pi/2,0}$, both the linear entropy and entanglement entropy attain their maximum values, while the concurrence vanishes, as shown in Fig. \ref{fig:4.88qubitavg}~(b). The average value of concurrence is maximum of $4.450511364\times10^{-2}$ for the states $\ket{1.02887,\pm\pi}$, $\ket{1.02887,\pm\pi/2}$, $\ket{1.02887,0}$, $\ket{2.11272,\pm\pi}$, $\ket{2.11272,\pm\pi/2}$, $\ket{2.11272,0}$, $\ket{2.59966,\pm\pi}$, $\ket{2.59666,\pm\pi/2}$, $\ket{2.59666,0}$, $\ket{0.54192,\pm\pi}$, $\ket{0.54912,\pm\pi/2}$ and $\ket{0.54912,0}$.
\subsection{Exact solution for $8$ qubit}
The unitary operator $\mathcal{U}$ can be expressed in two blocks $\mathcal{U}_{+}$ and $\mathcal{U}_{-}$ in $\ket{\phi}$ basis \cite{sharma2024exactly}. The two blocks are given as follows:
\begin{equation}
\mathcal{U}_{+}={\frac{1}{8}\left(
\begin{array}{ccccc}
 -1& 0 & -2\sqrt{7}  & 0 & -\sqrt{35}  \\
 0 & -6i & 0 & -2i \sqrt{7}  & 0 \\
-2 \sqrt{7}  & 0 & -4 & 0 & 2\sqrt{5}  \\
 0 & -2i \sqrt{7}  & 0 & 6i & 0 \\
 - \sqrt{35}  & 0 & 2 \sqrt{5}  & 0 & -3 \\
\end{array}
\right)}~~ \mbox{and}
\end{equation}
\begin{equation}
\mathcal{U}_{-}={\frac{1}{2 \sqrt{2}}\left(
\begin{array}{cccc}
 0 & 1 & 0 & \sqrt{7}  \\
i & 0 & i\sqrt{7}  & 0 \\
 0 & \sqrt{7}  & 0 & -1 \\
 i \sqrt{7}  & 0 & -i & 0 \\
\end{array}
\right)}.
\end{equation}
Applying the unitary operator $\mathcal{U}$ $n$ times on the state $\ket{\psi}$ we get,
\begin{eqnarray}
\ket{\psi_n}&=&\mathcal{U}^n\ket{\psi}\\ \nonumber
&=& f_{1n}\ket{\phi_0^+}+f_{2n}\ket{\phi_1^+}+f_{3n}\ket{\phi_2^+}+f_{4n}\ket{\phi_3^+} +f_{5n}\ket{\phi_4^+}\\ \nonumber&&+f_{6n}\ket{\phi_0^-}+f_{7n}\ket{\phi_1^-}+f_{8n}\ket{\phi_2^-}+f_{9n}\ket{\phi_3^-},\nonumber
\end{eqnarray}
where the coefficients $f_{jn}$ can be computed as follows:
\begin{equation}\label{Eq:qkt6}
f_{jn}=\sum_{q=1}^{\frac{N+2}{2}}\mathcal{U}^n_{j,q}~ a_q+\sum_{q=\frac{N+4}{2}}^{N+1}\mathcal{U}^n_{j,q}~ b_{q-\frac{N+2}{2}},1\leq j\leq N+1.
\end{equation}
The expressions of the coefficients $f_{jn}$ can be calculated using  Eqs.  (\ref{Eq:arbitaray}), (\ref{Eq:arbitaray1}), (\ref{Eq:arbitaray4}) and (\ref{Eq:qkt6}) for $8$ qubits. The detailed calculations  regarding the $n${th} time evolution of $\mathcal{U}$ and  the coefficient $f_{jn}$ are provided in the supplemental material \cite{supplementry2023}.
The single qubit RDM is given as,
\begin{equation}
\rho_1(n)=\frac{1}{2}\left(
\begin{array}{cc}
 y_n & \bar{v}_n \\
\bar{ v}_n^* & 2-y_n \\
\end{array}
\right),
\end{equation}
where the coefficient $y_n$ and ${m}_n$ are calculated as follows:
\begin{eqnarray} \nonumber
 y_n&=&\frac{1}{2}+f_{1n}f_{6n}^*+f_{6n}f_{1n}^*+\frac{3}{4}\left(f_{7n}f_{2n}^*+f_{2n}f_{7n}^*\right)+\\ \nonumber&&\frac{1}{2}\left(f_{8n}f_{3n}^*+f_{3n}f_{8n}^*\right)+\frac{1}{4}\left(f_{9n}f_{4n}^*+f_{4n}f_{9n}^*\right)~\mbox{and}\\ \nonumber
 \end{eqnarray}
 \begin{eqnarray} \nonumber
 m_n&=&\left[2\left(\left(f_{1n}+f_{6n}\right)\left(f_{2n}^*+f_{7n}^*\right)+\left(f_{7n}-f_{2n}\right)\right.\right.\\ \nonumber &&\left.\left.\left(f_{1n}^*-f_{6n}^*\right)\right)+\sqrt{7}\left(\left(f_{2n}+f_{7n}\right)\left(f_{3n}^*+f_{8n}^*\right)\right.\right.\\ \nonumber &&\left.\left.+\left(f_{3n}-f_{8n}\right)\left(f_{7n}^*-f_{2n}^*\right)\right)+{3}\left(\left(f_{3n}+f_{8n}\right)\right.\right.\\ \nonumber &&\left.\left.\left(f_{4n}^*+f_{9n}^*\right)+\left(f_{9n}-f_{4n}\right)\left(f_{8n}^*-f_{3n}^*\right)\right) +2\sqrt{7}\right.\\ \nonumber &&\left.\left(f_{4n}+f_{9n}\right)f_{5n}^*+\left(-f_{4n}^*+f_{9n}^*\right)f_{5n}\right]\Big{/}{4\sqrt{2}}.
\end{eqnarray}
The eigenvalues  of $\rho_1(n)$ are $\frac{1}{2} \left(1\pm\sqrt{1-{y}_n\left(2-{y}_n\right)-|{{m}_n}|^2}\right)$. The linear entropy of single qubit RDM is given as follows:
\begin{equation}
 S_{(\theta_0,\phi_0)}^{(8)}(n,1)=[y_n(2-y_n)-|{m_n}|^2]/2.
\end{equation}
The eigenvalues of $\mathcal{U}$ for the case $J=1$ and $\tau=\pi/4$ are $\left\lbrace \pm1,\pm1,\pm \exp({\frac{i\pi}{4}}),-\exp({\frac{3i\pi}{4}})\right\rbrace$, which implies that $\mathcal{U}^8=I$. We find that  the entanglement dynamics are periodic in nature having period $4$  for any arbitrary initial  state, except for the initial state $\ket{\pi/2,\pm\pi/2}$ where the period $2$. Which is shown in Fig. \ref{fig:4.795qubitavg}~(c) for various initial states. Thus, the time-averaged linear entropy for an arbitrary initial coherent states, is given  as follows:
\begin{figure}[htbp]\vspace{0.4cm}
\includegraphics[width=0.47\textwidth,height=0.15\textheight]{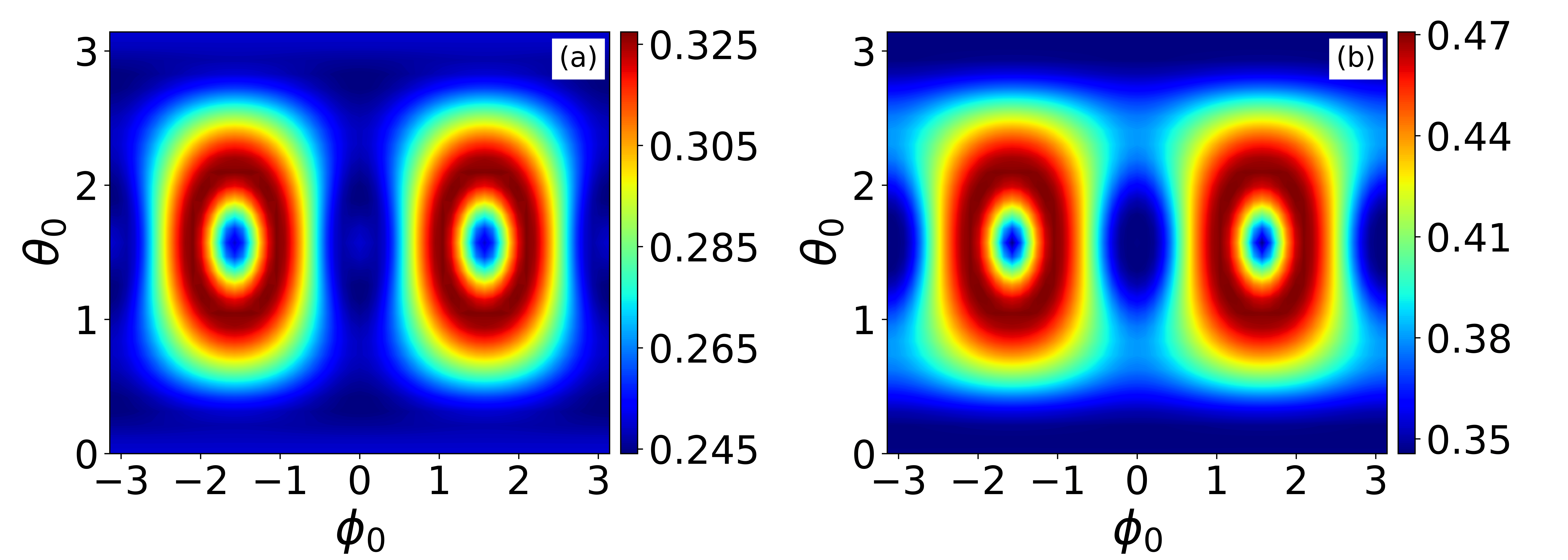}
\caption{Contour plot of time-averaged values of (a) linear entropy and (b) entanglement entropy for any arbitrary  initial states $\ket{\theta_{0},\phi{_0}}$ for $8$ qubits.}
\label{fig:8qubitavg}
\end{figure}
\begin{widetext}
\begin{eqnarray}\nonumber
 \langle S_{(\theta_0,\phi_0)}^{(8)}\rangle&=&\left[1169626257-100179152 \cos\left(2 \theta_0\right)-30918888 \cos\left(4 \theta_0\right)+23312016 \cos\left(6 \theta_0\right)+4417140
\cos\left(8 \theta_0\right)+5633712\right.\\ \nonumber && \left. \cos\left(10 \theta_0\right)+1548840 \cos\left(12 \theta_0\right)+274320 \cos\left(14 \theta_0\right)+27579 \cos\left(16 \theta_0\right)-224
(3335046+3546760 \cos\left(2 \theta_0\right)\right.\\ \nonumber && \left.+1015019 \cos\left(4 \theta_0\right)+431796 \cos\left(6 \theta_0\right)+45770 \cos\left(8 \theta_0\right)+14020 \cos\left(10
\theta_0\right)+197 \cos\left(12 \theta_0\right)) \cos\left(2 \phi_0 \right) \sin^{4}\left(\theta_0\right)\right.\\ \nonumber && \left.-65536 (1001 (3+\cos\left(2 \theta_0\right)) \cos\left(4 \phi_0 \right)+91 (5+3
\cos\left(2 \theta_0\right)) \cos\left(8 \phi_0 \right)+(5+23 \cos\left(2 \theta_0\right))\cos\left(12 \phi_0 \right)) \sin^{14}\left(\theta_0\right) \right.\\ && \left.+32768 (14 (169 \cos\left(6
\phi_0 \right)+27 \cos\left(10 \phi_0 \right)+\cos\left(14 \phi_0 \right))+\cos\left(16 \phi_0 \right)) \sin^{16}\left(\theta_0\right)\right]\Big{/}{4294967296}.
\end{eqnarray}
\end{widetext}
The analytically obtained expression for the time-averaged linear entropy lies within a small interval $[0.2438, 0.3275]$ and shown in Fig. \ref{fig:8qubitavg}~(a). We find that it attains its maximum values for the initial states,  $\ket{1.189485,\pm1.983328}$, $\ket{1.952105,\pm1.1582623}$, $\ket{1.952105,\pm1.983328}$, $\ket{1.189485,\pm1.1582623}$ and reaches a minimum for the initial states such as $\ket{0.361377,\pm\pi}$, $\ket{0.361377,0}$, $\ket{2.780216,\pm\pi}$, $\ket{2.780216,0}$, $\ket{1.209419,\pm\pi}$, $\ket{1.9321731,\pm\pi}$,$\ket{1.9321731,0}$ and $\ket{1.9321731,0}$ which can be seen from  Fig. \ref{fig:8qubitavg}~(a).  Using numerical methods, we computed the average values of concurrence for arbitrary initial states, and plotted in Fig. \ref{fig:4.781qubitavg}. We observe that the concurrence is zero for the states  $\ket{0,0}$,$\ket{\pi/2,\pm\pi/2}$ and $\ket{\pi/2,0}$, while both the linear entropy and entanglement entropy attains their maximum values \cite{sharma2024exactly}. The average value of concurrence is maximum of $4.32522158 \times 10^{-2}$ for the states $\ket{0.518363,\pm\pi}$, $\ket{0.518363,0}$ $\ket{1.0524335,\pm\pi}$, $\ket{1.0524335,0}$ $\ket{2.6232298,\pm\pi}$, $\ket{2.6232298,0}$ $\ket{2.0891591,\pm\pi}$, and $\ket{2.0891591,0}$.
\subsection{Exact solution for $9$ qubits}
In this set of basis, the unitary operator is block diagonalized in two blocks $\mathcal{U}_{+}~(\mathcal{U}_{-})$ having dimension $5\times5~(5\times5)$. The blocks are given as follows:
\begin{equation}
\mathcal{U}_{\pm}=\frac {e^ {\frac{{\mp}i \pi  }{4}}}{16}\left(
\begin{array}{ccccc}
1 & \mp 3i~ & {6} & \mp{2i} \sqrt{21}  & {3} \sqrt{14}  \\
 \pm 3 i & -{7}  & \pm {10 i}  & -2\sqrt{21}  & \pm {i}\sqrt{14}  \\
 6  & \mp {10 i} & 8  & 0 & -2{\sqrt{14}} \\
\pm {2i} \sqrt{21} & - 2\sqrt{21} & 0 & 8 & \mp {2i}\sqrt{6} \\
 3\sqrt{14}  & \mp i\sqrt{14}  & -2{\sqrt{14}}  & \pm {2i} \sqrt{6}  & {6}  \\
\end{array}
\right).
\end{equation}
The state $\ket{\psi_n}$ is obtained by applying the $n${th} iteration of unitary operator $\mathcal{U}$ to the initial state $\ket{\psi}$ and is expressed as,
\begin{eqnarray}\nonumber
\ket{\psi_n}&=&\mathcal{U}^n\ket{\psi}\\ \nonumber
&=& \bar{c}_{1n}\ket{\phi_1^+}+\bar{c}_{2n}\ket{\phi_2^+}+\bar{c}_{3n}\ket{\phi_3^+}+\bar{c}_{4n}\ket{\phi_4^+}+\\ \nonumber&&\bar{c}_{5n}\ket{\phi_5^+} +\bar{c}_{6n}\ket{\phi_1^-}+\bar{c}_{7n}\ket{\phi_2^-}+\bar{c}_{8n}\ket{\phi_3^-}+\\ &&\bar{c}_{9n}\ket{\phi_4^-}+\bar{c}_{10n}\ket{\phi_5^-},
\end{eqnarray}
where the coefficients  $\bar{c}_{jn}$ are  calculated as follows:
\begin{equation}\label{Eq:qkt53}
 \bar{c}_{jn}=\sum_{q=1}^{\frac{N+1}{2}}\mathcal{U}^n_{j,q}~ a_q+\sum_{q=\frac{N+3}{2}}^{N+1}\mathcal{U}^n_{j,q}~ b_{q-\frac{N+1}{2}},1\leq j\leq N+1.
 \end{equation}
 The expressions of the coefficients $\bar{c}_{jn}$ can be calculated using Eqs. (\ref{Eq:arbitaray2}), (\ref{Eq:arbitaray3}) and (\ref{Eq:qkt53}) for $9$ qubits. The detailed calculations  regarding the $n${th} time evolution of $\mathcal{U}$ and  the coefficient $\bar{c}_{jn}$ are provided in the supplemental material \cite{supplementry2023}.
The  single qubit $\rho_1(n)$, is given as,
\begin{equation}
\rho_1(n)=\frac{1}{2}\left(
\begin{array}{cc}
 \bar{h}_n & \bar{k}_n \\
 \bar{k}_n^* & 2-\bar{h}_n \\
\end{array}
\right),
\end{equation}
where the coefficient $\bar{h}_n$ and $\bar{k}_n$ are calculated as,
\begin{eqnarray} \nonumber
 \bar{h}_n&=&\frac{1}{2}+\bar{c}_{1n}\bar{c}_{6n}^*+\bar{c}_{6n}\bar{c}_{1n}^*+\left[{7}\left(\bar{c}_{7n}\bar{c}_{2n}^*+\bar{c}_{2n}\bar{c}_{7n}^*\right)+\right.\\ \nonumber &&\left.{5}\left(\bar{c}_{8n}\bar{c}_{3n}^*+\bar{c}_{3n}\bar{c}_{8n}^*\right)+{3}\left(\bar{c}_{9n}\bar{c}_{4n}^*+\bar{c}_{4n}\bar{c}_{9n}^*\right)+\right.\\ \nonumber && \left.\left(\bar{c}_{10n}\bar{c}_{5n}^*+\bar{c}_{5n}\bar{c}_{10n}^*\right)\right]\Big{/}{9} ~~~\mbox{and}\\ \nonumber
 \end{eqnarray}
 \begin{figure}[htbp]\vspace{0.4cm}
\includegraphics[width=0.47\textwidth,height=0.15\textheight]{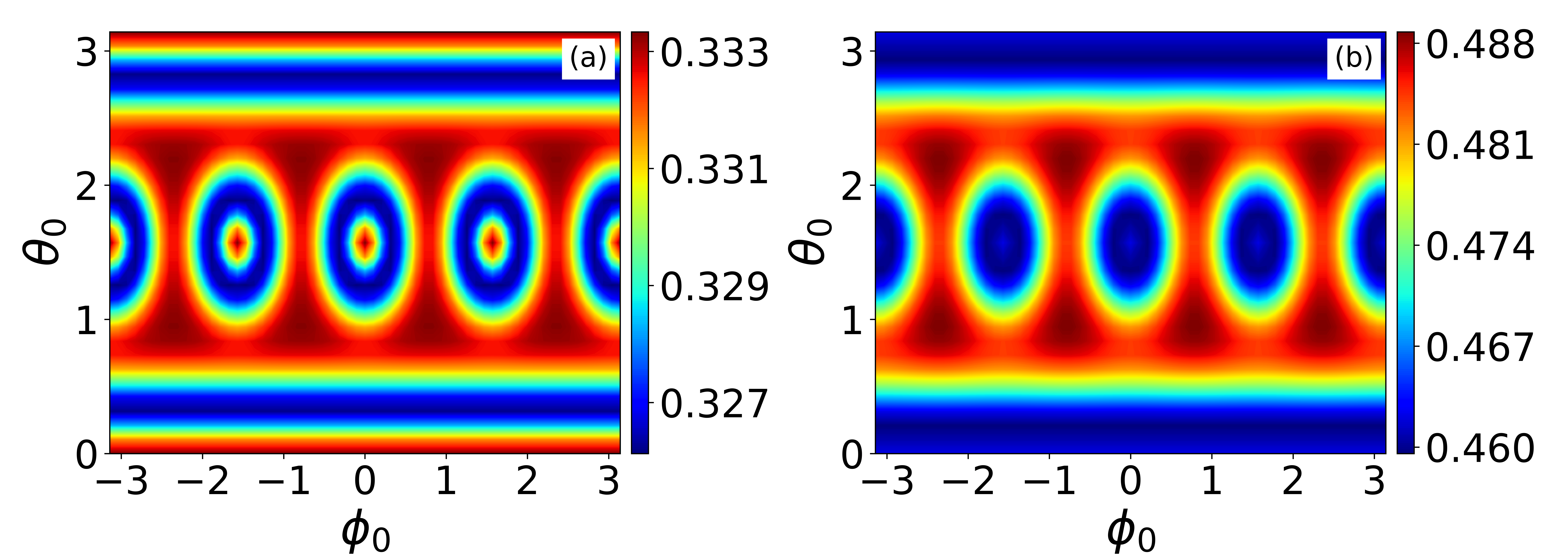}
\caption{Contour plot of time-averaged values of (a) linear entropy and (b) entanglement entropy for any arbitrary  initial states $\ket{\theta_{0},\phi{_0}}$ for $9$ qubits.}
\label{fig:9qubitavg}
\end{figure}
 \begin{eqnarray} \nonumber
 \bar{k}_n&=&\frac{1}{9}\left[3\left(\left(\bar{c}_{1n}+\bar{c}_{6n}\right)\left(\bar{c}_{2n}^*+\bar{c}_{7n}^*\right)+\left(\bar{c}_{7n}-\bar{c}_{2n}\right)\right.\right.\\ \nonumber &&\left.\left.\left(\bar{c}_{1n}^*-\bar{c}_{6n}^*\right)\right)+4(\left(\bar{c}_{2n}+\bar{c}_{7n}\right)\left(\bar{c}_{3n}^*+\bar{c}_{8n}^*\right)+\right. \\ \nonumber && \left.\left(\bar{c}_{3n}-\bar{c}_{8n}\right)\left(\bar{c}_{7n}^*-\bar{c}_{2n}^*\right))+\sqrt{21}\left(\left(c_{3n}+c_{8n}\right)\right.\right.\\ \nonumber &&\left.\left.\left(c_{4n}^*+c_{9n}^*\right)+\left(c_{9n}-c_{4n}\right)\left(\bar{c}_{3n}^*-\bar{c}_{8n}^*\right)\right)-5i \right.\\ \nonumber &&\left.\left(\left(\bar{c}_{5n}+\bar{c}_{10n}\right)\left(\bar{c}_{5n}^*-\bar{c}_{10n}^*\right)\right)\right]+\dfrac{4}{3\sqrt{7}} \left(\left(\bar{c}_{4n}+\bar{c}_{9n}\right)\right.\\ \nonumber &&\left.\left(\bar{c}_{5n}^*+\bar{c}_{10n}^*\right)+\left(\bar{c}_{5n}-\bar{c}_{10n}\right)\left(\bar{c}_{9n}^*-\bar{c}_{4n}^*\right)\right).
\end{eqnarray}
The eigenvalues of $\rho_1(n)$ are $\frac{1}{2} \left(1\pm\sqrt{1-\bar{h}_n\left(2-\bar{h}_n\right)-|\bar{k}_n}|^2\right)$. The linear entropy of single qubit RDM is given as follows:
\begin{equation}
 S_{\theta_0,\phi_0}^{(9)}(n,1)=[\bar{h}_n(2-\bar{h}_n)-|\bar{k}_n|^2]/2.
\end{equation}

The eigenvalues of $\mathcal{U_{+}}~\left(\mathcal{U_{-}}\right)$ is given as ${\exp({\frac{3i \pi }{4}})\left\{1,\exp({\pm\frac{2i \pi }{3}}),\exp({\pm\frac{2i \pi }{3}})\right\}}~\left(\exp({\frac{i \pi }{4}})\left\{-1,\exp({\pm\frac{i \pi }{3}})\right.\right.\\ \left.\left.,\exp({\pm\frac{i \pi }{3}})\right\}\right)$, which implies that $\mathcal{U}^{12}=I$. We find that  the entanglement dynamics are periodic in nature having period $6$  for any arbitrary initial coherent states, except for the initial state  $\ket{\pi/2,\pm\pi/2}$ where the period $3$ and plotted in Fig. \ref{fig:4.79qubitavg}~(c) for various initial states. Thus, the time-averaged linear entropy is given  as follows:
\begin{widetext}
\begin{eqnarray}\nonumber
 \langle S_{(\theta_0,\phi_0)}^{(9)}\rangle&=&\left[8507290602-8548826\cos\left(2 \theta_0 \right)-32246864\cos\left(4 \theta_0 \right)+35001876\cos\left(6 \theta_0 \right)+52655624
\cos\left(8 \theta_0 \right)+18504260\right.\\ \nonumber &&\left.\cos\left(10 \theta_0 \right)+14750928\cos\left(12 \theta_0 \right)+1880317\cos\left(14 \theta_0 \right)+626062\cos\left(16 \theta_0
\right)+20613\cos\left(18 \theta_0 \right)-65536 (104\right.\\ \nonumber &&\left. (155+153\cos\left(2 \theta_0 \right))\cos\left(4 \phi_0 \right)+28 (107+153\cos\left(2 \theta_0 \right))\cos\left(8 \phi_0
\right)+24 (3+17\cos\left(2 \theta_0 \right))\cos\left(12 \phi_0 \right)+\right.\\  &&\left.(-5+9\cos\left(2 \theta_0 \right))\cos\left(16 \phi_0 \right)) \sin^{16}\left(\theta_0 \right)\right]\Big{/}{25769803776}.
\end{eqnarray}
\end{widetext}
The time-averaged linear entropy lies within a small interval $[0.3261, 1/3]$, as shown in Fig. \ref{fig:9qubitavg}~(a). It reaches its maximum value for the states such as $\ket{0,0}$, $\ket{\pi/2,\pm \pi/2}$, and $\ket{\pi,\pm \pi}$, while it attains minimum values for states like $\ket{\pi/2,\pm0.339837}$, $\ket{\pi/2,\pm1.23096}$, $\ket{\pi/2,\pm1.910633}$, $\ket{\pi/2,\pm2.801756}$, $\ket{0.339837,-0.020757}$ and $\ket{2.801755,-0.01508}$, as can be observe in Fig. \ref{fig:9qubitavg}. We numerically obtain the average values of concurrence for arbitrary initial states and plotted them in Fig. \ref{fig:4.88qubitavg}. The average value of concurrence is maximum of $ 3.043055600896\times10^{-2}$ for the states $\ket{2.72533,\pm\pi}$, $\ket{2.72533,\pm\pi/2}$, $\ket{2.72533,0}$, $\ket{1.98706,\pm\pi}$, $\ket{1.98706,\pm\pi/2}$, $\ket{1.98706,0}$, $\ket{0.41626,\pm\pi}$, $\ket{0.41626,\pm\pi/2}$, $\ket{0.41626,0}$, $\ket{1.154535,\pm\pi}$, $\ket{1.154535,\pm\pi/2}$ and $\ket{1.154535,0}$. From the numerical results, we observe that for the states such as, $\ket{0,0}$, $\ket{\pi/2,\pm\pi/2}$, and $\ket{\pi/2,0}$, the  linear entropy and entanglement entropy reaches their maximum values, while the concurrence vanishes \cite{sharma2024exactly}. A similar periodic nature of the entanglement dynamics is observe for  any odd-$N$ (results are not shown here).  We also find that the qualitative structure of the contour plots  of linear entropy and entanglement entropy are  same for odd qubits, despite the different values of linear entropy and entanglement entropy corresponding to different arbitrary initial states $\ket{\theta_0,\phi_0}$.
\subsection{Exact solution for $10$ qubit}
The unitary operator in $\phi$ basis can be written in two blocks $\mathcal{U}_{+}$ and $\mathcal{U}_{-}$ with dimensions $6\times6$ and $5\times5$ \cite{sharma2024exactly}. The two blocks can be written as follows:
\begin{widetext}
\begin{equation}
\mathcal{U}_{+}=\frac{1}{8\sqrt{2}}\left(
\begin{array}{cccccc}
 0 & - \sqrt{{5}}~ e^{\frac{3 i \pi}{4}} & 0 & -2 \sqrt{{15}}~ e^{\frac{3 i \pi}{4}} & 0 & -{3}\sqrt{{7}}~
e^{\frac{3 i \pi}{4}} \\
  \sqrt{{5}} ~e^{\frac{-3 i \pi}{4}} & 0 & 9~ e^{\frac{-3 i \pi}{4}} & 0 &  \sqrt{42} ~e^{\frac{-3 i \pi}{4}} & 0 \\
 0 & -9~ e^{\frac{3 i \pi}{4}} & 0 & -2\sqrt{{3}}~ e^{\frac{3 i \pi}{4}} & 0 &  \sqrt{{35}}
~e^{\frac{3 i \pi}{4}} \\
 2 \sqrt{{15}}~ e^{\frac{-3 i \pi}{4}} & 0 &  2\sqrt{3} e^{\frac{-3 i \pi}{4}} & 0 & - 2\sqrt{14}~ e^{\frac{-3 i \pi}{4}} & 0 \\
 0 & -\sqrt{42}~ e^{\frac{3 i \pi}{4}} & 0 & 2\sqrt{14}~ e^{\frac{3 i \pi}{4}} & 0 & - \sqrt{30}~ e^{\frac{3 i \pi}{4}}
\\
 {3}\sqrt{{7}}~ e^{\frac{-3 i \pi}{4}} & 0 & - \sqrt{{35}}~ e^{\frac{-3 i \pi}{4}} & 0 &  \sqrt{30}~ e^{\frac{-3 i \pi}{4}} & 0 \\
\end{array}
\right) ~~ \mbox{and}
\end{equation}
\begin{equation}
\mathcal{U}_{-}={ \frac{1}{16}\left(
\begin{array}{ccccc}
 e^{\frac{3 i \pi}{4}} & 0 & {3} \sqrt{5}~ e^{\frac{3 i \pi}{4}} & 0 &  \sqrt{{210}}~ e^{\frac{3 i \pi}{4}} \\
 0 & -8~ e^{-\frac{3 i \pi}{4}} & 0 & -8 \sqrt{3}~ e^{-\frac{3 i \pi}{4}} & 0 \\
 {3} \sqrt{5}~ e^{\frac{3 i \pi}{4}} & 0 & {13}~ e^{\frac{3 i \pi}{4}} & 0 & - \sqrt{{42}}~ e^{\frac{3 i \pi}{4}}
\\
 0 & -8 \sqrt{3} ~e^{-\frac{3 i \pi}{4}} & 0 & 8~ e^{-\frac{3 i \pi}{4}} & 0 \\
 \sqrt{{210}}~ e^{\frac{3 i \pi}{4}} & 0 & -\sqrt{{42}}~ e^{\frac{3 i \pi}{4}} & 0 & 2~ e^{\frac{3 i \pi}{4}} \\
\end{array}
\right)}.
\end{equation}
\end{widetext}
Applying the unitary operator $\mathcal{U}$ $n$ times on the state $\ket{\psi}$ we get,
\begin{eqnarray}
\ket{\psi_n}&=&\mathcal{U}^n\ket{\psi_0}\\ \nonumber
&=& d_{1n}\ket{\phi_0^+}+d_{2n}\ket{\phi_1^+}+d_{3n}\ket{\phi_2^+}+d_{4n}\ket{\phi_3^+}\\ \nonumber&& +d_{5n}\ket{\phi_4^+}+d_{6n}\ket{\phi_5^+}+d_{7n}\ket{\phi_0^-}+d_{8n}\ket{\phi_1^-}\\ \nonumber&&+d_{9n}\ket{\phi_2^-}+d_{10n}\ket{\phi_3^-}+d_{11n}\ket{\phi_4^-},\nonumber
\end{eqnarray}
where the coefficients are given as follows:
\begin{equation}\label{Eq:qkt54}
d_{jn}=\sum_{q=1}^{\frac{N+2}{2}}\mathcal{U}^n_{j,q}~ a_q+\sum_{q=\frac{N+4}{2}}^{N+1}\mathcal{U}^n_{j,q}~ b_{q-\frac{N+2}{2}},1\leq j\leq N+1.
\end{equation}
The expressions of the coefficients $d_{jn}$ can be calculated using  Eqs.  (\ref{Eq:arbitaray}), (\ref{Eq:arbitaray1}), (\ref{Eq:arbitaray4}) and (\ref{Eq:qkt54}) for $10$ qubits. The detailed calculations  regarding the $n${th} time evolution of $\mathcal{U}$ and  the coefficient $d_{jn}$ are provided in the supplemental material \cite{supplementry2023}.
The single qubit RDM, $\rho_1(n)$, is given as,
\begin{equation}
\rho_1(n)=\frac{1}{2}\left(
\begin{array}{cc}
 x_n & \bar{z}_n \\
\bar{ z}_n^* & 2-x_n \\
\end{array}
\right),
\end{equation}
where the coefficients  $x_n$ and $\bar{z}_n$ are given as follows:
\begin{eqnarray} \nonumber
 x_n&=&\frac{1}{2}+d_{1n}d_{7n}^*+d_{7n}d_{1n}^*+\frac{1}{5}\left[4\left(d_{8n}d_{2n}^*+d_{2n}d_{8n}^*\right)+\right.\\ \nonumber&&\left.{3}\left(d_{9n}d_{3n}^*+d_{3n}d_{9n}^*\right)+2\left(d_{4n}d_{10n}^*+d_{10n}d_{4n}^*\right)+\right.\\ \nonumber&&\left.\left(d_{11n}d_{5n}^*+d_{5n}d_{11n}^*\right)\right]~~\mbox{and} \\ \nonumber
 \bar{z}_n&=&\left[\sqrt{2}\left(\left(d_{1n}+d_{7n}\right)\left(d_{2n}^*+d_{8n}^*\right)+\left(d_{2n}-d_{8n}\right)\right.\right.\\ \nonumber &&\left.\left.\left(d_{7n}^*-d_{1n}^*\right)\right)+3\left(\left(d_{2n}+d_{8n}\right)\left(d_{3n}^*+d_{9n}^*\right)+\right.\right.\\ \nonumber &&\left.\left.\left(d_{9n}-d_{3n}\right)\left(d_{2n}^*-d_{8n}^*\right)\right)+2\sqrt{3}\left(\left(d_{3n}+d_{9n}\right)\right.\right.\\ \nonumber &&\left.\left.\left(d_{4n}^*+d_{10n}^*\right)+\left(d_{4n}-d_{10n}\right)\left(d_{9n}^*-d_{3n}^*\right)\right)+\sqrt{14} \right.\\ \nonumber &&\left.\left(\left(d_{4n}+d_{10n}\right)\left(d_{5n}^*+d_{11n}^*\right)+\left(d_{11n}-d_{5n}\right)\right.\right.\\ \nonumber &&\left.\left.\left(d_{4n}^*-d_{10n}^*\right)\right) +\sqrt{30}\left(d_{5n}+d_{11n}\right)d_{6n}^*+\right.\\ \nonumber &&\left.\left(-d_{5n}^*+d_{11n}^*\right)d_{6n}\right]\Big{/}{5\sqrt{2}}.
\end{eqnarray}
The eigenvalues  of single qubit RDM, $\rho_1(n)$ are $\frac{1}{2} \left(1\pm\sqrt{1-{x}_n\left(2-{x}_n\right)-|{\bar{z}_n}|^2}\right)$. The linear entropy of single qubit RDM can be expressed as,
\begin{equation}
 S_{(\theta_0,\phi_0)}^{(10)}(n,1)=[x_n(2-x_n)-|{z_n}|^2]/2.
\end{equation}
\begin{figure}[htbp]\vspace{0.4cm}
\includegraphics[width=0.47\textwidth,height=0.15\textheight]{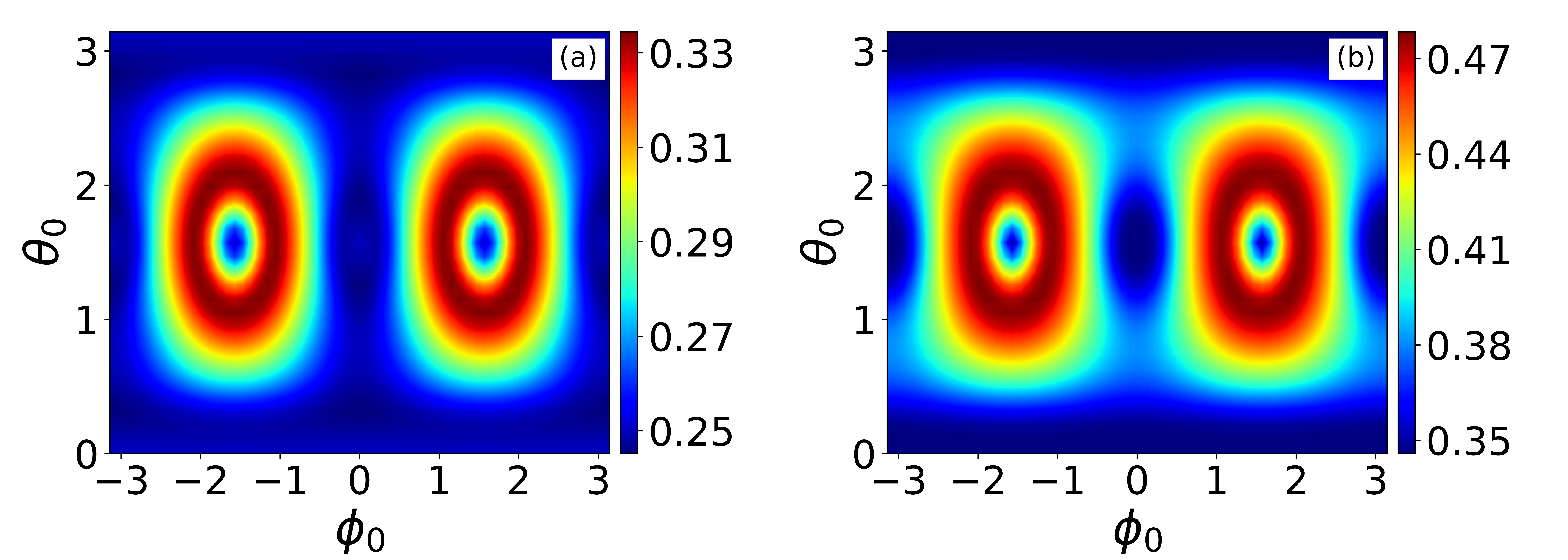}
\caption{Contour plot of time-averaged values of (a) linear entropy and (b) entanglement entropy for any arbitrary  initial states $\ket{\theta_{0},\phi{_0}}$ for $10$ qubits.}
\label{fig:10qubitavg}
\end{figure}
The eigenvalues of $\mathcal{U}$ for the case $J=1$ and $\tau=\pi/4$ are $\left\lbrace \pm i,\pm i,\pm i,\pm \exp({\frac{i\pi}{4}}),\pm \exp({\frac{3i\pi}{4}}), \exp({\frac{3i\pi}{4}})\right\rbrace$, which implies that $\mathcal{U}^8=I$. We find that  the entanglement dynamics are periodic in nature having period $4$ for arbitrary initial coherent states, except for the initial state $\ket{\pi/2,\pm\pi/2}$ where the  period $2$. Which is shown in Fig. \ref{fig:4.795qubitavg}~(d) for various initial states. Thus, the time-averaged linear entropy
for an arbitrary initial coherent states, is given  as follows:
\begin{widetext}
\begin{eqnarray}\nonumber
 \langle S_{(\theta_0,\phi_0)}^{(10)}\rangle&=&\left[150956244022-13735021960 \cos\left(2 \theta_0\right)-3693736098 \cos\left(4 \theta_0\right)+2555995152 \cos\left(6 \theta
\right)+64102104 \cos\left(8 \theta_0\right)+\right.\\ \nonumber && \left.897625296 \cos\left(10 \theta_0\right)+258145323 \cos\left(12 \theta_0\right)+108622404 \cos\left(14 \theta_0\right)+23856114 \cos\left(16
\theta_0\right)+2902628 \cos\left(18 \theta_0\right)\right.\\ \nonumber && \left.+218487 \cos\left(20 \theta_0\right)-96 (1151747025+1300334448 \cos\left(2 \theta_0\right)+481789848 \cos\left(4 \theta
\right)+224185680 \cos\left(6 \theta_0\right)+\right.\\ \nonumber && \left. 44903028\cos\left(8 \theta_0\right)+16642800 \cos\left(10 \theta_0\right)+1292712 \cos\left(12 \theta_0\right)+325712 \cos\left(14
\theta_0\right)+4219 \cos\left(16 \theta_0\right)) \cos\left(2 \phi_0 \right) \right.\\ \nonumber && \left.\sin^{20}\left(\theta_0\right)-131072 (51 (26 (145+47 \cos\left(2 \theta_0\right)) \cos\left(4 \phi_0 \right)+8
(115+53 \cos\left(2 \theta_0\right)) \cos\left(8 \phi_0 \right)+(65+63 \cos\left(2 \theta_0\right)) \right.\\ \nonumber && \left.\cos\left(12 \phi_0 \right))+2 (-5+77 \cos\left(2 \theta_0\right)) \cos\left(16
\phi_0 \right)) \sin^{18}\left(\theta_0\right)+262144 (40392 \cos\left(6 \phi_0 \right)+9384 \cos\left(10 \phi_0 \right)+834 \right.\\  && \left.\cos\left(14 \phi_0 \right)+18 \cos\left(18 \phi_0 \right)+\cos\left(20
\phi_0 \right)) \sin^{20}\left(\theta_0\right)\right]\Big{/}{549755813888}.
\end{eqnarray}
\end{widetext}
It takes values from the  narrow interval [0.2451, 0.3345] and shown in Fig. \ref{fig:10qubitavg}~(a). The maximum valued corresponds to  initial states as follows: $\ket{1.21692837,\pm1.949179}$, $\ket{1.924664,\pm1.949179}$, $\ket{1.21692837,\pm1.1924136}$ and $\ket{1.924664,\pm1.1924136}$ and the minimum associated with  the states $\ket{0.321751,\pm \pi}$, $\ket{2.819842,\pm\pi}$, $\ket{1.2490457,\pm \pi}$, $\ket{1.892547,\pm \pi}$, $\ket{0.321751,0}$, $\ket{2.819842,0}$, $\ket{1.2490457,0}$, $\ket{1.892547,0}$, which can be seen from same figure. We numerically obtain the average values of concurrence for any arbitrary initial state and plotted in Fig. \ref{fig:4.781qubitavg}~(d).  The average value of concurrence is maximum of $ 3.1299247\times10^{-2}$ for the states $\ket{ 0.424115,\pm\pi}$,$\ket{1.146681318,\pm\pi}$, $\ket{1.146681318,0}$, $\ket{ 0.424115,0}$, $\ket{ 2.717477,\pm\pi}$,$\ket{1.994911,\pm\pi}$, $\ket{1.994911,0}$ and $\ket{2.717477,0}$. We also observe that for the  states  $\ket{0,0}$, $\ket{\pi/2,\pm\pi/2}$ and $\ket{\pi/2,0}$, its concurrence vanishes, which indicating the multipartite nature of entanglement. We also observe that for these states the  linear and entanglement entropy are maximum \cite{sharma2024exactly}.

As the number of qubits is increased, performing analytical calculations becomes  challenging due to the complexity and large size of the expressions. Therefore, we rely on numerical methods. We  numerically find that the contour plots for both linear and entanglement entropy exhibit qualitatively similar behavior for any even number of qubits $N$, despite variations in the entropy values corresponding to different initial conditions. We  observe that the average value of concurrence tends to  zero with $N$. This suggest that  entanglement  is now shared globally (multipartite) rather then bipartite manner for arbitrary initial states. We also observe the periodic behavior of entanglement measures for arbitrary initial states at any $N$ (results are not shown here). In our recent work \cite{sharma2024exactly}, we demonstrated that for  the parameter $J=1$ and $\tau=\pi/4$ the time-evolved unitary operator is periodic in nature and the spectrum  of $\mathcal{U}$ is  highly degenerate. Thus, we  conclude that, for these parameter values, the  system  exhibits signatures of the quantum integrability for arbitrary initial states at any $N$.
\section{The case $J=1/2$}\label{sec:example-section4}
In this section, we extend our analysis to the case where the interaction parameter is set to $J=1/2$. Following the same approach as in the previous section \ref{sec:example-section3}, we analytically calculated the linear entropy and entanglement entropy for arbitrary initial states for even qubits ranging from $4$ to $10$. In Ref. \cite{sharma2024signatures}, we have shown that our model exhibits quantum integrability only for even-N for the initial state $\ket{0,0}$ and $\ket{\pi/2,-\pi/2}$ for  $J=1/2$ and $\tau=\pi/4$. In contrast, the signatures of QI are absent for odd $N$. We find the signatures of QI for arbitrary initial states under this modified parameter. Similar to the previous section \ref{sec:example-section3}, we calculated the expression of the time-average linear entropy analytically and average concurrence numerically for an arbitrary initial state.
\subsection{Exact solution for $4$ qubit}
\begin{figure}[htbp]\vspace{0.4cm}
\includegraphics[width=0.47\textwidth,height=0.23\textheight]{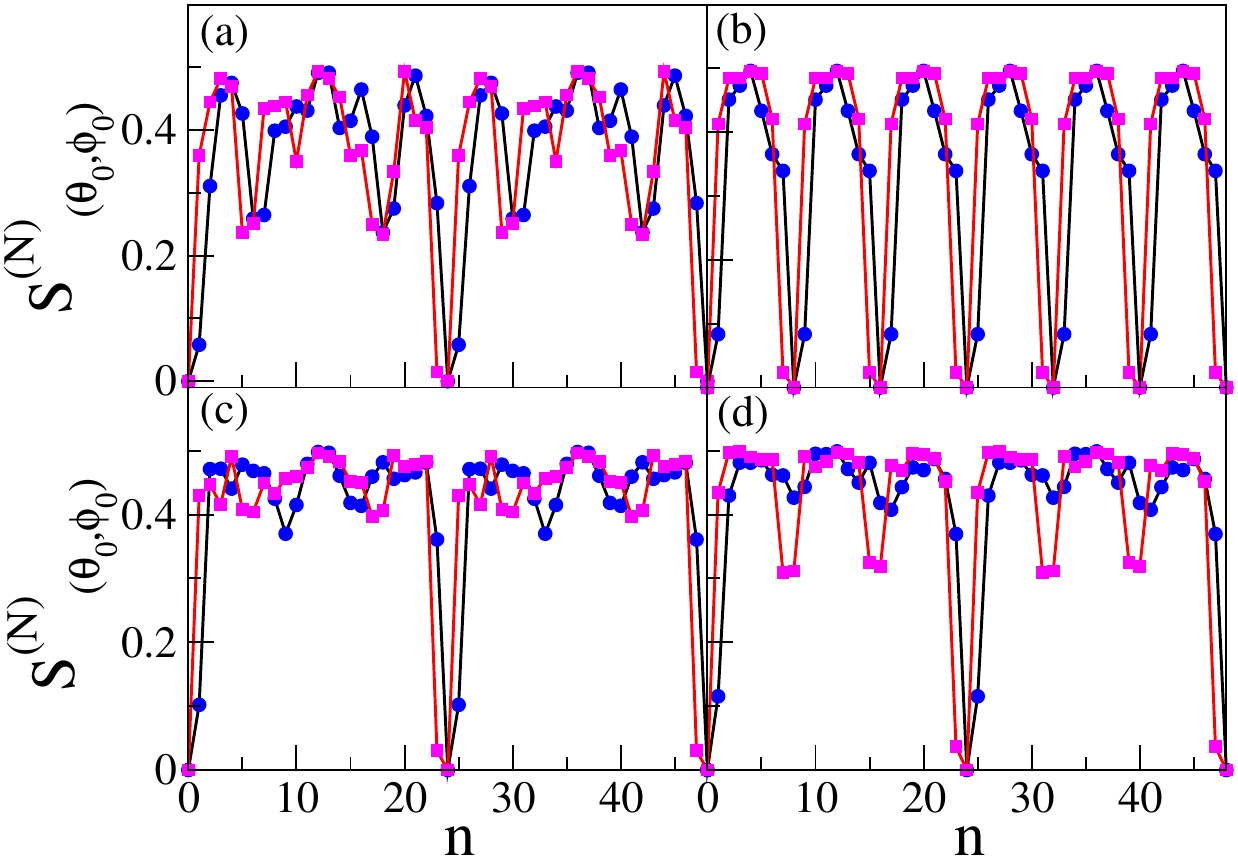}
\caption{The linear entropy for various initial states $\ket{\theta_0,\phi_0}$ = $\ket{2\pi/3,-\pi/12}$ (black line with circle) and $\ket{\pi/8,-\pi/8}$ (red line with square)) are plotted for (a) $4$ qubits  (b) $6$ qubits (c) $8$ qubits and (d) $10$ qubits.}
\label{fig:4.25qubitavg}
\end{figure}

\begin{figure}[htbp]\vspace{0.4cm}
\includegraphics[width=0.45\textwidth,height=0.25\textheight]{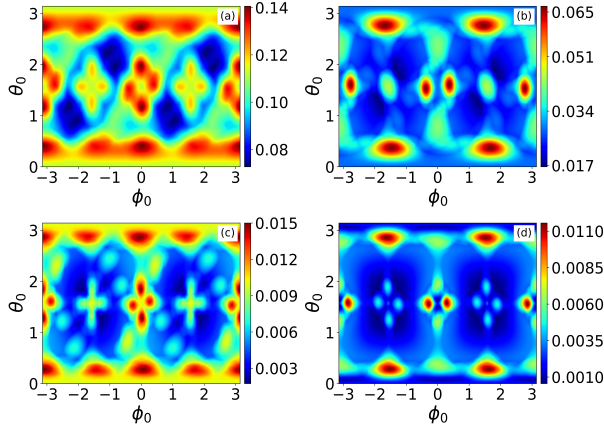}
\caption{The  time-average concurrence are plotted  for $J=1/2$ (a) 4 qubits (b) $6$ qubits (c) $8$ qubits (d) $10$ qubits  for the various  initial states $\ket{\theta_0,\phi_0}$.}
\label{fig:4.26qubitavg}
\end{figure}

Using Eq. (\ref{Eq:QKT1}), the unitary operator $\mathcal{U}$ for $4$ qubits in $\ket{\phi}$  basis, for the parameter $J=1/2$ and $\tau=\pi/4$, can be written as follows:
\begin{equation}
\mathcal{U}= \left(
\begin{array}{ccccc}
  -1&  0 & 0 & 0&0 \\
 0 & e^{\frac{-3i\pi}{4}}/2 & \sqrt{3}~ e^{\frac{-3i\pi}{4}}/2 & 0 &0\\
 0 &  \sqrt{3}~ e^{\frac{i\pi}{4}}/2 & - e^{\frac{i\pi}{4}}/2 & 0&0 \\
 0 & 0 & 0  & 0 &1 \\
 0 & 0 & 0  & - e^{\frac{-3i\pi}{4}}&0 \\
\end{array}
\right).
 \end{equation}
 The state $\ket{\psi_n}$ can be calculated by  $n$ implementations of the unitary operator $\mathcal{U}$ on the state $\ket{\psi}$ we get,
\begin{eqnarray}
\ket{\psi_n}&=&\mathcal{U}^n\ket{\psi}\\ \nonumber
&=& \bar{p}_{1n}\ket{\phi_0^+}+\bar{p}_{2n}\ket{\phi_1^+}+\bar{p}_{3n}\ket{\phi_2^+}+\bar{p}_{4n}\ket{\phi_0^-} +\bar{p}_{5n}\ket{\phi_1^-},\nonumber
\end{eqnarray}
where the coefficients  are given as follows:
\begin{equation}\label{Eq:qkt55}
\bar{p}_{jn}=\sum_{q=1}^{\frac{N+2}{2}}\mathcal{U}^n_{j,q}~ a_q+\sum_{q=\frac{N+4}{2}}^{N+1}\mathcal{U}^n_{j,q}~ b_{q-\frac{N+2}{2}},1\leq j\leq N+1.
\end{equation}
\begin{figure}[htbp]\vspace{0.4cm}
\includegraphics[width=0.5\textwidth,height=0.15\textheight]{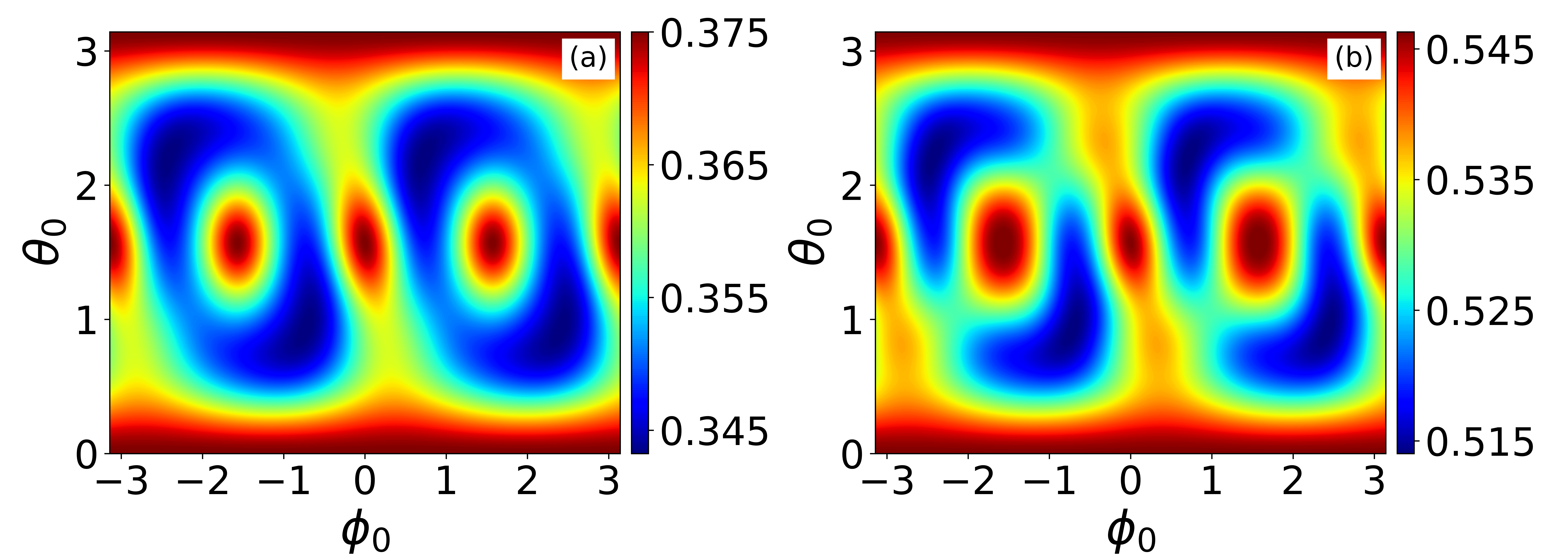}
\caption{ Contour plot of time-averaged values of (a) linear entropy and (b) entanglement entropy for any arbitrary  initial states $\ket{\theta_{0},\phi{_0}}$ for $4$ qubits.}
\label{fig:4.2qubitavg}
\end{figure}
The expressions of the coefficients $p_{jn}$ can be calculated using  Eqs.  (\ref{Eq:arbitaray}), (\ref{Eq:arbitaray1}), (\ref{Eq:arbitaray4}) and (\ref{Eq:qkt55}) for $4$ qubits. The detailed calculations  regarding the $n${th} time evolution of $\mathcal{U}$ and  the coefficient $p_{jn}$ are provided in the supplemental material \cite{supplementry2023}.
The single qubit RDM, $\rho_1(n)$, is given as,
\begin{equation}
\rho_1(n)=\frac{1}{2}\left(
\begin{array}{cc}
 {t}_{n}^{\prime} & {v}_{n}^{\prime} \\
({v}_{n}^{\prime})^* & 2-{t}_{n}^{\prime} \\
\end{array}
\right),
\end{equation}
where the coefficient ${t}_{n}^{\prime}$ and ${v}_{n}^{\prime}$ are given as follows:
\begin{eqnarray} \nonumber
 {t}_{n}^{\prime}&=&\frac{1}{2}+\bar{p}_{2n}\bar{p}_{5n}^*+\bar{p}_{5n}\bar{p}_{2n}^*+\frac{1}{2}\left(\bar{p}_{1n}\bar{p}_{4n}^*+\bar{p}_{4n}\bar{p}_{1n}^*\right)~~ \mbox{and}\\ \nonumber
 {v}_{n}^{\prime}&=&\left[\left(\left(\bar{p}_{2n}+\bar{p}_{5n}\right)\left(\bar{p}_{1n}^*+\bar{p}_{4n}^*\right)+\left(\bar{p}_{4n}-\bar{p}_{1n}\right)\left(\bar{p}_{2n}^*-\bar{p}_{5n}^*\right)\right)\right.\\ \nonumber &&\left.+{\sqrt{3}}(\left(\bar{p}_{1n}+\bar{p}_{4n}\right)\bar{p}_{3n}^*+\left(\bar{p}_{4n}^*-\bar{p}_{1n}^*\right))\bar{p}_{3n}\right]\Big{/}{2}.
\end{eqnarray}
The eigenvalues  of single qubit RDM, $\rho_1(n)$, are $\frac{1}{2} \left(1\pm\sqrt{1-{t}_{n}^{\prime}\left(2-{t}_{n}^{\prime}\right)-|{v}_{n}^{\prime}|^2}\right)$. The  linear entropy of single qubit  RDM, can be calculated  as follows:
\begin{equation}
 S_{(\theta_0,\phi_0)}^{(4)}(n,1/2)=[{t}_{n}^{\prime}(2-{t}_{n}^{\prime})-|{{v}_{n}^{\prime}}|^2]/2.
\end{equation}
The eigenvalues of $\mathcal{U}$ for the case $J=1/2$ and $\tau=\pi/4$ are  $\left\{\pm \exp({-\frac{i  \pi}{8}}), \exp({-\frac{5i  \pi}{12}}), \exp({\frac{11i  \pi}{12}}),-1\right\}$, which implies that $\mathcal{U}^{48}=I$. We find that  the entanglement dynamics are periodic in nature having period $24$  for arbitrary initial  state, except for the initial state $\ket{\pi/2,\pm\pi/2}$, where the period $12$.  Which is shown in Fig. \ref{fig:4.25qubitavg}~(a) for various initial states. Thus, the time-averaged linear entropy
for an arbitrary initial coherent states, is given  as follows:
\begin{widetext}
\begin{eqnarray}\nonumber
 \langle S_{(\theta_0,\phi_0)}^{(4)}\rangle&=&\left[47043+776\cos\left(2 \theta_0 \right)+1052\cos\left(4 \theta_0 \right)+184\cos\left(6 \theta_0 \right)+97\cos\left(8 \theta_0 \right)-8
(3+\cos\left(2 \theta_0 \right)) (-61-68\cos\left(2 \theta_0 \right)+\right.\\ \nonumber && \left.\cos\left(4 \theta_0 \right))\cos\left(2 \phi_0 \right) \sin^2\left(\theta_0 \right)+32 (4\cos\left(2 \theta_0
\right)+15 (3+\cos\left(4 \theta_0 \right)))\cos\left(4 \phi_0 \right) \sin^4\left(\theta_0 \right)+64 (3+\cos\left(2 \theta_0 \right))\cos\left(6 \phi_0 \right) \right.\\ \nonumber && \left.\sin^6\left(\theta_0 \right)+128
\cos\left(8 \phi_0 \right) \sin^8\left(\theta_0 \right)+16 (202\cos\left(\theta_0 \right)+49\cos\left(3 \theta_0 \right)+5\cos\left(5 \theta_0 \right)) \sin^2\left(\theta_0 \right) \sin\left(2
\phi_0 \right)+512\cos\left(\theta_0 \right)\right.\\  && \left. (3+\cos\left(2 \theta_0 \right)) \sin^4\left(\theta_0 \right) \sin\left(4 \phi_0 \right)+256\cos\left(\theta_0 \right) \sin^6\left(\theta_0 \right)
\sin\left(6 \phi_0 \right)\right]\Big{/}{131072}.
\end{eqnarray}
\end{widetext}
It takes values from the  narrow interval $[0.34323, 0.375]$ and shown in Fig. \ref{fig:4.2qubitavg}~(a). The maximum values corresponds to various initial states as follows: $\ket{\pi,\phi_0}$, $\ket{0,\phi_0}$, $\ket{\pi/2,\pm \pi/2}$, $\ket{\pi/2,\pm \pi}$, $\ket{\pi/2,0}$, while the minimum values associated with  the initial states $\ket{0.928472,2.41608527}$, $\ket{2.21312,0.7255075}$, $\ket{0.928472,0.7255705}$ and  $\ket{2.21312,-2.416085}$. Which can be seen from  Fig. \ref{fig:4.2qubitavg}. We numerically obtain the average values of concurrence for any arbitrary initial state and plotted in Fig. \ref{fig:4.26qubitavg}~(a). We numerically observe that, the average value of concurrence is maximum of $0.14057897625$  for the states $\ket{2.71748,-0.01571}$, $\ket{2.71748,3.12588}$, $\ket{0.424115,-3.12588}$, $\ket{0.424115,-0.01571}$ and $\ket{0.41888,0}$, whereas  minimum of $7.26644726\times10^{-2}$ for the states $\ket{0.8875,-2.0813}$, $\ket{0.8875,1.060287}$, $\ket{2.25409,-1.60287}$ and $\ket{2.25409,2.0813}$.
\subsection{Exact solution for $6$ qubits}
\begin{figure}[t]\vspace{0.4cm}
\includegraphics[width=0.45\textwidth,height=0.15\textheight]{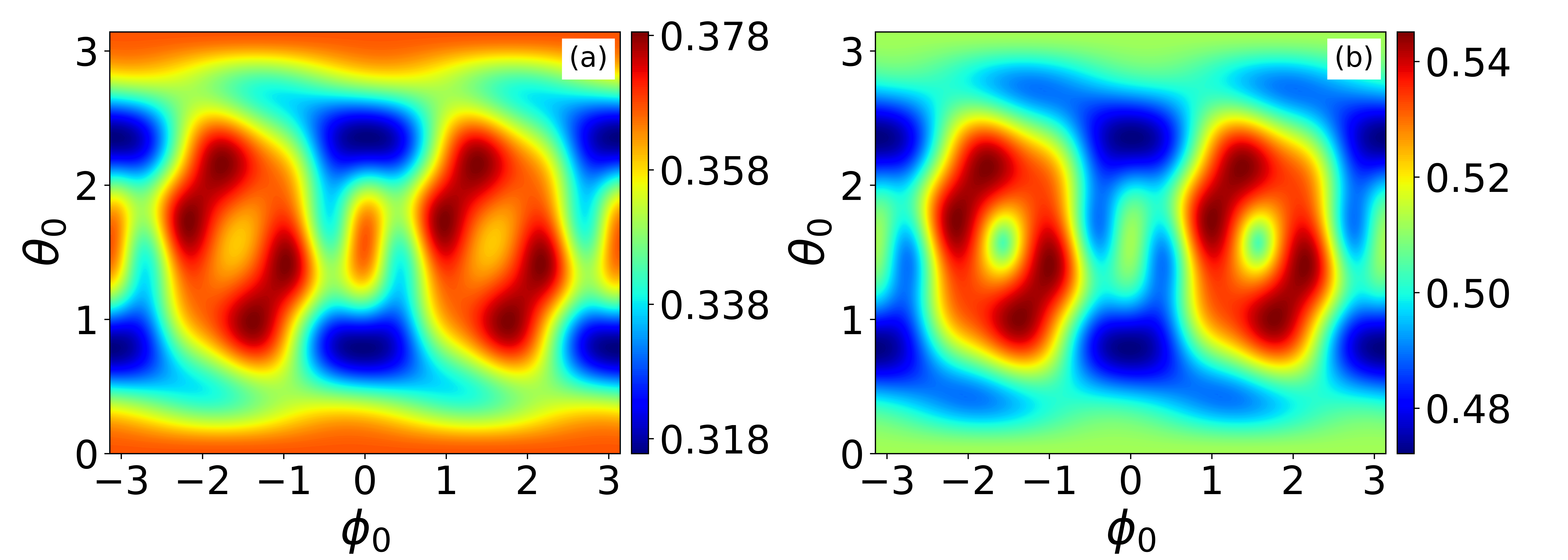}
\caption{ Contour plot of time-averaged values of (a) linear entropy and (b) entanglement entropy for any arbitrary  initial states $\ket{\theta_{0},\phi{_0}}$ for $6$ qubits.}
\label{fig:6.2qubitavg}
\end{figure}

In $\ket{\phi}$ basis, the unitary operator $\mathcal{U}$ can be expressed in two blocks $\mathcal{U}_{+}$ and $\mathcal{U}_{-}$ \cite{sharma2024signatures}. The two blocks are given as follows:
\begin{equation}
\mathcal{U_+}= \frac{-e^{\frac{i \pi }{8}}}{2\sqrt{2}} \left(
\begin{array}{cccc}
 0 &  \sqrt{3} & 0 & \sqrt{5} \\
 \sqrt{3}~e^{\frac{i \pi }{4}} & 0 & \sqrt{5}~e^{\frac{i \pi }{4}} & 0 \\
 0 &  \sqrt{5} & 0 & - \sqrt{3} \\
 -\sqrt{5}~e^{\frac{i \pi }{4}} & 0 & \sqrt{3}~e^{\frac{i \pi }{4}}  & 0 \\
\end{array}
\right)
 \end{equation}
 \begin{equation}
\mbox{and}~~\mathcal{U_-}= \frac{ e^{\frac{i \pi }{8}}}{4} \left(
\begin{array}{ccc}
 1 & 0 &  \sqrt{15} \\
 0 & 4~e^{\frac{i \pi }{4}} & 0 \\
  \sqrt{15} & 0 & -1 \\
\end{array}
\right).
 \end{equation}
Applying the unitary operator $\mathcal{U}$ $n$ times on the state $\ket{\psi}$ we get,
\begin{eqnarray}
\ket{\psi_n}&=&\mathcal{U}^n\ket{\psi}\\ \nonumber
&=& \bar{g}_{1n}\ket{\phi_0^+}+\bar{g}_{2n}\ket{\phi_1^+}+\bar{g}_{3n}\ket{\phi_2^+}+\bar{g}_{4n}\ket{\phi_3^+}\\ \nonumber&& +\bar{g}_{5n}\ket{\phi_0^-}+\bar{g}_{6n}\ket{\phi_1^-}+\bar{g}_{7n}\ket{\phi_2^-},\nonumber
\end{eqnarray}
where the coefficients are given as follows:
\begin{equation}\label{Eq:qkt56}
\bar{g}_{jn}=\sum_{q=1}^{\frac{N+2}{2}}\mathcal{U}^n_{j,q}~ a_q+\sum_{q=\frac{N+4}{2}}^{N+1}\mathcal{U}^n_{j,q}~ b_{q-\frac{N+2}{2}},1\leq j\leq N+1.
\end{equation}
The expressions of the coefficients $\bar{g}_{jn}$ can be calculated using Eqs.  (\ref{Eq:arbitaray}), (\ref{Eq:arbitaray1}), (\ref{Eq:arbitaray4}) and (\ref{Eq:qkt56}) for $6$ qubits. The detailed calculations regarding the $n${th} time evolution of $\mathcal{U}$ and  the coefficient $\bar{g}_{jn}$ are provided in the supplemental material \cite{supplementry2023}.
The single qubit RDM, $\rho_1(n)$, is given as,
\begin{equation}
\rho_1(n)=\frac{1}{2}\left(
\begin{array}{cc}
 \bar{y}_n & \bar{m}_n \\
\bar{m}_n^* & 2-\bar{y}_n \\
\end{array}
\right),
\end{equation}
where the coefficient $\bar{y}_n$ and $\bar{m}_n$ are given as follows:
\begin{eqnarray} \nonumber
 \bar{y}_n&=&\frac{1}{2}+\bar{g}_{1n}\bar{g}_{5n}^*+\bar{g}_{5n}\bar{g}_{1n}^*+\frac{2}{3}\left(\bar{g}_{6n}\bar{g}_{2n}^*+\bar{g}_{2n}\bar{g}_{6n}^*\right)\\ \nonumber&&+\frac{1}{3}\left(\bar{g}_{7n}\bar{g}_{3n}^*+\bar{g}_{3n}\bar{g}_{7n}^*\right)~~~ \mbox{and} \\ \nonumber
 \bar{m}_n&=&\left[{\sqrt{3}}\left(\left(\bar{g}_{1n}+\bar{g}_{5n}\right)\left(\bar{g}_{2n}^*+\bar{g}_{6n}^*\right)+\left(\bar{g}_{2n}-\bar{g}_{6n}\right)\right.\right.\\ \nonumber &&\left.\left.\left(\bar{g}_{5n}^*-\bar{g}_{1n}^*\right)\right)+{\sqrt{5}}\left(\left(\bar{g}_{2n}+\bar{g}_{6n}\right)\left(\bar{g}_{3n}^*+\bar{g}_{7n}^*\right)+\right.\right.\\ \nonumber &&\left.\left.\left(\bar{g}_{7n}-\bar{g}_{3n}\right)\left(\bar{g}_{2n}^*-\bar{g}_{6n}^*\right)\right)+2\sqrt{3}\left(\bar{g}_{3n}+\bar{g}_{7n}\right)\bar{g}_{4n}^*\right.\\ \nonumber &&\left.+\left(-\bar{g}_{3n}^*+\bar{g}_{7n}^*\right)\bar{g}_{4n}\right]\Big{/}{3\sqrt{2}}
\end{eqnarray}
The eigenvalues of single qubit RDM, $\rho_1(n)$, are $\frac{1}{2} \left(1\pm\sqrt{1-\bar{y}_n\left(2-\bar{y}_n\right)-|\bar{m}_n|^2}\right)$. The linear entropy of single qubit RDM is given as follows:
\begin{equation}
 S_{(\theta_0,\phi_0)}^{(6)}(n,1/2)=[\bar{y}_n(2-\bar{y}_n)-|{\bar{m}_n}|^2]/2.
\end{equation}
The eigenvalues of $\mathcal{U}$ for the case $J=1/2$ and $\tau=\pi/4$ are $\left\lbrace \pm(-1)^{1/4}, \pm(-1)^{3/4},(-1)^{3/8}, \pm(-1)^{1/8}\right\rbrace$, which implies that $\mathcal{U}^{16}=I$. We find that  the entanglement dynamics are periodic in nature having period $8$ for arbitrary initial coherent state, except for the initial state $\ket{\pi/2,\pm\pi/2}$, where the period $4$. Which is shown in Fig. \ref{fig:4.25qubitavg}~(b) for various initial states. Thus, the time-averaged linear entropy
for an arbitrary initial coherent states, is given  as follows:
\begin{widetext}
\begin{eqnarray}\nonumber
 \langle S_{(\theta_0,\phi_0)}^{(6)}\rangle&=&\left[94703502-664344 \cos\left(2 \theta_0\right)+3016975 \cos\left(4 \theta_0\right)+1333540 \cos\left(6 \theta_0\right)+74370 \cos\left(8
\theta_0\right)+91636 \cos\left(10 \theta_0\right)+  \right.\\ \nonumber && \left.10465\cos\left(12 \theta_0\right)+8 (53222+1607230 \cos\left(2 \theta_0\right)+1942680 \cos\left(4 \theta_0\right)+679195 \cos\left(6
\theta_0\right)+38530 \cos\left(8 \theta_0\right)+ \right.\\ \nonumber && \left.4519 \cos\left(10 \theta_0\right)) \cos\left(2 \phi_0 \right) \sin^2\left(\theta_0\right)-8 (233363+343624 \cos\left(2 \theta
\right)+214748 \cos\left(4 \theta_0\right)+23928 \cos\left(6 \theta_0\right)+3537   \right.\\ \nonumber && \left.\cos\left(8 \theta_0\right))\cos\left(4 \phi_0 \right) \sin^4\left(\theta_0\right)+64 (67002+42365
\cos\left(2 \theta_0\right)+20406 \cos\left(4 \theta_0\right)-749 \cos\left(6 \theta_0\right)) \cos\left(6 \phi_0 \right) \sin^6\left(\theta_0\right)  \right.\\ \nonumber && \left.+256(6333+2436 \cos\left(2
\theta_0\right)+511 \cos\left(4 \theta_0\right)) \cos\left(8 \phi_0 \right) \sin^8\left(\theta_0\right)+1024 (69-97 \cos\left(2 \theta_0\right)) \cos\left(10 \phi_0 \right) \sin^{10}\left(\theta
\right)+  \right.\\ \nonumber && \left.17408 \cos\left(12 \phi_0 \right)\sin^{12}\left(\theta_0\right)-64 \cos\left(\theta_0\right) (110947+39176 \cos\left(2 \theta_0\right)+80284  \cos\left(4 \theta_0\right)-18760
\cos\left(6 \theta_0\right)+1345 \right.\\ \nonumber && \left.\cos\left(8 \theta_0\right)) \sin^{2}\left(\theta_0\right)  \sin\left(2 \phi_0 \right)+12288 \cos\left(\theta_0\right) (3+\cos\left(2 \theta_0\right)) (133+196
\cos\left(2 \theta_0\right)-9 \cos\left(4 \theta_0\right)) \sin^{4}\left(\theta_0\right) \sin\left(4 \phi_0 \right)\right.\\ \nonumber && \left.-32256 \cos\left(\theta_0\right) (131+124 \cos\left(2 \theta_0\right)+\cos\left(4
\theta_0\right)) \sin^{6}\left(\theta_0\right)  \sin\left(6 \phi_0 \right)+245760 \cos\left(\theta_0\right) (3+\cos\left(2 \theta_0\right)) \sin^8\left(\theta_0\right)\right.\\  && \left.\sin\left(8 \phi_0
\right)-53248  \cos\left(\theta_0\right) \sin^{10}\left(\theta_0\right) \sin\left(10 \phi_0 \right)\right]\Big{/}{268435456}.
\end{eqnarray}
\end{widetext}
The  time-averaged linear entropy lies within a small interval $[0.31578, 0.37855]$ and is shown in Fig. \ref{fig:6.2qubitavg}~(a). It attains the   maximum value for  eight initial states as follows:  $\ket{1.730825,-2.1632874}$, $\ket{2.154710,1.37862}$, $\ket{1.410768,2.163287}$, $\ket{2.154710,-1.762968}$, $\ket{0.986882,-1.378624}$, $\ket{1.410768,-0.978305}$, $\ket{1.730824,0.978305}$ and $\ket{0.986882,1.762968}$, while  minimum value  corresponds to  the states $\ket{\pi/4,-0.0262045}$, $\ket{3\pi/4,0.0262045}$, $\ket{3\pi/4,-3.115387}$ and $\ket{\pi/4,3.115387}$. Which can be observe from same figure. We numerically obtaine the average values of concurrence for  arbitrary initial state and find that it is  maximum of $6.68085503\times{10^{-2}}$ for the states $\ket{2.764015,1.46084}$ and $\ket{2.764015,-1.680752}$, $\ket{0.376911,1.680752}$ and $\ket{0.376991,-1.46084}$. The minimum value of $1.6581305\times{10^{-2}}$  for the states $\ket{1.24092,1.3744467}$ and $\ket{1.24092,-1.76715}$, $\ket{1.9,-1.3744467}$ and $\ket{1.9,1.767145}$. These numerical results are presented in  Fig.  \ref{fig:4.26qubitavg}~(b).
\subsection{Exact solution for $8$ qubits}
The unitary operator $ \mathcal{U}$ can be written in two blocks $\mathcal{U}_{+}~(\mathcal{U}_{-})$ in $\ket{\phi}$ basis \cite{sharma2024signatures},  as follows:
\begin{equation}
\mathcal{U}_{+}=\frac{1}{8}\left(
\begin{array}{ccccc}
 i & 0 & 2~i \sqrt{7} & 0 & i~ \sqrt{35} \\
 0 & -6 ~e^{\frac{i \pi }{4}} & 0 & -2 \sqrt{7}~ e^{\frac{i \pi }{4}} & 0 \\
 -2~i \sqrt{7} & 0 & -4~i & 0 & 2~i \sqrt{5} \\
 0 & -2 \sqrt{7}~ e^{\frac{i \pi }{4}} & 0 & 6~ e^{\frac{i \pi }{4}} & 0 \\
 i~ \sqrt{35} & 0 & -2~i \sqrt{5} & 0 & 3 ~i \\
\end{array}
\right)
\end{equation}
\begin{equation}
\mbox{and}~~\mathcal{U}_{-}=\frac{-1}{2\sqrt{2}}\left(
\begin{array}{cccc}
 0 & i & 0 &  i ~\sqrt{7} \\
 -e^{\frac{i \pi }{4}} & 0 & -\sqrt{7}~ e^{\frac{i \pi }{4}} & 0 \\
 0 & - i ~\sqrt{7} & 0 & i \\
 - \sqrt{7}~ e^{\frac{i \pi }{4}} & 0 & e^{\frac{i \pi }{4}} & 0 \\
\end{array}
\right).
\end{equation}
The state $\ket{\psi_n}$ can be calculated by  applying  unitary operator $\mathcal{U}$ $n$ times  on the state $\ket{\psi}$ we get,
\begin{eqnarray}
\ket{\psi_n}&=&\mathcal{U}^n\ket{\psi_0}\\ \nonumber
&=& \bar{f}_{1n}\ket{\phi_0^+}+\bar{f}_{2n}\ket{\phi_1^+}+\bar{f}_{3n}\ket{\phi_2^+}+\bar{f}_{4n}\ket{\phi_3^+} +\bar{f}_{5n}\ket{\phi_4^+}\\ \nonumber&&+\bar{f}_{6n}\ket{\phi_0^-}+\bar{f}_{7n}\ket{\phi_1^-}+\bar{f}_{8n}\ket{\phi_2^-}+\bar{f}_{8n}\ket{\phi_3^-},\nonumber
\end{eqnarray}
where the coefficients are  given as follows:
\begin{equation}\label{Eq:qkt57}
\bar{f}_{jn}=\sum_{q=1}^{\frac{N+2}{2}}\mathcal{U}^n_{j,q}~ a_q+\sum_{q=\frac{N+4}{2}}^{N+1}\mathcal{U}^n_{j,q}~ b_{q-\frac{N+2}{2}},1\leq j\leq N+1.
\end{equation}
The expressions of the coefficients $\bar{f}_{jn}$ can be calculated using Eqs. (\ref{Eq:arbitaray}), (\ref{Eq:arbitaray1}), (\ref{Eq:arbitaray4}) and (\ref{Eq:qkt57}) for $8$ qubits. The detailed calculations  regarding the $n${th} time evolution of $\mathcal{U}$ and  the coefficient $\bar{f}_{jn}$ are provided in the supplemental material \cite{supplementry2023}.
\begin{figure}[t]\vspace{0.4cm}
\includegraphics[width=0.47\textwidth,height=0.15\textheight]{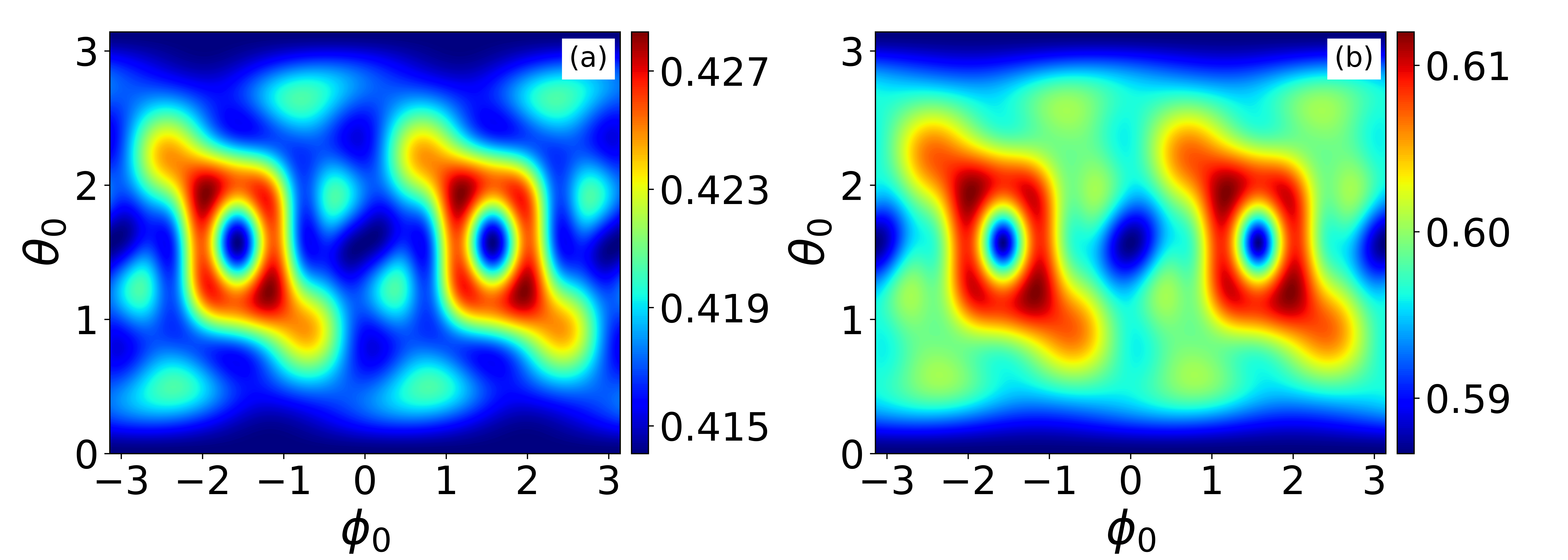}
\caption{Contour plot of time-averaged values of (a) linear entropy and (b) entanglement entropy for any arbitrary  initial states $\ket{\theta_{0},\phi{_0}}$ for $8$ qubits.}
\label{fig:8.2qubitavg}
\end{figure}
The single qubit RDM, $\rho_1(n)$, is given as,
\begin{equation}
\rho_1(n)=\frac{1}{2}\left(
\begin{array}{cc}
 l_n & \bar{o}_n \\
\bar{ o}_n^* & 2-l_n \\
\end{array}
\right),
\end{equation}
where the coefficient ${l}_n$ and $\bar{o}_n$ can be computed as follows:
\begin{eqnarray} \nonumber
 l_n&=&\frac{1}{2}+\bar{f}_{1n}\bar{f}_{6n}^*+\bar{f}_{6n}\bar{f}_{1n}^*+\frac{3}{4}\left(\bar{f}_{7n}\bar{f}_{2n}^*+\bar{f}_{2n}\bar{f}_{7n}^*\right)+\\ \nonumber&&\frac{1}{2}\left(\bar{f}_{8n}\bar{f}_{3n}^*+\bar{f}_{3n}\bar{f}_{8n}^*\right)+\frac{1}{4}\left(\bar{f}_{9n}\bar{f}_{4n}^*+\bar{f}_{4n}\bar{f}_{9n}^*\right)~ \mbox{and}\\ \nonumber
 \bar{o}_n&=&\left[2\left(\left(\bar{f}_{1n}+\bar{f}_{6n}\right)\left(\bar{f}_{2n}^*+\bar{f}_{7n}^*\right)+\left(\bar{f}_{7n}-\bar{f}_{2n}\right)\right.\right.\\ \nonumber &&\left.\left.\left(\bar{f}_{1n}^*-\bar{f}_{6n}^*\right)\right)+\sqrt{7}\left(\left(\bar{f}_{2n}+\bar{f}_{7n}\right)\left(\bar{f}_{3n}^*+\bar{f}_{8n}^*\right)+\right.\right.\\ \nonumber &&\left.\left.\left(\bar{f}_{3n}-\bar{f}_{8n}\right)\left(\bar{f}_{7n}^*-\bar{f}_{2n}^*\right)\right)+{3}\left(\left(\bar{f}_{3n}+\bar{f}_{8n}\right)\right.\right.\\ \nonumber &&\left.\left.\left(\bar{f}_{4n}^*+\bar{f}_{9n}^*\right)+\left(\bar{f}_{9n}-\bar{f}_{4n}\right)\left(\bar{f}_{3n}^*-\bar{f}_{8n}^*\right)\right)+2\sqrt{7} \right.\\ \nonumber &&\left.\left(\bar{f}_{4n}+\bar{f}_{9n}\right)\bar{f}_{5n}^*+\left(-\bar{f}_{4n}^*+\bar{f}_{9n}^*\right)\bar{f}_{5n}\right]\Big{/}{4\sqrt{2}}.
\end{eqnarray}
The eigenvalues of $\rho_1(n)$ are $\frac{1}{2} \left(1\pm\sqrt{1-{l}_n\left(2-{l}_n\right)-|\bar{o}_n}|^2\right)$. The linear entropy of single qubit  is given as follows:
\begin{equation}
 S_{(\theta_0,\phi_0)}^{(8)}(n,1/2)=[l_n(2-l_n)-|{ \bar{o}_n}|^2]/2.
\end{equation}
The eigenvalues of $\mathcal{U}$ for the case $J=1/2$ and $\tau=\pi/4$ are $\left\lbrace i,\exp({\frac{7i \pi }{6}}), \exp({\frac{- i \pi }{6}}), \exp({\frac{5 i \pi }{4}}), \exp({\frac{ i \pi }{4}}), \exp({\frac{7i \pi }{8}}), \exp({\frac{-i \pi }{8}})\right.\\ \left., \exp({\frac{-5i \pi }{8}}), \exp({\frac{3i \pi }{8}})\right\rbrace$, which implies that $\mathcal{U}^{48}=I$. We find that  the entanglement dynamics are periodic in nature having period $24$  for any arbitrary initial coherent states, except for the initial state $\ket{\pi/2,\pm\pi/2}$, where the period $12$. Which is shown in Fig. \ref{fig:4.25qubitavg}~(c) for various initial states. Thus, the time-averaged linear entropy for an arbitrary initial coherent states, is given  as follows:
\begin{widetext}
\begin{eqnarray}\nonumber
\langle S_{(\theta_0,\phi_0)}^{(8)}\rangle&=&\left[172521535907-1076481488 \cos\left(2\theta_0 \right)-483908600 \cos\left(4\theta_0 \right)+204246672 \cos\left(6\theta_0
\right)-519456868 \cos\left(8\theta_0 \right)\right.\\ &&\nonumber \left.+50430128 \cos\left(10\theta_0 \right)+23220152 \cos\left(12\theta_0 \right)+
4554384 \cos\left(14\theta_0 \right)+809729 \cos\left(16\theta_0 \right)+\cos\left(2 \phi_0 \right) \sin^2\left(\theta_0 \right)\right.\\ &&\nonumber \left.\left\lbrace \left(-704643072  -11509170176 \sqrt{2}+1409286144 \cos\left(2\theta_0 \right)+23018340352 \sqrt{2} \cos\left(2\theta_0 \right)+1526726656  \cos\left(4\theta_0 \right)\right.\right.\right.\\ &&\nonumber \left.\left. \left.
-117440512 \sqrt{2}  \cos\left(4\theta_0 \right)\right) \cos\left(\theta_0 \right)+
4298888426 \cos\left(2\theta_0 \right) -
\left(1468006400  +11450449920 \sqrt{2}\right) \cos\left(3\theta_0 \right) +\right.\right.\\ &&\nonumber \left.\left. 2239395284\cos\left(4\theta_0 \right) -763363328 \cos\left(5\theta_0 \right) +58720256 \sqrt{2} \cos\left(5\theta_0 \right) +1005475086 \cos\left(6\theta_0 \right) +
1147266680 \right.\right.\\ &&\nonumber \left.\left.\cos\left(8\theta_0 \right) +391988842 \cos\left(10\theta_0 \right) +14079212 \cos\left(12\theta_0 \right) +12190 \cos\left(14\theta_0 \right) \right\rbrace +
\cos\left(4 \phi_0 \right) \sin^4\left(\theta_0 \right)\right.\\ &&\nonumber \left.\left[-5383084896 -7342837376 \cos\left(2\theta_0 \right) -1820455728 \cos\left(4
\theta_0 \right) +874776000 \cos\left(6\theta_0 \right) +
53986912 \cos\left(8\theta_0 \right)-\right.\right.\\ &&\nonumber \left.\left.8714048 \cos\left(10\theta_0 \right)+3229744 \cos\left(12\theta_0 \right) \right]+\cos\left(6 \phi_0 \right) \sin^6\left(\theta_0 \right)\left[221898944 -
560204352 \cos\left(2\theta_0 \right) + \right.\right.\\ &&\nonumber \left.\left. 739468032\cos\left(4\theta_0 \right) -197632416\cos\left(6\theta_0 \right) -25193408 \cos\left(8\theta_0 \right) -
2176032 \cos\left(10\theta_0 \right)\right]+\cos\left(6 \phi_0 \right) \sin^8\left(\theta_0 \right)\right.\\ &&\nonumber \left.\left[-3043660032 -3938801664 \cos\left(2
\theta_0 \right) -1292956672 \cos\left(4\theta_0 \right) +
123475968 \cos\left(6\theta_0 \right) -10173184 \cos\left(8\theta_0 \right)\right]\right.\\ &&\nonumber \left. +\cos\left(8 \phi_0 \right) \sin^8\left(\theta_0 \right)\left[-223133696 +959709184\cos\left(2\theta_0 \right) +
165609472 \cos\left(4\theta_0 \right) +18087936 \cos\left(6\theta_0 \right) +\right.\right.\\ &&\nonumber \left.\left.10862592 \cos\left(8\theta_0 \right) \right] +\sin^{10}\left(\theta_0 \right)\left[-106975232 \cos\left(8 \phi_0 \right) +
69056512 \cos\left(2\theta_0 \right) \cos\left(8 \phi_0 \right) +76926976\cos\left(4\theta_0 \right) \right.\right.\\ &&\nonumber \left. \left. \cos\left(8 \phi_0 \right) -24328192 \cos\left(6\theta_0 \right) \cos\left(8 \phi_0 \right) +149875712 \cos\left(10 \phi_0 \right)-
156817920 \cos\left(2\theta_0 \right)\cos\left(10 \phi_0 \right) +\right.\right.\\ &&\nonumber \left.\left.26457088 \cos\left(4\theta_0 \right) \cos\left(10 \phi_0 \right)  -8504832 \cos\left(6\theta_0 \right) \cos\left(10 \phi_0 \right) \right]+\sin^{12}\left(\theta \right)\left[120508416 \cos\left(10 \phi_0 \right) +307249152
\right.\right.\\ &&\nonumber \left.\left. \cos\left(2\theta_0 \right) \cos\left(10 \phi_0 \right) +1634304 \cos\left(4\theta_0 \right) \cos\left(10 \phi_0 \right) +62222336 \cos\left(12 \phi_0 \right) -56508416 \cos\left(2\theta_0 \right) \cos\left(12 \phi_0 \right) +
\right.\right.\\ &&\nonumber \left.\left.14471168 \cos\left(4\theta_0 \right) \cos\left(12 \phi_0 \right)\right]+\cos\left(14 \phi_0 \right) \sin^{14}\left(\theta_0 \right)\left(4104192 -1646592 \cos\left(2
\theta_0 \right) \right)-\left(622592 \cos\left(14 \phi_0 \right) -\right.\right.\\ &&\nonumber \left.\left.1081344 \cos\left(16 \phi_0 \right)\right) \sin^{16}\left(\theta_0
\right)+\cos\left(\theta_0 \right) \sin^2\left(\theta_0 \right) \sin\left(2 \phi_0 \right)\left[
1328256288 +4967799168  \cos\left(2\theta_0 \right) +4227791376 \right. \right.\\ &&\nonumber \left.\left.\cos\left(4\theta_0 \right) -118115904  \cos\left(6\theta_0 \right) -
307365408  \cos\left(8\theta_0 \right) -32246592  \cos\left(10\theta_0
\right) +210672  \cos\left(12\theta_0 \right) \right]+\cos\left(\theta_0 \right)  \right.\\ &&\nonumber \left.\sin^4\left(\theta_0 \right) \sin\left(4 \phi_0 \right)\left[-244016640
-3233395200 \cos\left(2\theta_0 \right) -1050685440  \cos^2\left(2\theta_0
\right) +448874496  \cos\left(4\theta_0 \right)+
  \right.\right.\\ &&\nonumber \left.\left.149624832 \cos\left(2\theta_0 \right) \cos\left(4\theta_0 \right) -252112896  \cos\left(6\theta_0 \right) -84037632  \cos\left(2\theta_0 \right) \cos\left(6\theta_0 \right) +
28417536 \cos\left(\theta_0 \right)\right.\right.\\ &&\nonumber \left.\left. \cos\left(8\theta_0 \right)+9472512  \cos\left(2\theta_0
\right) \cos\left(8\theta_0 \right) \right]+\sin^6\left(\theta_0 \right) \sin\left(6 \phi_0 \right)\left[624576768 \cos\left(\theta_0 \right) -567144960
\cos\left(3\theta_0 \right)  \right.\right.\\ &&\nonumber \left.\left.+
29308416 \cos\left(5\theta_0 \right)+81726336 \cos\left(7\theta_0 \right) +7694208 \cos\left(9\theta_0 \right) \right]+\cos\left(\theta_0 \right) \sin^8\left(\theta_0 \right) \sin\left(8 \phi_0 \right)\left[381911040 +\right.\right.\\ &&\nonumber \left.\left.432832512  \cos\left(2\theta_0 \right) +101842944 \cos^2\left(2\theta_0
\right) +10911744  \cos\left(4\theta_0 \right) +
3637248  \cos\left(2\theta_0 \right) \cos\left(4\theta_0 \right) \right]+\sin^{10} \left(\theta_0 \right) \right.\\ &&\nonumber \left.\sin\left(10 \phi_0 \right)\left[-24858624 \cos\left(\theta_0
\right) +52426752\cos\left(3\theta_0 \right) -16558080 \cos\left(5
\theta_0 \right) \right]+\sin^{12}\left(\theta_0 \right) \sin\left(12 \phi_0 \right)\left[
-229376 -\right.\right.\\ \nonumber && \left.\left.\cos\left(\frac{\theta_0 }{48}\right) -12959744 \cos\left(\theta_0 \right) 2523136 \cos\left(\frac{\theta_0 }{48}\right) \cos\left(2\theta_0 \right) -802816
\cos\left(3\theta_0 \right) \right]+
2457600 \cos\left(\theta_0 \right) \sin^{14}\left(\theta_0 \right) \right.\\ && \left.\sin\left(14 \phi_0 \right)\right]/{412316860416}.
\end{eqnarray}
\end{widetext}
The time-averaged linear entropy takes values from the  narrow interval $[0.41406, 0.4284]$ and shown in Fig. \ref{fig:8.2qubitavg}~(a). The maximum value corresponds to initial states:  $\ket{1.934486,1.181507}$ and $\ket{1.934486,-1.961442}$, while the minimum value for the states $\ket{1.553926,3.100515}$, $\ket{1.553926,-0.041063}$, which can be observe from same figure. We numerically obtain the average values of concurrence for any arbitrary initial state and plotted in Fig. \ref{fig:4.26qubitavg}~(c). The average value of concurrence is maximum of $1.50011786\times10^{-2}$ for the states  $\ket{1.29591,-0.01571}$, $\ket{1.8456885,-3.125885}$, $\ket{1.29591,3.125885}$ and $\ket{1.8456885,0.01571}$, while it is minimum   of $1.73226016\times10^{-3}$  for the states $\ket{1.829977,1.303761}$, $\ket{1.311615,-1.30371}$, $\ket{1.3116145,1.837832}$ and $\ket{1.829977,-1.837832}$ .
\subsection{Exact solution for $10$ qubit}
In this set of basis, the unitary operator is block diagonalized in two blocks $\mathcal{U}_{+}~(\mathcal{U}_{-})$ having dimension $6\times6~(5\times5)$ \cite{sharma2024signatures}. The blocks are given as follows:
\begin{widetext}
\begin{equation}
\mathcal{U}_{+}=-\frac{e^{\frac{3i \pi }{8}}}{8\sqrt{2}} \left(
\begin{array}{cccccc}
 0 &  \sqrt{5} & 0 &  2\sqrt{15} & 0 & 3\text{  }\sqrt{7} \\
 -e^{\frac{i \pi }{4}} \sqrt{5} & 0 & -9 ~e^{\frac{i \pi }{4}} & 0 & -e^{\frac{i \pi }{4}} \sqrt{42} & 0 \\
 0 & 9  & 0 & 2\text{  }\sqrt{3} & 0 & - \sqrt{35} \\
 2~ e^{\frac{i \pi }{4}}\sqrt{15} & 0 & 2~ e^{\frac{i \pi }{4}}\sqrt{3} & 0 & 2~ e^{\frac{i \pi }{4}} \sqrt{14} & 0 \\
 0 &  \sqrt{42} & 0 & - 2\sqrt{14} & 0 &  \sqrt{30} \\
 -3~ e^{\frac{i \pi }{4}} \sqrt{7} & 0 & e^{\frac{i \pi }{4}} \sqrt{35} & 0 & -e^{\frac{i \pi }{4}} \sqrt{30} & 0 \\
\end{array}
\right)~~~\mbox{and}
\end{equation}
\end{widetext}
\begin{equation}
\mathcal{U}_{-}=\frac{e^{\frac{3i \pi }{8}}}{16} \left(
\begin{array}{ccccc}
 1 & 0 & 3\text{  }\sqrt{5} & 0 & \sqrt{210} \\
 0 & -8~ e^{\frac{i \pi }{4}} & 0 & -8 \sqrt{3}~ e^{\frac{i \pi }{4}}  & 0 \\
 3\text{  }\sqrt{5} & 0 & 13  & 0 & - \sqrt{42} \\
 0 & 8\sqrt{3}~ e^{\frac{i \pi }{4}}  & 0 & -8~ e^{\frac{i \pi }{4}} & 0 \\
  \sqrt{210} & 0 & -\sqrt{42} & 0 & 2 \\
\end{array}
\right).
\end{equation}
The initial state $\ket{\psi}$ after the $n${th} implementations of the unitary operator $\mathcal{U}$ can be expressed as follows:
\begin{eqnarray}
\ket{\psi_n}&=&\mathcal{U}^n\ket{\psi}\\ \nonumber
&=& \bar{d}_{1n}\ket{\phi_0^+}+\bar{d}_{2n}\ket{\phi_1^+}+\bar{d}_{3n}\ket{\phi_2^+}+\bar{d}_{4n}\ket{\phi_3^+}\\ \nonumber&& +\bar{d}_{5n}\ket{\phi_4^+}+\bar{d}_{6n}\ket{\phi_5^+}+\bar{d}_{7n}\ket{\phi_0^-}+\bar{d}_{8n}\ket{\phi_1^-}\\ \nonumber&&+\bar{d}_{9n}\ket{\phi_2^-}+\bar{d}_{10n}\ket{\phi_3^-}+\bar{d}_{11n}\ket{\phi_4^-},\nonumber
\end{eqnarray}
where the coefficients are given as follows:
\begin{equation}\label{Eq:qkt58}
\bar{d}_{jn}=\sum_{q=1}^{\frac{N+2}{2}}\mathcal{U}^n_{j,q}~ a_q+\sum_{q=\frac{N+4}{2}}^{N+1}\mathcal{U}^n_{j,q}~ b_{q-\frac{N+2}{2}},1\leq j\leq N+1.
\end{equation}
The expressions of the coefficients $\bar{d}_{jn}$ can be calculated using  Eqs.  (\ref{Eq:arbitaray}), (\ref{Eq:arbitaray1}), (\ref{Eq:arbitaray4}) and (\ref{Eq:qkt58}) for $10$ qubits. The detailed calculations  regarding the $n${th} time evolution of $\mathcal{U}$ and  the coefficient $\bar{d}_{jn}$ are provided in the supplemental material \cite{supplementry2023}.
The single qubit RDM, $\rho_1(n)$, is given as,
\begin{equation}
\rho_1(n)=\frac{1}{2}\left(
\begin{array}{cc}
 \bar{l}_n & o_n \\
o_n^* & 2-\bar{l}_n \\
\end{array}
\right),
\end{equation}
where the coefficient $\bar{l}_n$ and ${o}_n$ are given as follows:
\begin{eqnarray} \nonumber
 \bar{l}_n&=&\frac{1}{2}+\bar{d}_{1n}\bar{d}_{7n}^*+\bar{d}_{7n}\bar{d}_{1n}^*+\frac{1}{5}\left[4\left(\bar{d}_{8n}\bar{d}_{2n}^*+\bar{d}_{2n}\bar{d}_{8n}^*\right)\right.\\ \nonumber &&\left.+{3}\left(\bar{d}_{9n}\bar{d}_{3n}^*+\bar{d}_{3n}\bar{d}_{9n}^*\right)+{2}\left(\bar{d}_{4n}\bar{d}_{10n}^*+\bar{d}_{10n}\bar{d}_{4n}^*\right)\right.\\ \nonumber &&\left.+\left(\bar{d}_{11n}\bar{d}_{5n}^*+\bar{d}_{5n}\bar{d}_{11n}^*\right)\right]~~ \mbox{and} \\ \nonumber
 \end{eqnarray}
 \begin{eqnarray} \nonumber
 o_n&=&\left[\sqrt{2}\left(\left(\bar{d}_{1n}+\bar{d}_{7n}\right)\left(\bar{d}_{2n}^*+\bar{d}_{8n}^*\right)+\left(\bar{d}_{2n}-\bar{d}_{8n}\right)\right.\right.\\ \nonumber &&\left.\left.\left(\bar{d}_{7n}^*-\bar{d}_{1n}^*\right)\right)+3\left(\left(\bar{d}_{2n}+\bar{d}_{8n}\right)\left(\bar{d}_{3n}^*+\bar{d}_{9n}^*\right)+\right.\right.\\ \nonumber &&\left.\left.\left(\bar{d}_{9n}-\bar{d}_{3n}\right)\left(\bar{d}_{2n}^*-\bar{d}_{8n}^*\right)\right)+2\sqrt{3}\left(\left(\bar{d}_{3n}+\bar{d}_{9n}\right)\right.\right.\\ \nonumber &&\left.\left.\left(\bar{d}_{4n}^*+\bar{d}_{10n}^*\right)+\left(\bar{d}_{4n}-\bar{d}_{10n}\right)\left(\bar{d}_{9n}^*-\bar{d}_{3n}^*\right)\right)+ \right.\\ \nonumber &&\left.\sqrt{14}\left(\left(\bar{d}_{4n}+\bar{d}_{10n}\right)\left(\bar{d}_{5n}^*+\bar{d}_{11n}^*\right)+\left(\bar{d}_{11n}-\bar{d}_{5n}\right)\right.\right.\\ \nonumber &&\left.\left.\left(\bar{d}_{4n}^*-\bar{d}_{10n}^*\right)\right) +\sqrt{30}\left(\bar{d}_{5n}+\bar{d}_{11n}\right)\bar{d}_{6n}^*+\right.\\ \nonumber &&\left.\left(-\bar{d}_{5n}^*+\bar{d}_{11n}^*\right)\bar{d}_{6n}\right]\Big{/}{5\sqrt{2}}.
\end{eqnarray}
The eigenvalues  of $\rho_1(n)$ are $\frac{1}{2} \left(1\pm\sqrt{1-\bar{l}_n\left(2-\bar{l}_n\right)-|{o}_n}|^2\right)$. The linear entropy of single qubit is given as follows:
\begin{equation}
 S_{(\theta_0,\phi_0)}^{(10)}(n,1/2)=[\bar{l}_n(2-\bar{l}_n)-|{o_n}|^2]/2.
\end{equation}
\begin{figure}[htbp!]\vspace{0.2cm}
\includegraphics[width=0.47\textwidth,height=0.15\textheight]{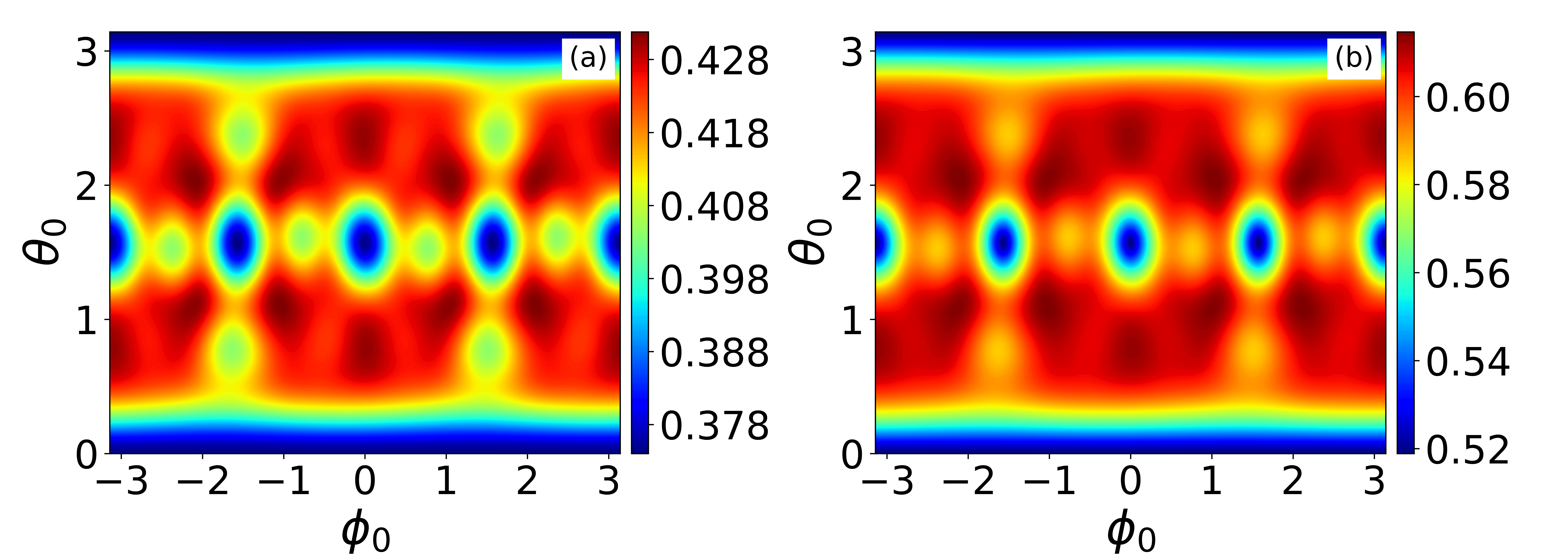}
\caption{Contour plot of time-averaged values of (a) linear entropy and (b) entanglement entropy for any arbitrary  initial states $\ket{\theta_{0},\phi{_0}}$ for $10$ qubits.}
\label{fig:10.2qubitavg}
\end{figure}
The eigenvalues of $\mathcal{U}$ for the case $J=1/2$ and $\tau=\pi/4$ are $\left\lbrace \pm1, \pm1, \pm i, \pm(-1)^{3/8},-(-1)^{23/24},(-1)^{3/8},-(-1)^{7/24}\right\rbrace$, which implies that $\mathcal{U}^{48}=I$. We find that  the entanglement dynamics are periodic in nature having period $24$  for any arbitrary initial coherent states, except for the initial state $\ket{\pi/2,\pm\pi/2}$, where the period $12$. Which is shown in Fig. \ref{fig:4.25qubitavg}~(d) for various initial states. Thus, the time-averaged linear entropy
for an arbitrary initial coherent states, is given  as follows:
\begin{widetext}
\begin{eqnarray}\nonumber
 \langle S_{(\theta_0,\phi_0)}^{(10)}\rangle&=&\left[115848776942454-4817176106370 \cos\left(4 \theta_0 \right)-740033340336 \cos\left(6 \theta_0 \right)-1927392828968 \cos\left(8 \theta_0 \right)+  \right.\\ \nonumber &&\left.31230215\cos\left(20
\theta_0 \right)-95285484165 \cos\left(12 \theta_0 \right)+402136020 \cos\left(14 \theta_0 \right)+1682227634 \cos\left(16 \theta_0 \right)+74611316 \right.\\ \nonumber &&\left.\cos\left(18 \theta_0 \right)-69994693616 \cos\left(10 \theta_0 \right)+4180631963664
\cos\left(2 \phi_0 \right) \sin^2\left(\theta_0 \right)+{8 }(-348175922645 \cos\left(2 \theta_0 \right)+\right.\\ \nonumber &&\left.(669867579870 \cos\left(2 \theta_0 \right)+3 (12404991056 \cos\left(4 \theta_0 \right)-3803296500 \cos\left(6 \theta_0 \right)+37127377656 \cos\left(8 \theta_0 \right)+\right.\\ \nonumber &&\left.17776621148
\cos\left(10 \theta_0 \right)-337165520 \cos\left(12 \theta_0 \right)+329527115 \cos\left(14 \theta_0 \right)+8228146 \cos\left(16 \theta_0 \right))+14044681 \right.\\ \nonumber &&\left.\cos\left(18 \theta_0 \right)) \cos\left(2 \phi_0 \right) \sin^2\left(\theta_0 \right)-
6 (144497518439+117479206160 \cos\left(2 \theta_0 \right)+46693202344 \cos\left(4 \theta_0 \right)-\right.\\ \nonumber &&\left. 11922823120 \cos\left(6 \theta_0 \right)+8270640364\cos\left(8
\theta_0 \right)+4519785872 \cos\left(10 \theta_0 \right)+761948952 \cos\left(12 \theta_0 \right)+10990896 \right.\\ \nonumber &&\left.\cos\left(14 \theta_0 \right)+917229 \cos\left(16 \theta_0 \right)) \cos\left(4 \phi_0 \right) \sin^4\left(\theta_0 \right)+96 (23979894604+33384711527 \cos\left(2 \theta_0 \right)+11393273102 \right.\\ \nonumber &&\left.\cos\left(4 \theta_0 \right)+1595029997 \cos\left(6 \theta_0 \right)+42380692 \cos\left(8 \theta_0
\right)-435801 \cos\left(10 \theta_0 \right)+1945106 \cos\left(12 \theta_0 \right)+399109 \right.\\ \nonumber &&\left.\cos\left(14 \theta_0 \right)) \cos^6\left(6 \phi_0 \right) \sin\left(\theta_0 \right)+128 (-6577422+2579017752 \cos\left(2 \theta_0 \right)+138806961 \cos\left(4 \theta_0 \right)+506044252  \right.\\ \nonumber &&\left.\cos\left(6
\theta_0 \right)+41317118 \cos\left(8 \theta_0 \right)-7105396\cos\left(10 \theta_0 \right)+1179487 \cos\left(12 \theta_0 \right))\cos\left(8 \phi_0 \right) \sin^8\left(\theta_0 \right)+512\right.\\ \nonumber &&\left. (564503774+733829718 \cos\left(2 \theta_0 \right)+206880888 \cos\left(4 \theta_0 \right)+17813079 \cos\left(6
\theta_0 \right)+766378 \cos\left(8 \theta_0 \right)+180307 \right.\\ \nonumber &&\left.\cos\left(10 \theta_0 \right)) \cos\left(10 \phi_0 \right) \sin^{10}\left(\theta_0 \right)+768 (112119405+148245944 \cos\left(2 \theta_0 \right)+45407524 \cos\left(4 \theta_0 \right)+3360392 \right.\\ \nonumber &&\left.\cos\left(6 \theta_0 \right)+393263 \cos\left(8 \theta_0 \right)) \cos\left(12
\phi_0 \right) \sin^{12}\left(\theta_0 \right)-6144 (498554+373309 \cos\left(2 \theta_0 \right)+91894 \cos\left(4 \theta_0 \right)+ \right.\\ \nonumber &&\left.15187\cos\left(6 \theta_0 \right)) \cos\left(14 \phi_0 \right) \sin^{14}\left(\theta_0 \right)+8192
(51255+50956 \cos\left(2 \theta_0 \right)+30077 \cos\left(4 \theta_0 \right)) \cos\left(16 \phi_0 \right) \sin^{16}\left(\theta_0 \right)\right.\\ \nonumber &&\left.+32768(-27+95 \cos\left(2 \theta_0 \right)) \cos\left(18 \phi_0 \right) \sin^{18}\left(\theta_0 \right)+8421376 \cos\left(20 \phi_0 \right) \sin^{20}\left(\theta_0 \right)-4 (89519598618 \cos\left(\theta_0 \right)+\right.\\ \nonumber &&\left.59204770728 \cos\left(3 \theta_0 \right)+37915640328 \cos\left(5 \theta_0 \right)+3680215764 \cos\left(7 \theta_0 \right)+4172266996
\cos\left(9 \theta_0 \right)-4272505992\right.\\ \nonumber &&\left. \cos\left(11 \theta_0 \right)+990421848 \cos\left(13 \theta_0 \right)-87026217 \cos\left(15 \theta_0 \right)+2662599  \cos\left(17 \theta_0 \right))\sin^{2}\left(\theta_0 \right) \sin\left(2 \phi_0 \right)+3072\right.\\ \nonumber &&\left. \cos\left(\theta_0 \right) (3+\cos\left(2 \theta_0 \right)) (36681846+55081352 \cos\left(2 \theta_0 \right)+5009923 \cos\left(4 \theta_0 \right) +7022388\cos\left(6
\theta_0 \right)+831322\right.\\ \nonumber &&\left. \cos\left(8 \theta_0 \right)+240452 \cos\left(10 \theta_0 \right)-9683 \cos\left(12 \theta_0 \right)) \sin^{4}\left(\theta_0 \right) \sin\left(4 \phi_0 \right)-384 (3759046020 \cos\left(\theta_0 \right)+2190634743 \right.\\ \nonumber &&\left.\cos\left(3 \theta_0 \right)+715229299 \cos\left(5 \theta_0 \right)+138377062 \cos\left(7 \theta_0 \right)-11543754
\cos\left(9 \theta_0 \right)+3113237 \cos\left(11 \theta_0 \right)-84127\right.\\ \nonumber &&\left. \cos\left(13 \theta_0 \right)) \sin^{6}\left(\theta_0 \right) \sin\left(6 \phi_0 \right)+16384 \cos\left(\theta_0 \right) (3+\cos\left(2 \theta_0 \right)) (11528151+16707368 \cos\left(2 \theta_0 \right) +4016556\right.\\ \nonumber &&\left.\cos\left(4 \theta_0 \right)-48744 \cos\left(6
\theta_0 \right)-8771 \cos\left(8 \theta_0 \right)) \sin^{8}\left(\theta_0 \right) \sin\left(8 \phi_0 \right)-4096 \cos\left(\theta_0 \right) (76447379+90531976 \right.\\ \nonumber &&\left.\cos\left(2 \theta_0 \right)+22418780 \cos\left(4 \theta_0 \right)+1041208 \cos\left(6 \theta_0 \right)+57425 \cos\left(8
\theta_0 \right)) \sin^{10}\left(\theta_0 \right) \sin\left(10 \phi_0 \right)+786432 \cos\left(\theta_0 \right) \right.\\ \nonumber &&\left.(3+\cos\left(2 \theta_0 \right)) (24323+16476 \cos\left(2 \theta_0 \right)+1121 \cos\left(4 \theta_0 \right)) \sin^{12}\left(\theta_0 \right)
\sin\left(12 \phi_0 \right)-49152 \cos\left(\theta_0 \right) (112287\right.\\ \nonumber &&\left.+92204 \cos\left(2 \theta_0 \right)+2869 \cos\left(4 \theta_0 \right)) \sin^{14}\left(\theta_0 \right) \sin\left(14\phi_0 \right)+78643200 \cos\left(\theta_0 \right) (3+\cos\left(2 \theta_0 \right)) \sin^{16}\left(\theta_0 \right) \sin\left(16 \phi_0 \right)\right.\\  &&\left.-11665408 \cos\left(\theta_0
\right) \sin^{18}\left(\theta_0 \right) \sin\left(18 \phi_0 \right)\right]\Big{/}{281474976710656}.
\end{eqnarray}
\end{widetext}
It lies within the  small interval $[0.37402, 0.4318]$ and shown in Fig. \ref{fig:10.2qubitavg}~(a). The maximum  value corresponds to the initial state:  $\ket{1.1056653,-1.045053}$,$\ket{2.0359276,1.045053}$, $\ket{1.1056653,2.09653933}$ and $\ket{2.0359276,-2.09653933}$, while it  is minimum  for the states $\ket{\pi/2,\pm\pi/2}$, which can be seen from  Fig. \ref{fig:10.2qubitavg}. The  numerically obtained time-average value of concurrence is maximum of $1.155818823\times10^{-2}$  for the states $\ket{2.8501,1.523672}$, $\ket{0.290597,-1.523672}$, $\ket{2.8501,-1.617902}$, $\ket{0.290597,1.6179202}$, while it attains minimum  value of $5.49013879\times10^{-4}$ for the states  $\ket{1.861394,\pm1.8692476}$, $\ket{1.2802,\pm1.272345}$$\ket{1.861394,\pm1.272345}$, $\ket{1.2802,\pm1.869247}$,  as plotted in Fig. \ref{fig:4.26qubitavg}~(d). Based on these numerical results, we find that the time-averaged concurrence tends to zero for arbitrary initial states  with $N$. It indicates that  bipartite entanglement is now become multipartite entanglement for arbitrary initial states. For the parameters $J=1/2$ and $\tau=\pi/4$, we numerically observe the periodic nature of entanglement dynamics for arbitrary initial states at any even $N$, while this periodicity disappear for any odd $N$. Thus, we conclude that, for the these parameters the system exhibits quantum integrablity only for even-$N$ with arbitrary initial states.
\section{Impact of Ising Strength ($J$) on Average Linear Entropy of An Arbitrary Initial States}\label{sec:example-section5}
In previous sections \ref{sec:example-section3} and \ref{sec:example-section4}, we have analytically calculated the linear entropy and its average values for any arbitrary initial states for the specific values of the parameters $J=1$, $1/2$ and $\tau=\pi/4$. We have shown that it exhibits the signatures of quantum integrability for $J=1$ with qubits ranging from $4$ to $10$ for arbitrary initial states. In contrast, for $J=1/2$, the signatures of integrability are observed only for even-$N$ ($4,6,8,$ and $10$) for arbitrary initial states. Using our procedure, in principle one can solve analytically for any finite $N$. However, solving them becomes more and more challenging and cumbersome. Therefore, for $N>10$, we opt for the method of numerical simulations. In this section, we numerically investigate the impact of Ising strength ($J$) on the normalized average linear entropy $\langle S\rangle/S_{{Max}}$ for arbitrary initial states, while varying $N$.
\begin{figure}[t]\vspace{0.4cm}
\includegraphics[width=0.47\textwidth,height=0.25\textheight]{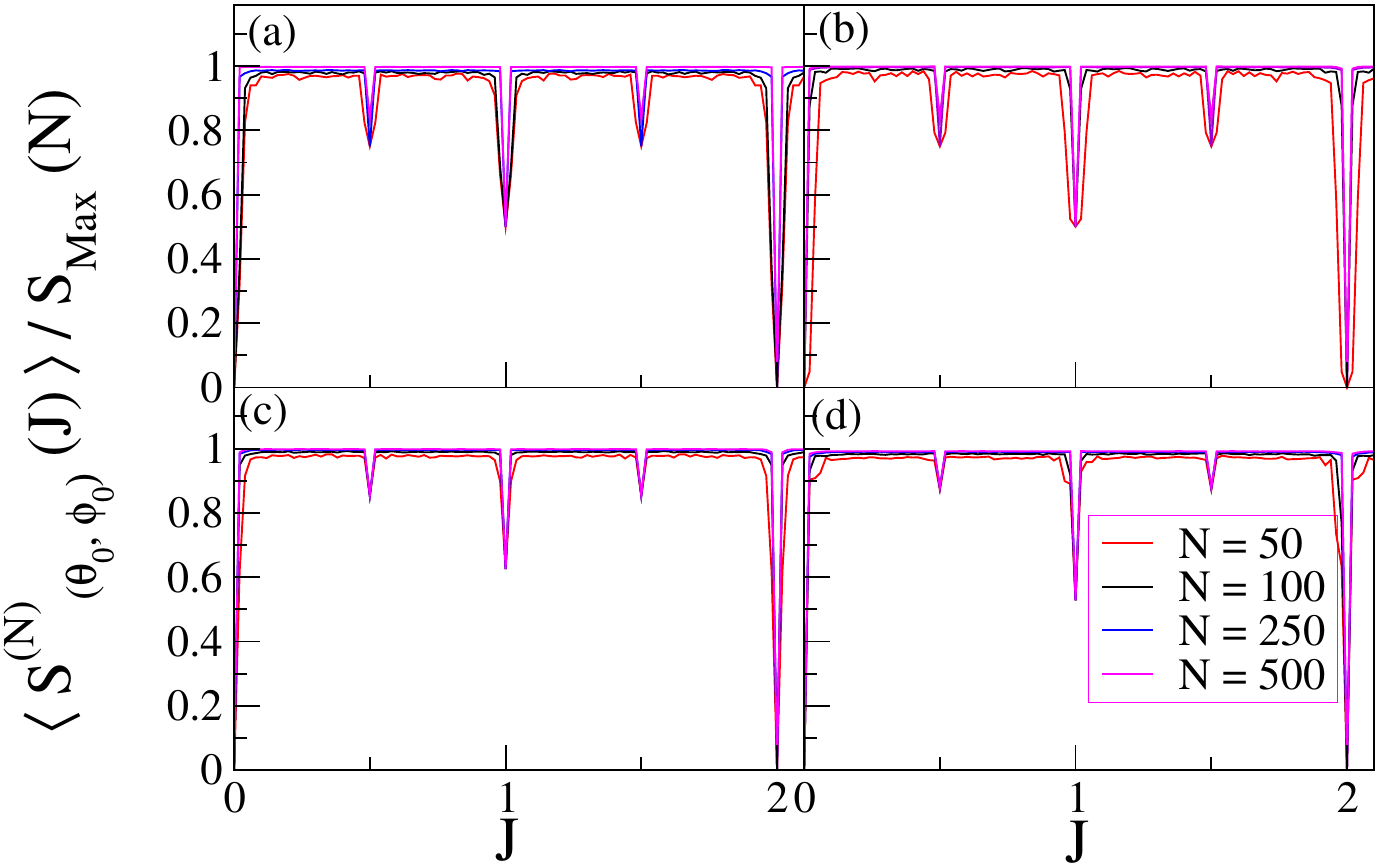}
\caption{ Normalized average single-qubit linear entropy vs Ising strength (J), when the initial states are (a) $\ket{0,0}$ (b) $\ket{\pi/2,-\pi/2}$ (c) $\ket{\pi/4,-\pi/2}$ and (d) $\ket{2\pi/3,-\pi/8}$ with even-$N$.}
\label{fig:4.94qubitavg}
\end{figure}
The average entanglement dynamics play a crucial role in distinguishing integrability and chaos in many-body quantum systems \cite{kumari2022eigenstate,madhok2015comment,ghose2008chaos}. The dips in time-averaged linear entropy and entanglement entropy indicate the presence of periodic orbits, implying that regular regions exhibit lower average entanglement than the chaotic ones \cite{madhok2015comment,ghose2008chaos}. Numerous studies have reported that, in integrable systems, the average entanglement entropy significantly deviates from its maximum value (converges to a value less than $1$). Whereas for non-integrable systems, $\langle S\rangle/S_{{Max}}\rightarrow 1$, in the thermodynamic limit \cite{vidmar2017entanglement,hackl2019average,lydzba2020eigenstate,kumari2022eigenstate}. Recently \cite{kumari2022eigenstate}, in a bipartite system, it has been analytically shown that the ratio $\langle S\rangle/S_{Max}$ is strictly less than  $1$  for various integrable models. For instance, in the case of free fermions \cite{vidmar2017entanglement} and $XY$ chain \cite{hackl2019average} the ratio $\langle S\rangle/S_{Max}$ lies within range  $[0.52,0.59]$, for random quadratic model around $0.557$  \cite{lydzba2020eigenstate} and for the Dicke basis and LMG model, it is around $0.7213$ and $0.5$ respectively  \cite{kumari2022eigenstate}.
\begin{figure}[htbp!]\vspace{0.4cm}
\includegraphics[width=0.47\textwidth,height=0.25\textheight]{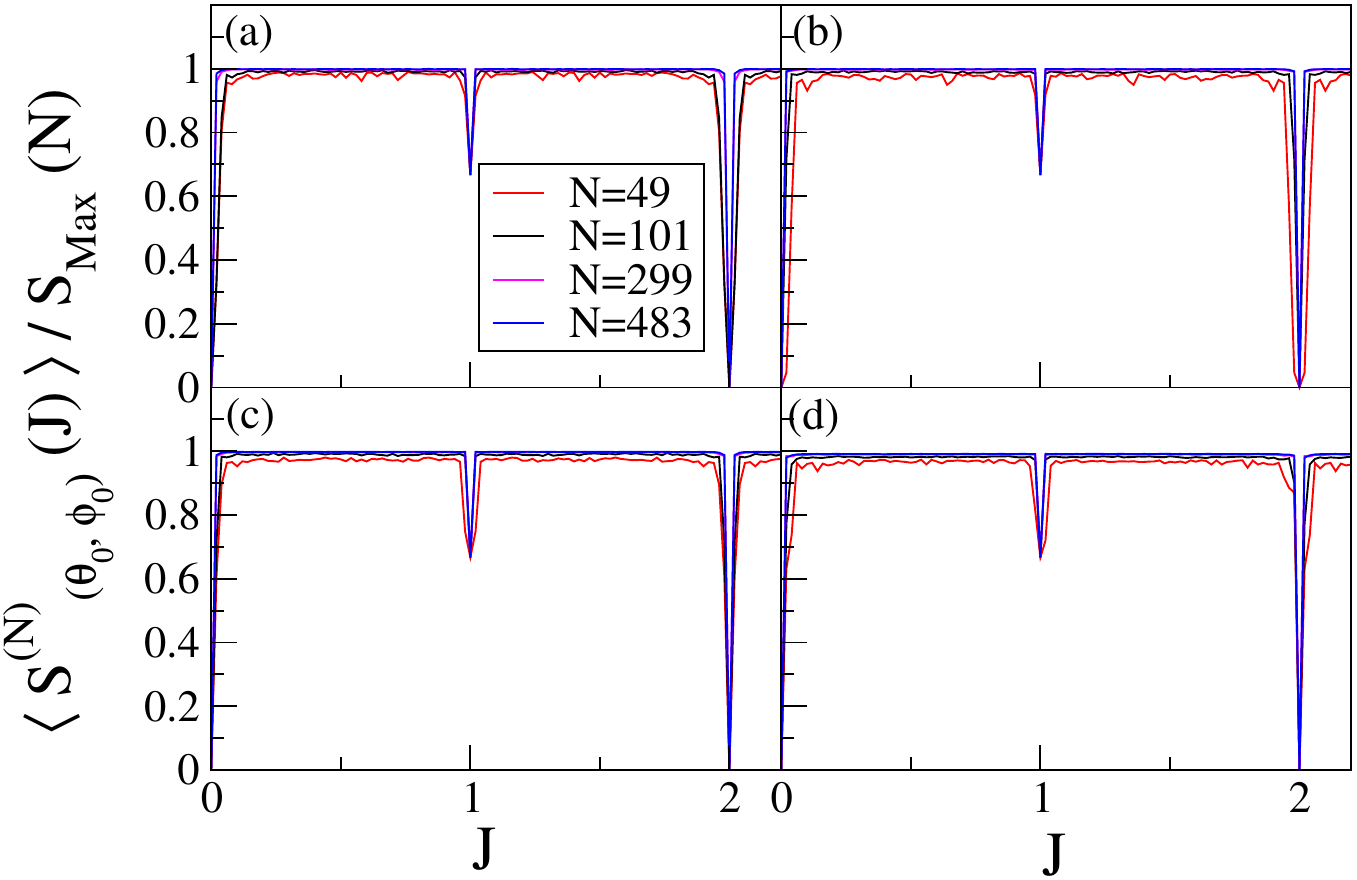}
\caption{ Normalized average single-qubit linear entropy vs Ising strength (J), when the initial states are (a) $\ket{0,0}$ (b) $\ket{\pi/2,-\pi/2}$ (c) $\ket{\pi/4,-\pi/2}$ and (d) $\ket{2\pi/3,-\pi/8}$ with odd-$N$.}
\label{fig:4.95qubitavg}
\end{figure}

\begin{figure}[htbp!]\vspace{0.4cm}
\includegraphics[width=0.45\textwidth,height=0.22\textheight]{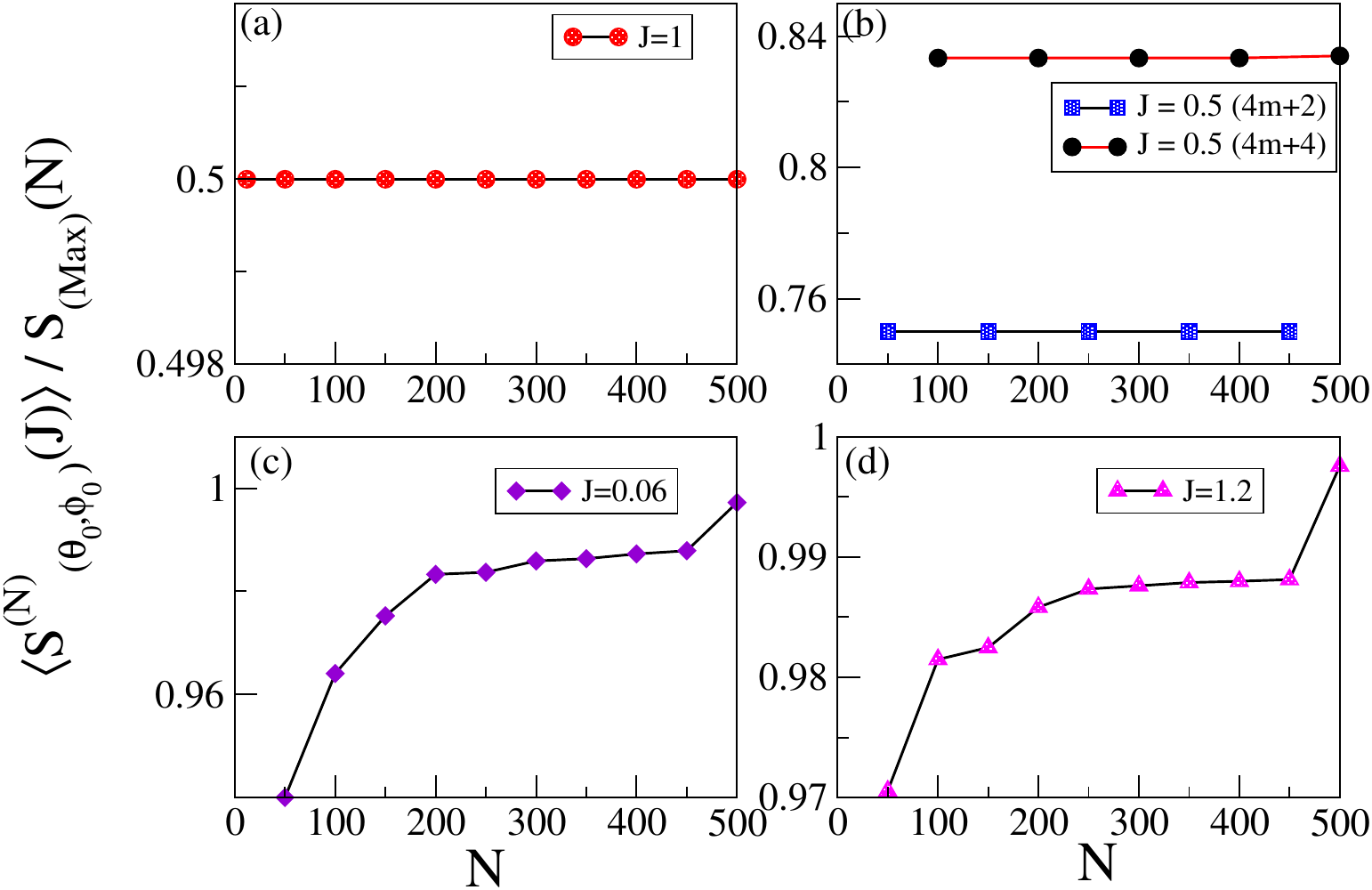}
\caption{Normalized average single-qubit linear entropy vs even-$N$ for the initial state $\ket{0,0}$.}
\label{fig:4.77qubitavg}
\end{figure}

\begin{figure}[htbp!]\vspace{0.4cm}
\includegraphics[width=0.45\textwidth,height=0.22\textheight]{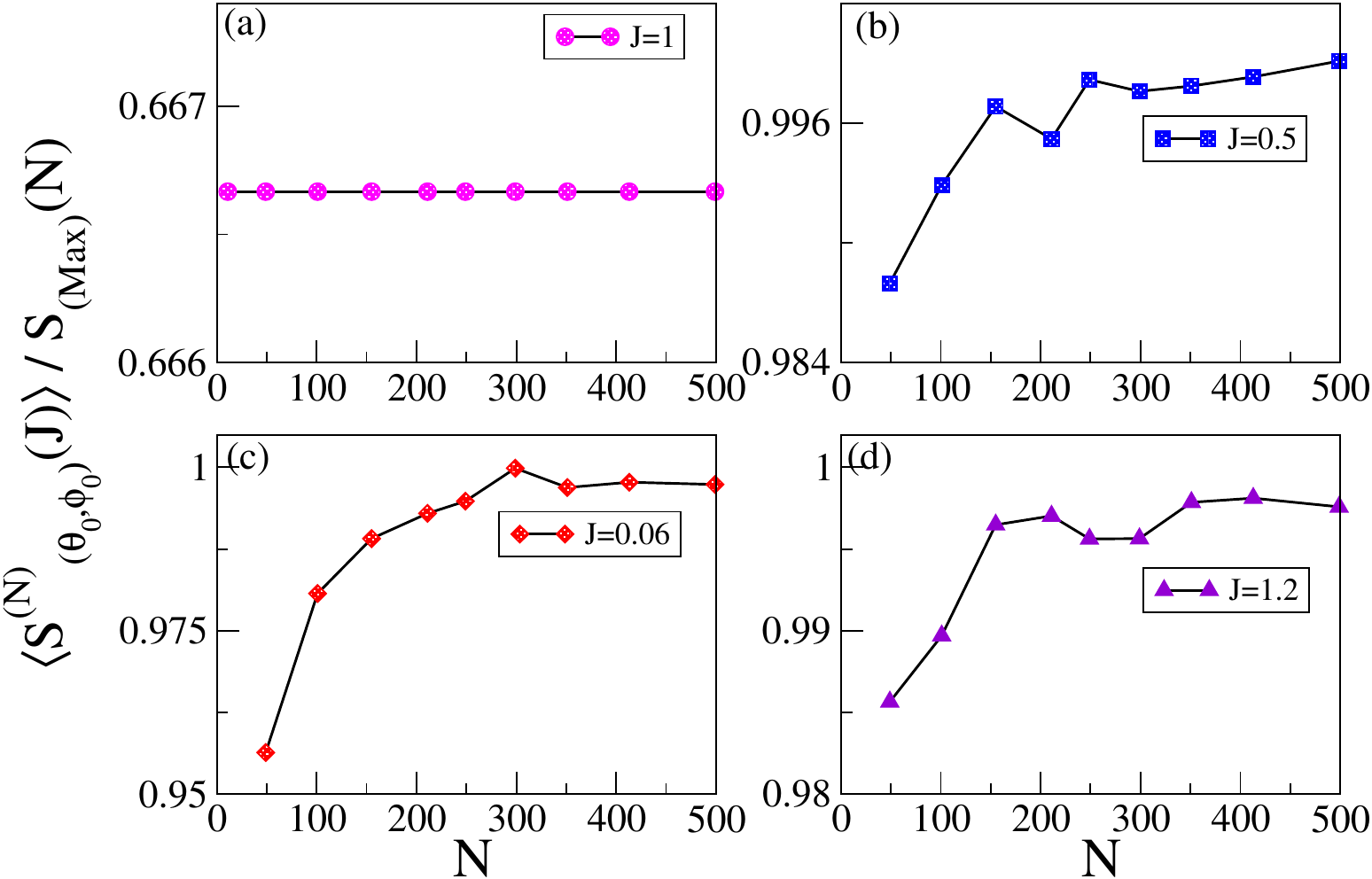}
\caption{Normalized average single-qubit linear entropy vs odd-$N$ for the initial state $\ket{0,0}$.}
\label{fig:4.78qubitavg}
\end{figure}

We numerically plotted the normalized average single qubit linear entropy $\langle S\rangle/S_{Max}$ and examined its behavior with the Ising strength ($J$) for various initial states, while varying  $N$. The results are plotted in Figs. \ref{fig:4.94qubitavg}, \ref{fig:4.95qubitavg}, \ref{fig:4.77qubitavg} and \ref{fig:4.78qubitavg}. We observe that for the specific values of ising strength, $J=1,1/2$ (for even qubits) and $J=1$ (for odd qubits),  the ratio $\langle S\rangle/S_{{Max}}$ in the limit $N\rightarrow\infty$ converges to a value less than $1$ for arbitrary initial states, which implies integrable nature.  From the Figs. \ref{fig:4.94qubitavg} and \ref{fig:4.95qubitavg}, we observe that the occurrence of these dips is independent of the initial state. However, the depth of these dips in general depends on the arbitrary initial state and  $N$. In contrast, for other values of $J$, this ratio tends towards $1$ implying non-integrable nature.

We have plotted  $\langle S\rangle/S_{Max}$ as a function of $N$ for different values of $J$ and the  initial state $\ket{0,0}$ (see  Figs. \ref{fig:4.77qubitavg} and \ref{fig:4.78qubitavg}). For the case $J=1$ the system is integrable for both even and odd-$N$. We observe that for this case the ratio  $\langle S\rangle/S_{{Max}}$ asymptotically approaches  to a value less than $1$ with $N$ (see (a) of  Figs. \ref{fig:4.77qubitavg} and \ref{fig:4.78qubitavg}). Additionally, we observe that the asymptotic value of the ratio $\langle S\rangle/S_{{Max}}$ depends on the parity of $N$. A few points are worth noting here: we observe that for the integrable case, the depth of the dips is independent of $N$ for the special class of the initial state, as shown in Figs. \ref{fig:4.77qubitavg} (a), \ref{fig:4.77qubitavg} (b) and \ref{fig:4.78qubitavg} (a). While for most of the initial states, the depth of dips saturates very fast for small $N$ of the order of $12$ itself. Thus, the increase in the values of average linear entropy is too small to observe in the figures. However, it can be observed in the numerical data from the third decimal place onwards (results are not shown here).

For the case $J = {1}/{2}$, there are two cases for even-$N$ depending on whether  $N=4m+2$ or $N=4m+4$. For both the cases, we observed that the ratio  $\langle S\rangle/S_{{Max}}$  converges to
different value less than $1$. This can be seen in  Fig. \ref{fig:4.77qubitavg} (b). Similarly, for $J=1/2$ and odd $N$, the ratio tends toward $1$ implying non-integrability, as shown in Fig. \ref{fig:4.78qubitavg} (b).
Furthermore, we observed that for other values of $J$, the ratio tends toward $1$ with $N$ (see (c) and (d) of  Figs. \ref{fig:4.77qubitavg} and \ref{fig:4.78qubitavg}). We find that for the integrable cases,
the depth of the dips in the average entanglement increases with $N$ and for large $N$, it saturates to a specific value less than $1$, as shown in Figs. \ref{fig:4.77qubitavg} (a), \ref{fig:4.77qubitavg} (b) and \ref{fig:4.78qubitavg} (a).

The convergence of the ratio $\langle S\rangle/S_{Max}$ to a value less than $1$ serves as a good indicator of the absence or lack of quantum chaos in many-body quantum systems in the thermodynamic limit \cite{vidmar2017entanglement,hackl2019average,lydzba2020eigenstate,kumari2022eigenstate}. We observe a similar signature for specific values of $J$, even though the thermodynamic limit does not exist in our model \cite{sharma2024exactly,sharma2024signatures}. We have demonstrated that, for these specific values, the system exhibits signatures of quantum integrability for arbitrary initial states. Based on these numerical findings, we conjecture that the normalized linear entropy converging to a value other than $1$ indicates the absence of chaos for large $N$ limit, serving as a potential signature of integrability in systems where the thermodynamic limit does not exist.
\section{Conclusions}\label{sec:example-section6}
In this paper, we have analytically calculated the single qubit reduced density matrix and its eigenvalues and entanglement dynamics for arbitrary initial states for the parameters $J=1,1/2$ and $\tau =\pi/4$. In Refs. \cite{sharma2024exactly,sharma2024signatures}, we have shown that our model, which consists of all-to-all ising interaction, exhibits signatures of integrability for the aforementioned parameters. In our previous works \cite{sharma2024exactly,sharma2024signatures}, we calculated these signatures of QI for the initial states $\ket{0,0}$ and  $\ket{\pi/2,-\pi/2}$. As a byproduct of these studies, we have now extended these signatures of QI to arbitrary initial states. We calculated linear entropy, von Neumann entropy analytically, and concurrence numerically to measure the entanglement dynamics for arbitrary initial states for $4$ to $10$ qubits. We have found that, for any initial state, the entanglement dynamics exhibit a periodic nature. We analytically calculated the expression for average linear entropy for arbitrary initial states for the said values of parameters. We have identified the initial states, where the average entanglement dynamics attain their maximum and minimum values. For $J=1$, we observe that the contour plots for both linear entropy and entanglement entropy exhibit a qualitatively similar structure for even $N$. Likewise, a consistent qualitative structure is observed across all odd numbers of qubits. However, the entropy values vary with different initial conditions. In contrast, for $J=1/2$, this behavior is not observed.

We numerically observe that the average linear entropy tends to its maximum values ($\langle S\rangle/S_{Max}\rightarrow1$) for the other values of Ising strength ($J\neq1,1/2$)  for arbitrary initial state. We have numerically shown that $\langle S\rangle/S_{Max}$ converges to a value less than $1$ for $J=1,1/2$ for any even number of qubits and $J=1$ for any odd number of qubits for any arbitrary initial states. We also observe that the depth of the dips varies for different initial states and $N$, the values of $\langle S\rangle/S_{Max}$ are different, but they always diverge from $1$ for the aforementioned parameters. In various studies, it has been shown that for the integrable systems, average entanglement entropy and linear entropy are far away from their maximal in the thermodynamic limit. In our model, the thermodynamic limit does not exist, but we still found similar behavior. This could serve as a good indicator in distinguishing the integrable and nonintegrable systems with or without thermodynamic limits. We also observe that for the same parameters,  time-average concurrence decreases with $N$, indicating the multipartite nature of the system for arbitrary initial state.

The nature of our model is disorder-free (clean), and the integrability exists only for the special values of the parameters $J$ and $\tau$ \cite{sharma2024exactly,sharma2024signatures}. The signatures of quantum integrability disappear in our model applying minor perturbations in either of the parameters, and the exact solution is no longer possible as far as we know. Our results could be experimentally verified in various setups like NMR \cite{Krithika2019}, superconducting qubits \cite{neill2016ergodic}, and laser-cooled atoms \cite{Chaudhury09}, where the QKT has been implemented. However for higher number of qubits, one can use ion trap \cite{monroe2021programmable,defenu2023long}. Recently, an exact solution for quantum-strong-long-range Ising chains \cite{roman2023exact} was obtained by applying the Hubbard-Stratonovich transformation. While our research successfully identified the integrability for arbitrary initial states for the parameter $J=1,1/2$ and $\tau=\pi/4$. We hope our work raises several open questions. A few of them are as follows:
~1) Are there other possible values of $\tau$ or the combination of $J$ and $\tau$ that exhibit integrability within this framework?;
~2) Since our model is disorder-free and shows integrability, a question arises:  is it possible for our model to remain integrable even if the disorder is introduced into the system?;
~3) Are there additional signatures, beyond the known ones, to identify integrability in such systems?
\section{Acknowledgment}
The authors express their gratitude to the Department of Science and Technology (DST) for their invaluable financial support, which made this research possible through the approved project SR/FST/PSI/2017/5(C), granted to the Department of Physics at VNIT, Nagpur. We also thank Avadhut Purohit for his insightful discussions and suggestions on the manuscript.
\let\oldaddcontentsline\addcontentsline
\renewcommand{\addcontentsline}[3]{}
\bibliography{refrence11,ref18}
\newpage
\onecolumngrid
\setlength{\belowcaptionskip}{0pt}
\setcounter{secnumdepth}{4}
\setcounter{tocdepth}{5}
\setcounter{page}{1}
\setcounter{figure}{0}
\setcounter{table}{0}
\setcounter{equation}{0}
\setcounter{section}{0}
\renewcommand{\thepage}{S\arabic{page}}
\renewcommand{\thesection}{S\arabic{section}}
\renewcommand{\thesection}{S\arabic{section}}
\renewcommand{\thetable}{S\arabic{table}}
\renewcommand{\thefigure}{S\arabic{figure}}
\renewcommand{\theequation}{S\arabic{equation}}
\counterwithout*{equation}{section}

\titleformat{\paragraph}[runin]{\normalfont\normalsize\bfseries}{}{0em}{}[.]
\titlespacing*{\paragraph}{0pt}{3.25ex plus 1ex minus .2ex}{1em}

\setlength{\belowcaptionskip}{0pt}
\newpage
\begin{center}
    \large\textbf{Supplemental Materials for\\``\textit{Exact Solvability Of Entanglement For Arbitrary Initial State in an Infinite-Range Floquet System''}}
\end{center}


In this supplement, we provide the analytical calculations of the eigenvalues, eigenvectors, and the $n$th time evolution of unitary operator.  We also provide the  expressions for any
arbitrary  initial coherent state $\ket{\psi}$  in $\ket{\phi}$ basis for any $N$. Additionally, we calculated the expressions for the  coefficients of $\ket{\psi_n}=\mathcal{U}^n\ket{\psi}$  used in the main text for various numbers of qubits ($N$) and Ising strengths ($J$). The first section contains  the expressions for $N=5$ to $10$ qubits with  parameters $J=1$ and $\tau=\pi/4$. In the second section,  the expressions for an even number of qubits from $N=4$ to $10$ with $J=1/2$ and $\tau=\pi/4$ are provided.
\section{The case for $J=1$}
\subsection{Expressions for the Coefficients for the case of $5$ qubit}
The general basis $\ket{\phi}$, for any odd number of qubits \cite{sharma2024exactly} is given  as follows:
\begin{equation}\label{Eq:oddBasis115}
\ket{\phi_q^{\pm}}=\frac{1}{\sqrt{2}}\left(\ket{w_q}\pm {i^{\left(N-2q\right)}} \ket{\overline{w_q}}\right),
0\leq q\leq \dfrac{N-1}{2},
\end{equation}
where $\ket{w_q}=\left({1}/{\sqrt{\binom{N}{q}}}\right)\sum_\mathcal{P}\left(\otimes^q \ket{1} \otimes^{(N-q)}\ket{0}\right)_\mathcal{P}$ and
$\ket{\overline{w_q}}=\left({1}/{\sqrt{\binom{N}{q}}}\right)\sum_\mathcal{P}\left(\otimes^{q}\ket{0}\otimes^{(N-q)}\ket{1}\right)_\mathcal{P}$,
both being definite particle states \cite{Vikram11}. The $\sum_\mathcal{P}$ denotes the sum over all possible permutations.
Using Eq. (3), from the main text the unitary operator $\mathcal{U}$ can be expressed in two blocks   $\mathcal{U}_{\pm}$ in $\ket{\phi}$ basis \cite{sharma2024exactly} as follows:
\begin{equation}
  \mathcal{U}_{\pm}= \frac{e^{\pm{\frac{ i \pi }{4}}}}{4}
\begin{pmatrix}
 \mp{1} &  i\sqrt{5}  &  \mp{\sqrt{10}} \\
- i\sqrt{5} & \pm 3 & - i\sqrt{2} \\
  \pm \sqrt{10} & - i\sqrt{2} & \mp 2 \\
\end{pmatrix}.
\end{equation}
The eigenvalues  of $\mathcal{U_+}\left(\mathcal{U_-}\right)$ are ${e^{\frac{i \pi }{4}}}\left\lbrace 1,e^{\frac{2 i \pi }{3}},e^{-\frac{2 i \pi }{3}}\right\rbrace \left(e^{\frac{3i \pi }{4}}\left\lbrace1,e^{-\frac{2 i \pi }{3}},e^{\frac{2 i \pi }{3}}\right\rbrace\right)$ and  the eigenvectors are $\left\lbrace\left[\frac{\pm i}{\sqrt{5}},1,0\right]^T, \left[\pm i\sqrt{\frac{5}{6}},-\frac{1}{\sqrt{6} },1\right]^T, \right.\\ \left. \left[\mp i\sqrt{\frac{5}{6}},\frac{1}{\sqrt{6} },1\right]^T\right\rbrace$.   The time  evolution of the two blocks  $\mathcal{U}_+$ and $\mathcal{U}_{-}$ given  as follows:
\begin{equation}
\mathcal{U}_{\pm}^n =(\pm 1)^n e^{\pm{\frac{ in \pi }{4}}}
{\left[
\begin{array}{ccc}
 \frac{1}{6} \left(1+5 \cos \left(\frac{2 n \pi }{3}\right)\right) &\pm \frac{1}{3} i \sqrt{5} \sin^2\left(\frac{n \pi }{3}\right) & -\sqrt{\frac{5}{6}}
\sin \left(\frac{2 n \pi }{3}\right) \\
 \mp \frac{1}{3} i \sqrt{5} \sin^2 \left(\frac{n \pi }{3}\right) & \frac{1}{6} \left(5+\ \cos \left(\frac{2 n\pi }{3}\right)\right) &\mp {i
\sin \left(\frac{2 n \pi }{3}\right)}/{\sqrt{6}} \\
 \sqrt{\frac{5}{6}} \sin \left(\frac{2 n \pi }{3}\right) & \mp {i \sin \left(\frac{2 n \pi }{3}\right)}/{\sqrt{6}} & \cos \left(\frac{2n
 \pi }{3}\right) \\
\end{array}
\right]}.
\end{equation}
The analytical calculations of  $\ket{\phi}$ basis, eigenvalues, eigenvectors, and the $n$th time evolution of unitary operator for the parameters $J=1$ and $\tau=\pi/4$ are shown in the supplementary material of Ref. \cite{sharma2024exactly}. Here, we rewrite  them  for  better understanding and clarity for the results. The coherent state in the computational basis is given as follows:
\begin{equation}
  \ket{\psi_0}=|\theta_0,\phi_0\rangle = \cos(\theta_0/2) |0\rangle + e^{-i \phi_0} \sin(\theta_0/2) |1\rangle.
\end{equation}
The arbitrary initial state $\ket{\psi}=\otimes^{N}\ket{\psi_0}$ for any  odd  number of qubits  in $\ket{\phi}$ basis can be expressed as follows:
\begin{equation}
 \ket{\psi}= \sum_{q=1}^{(N+1)/2}\frac{1}{\sqrt{2}}\left( a_q \ket{\phi_{q-1}^+} +b_q \ket{\phi_{q-1}^-}\right),
\end{equation}
where the coefficients $a_q$ and $b_q$ are given as follows:
\begin{eqnarray}\label{Eq:arbitaray22}
a_q&=&\sqrt{\binom{N}{q-1}}\left(\cos^{N-(q-1)}\left(\theta_0/2\right) e^{-i (q-1)\phi_0} \sin^{(q-1)}\left(\theta_0/2\right) -~i^{N-2(q-1)}\cos^{(q-1)}\left(\theta_0/2\right) e^{-i (N-(q-1))\phi_0}\right. \\ \nonumber  &&\left.\sin^{N-(q-1)}\left(\theta_0/2\right)\right),~  1\leq q\leq \dfrac{N+1}{2}~~~~~\mbox{and}\\  \label{Eq:arbitaray32}
b_q&=&\sqrt{\binom{N}{(q-1)}}\left(\cos^{N-(q-1)}\left(\theta_0/2\right) e^{-i (q-1) \phi_0} \sin^{q-1}\left(\theta_0/2\right) +~i^{N-2(q-1)}\cos^{(q-1)}\left(\theta_0/2\right) e^{-i (N-(q-1))\phi_0}\right. \\ \nonumber  &&\left.\sin^{N-(q-1)}\left(\theta_0/2\right)\right),~ 1\leq q\leq \dfrac{N+1}{2}.
\end{eqnarray}
Applying the unitary operator $\mathcal{U}$ $n$ times on the state $\ket{\psi}$ we obtain,
\begin{eqnarray}
\ket{\psi_n}&=&\mathcal{U}^n\ket{\psi}\\ \nonumber
&=& a_{1n}\ket{\phi_0^+}+a_{2n}\ket{\phi_1^+}+a_{3n}\ket{\phi_2^+}+a_{4n}\ket{\phi_0^-} +a_{5n}\ket{\phi_1^-}+a_{6n}\ket{\phi_2^-},\nonumber
\end{eqnarray}
where the coefficients are given as follows:
\begin{equation}\label{Eq:qkt59}
 a_{jn}=\sum_{q=1}^{\frac{N+1}{2}}\mathcal{U}^n_{j,q}~ a_q+\sum_{q=\frac{N+3}{2}}^{N+1}\mathcal{U}^n_{j,q}~ b_{q-\frac{N+1}{2}},1\leq j\leq N+1.
\end{equation}
The expressions of the coefficients $a_{jn}$ for $N=5$ qubits can be calculated using   Eqs. (\ref{Eq:arbitaray22}),  (\ref{Eq:arbitaray32}) and (\ref{Eq:qkt59}), as follows:
\begin{eqnarray}\nonumber
 a_{1n}&=&\frac{e^{-5 i \phi_0 }}{6 \sqrt{2}}\left[e^{5 i \phi_0 } \left(1+5 \cos\left(\frac{2 n \pi }{3}\right)\right) \cos^5\left(\frac{\theta_0 }{2}\right)-i \sin^5\left(\frac{\theta_0
}{2}\right)-5 i \cos\left(\frac{2 n \pi }{3}\right) \sin^5\left(\frac{\theta_0 }{2}\right)+\frac{5~e^{i \phi_0 }}{4}  \sin\left(\theta_0
\right) \left(\sin^2\left(\frac{n \pi }{3}\right) \right.\right. \\ \nonumber&&\left.\left.\left(4 i e^{3 i \phi_0 } \cos^3\left(\frac{\theta_0 }{2}\right)-
4 \sin^3\left(\frac{\theta_0
}{2}\right)\right)-2 \sqrt{3} e^{i \phi_0 }\sin\left(\frac{2 n \pi }{3}\right)\left(e^{i \phi_0 } \cos\left(\frac{\theta_0 }{2}\right)-i
\sin\left(\frac{\theta_0 }{2}\right)\right) \sin(\theta_0)\right)\right],\\ \nonumber
a_{2n}&=&\frac{e^{-4 i \phi_0 }}{6} \sqrt{\frac{5}{2}}  \left[ \left(5+\cos\left(\frac{2 n \pi }{3}\right)\right)\left(i\sin^4\left(\frac{\theta_0
}{2}\right) \cos\left(\frac{\theta_0 }{2}\right)+ e^{3 i \phi_0 }  \sin\left(\frac{\theta_0
}{2}\right)\cos^4\left(\frac{\theta_0 }{2}\right)\right)-2e^{-i \phi_0 } \sin^2\left(\frac{n \pi }{3}\right)\sin^5\left(\frac{\theta_0 }{2}\right)\right.  \\ \nonumber &&\left.-2 i e^{4
i \phi_0 } \cos^5\left(\frac{\theta_0 }{2}\right) \sin^2\left(\frac{n \pi }{3}\right)-\right.
\left.2 \sqrt{3} e^{ i \phi_0 } \cos^2\left(\frac{\theta_0 }{2}\right) \sin^3\left(\frac{\theta_0 }{2}\right)\sin\left(\frac{2
n \pi }{3}\right)-2 i \sqrt{3} e^{2 i \phi_0 } \sin^2\left(\frac{\theta_0 }{2}\right)\cos^3\left(\frac{\theta_0 }{2}\right) \sin\left(\frac{2
n \pi }{3}\right)\right], \\ \nonumber
a_{3n}&=&\frac{\sqrt{5} e^{-5 i \phi_0 }}{6}  \left(e^{i \phi_0 } \cos\left(\frac{\theta_0 }{2}\right)-i \sin\left(\frac{\theta_0
}{2}\right)\right) \left[\sqrt{3} \sin\left(\frac{2 n \pi }{3}\right)\left(e^{4 i \phi_0 } \cos^4\left(\frac{\theta_0 }{2}\right) +\sin^4\left(\frac{\theta_0 }{2}\right)\right)+\frac{3~e^{2 i \phi_0 }}{2}  \cos\left(\frac{2 n \pi }{3}\right) \sin^2(\theta_0)\right], \\ \nonumber
\vspace{2cm}
a_{4n}&=&\frac{ e^{\frac{i n \pi }{2}-5 i \phi_0}}{6 \sqrt{2}} \left[\left(1+5 \cos\left(\frac{2 n \pi }{3}\right)\right) \left(i \sin^5\left(\frac{\theta_0 }{2}\right)+e^{5 i \phi_0 } \cos^5\left(\frac{\theta_0
}{2}\right)\right)-10 e^{i \phi_0 }\sin^2\left(\frac{n \pi}{3}\right)\left( \cos\left(\frac{\theta_0 }{2}\right) \sin^4\left(\frac{\theta_0 }{2}\right)-\right.\right.\\ \nonumber &&
\left.\left.i e^{3 i \phi_0 } \cos^4\left(\frac{\theta_0 }{2}\right)\sin\left(\frac{\theta_0 }{2}\right)\right) -10 \sqrt{3} e^{2 i \phi_0 }\sin\left(\frac{2n \pi }{3}\right)\left( i\cos^2\left(\frac{\theta_0 }{2}\right) \sin^3\left(\frac{\theta_0 }{2}\right)-e^{ i \phi_0 } \cos^3\left(\frac{\theta_0 }{2}\right) \sin^2\left(\frac{\theta_0 }{2}\right)\right)\right], \\ \nonumber
a_{5n}&=&\frac{ e^{\frac{i n \pi }{2}-5 i \phi_0}}{12} \sqrt{\frac{5}{2}} \left(2 e^{i \phi_0 } \left(5+\cos\left(\frac{2 n \pi }{3}\right)\right)
\cos\left(\frac{\theta_0 }{2}\right) \sin\left(\frac{\theta_0 }{2}\right) \left(e^{3 i \phi_0 } \cos^3\left(\frac{\theta_0 }{2}\right)-i
\sin^3\left(\frac{\theta_0 }{2}\right)\right)+ \left(e^{5 i \phi_0 } \cos^4\left(\frac{\theta_0
}{2}\right)+i \sin^5\left(\frac{\theta_0 }{2}\right)\right)\right.  \\ \nonumber&&\left.4 i \sin^2\left(\frac{n \pi }{3}\right)+i \sqrt{3} e^{2 i \phi_0 } \left(i \sin\left(\frac{\theta_0 }{2}\right)+e^{i
\phi_0 } \cos\left(\frac{\theta_0 }{2}\right)\right) \sin\left(\frac{2 n \pi }{3}\right)\sin^2\left(\theta_0 \right)\right) ~\mbox{and}\\ \nonumber
a_{6n}&=&\frac{\sqrt{5}  e^{\frac{i n \pi }{2}-5 i \phi_0}}{6}  \left(e^{i \phi_0 }
\cos\left(\frac{\theta_0 }{2}\right)+i
\sin\left(\frac{\theta_0 }{2}\right)\right) \left[\sqrt{3}\sin\left(\frac{2 n
\pi }{3}\right)\left( e^{4 i \phi_0 } \cos^4\left(\frac{\theta_0 }{2}\right)+ \sin^4\left(\frac{\theta_0 }{2}\right)\right)+\frac{3~e^{2 i \phi_0 }}{2}  \cos\left(\frac{2
n \pi }{3}\right) \sin^2\left(\theta_0 \right)\right].
\end{eqnarray}
\subsection{Expressions for the Coefficients for the case of  $6$ qubit}
The general basis for any even number of qubits \cite{sharma2024exactly} is given as follows:
\begin{eqnarray}\label{Eq:evenBasis11}
\begin{split}
\ket{\phi_q^{\pm}}&=\frac{1}{\sqrt{2}}\left(\ket{w_q}\pm {(-1)^{\left(N/2-q\right)}} \ket{\overline{w_q}}\right), 0\leq q\leq N/2-1\\
\mbox{and} &\;\;
\ket{\phi_{N/2}^+}=\left({1}/{\sqrt{\binom{N}{N/2}}}\right)\sum_{\mathcal{P}}\left(\otimes^{N/2}\ket{0}\otimes^{N/2}\ket{1}\right)_{\mathcal{P}}
\end{split}
\end{eqnarray}
  where
$\ket{w_q}=\left({1}/{\sqrt{\binom{N}{q}}}\right)\sum_\mathcal{P}\left(\otimes^q \ket{1} \otimes^{(N-q)}\ket{0}\right)_\mathcal{P}$ and
$\ket{\overline{w_q}}=\left({1}/{\sqrt{\binom{N}{q}}}\right)\sum_\mathcal{P}\left(\otimes^{q}\ket{0}\otimes^{(N-q)}\ket{1}\right)_\mathcal{P}$,
both being definite particle states \cite{vikram11}. The $\sum_\mathcal{P}$ denotes the sum over all possible permutations.
The unitary operator $\mathcal{U}$ is block diagonal in two blocks $\mathcal{U}_{+}~(\mathcal{U}_{-})$ in $\ket{\phi}$ basis  with dimension $4\times4~(3\times3)$ \cite{sharma2024exactly}. The blocks are given as follows:
\begin{equation}
\mathcal{U_+}= \frac{e^{\frac{ i \pi}{4}}}{2\sqrt{2}}\left(
\begin{array}{cccc}
 0 & -\sqrt{3} & 0 & -\sqrt{5} \\
i\sqrt{3} & 0 & i\sqrt{5}& 0 \\
 0 & -\sqrt{5} & 0 & \sqrt{3} \\
 i\sqrt{5} & 0 & -i\sqrt{3} & 0 \\
\end{array}
\right)~~~~\mbox{and}
 \end{equation}
 \begin{equation}
\mathcal{U_-}= \frac{e^{\frac{ i \pi}{4}}}{4}{\left(
\begin{array}{ccc}
  1 & 0 &  \sqrt{15}  \\
 0 & -4i & 0 \\
 \sqrt{15}  & 0 & - 1 \\
\end{array}
\right)}.
 \end{equation}
The eigenvalues of the $\mathcal{U_+} \left( \mathcal{U_-}\right)$  are ${\left\{-1,-1,1,1\right\}}\left(\left\{-(-1)^{1/4},-(-1)^{3/4},(-1)^{1/4}\right\}\right)$  and the eigenvectors are  $\left\lbrace\left[\frac{2+2 i}{\sqrt{5}},\sqrt{\frac{3}{5}},0,1\right]^T\right.\\ \left., \left[\sqrt{\frac{3}{5}},\frac{2-2 i}{\sqrt{5}},1,0\right]^T, \left[-\frac{2+2 i}{\sqrt{5}},\sqrt{\frac{3}{5}},0,1\right]^T, \left[\sqrt{\frac{3}{5}},-\frac{2-2 i}{\sqrt{5}},1,0\right]^T\right\rbrace\left(\left\lbrace\left[-\sqrt{\frac{3}{5}},0,1\right]^T, \left[0,1,0 \right]^T, \left[\sqrt{\frac{5}{3}},0,1\right]^T\right\rbrace\right)$. The $n$th time evolution of two blocks $\mathcal{U_{\pm}}$ is given as follows:
 \begin{equation}
\mathcal{U}_{+}^n= \frac{1}{8}{\left[
\begin{array}{cccc}
 4 \left(1+e^{i n \pi }\right) & {\sqrt{6}}~ e^{\frac{i  \pi }{4}} \left(-1+e^{i n \pi }\right) & 0 & {\sqrt{10}}~ e^{\frac{i  \pi }{4}} \left(-1+e^{i n \pi }\right) \\
 {\sqrt{6}}~ e^{-\frac{i  \pi }{4}} \left(-1+e^{i n \pi }\right) & 4 \left(1+e^{in \pi }\right) & {\sqrt{10}}~ e^{-\frac{i  \pi }{4}} \left(-1+e^{i n \pi }\right) & 0 \\
 0 & {\sqrt{10}}~ e^{\frac{i  \pi }{4}} \left(-1+e^{i n \pi }\right) &  4\left(1+e^{i n \pi }\right) & -{\sqrt{6}}~ e^{\frac{i  \pi }{4}} \left(-1+e^{i m \pi }\right) \\
\sqrt{10}~ e^{-\frac{i  \pi }{4}} \left(-1+e^{i n \pi } \right) & 0 & -\sqrt{6}~  e^{-\frac{i  \pi }{4}} \left(-1+e^{i n
\pi }\right) &  4\left(1+e^{i n \pi }\right) \\
\end{array}
\right]}~~ \mbox{and}
 \end{equation}
 \begin{equation}
 \mathcal{U}_{-}^n =\frac{e^{\frac{i n \pi }{4}}}{8}{\left[
\begin{array}{ccc}
 5 + 3~e^{i n \pi } & 0 &  \sqrt{15}\left( 1-e^{{ i n \pi }}\right) \\
 0 & 8 ~ e^{\frac{3 i m \pi }{2}} & 0 \\
 \sqrt{15}\left( 1-e^{{ i n \pi }}\right) & 0 & 3 + 5~ e^{i n \pi } \\
\end{array}
\right]}.
 \end{equation}
 The arbitrary initial state $\ket{\psi}$ for any even number of qubits  in $\ket{\phi}$ basis can be expressed as,
\begin{equation}
 \ket{\psi}= \sum_{q=1}^{N/2}\frac{1}{\sqrt{2}}\left( a_q \ket{\phi_{q-1}^+} +b_q \ket{\phi_{q-1}^-}\right)+a_{\frac{N+2}{2}} \ket{\phi_{\frac{N}{2}}^+},
\end{equation}
where the coefficients $a_q$, $b_q$ and $a_{\frac{N+2}{2}}$ are given as follows:
\begin{eqnarray} \label{Eq:arbitaray12}
a_q&=&\sqrt{\binom{N}{q-1}}\left(\cos^{N-(q-1)}\left(\theta_0/2\right) e^{-i (q-1)\phi_0} \sin^{(q-1)}\left(\theta_0/2\right) +~i^{N-2(q-1)}\cos^{(q-1)}\left(\theta_0/2\right) e^{-i (N-(q-1))\phi_0}\right. \\ \nonumber  &&\left.\sin^{N-(q-1)}\left(\theta_0/2\right)\right),1\leq q\leq \dfrac{N}{2};\\  \label{Eq:arbitaray15}
b_q&=&\sqrt{\binom{N}{(q-1)}}\left(\cos^{N-(q-1)}\left(\theta_0/2\right) e^{-i (q-1) \phi_0} \sin^{q-1}\left(\theta_0/2\right) -~i^{N-2(q-1)}\cos^{(q-1)}\left(\theta_0/2\right) e^{-i (N-(q-1))\phi_0}\right. \\ \nonumber  &&\left.\sin^{N-(q-1)}\left(\theta_0/2\right)\right),1\leq q\leq \dfrac{N}{2}; ~~~\mbox{and} \\ \label{Eq:arbitaray42}
a_{\frac{N+2}{2}}&=&\sqrt{\binom{N}{\frac{N}{2}}}\left(e^{-i\frac{N}{2} \phi_0}\cos^{\frac{N}{2}}\left(\theta_0/2\right)  \sin^{\frac{N}{2}}\left(\theta_0/2\right)\right).
\end{eqnarray}
 The state $\ket{\psi_n}$ can be obtain after the $n$ implementations of $\mathcal{U}$ on the state $\ket{\psi}$. Thus,
\begin{eqnarray}
\ket{\psi_n}&=&\mathcal{U}^n\ket{\psi}\\ \nonumber
&=& g_{1n}\ket{\phi_0^+}+g_{2n}\ket{\phi_1^+}+g_{3n}\ket{\phi_2^+}+g_{4n}\ket{\phi_3^+}+g_{5n}\ket{\phi_0^-} + ~g_{6n}\ket{\phi_1^-}+g_{7n}\ket{\phi_2^-},
\end{eqnarray}
where the coefficients are given as follows:
\begin{equation}\label{Eq:qkt501}
g_{jn}=\sum_{q=1}^{\frac{N+2}{2}}\mathcal{U}^n_{j,q}~ a_q+\sum_{q=\frac{N+4}{2}}^{N+1}\mathcal{U}^n_{j,q}~ b_{q-\frac{N+2}{2}},~1\leq j\leq N+1.
\end{equation}
 The expressions of the coefficients $g_{jn}$ for the special case of  $N=6$ qubits can be calculated  using  Eqs. (\ref{Eq:arbitaray12}), (\ref{Eq:arbitaray15}), (\ref{Eq:arbitaray42}) and (\ref{Eq:qkt501}), as follows:
\begin{eqnarray}\nonumber
g_{1n}&=&\frac{(1+i)\left(-1+e^{i n \pi }\right)}{4}\left[5 e^{-3 i \phi_0 }  \cos^3\left(\frac{\theta_0 }{2}\right)
\sin^3\left(\frac{\theta_0 }{2}\right)+\frac{3}{2}  \left( e^{-i \phi_0 }
\cos^5\left(\frac{\theta_0 }{2}\right) \sin\left(\frac{\theta_0 }{2}\right)+ e^{-5 i \phi_0 } \cos\left(\frac{\theta_0 }{2}\right)\sin^5\left(\frac{\theta_0 }{2}\right)\right)\right]
\\ \nonumber&&+\frac{\left(1+e^{i n \pi }\right)}{2\sqrt{2}}  \left(\cos^6\left(\frac{\theta_0 }{2}\right)-e^{-6
i \phi_0 } \sin^6\left(\frac{\theta_0 }{2}\right)\right),\\ \nonumber
g_{2n}&=&\frac{\sqrt{3}(1-i)\left(-1+e^{i n \pi }\right)}{8\sqrt{2}}    \left[5\left(e^{-2 i \phi_0 } \cos^4\left(\frac{\theta_0
}{2}\right) \sin^2\left(\frac{\theta_0 }{2}\right)- e^{-4 i \phi_0 } \cos^2\left(\frac{\theta_0 }{2}\right) \sin^4\left(\frac{\theta_0
}{2}\right)\right)+ \cos^6\left(\frac{\theta_0 }{2}\right)-e^{-6
i \phi_0 } \sin^6\left(\frac{\theta_0 }{2}\right)\right]  \\ \nonumber&&+\frac{\sqrt{3}}{2} \left(1+e^{i n \pi }\right) \left(e^{-i \phi_0 } \cos^5\left(\frac{\theta_0 }{2}\right)\sin\left(\frac{\theta_0
}{2}\right)+ e^{-5 i \phi_0 } \cos\left(\frac{\theta_0 }{2}\right) \sin^5\left(\frac{\theta_0 }{2}\right)\right),\\ \nonumber
g_{3n}&=&\frac{\sqrt{15}(1+i)~\left(-1+e^{i n \pi }\right)}{4}\left[-  e^{-3 i \phi_0 }  \cos^3\left(\frac{\theta_0
}{2}\right) \sin^3\left(\frac{\theta_0 }{2}\right)+\frac{1}{2} \left( e^{-i \phi_0 } \cos^5\left(\frac{\theta_0 }{2}\right)
\sin\left(\frac{\theta_0 }{2}\right)+ e^{-5 i \phi_0 } \cos\left(\frac{\theta_0 }{2}\right) \sin^5\left(\frac{\theta_0 }{2}\right)\right)\right] \\ \nonumber&&+\frac{1}{2}\sqrt{\frac{15}{2}} \left(1+e^{i n \pi }\right) \left( e^{-2 i \phi_0 } \cos^4\left(\frac{\theta_0
}{2}\right) \sin^2\left(\frac{\theta_0 }{2}\right)- e^{-4 i \phi_0 } \cos^2\left(\frac{\theta_0 }{2}\right)\sin^4\left(\frac{\theta_0
}{2}\right)\right),\\ \nonumber
g_{4n}&=&\sqrt{5} e^{-3 i \phi_0 } \left(1+e^{i n \pi }\right) \cos^3\left(\frac{\theta_0 }{2}\right) \sin^3\left(\frac{\theta_0
}{2}\right)-\left(\frac{3}{8}-\frac{3~i}{8}\right) \sqrt{5} \left(-1+e^{i n \pi }\right) \left(e^{-2 i \phi_0 } \cos^4\left(\frac{\theta_0
}{2}\right) \sin^2\left(\frac{\theta_0 }{2}\right)- e^{-4 i \phi_0 } \cos^2\left(\frac{\theta_0 }{2}\right) \right. \\ \nonumber&&\left.\sin^4\left(\frac{\theta_0
}{2}\right)\right)+\left(\frac{1}{8}-\frac{i}{8}\right) \sqrt{\frac{5}{2}} \left(-1+e^{i n \pi }\right) \left(\cos^6\left(\frac{\theta_0 }{2}\right)-e^{-6
i \phi_0 } \sin^6\left(\frac{\theta_0 }{2}\right)\right),\\ \nonumber
g_{5n}&=&\frac{ e^{-\frac{3}{4} i n \pi }}{8}  \left[\frac{ \left(3+5 e^{i n \pi }\right)}{\sqrt{2}} \left(\cos^6\left(\frac{\theta_0 }{2}\right)+e^{-6
i \phi_0 } \sin^6\left(\frac{\theta_0 }{2}\right)\right)+\frac{15\left(-1+e^{i n \pi }\right)}{\sqrt{2}} \left( e^{-2 i \phi_0 } \cos^4\left(\frac{\theta_0
}{2}\right) \sin^2\left(\frac{\theta_0 }{2}\right)+\right.\right.\\ \nonumber && \left.\left. e^{-4 i \phi_0 } \cos^2\left(\frac{\theta_0 }{2}\right) \sin^4\left(\frac{\theta_0
}{2}\right)\right)\right],~~~g_{6n}=\sqrt{3} ~e^{-\frac{1}{4} i n \pi } \left[e^{-i \phi_0 } \cos^5\left(\frac{\theta_0 }{2}\right) \sin\left(\frac{\theta_0
}{2}\right)-e^{-5 i \phi_0 } \cos\left(\frac{\theta_0 }{2}\right) \sin^5\left(\frac{\theta_0 }{2}\right)\right]~~ \mbox{and}\\ \nonumber
g_{7n}&=&\frac{\sqrt{15} e^{-\frac{3}{4} i n \pi }}{8} \left[ \sqrt{\frac{15}{2}}\left(5+3 e^{i n \pi }\right) \left( e^{-2 i \phi_0 } \cos^4\left(\frac{\theta_0
}{2}\right) \sin^2\left(\frac{\theta_0 }{2}\right)+ e^{-4 i \phi_0 } \cos^2\left(\frac{\theta_0 }{2}\right) \sin^4\left(\frac{\theta_0
}{2}\right)\right) -\frac{\left(1-e^{i n \pi }\right)}{\sqrt{2}}\right.\\ \nonumber && \left. \left(\cos^6\left(\frac{\theta_0 }{2}\right)+e^{-6
i \phi_0 } \sin^6\left(\frac{\theta_0 }{2}\right)\right)\right].
\end{eqnarray}
\subsection{Expressions for the Coefficients for the case of  $7$ qubit}
In the $\ket{\phi}$ basis, the unitary operator $\mathcal{U}$ is block diagonal in two blocks $\mathcal{U}_{+}~(\mathcal{U}_{-})$ having dimension $4\times4~(4\times4)$  \cite{sharma2024exactly}. The blocks are given as follows:
\begin{equation}
\mathcal{U}_{+} ={\frac{1}{8}\left(
\begin{array}{cccc}
 -1 & -i \sqrt{7} &  -\sqrt{21}  &- i \sqrt{35}
\\
 -i \sqrt{7}  & - 5  & - 3i\sqrt{3}  & - \sqrt{5} \\
  \sqrt{21}  &  3 i \sqrt{3}  &  1 & -i\sqrt{15}
\\
 i \sqrt{35}   &  \sqrt{5}  & -i \sqrt{15}  & -3  \\
\end{array}
\right)}~~~ \mbox{and}
\end{equation}
\begin{equation}
\mathcal{U}_{-} ={\frac{1}{8}\left(
\begin{array}{cccc}
 i &  \sqrt{7} &  i\sqrt{21}  & \sqrt{35}
\\
  \sqrt{7}  &  5i  & 3\sqrt{3}  & i\sqrt{5} \\
  -i\sqrt{21}  &  -3 \sqrt{3}  &  -i & \sqrt{15}
\\
 -\sqrt{35}   &  -i\sqrt{5}  & \sqrt{15}  & 3i  \\
\end{array}
\right)}.
\end{equation}
The eigenvalues of $\mathcal{U_{+}}$~($\mathcal{U_{-}}$) are
$\left\lbrace-1,-1,e^{\frac{i \pi }{3}},e^{-\frac{i \pi }{3}}\right\rbrace$ $\left(i\left\lbrace1,1,e^{\frac{2i \pi }{3}},e^{-\frac{2i \pi }{3}}\right\rbrace\right)$ and  the eigenvectors are  $\left\lbrace\left[ \pm i \sqrt{\frac{5}{7}},0,0,1\right]^T,\right.\\ \left. \left[0,\pm i \sqrt{3},1,0 \right]^T,\left[\mp i \sqrt{\frac{7}{5}},\pm i \sqrt{\frac{3}{5}},-\frac{3}{\sqrt{5}},1 \right]^T, \left[\mp i \sqrt{\frac{7}{5}},\mp i \sqrt{\frac{3}{5}},\frac{3}{\sqrt{5}},1 \right]^T \right\rbrace$. The $n$th time evolution of $\mathcal{U}_{\pm}$ is given as follows:

\begin{equation}
\mathcal{U}_{+}^n=
{e^{i n \pi }\left[
\begin{array}{cccc}
 \frac{1}{12} \left(5+7 \cos\left(\frac{2 n \pi }{3}\right)\right) & \frac{1}{4} i \sqrt{\frac{7}{3}} \sin\left(\frac{2 n \pi }{3}\right)
& \frac{1}{4} \sqrt{7} \sin\left(\frac{2 n \pi }{3}\right) & \frac{1}{6} i \sqrt{35} \sin^2\left(\frac{n \pi }{3}\right) \\
 \frac{1}{4} i \sqrt{\frac{7}{3}} \sin\left(\frac{2 n \pi }{3}\right) & \frac{1}{4} \left(3+\cos\left(\frac{2 n \pi }{3}\right)\right)
& \frac{1}{2} i \sqrt{3} \sin^2\left(\frac{n \pi }{3}\right) & \frac{1}{4} \sqrt{\frac{5}{3}} \sin\left(\frac{2 n \pi }{3}\right) \\
 -\frac{1}{4} \sqrt{7} \sin\left(\frac{2 n \pi }{3}\right) & -\frac{1}{2} i \sqrt{3} \sin^2\left(\frac{n \pi }{3}\right) & \frac{1}{4}
\left(1+3 \cos\left(\frac{2 n \pi }{3}\right)\right) & \frac{1}{4} i \sqrt{5} \sin\left(\frac{2 n \pi }{3}\right) \\
- \frac{1}{6} i \sqrt{35} \sin^2\left(\frac{n \pi }{3}\right) & -\frac{1}{4} \sqrt{\frac{5}{3}} \sin\left(\frac{2 n \pi }{3}\right) &
\frac{1}{4} i \sqrt{5} \sin\left(\frac{2 n \pi }{3}\right) & \frac{1}{12} \left(7+5 \cos\left(\frac{2 n \pi }{3}\right)\right) \\
\end{array}
\right]}~~~\mbox{and}
\end{equation}
\begin{equation}
\mathcal{U}_{-}^n=
e^{\frac{in \pi }{2}}\left[
\begin{array}{cccc}
 \frac{1}{12} \left(5+7 \cos\left(\frac{2 n \pi }{3}\right)\right) & -\frac{i}{4}  \sqrt{\frac{7}{3}} \sin\left(\frac{2 n \pi }{3}\right)
& \frac{\sqrt{7}}{4}  \sin\left(\frac{2 n \pi }{3}\right) & -\frac{1}{6} i \sqrt{35} \sin^2\left(\frac{n \pi }{3}\right) \\
 -\frac{1}{4} i \sqrt{\frac{7}{3}} \sin\left(\frac{2 n \pi }{3}\right) & \frac{1}{4} \left(3+\cos\left(\frac{2 n \pi }{3}\right)\right)
& -\frac{1}{2} i \sqrt{3} \sin^2\left(\frac{n \pi }{3}\right) & \frac{1}{4} \sqrt{\frac{5}{3}} \sin\left(\frac{2 n \pi }{3}\right) \\
 -\frac{1}{4} \sqrt{7} \sin\left(\frac{2 n \pi }{3}\right) & \frac{1}{2} i \sqrt{3} \sin^2\left(\frac{n\pi }{3}\right) & \frac{1}{4}
\left(1+3 \cos\left(\frac{2 n \pi }{3}\right)\right) & -\frac{1}{4} i \sqrt{5} \sin\left(\frac{2 n\pi }{3}\right) \\
 \frac{1}{6} i \sqrt{35} \sin^2\left(\frac{n \pi }{3}\right) & -\frac{1}{4} \sqrt{\frac{5}{3}} \sin\left(\frac{2 n \pi }{3}\right) &
-\frac{1}{4} i \sqrt{5} \sin\left(\frac{2 n \pi }{3}\right) & \frac{1}{12} \left(7+5 \cos\left(\frac{2 n \pi }{3}\right)\right) \\
\end{array}
\right].
\end{equation}
The initial state $\ket{\psi}$ after the $n${th} implementations of the unitary operator $\mathcal{U}$ can be expressed as follows:
\begin{eqnarray}
\ket{\psi_n}&=&\mathcal{U}^n\ket{\psi}\\ \nonumber
&=& b_{1n}\ket{\phi_1^+}+b_{2n}\ket{\phi_2^+}+b_{3n}\ket{\phi_3^+}+b_{4n}\ket{\phi_4^+} +b_{5n}\ket{\phi_1^-}+b_{6n}\ket{\phi_2^-}+b_{7n}\ket{\phi_3^-}+b_{8n}\ket{\phi_4^-},\nonumber
\end{eqnarray}
where the coefficients can be expressed as follows:
\begin{equation}\label{Eq:qkt525}
 b_{jn}=\sum_{q=1}^{\frac{N+1}{2}}\mathcal{U}^n_{j,q}~ a_q+\sum_{q=\frac{N+3}{2}}^{N+1}\mathcal{U}^n_{j,q}~ b_{q-\frac{N+1}{2}},1\leq j\leq N+1.
 \end{equation}
The expressions of the coefficients $b_{jn}$ for $7$ qubits can be calculated  using   Eqs. (\ref{Eq:arbitaray22}), (\ref{Eq:arbitaray32}) and (\ref{Eq:qkt525}), as follows:
\begin{eqnarray}\nonumber
 b_{1n}&=&\frac{35~i }{12\sqrt{2}}  \left(e^{i n \pi }-\cos\left(\frac{n \pi }{3}\right)\right) \left(e^{-3
i \phi_0 } \cos^4\left(\frac{\theta_0 }{2}\right) \sin^3\left(\frac{\theta_0 }{2}\right)-i  e^{-4 i \phi_0 } \cos^3\left(\frac{\theta_0
}{2}\right) \sin^4\left(\frac{\theta_0 }{2}\right)\right)-\frac{7}{4} \sqrt{\frac{3}{2}} \sin\left(\frac{n \pi }{3}\right)\\  \nonumber &&\left(
e^{-2 i \phi_0 } \cos^5\left(\frac{\theta_0 }{2}\right) \sin^2\left(\frac{\theta_0 }{2}\right)+i  e^{-5 i \phi_0 } \cos^2\left(\frac{\theta_0
}{2}\right)\sin^5\left(\frac{\theta_0 }{2}\right)\right)-
\frac{7~i}{4\sqrt{6}} \sin\left(\frac{n \pi }{3}\right) \left( e^{-i \phi_0 } \cos^6\left(\frac{\theta_0
}{2}\right) \sin\left(\frac{\theta_0 }{2}\right)-\right.\\  \nonumber &&\left.i  e^{-6 i \phi_0 } \cos\left(\frac{\theta_0 }{2}\right) \sin^6\left(\frac{\theta_0
}{2}\right)\right)+\frac{1}{12{\sqrt{2}}} \left(5 e^{i n \pi }+7 \cos\left(\frac{n \pi }{3}\right)\right) \left(\cos^7\left(\frac{\theta_0}{2}\right)+i e^{-7 i \phi_0 } \sin^7\left(\frac{\theta_0 }{2}\right)\right),\\ \nonumber
b_{2n}&=&-\frac{5}{4} \sqrt{\frac{7}{6}} \sin\left(\frac{n \pi }{3}\right) \left( e^{-3 i \phi_0 } \cos^4\left(\frac{\theta_0
}{2}\right) \sin^3\left(\frac{\theta_0 }{2}\right)-i  e^{-4 i \phi_0 } \cos^3\left(\frac{\theta_0 }{2}\right) \sin^4\left(\frac{\theta_0
}{2}\right)\right)+\frac{3~i}{4}\sqrt{\frac{7}{2}} \left(e^{i n \pi }-\cos\left(\frac{n \pi }{3}\right)\right) \\  \nonumber &&\left( e^{-2 i \phi_0
} \cos^5\left(\frac{\theta_0 }{2}\right) \sin^2\left(\frac{\theta_0 }{2}\right)+i e^{-5 i \phi_0 } \cos\left(\frac{\theta_0
}{2}\right)^2 \sin^5\left(\frac{\theta_0 }{2}\right)\right)+
\frac{\sqrt{7}}{4\sqrt{2}} \left(3 e^{i n \pi }+\cos\left(\frac{n \pi }{3}\right)\right) \left( e^{-i \phi_0 } \cos^6\left(\frac{\theta_0
}{2}\right) \sin\left(\frac{\theta_0 }{2}\right)-\right.\\  \nonumber &&\left.i  e^{-6 i \phi_0 } \cos\left(\frac{\theta_0 }{2}\right) \sin^6\left(\frac{\theta_0
}{2}\right)\right)-\frac{i}{4\sqrt{2}}  \sqrt{\frac{7}{3}} \sin\left(\frac{n \pi }{3}\right) \left(\cos^7\left(\frac{\theta_0}{2}\right)+i e^{-7 i \phi_0 } \sin^7\left(\frac{\theta_0 }{2}\right)\right), \\ \nonumber
\end{eqnarray}
\begin{eqnarray}\nonumber
b_{3n}&=&-\frac{5~i}{4} \sqrt{\frac{7}{2}} \sin\left(\frac{n \pi }{3}\right) \left( e^{-3 i \phi_0 } \cos^4\left(\frac{\theta_0
}{2}\right) \sin^3\left(\frac{\theta_0 }{2}\right)-i  e^{-4 i \phi_0 } \cos^3\left(\frac{\theta_0 }{2}\right) \sin^4\left(\frac{\theta_0
}{2}\right)\right)+\frac{1}{4}\sqrt{\frac{21}{2}} \left(e^{i n \pi }+3 \cos\left(\frac{n \pi }{3}\right)\right)\\  \nonumber && \left( e^{-2 i \phi_0 } \cos^5\left(\frac{\theta_0
}{2}\right) \sin^2\left(\frac{\theta_0 }{2}\right)+i  e^{-5 i \phi_0 } \cos^2\left(\frac{\theta_0 }{2}\right) \sin^5\left(\frac{\theta_0
}{2}\right)\right)+
\frac{\sqrt{3}}{4}  \left(i \left(\cos\left(\frac{n \pi }{3}\right)-\cos(n \pi )\right)+\sin(n \pi)\right)\\  \nonumber && \sqrt{\frac{7}{2}}\left(
e^{-i \phi_0 } \cos^6\left(\frac{\theta_0 }{2}\right) \sin\left(\frac{\theta_0 }{2}\right)-i  e^{-6 i \phi_0 } \cos\left(\frac{\theta_0
}{2}\right) \sin^6\left(\frac{\theta_0 }{2}\right)\right)+\frac{\sqrt{7}}{4\sqrt{2}}  \sin\left(\frac{n \pi }{3}\right) \left(\cos^7\left(\frac{\theta_0}{2}\right)+i e^{-7 i \phi_0 } \sin^7\left(\frac{\theta_0 }{2}\right)\right),\\ \nonumber
b_{4n}&=&\frac{1}{12}\sqrt{\frac{35}{2}} \left(7 e^{i n \pi }+5 \cos\left(\frac{n \pi }{3}\right)\right) \left( e^{-3 i \phi_0
} \cos^4\left(\frac{\theta_0 }{2}\right) \sin^3\left(\frac{\theta_0 }{2}\right)-i e^{-4 i \phi_0 } \cos^4\left(\frac{\theta_0}{2}\right) \sin^4\left(\frac{\theta_0 }{2}\right)\right)-\frac{i}{4}\sqrt{\frac{105}{2}} \sin\left(\frac{n \pi }{3}\right) \\  \nonumber && \left(
e^{-2 i \phi_0 } \cos^5\left(\frac{\theta_0 }{2}\right) \sin^2\left(\frac{\theta_0 }{2}\right)+i  e^{-5 i \phi_0 } \cos^2\left(\frac{\theta_0
}{2}\right) \sin^5\left(\frac{\theta_0 }{2}\right)\right)+
\frac{1}{4} \sqrt{\frac{35}{6}} \sin\left(\frac{n \pi }{3}\right) \left(e^{-i \phi_0 } \cos^6\left(\frac{\theta_0 }{2}\right)
\sin\left(\frac{\theta_0 }{2}\right)-\right.\\  \nonumber && \left. i e^{-6 i \phi_0 } \cos\left(\frac{\theta_0 }{2}\right) \sin^6\left(\frac{\theta_0
}{2}\right)\right)+\frac{\sqrt{35}}{12\sqrt{2}}  \left(i \left(\cos\left(\frac{n \pi }{3}\right)-\cos[n \pi ]\right)+\sin[n \pi ]\right)
\left(\cos^7\left(\frac{\theta_0}{2}\right)+i e^{-7 i \phi_0 } \sin^7\left(\frac{\theta_0 }{2}\right)\right), \\ \nonumber
b_{5n}&=&\frac{35~i  e^{-\frac{5}{6} i n \pi }}{24\sqrt{2}}  \left(1+e^{\frac{2 i n \pi }{3}}-2 e^{\frac{4 i n \pi }{3}}\right) \left(
e^{-3 i \phi_0 } \cos^4\left(\frac{\theta_0 }{2}\right) \sin^3\left(\frac{\theta_0 }{2}\right)+i  e^{-4 i \phi_0 } \cos^3\left(\frac{\theta_0
}{2}\right) \sin^4\left(\frac{\theta_0 }{2}\right)\right)+
\frac{7~i  e^{-\frac{5}{6} i n \pi }}{8} \sqrt{\frac{3}{2}} \left(-1+e^{\frac{2 i n \pi }{3}}\right) \\  \nonumber && \left( e^{-2 i \phi_0 } \cos^5\left(\frac{\theta_0
}{2}\right) \sin^2\left(\frac{\theta_0 }{2}\right)-i e^{-5 i \phi_0 } \cos^2\left(\frac{\theta_0 }{2}\right) \sin^5\left(\frac{\theta_0
}{2}\right)\right)+\frac{7~e^{-\frac{5}{6} i n \pi }}{8\sqrt{2}}   \left(-1+e^{\frac{2 i n \pi }{3}}\right) \left( e^{-i
\phi_0 } \cos^6\left(\frac{\theta_0 }{2}\right) \sin\left(\frac{\theta_0 }{2}\right)\right.\\  \nonumber && \left.+i e^{-6 i \phi_0 } \cos\left(\frac{\theta_0
}{2}\right) \sin^6\left(\frac{\theta_0 }{2}\right)\right)+\frac{e^{-\frac{5}{6} i n \pi }}{24\sqrt{2}}  \left(7+7 e^{\frac{2 i n \pi }{3}}+10 e^{\frac{4 i n \pi }{3}}\right) \left(\cos^7\left(\frac{\theta_0}{2}\right)+i e^{-7 i \phi_0 } \sin^7\left(\frac{\theta_0 }{2}\right)\right),\\ \nonumber
b_{6n}&=&\frac{\sqrt{7}~e^{-\frac{5}{6} i n \pi }}{8\sqrt{2}}  \left[5i\sqrt{3}  \left(-1+e^{\frac{2 i n \pi }{3}}\right) \left(
e^{-3 i \phi_0 } \cos^4\left(\frac{\theta_0 }{2}\right) \sin^3\left(\frac{\theta_0 }{2}\right)+i  e^{-4 i \phi_0 } \cos^3\left(\frac{\theta_0
}{2}\right) \sin^4\left(\frac{\theta_0 }{2}\right)\right)+
3i  \left(1+e^{\frac{2 i n \pi }{3}}-2 e^{\frac{4 i n \pi }{3}}\right)\right. \\  \nonumber && \left. \left(
e^{-2 i \phi_0 } \cos^5\left(\frac{\theta_0 }{2}\right) \sin^2\left(\frac{\theta_0 }{2}\right)-i  e^{-5 i \phi_0 } \cos^2\left(\frac{\theta_0
}{2}\right) \sin^5\left(\frac{\theta_0 }{2}\right)\right)+ \left(1+e^{\frac{2 i n \pi }{3}}+6 e^{\frac{4 i n \pi }{3}}\right) \left(e^{-i \phi_0
} \cos^6\left(\frac{\theta_0 }{2}\right) \sin\left(\frac{\theta_0 }{2}\right)+\right.\right.\\  \nonumber && \left.\left.i  e^{-6 i \phi_0 } \cos\left(\frac{\theta_0
}{2}\right) \sin^6\left(\frac{\theta_0 }{2}\right)\right)+ \frac{\left(-1+e^{\frac{2 i n \pi
}{3}}\right)}  {\sqrt{3}}\left(\cos^7\left(\frac{\theta_0}{2}\right)+i e^{-7 i \phi_0 } \sin^7\left(\frac{\theta_0 }{2}\right)\right)\right],\\ \nonumber
b_{7n}&=&\frac{\sqrt{7} e^{-\frac{5}{6} i n \pi }}{8} \left[\frac{5}{\sqrt{2}} \left(-1+e^{\frac{2 i n \pi }{3}}\right) \left( e^{-3
i \phi_0 } \cos^4\left(\frac{\theta_0 }{2}\right) \sin^3\left(\frac{\theta_0 }{2}\right)+i e^{-4 i \phi_0 } \cos^3\left(\frac{\theta_0
}{2}\right) \sin^4\left(\frac{\theta_0 }{2}\right)\right)+
\sqrt{\frac{3}{2}} \left(3+3 e^{\frac{2 i n \pi }{3}}+2 e^{\frac{4 i n \pi }{3}}\right) \right.\\  \nonumber &&\left. \left( e^{-2 i
\phi_0 } \cos^5\left(\frac{\theta_0 }{2}\right) \sin^2\left(\frac{\theta_0 }{2}\right)-i  e^{-5 i \phi_0 } \cos^2\left(\frac{\theta_0
}{2}\right) \sin^5\left(\frac{\theta_0 }{2}\right)\right)-
i\sqrt{\frac{3}{2}}\left(1+e^{\frac{2 i n \pi }{3}}-2 e^{\frac{4 i n \pi }{3}}\right) \left(
e^{-i \phi_0 } \cos^6\left(\frac{\theta_0 }{2}\right) \sin\left(\frac{\theta_0 }{2}\right)+\right.\right.\\  \nonumber && \left.\left.i e^{-6 i \phi_0 } \cos\left(\frac{\theta_0
}{2}\right) \sin^6\left(\frac{\theta_0 }{2}\right)\right)- \frac{i  \left(-1+e^{\frac{2 i n \pi }{3}}\right)}{\sqrt{2}}
\left(\cos^7\left(\frac{\theta_0}{2}\right)+i e^{-7 i \phi_0 } \sin^7\left(\frac{\theta_0 }{2}\right)\right)\right]~~ \mbox{and}\\ \nonumber
b_{8n}&=&\frac{\sqrt{5} e^{-\frac{5}{6} i n \pi }}{8}\left[\frac{1}{3} \sqrt{\frac{7}{2}} \left(5+5 e^{\frac{2 i n \pi }{3}}+14 e^{\frac{4 i n \pi }{3}}\right) \left(
e^{-3 i \phi_0 } \cos^4\left(\frac{\theta_0 }{2}\right) \sin^3\left(\frac{\theta_0 }{2}\right)+i  e^{-4 i \phi_0 } \cos^3\left(\frac{\theta_0
}{2}\right) \sin^4\left(\frac{\theta_0 }{2}\right)\right)-\sqrt{\frac{21}{2}}\left(1-e^{\frac{2 i n \pi }{3}}\right)\right. \\  \nonumber && \left. \left( e^{-2 i \phi_0 } \cos^5\left(\frac{\theta_0
}{2}\right) \sin^2\left(\frac{\theta_0 }{2}\right)-i  e^{-5 i \phi_0 } \cos^2\left(\frac{\theta_0 }{2}\right) \sin^5\left(\frac{\theta_0
}{2}\right)\right)- i \sqrt{\frac{7}{6}}  \left(-1+e^{\frac{2 i n \pi }{3}}\right) \left(
e^{-i \phi_0 } \cos^6\left(\frac{\theta_0 }{2}\right) \sin\left(\frac{\theta_0 }{2}\right)+\right.\right.\\  \nonumber && \left.\left.i  e^{-6 i \phi_0 } \cos\left(\frac{\theta_0
}{2}\right) \sin^6\left(\frac{\theta_0 }{2}\right)\right)-
\frac{ \sqrt{7}~i}{3\sqrt{2}} \left(1+e^{\frac{2 i n \pi }{3}}-2 e^{\frac{4 i n \pi }{3}}\right) \left(\cos^7\left(\frac{\theta_0}{2}\right)+i e^{-7 i \phi_0 } \sin^7\left(\frac{\theta_0 }{2}\right)\right)\right].
\end{eqnarray}
\subsection{Expressions for the Coefficients for the case of  $8$ qubit}
In  the $\ket{\phi}$ basis, the unitary operator $\mathcal{U}$ is block diagonal in two blocks $\mathcal{U}_{+}~(\mathcal{U}_{-})$ with dimension $5\times5~(4\times4)$  \cite{sharma2024exactly}. The blocks are given as follows:
\begin{equation}
\mathcal{U}_{+}={\frac{1}{8}\left(
\begin{array}{ccccc}
 -1& 0 & -2\sqrt{7}  & 0 & -\sqrt{35}  \\
 0 & -6i & 0 & -2i \sqrt{7}  & 0 \\
-2 \sqrt{7}  & 0 & -4 & 0 & 2\sqrt{5}  \\
 0 & -2i \sqrt{7}  & 0 & 6i & 0 \\
 - \sqrt{35}  & 0 & 2 \sqrt{5}  & 0 & -3 \\
\end{array}
\right)}~~~ \mbox{and}
\end{equation}
\begin{equation}
\mathcal{U}_{-}={\frac{1}{2 \sqrt{2}}\left(
\begin{array}{cccc}
 0 & 1 & 0 & \sqrt{7}  \\
i & 0 & i\sqrt{7}  & 0 \\
 0 & \sqrt{7}  & 0 & -1 \\
 i \sqrt{7}  & 0 & -i & 0 \\
\end{array}
\right)}.
\end{equation}
The eigenvalues of $\mathcal{U_{+}}$ $\left(\mathcal{U_{-}}\right)$ are $\left\lbrace-1,-1,i,-i,1\right\rbrace$ $\left(e^{\frac{i \pi }{4}}\left\lbrace-1,-1,1,1\right\rbrace\right)$ and the eigenvectors are\\
 $\left\lbrace\left[\sqrt{\frac{5}{7}},0,0,0,1 \right]^T, \left[\frac{2}{\sqrt{7}},0,1,0,0 \right]^T,\left[0, -\frac{1}{\sqrt{7}},0,1,0 \right]^T,\left[0,\sqrt{7},0,1,0 \right]^T,\left[-\sqrt{\frac{7}{5}},0,\frac{2}{\sqrt{5}},0,1 \right]^T \right\rbrace \left(\left\lbrace\left[-\frac{2-2 i}{\sqrt{7}},\frac{1}{\sqrt{7}},0,1 \right]^T\right.\right.$\\ $\left.\left.,\left[\frac{1}{\sqrt{7}},-\frac{2+2 i}{\sqrt{7}},1,0 \right]^T,\left[\frac{2-2 i}{\sqrt{7}},\frac{1}{\sqrt{7}},0,1 \right]^T,\left[\frac{1}{\sqrt{7}},\frac{2+2 i}{\sqrt{7}},1,0 \right]^T\right\rbrace \right)$. The $n$th time evolution of $\mathcal{U}_{+}$ and $\mathcal{U}_{-}$ is given as follows:
\begin{equation}
 \mathcal{U}_{+}^n=\frac{1}{16}{\left(
\begin{array}{ccccc}
  \left(7+9 e^{i n \pi }\right) & 0 & 2 \sqrt{7} \left(-1+e^{i n \pi }\right) & 0 &\sqrt{35} \left(-1+e^{i n\pi }\right) \\
 0 & 2~ e^{\frac{i n \pi }{2}} \left(1+7 e^{i n \pi }\right) & 0 & 2 \sqrt{7} e^{\frac{i n \pi }{2}} \left(-1+e^{i n \pi }\right)
& 0 \\
 2 \sqrt{7} \left(-1+e^{i n \pi }\right) & 0 & 4 \left(1+3 e^{i n \pi }\right) & 0 & -2\sqrt{5} \left(-1+e^{i n \pi
}\right) \\
 0 & 2 \sqrt{7} e^{\frac{i n \pi }{2}} \left(-1+e^{i n \pi }\right) & 0 & 2~ e^{\frac{i n \pi }{2}} \left(7+e^{i n \pi }\right)
& 0 \\
  \sqrt{35} \left(-1+e^{i n \pi }\right) & 0 & -2 \sqrt{5} \left(-1+e^{i n \pi }\right) & 0 &  \left(5+11 e^{i
n \pi }\right) \\
\end{array}
\right)} ~~\mbox{and}
\end{equation}
\begin{equation}
 \mathcal{U}_{-}^n=\frac{e^{\frac{i n \pi }{4}}}{2}{\left(
\begin{array}{cccc}
   \left(1+e^{i n \pi }\right) & -\frac{e^{\frac{-i n \pi}{4}}}{2\sqrt{2}} \left(-1+e^{i n \pi
}\right) & 0 &- \frac{e^{-\frac{i n \pi}{4}}}{2\sqrt{2}}\sqrt{7}  \left(-1+e^{i n \pi }\right) \\
- \frac{e^{\frac{i n \pi}{4}}}{2\sqrt{2}} \left(-1+e^{i n \pi }\right) &  \left(1+e^{i n \pi
}\right) & -\frac{e^{\frac{i n \pi}{4}}}{2\sqrt{2}} \sqrt{7}  \left(-1+e^{i n \pi }\right) & 0 \\
 0 &-\frac{e^{\frac{-i n \pi}{4}}}{2\sqrt{2}} \sqrt{7}  \left(-1+e^{i n \pi }\right) & \left(1+e^{i
n \pi }\right) & \frac{e^{\frac{-i n \pi}{4}}}{2\sqrt{2}}  \left(-1+e^{i n \pi }\right) \\
 -\frac{e^{\frac{i n \pi}{4}}}{2\sqrt{2}} \sqrt{7}  \left(-1+e^{i n \pi }\right) & 0 & \frac{e^{\frac{i n \pi}{4}}}{2\sqrt{2}}  \left(-1+e^{i n \pi }\right) &  \left(1+e^{i n \pi }\right) \\
\end{array}
\right)}.
\end{equation}
Applying the unitary operator $\mathcal{U}$ $n$ times on the state $\ket{\psi}$ we get,
\begin{eqnarray}
\ket{\psi_n}&=&\mathcal{U}^n\ket{\psi}\\ \nonumber
&=& f_{1n}\ket{\phi_0^+}+f_{2n}\ket{\phi_1^+}+f_{3n}\ket{\phi_2^+}+f_{4n}\ket{\phi_3^+} +f_{5n}\ket{\phi_4^+}+f_{6n}\ket{\phi_0^-}+f_{7n}\ket{\phi_1^-}+f_{8n}\ket{\phi_2^-}+f_{9n}\ket{\phi_3^-},\nonumber
\end{eqnarray}
where the coefficients $f_{jn}$ can be computed as follows:
\begin{equation}\label{Eq:qkt62}
f_{jn}=\sum_{q=1}^{\frac{N+2}{2}}\mathcal{U}^n_{j,q}~ a_q+\sum_{q=\frac{N+4}{2}}^{N+1}\mathcal{U}^n_{j,q}~ b_{q-\frac{N+2}{2}},1\leq j\leq N+1.
\end{equation}
The expressions of the coefficients $f_{jn}$ for $8$ qubits can be calculated using  Eqs. (\ref{Eq:arbitaray12}), (\ref{Eq:arbitaray15}), (\ref{Eq:arbitaray42}) and (\ref{Eq:qkt62}), as follows:
\begin{eqnarray}\nonumber
f_{1n}&=&\frac{{e^{-8 i \phi_0 }}}{128 \sqrt{2}} \left[8 \left(7+9 (-1)^n\right) \left(e^{8 i \phi_0 } \cos^8\left(\frac{\theta_0 }{2}\right)+ \sin^8\left(\frac{\theta_0 }{2}\right)\right)+7 \left(-1+(-1)^n\right) e^{2 i \phi_0 } \sin^2\left(\theta_0 \right) \left(8 e^{4 i \phi_0 } \cos^4\left(\frac{\theta_0
}{2}\right)+\right.\right. \\ \nonumber && \left.\left.8 \sin^4\left(\frac{\theta_0 }{2}\right)+5 e^{2 i \phi_0 } \sin^2\left(\theta_0 \right)\right)\right],\\ \nonumber
\end{eqnarray}
\begin{eqnarray}\nonumber
f_{2n}&=&\frac{e^{-\frac{1}{2} i (n \pi +14 \phi_0 )}}{8}  \left(- \sin^2\left(\frac{\theta_0 }{2}\right) +e^{2 i \phi_0 } \cos^2\left(\frac{\theta_0
}{2}\right)\right) \left[7\sin^4\left(\frac{\theta_0 }{2}\right) +e^{i n \pi }\sin^4\left(\frac{\theta_0 }{2}\right) +e^{2 i \phi_0 } \cos^2\left(\frac{\theta_0
}{2}\right)\left(14\sin^2\left(\frac{\theta_0 }{2}\right) \right.\right. \\ \nonumber && \left.\left.-6 e^{i n \pi }\sin^2\left(\frac{\theta_0 }{2}\right)+e^{2 i \phi_0 }
\left(7+e^{i n \pi }\right) \cos^2\left(\frac{\theta_0 }{2}\right)\right)\right]\sin(\theta_0),\\ \nonumber
f_{3n}&=&\frac{e^{-8 i \phi_0 }}{8} \sqrt{\frac{7}{2}} \left(\left(-1+(-1)^n\right)\left( e^{8 i \phi_0 } \cos^8\left(\frac{\theta_0 }{2}\right)+
\sin^8\left(\frac{\theta_0 }{2}\right)\right)+\left(1+3 (-1)^n\right) e^{2 i \phi_0 } \sin^4\left(\frac{\theta_0 }{2}\right) \sin^2\left(\theta_0 \right)+\frac{e^{4 i \phi_0 }}{2}
 \left(5\sin^2\left(\frac{\theta_0 }{2}\right)\right.\right. \\ \nonumber && \left.\left.-5 (-1)^n\sin^2\left(\frac{\theta_0 }{2}\right)+2 \left(1+3 (-1)^n\right) e^{2
i \phi_0 } \cos^2\left(\frac{\theta_0 }{2}\right)\right) \cos^2\left(\frac{\theta_0 }{2}\right)\sin^2\left(\theta_0 \right)\right),\\ \nonumber
f_{4n}&=&\frac{\sqrt{7} \sin(\theta_0) e^{-7 i \phi_0 }}{4}  \left(-\sin^2\left(\frac{\theta_0 }{2}\right)+e^{2 i \phi_0 } \cos^2\left(\frac{\theta_0
}{2}\right)\right) \left(\left((-i)^n+3 i^n\right) e^{2 i \phi_0 } \cos^2\left(\frac{\theta_0 }{2}\right)\sin^2\left(\frac{\theta_0 }{2}\right)-i
\sin\left(\frac{n \pi }{2}\right)\right.\\ \nonumber && \left.\left(\sin^4\left(\frac{\theta_0 }{2}\right)-i e^{4 i \phi_0 } \cos^4\left(\frac{\theta_0 }{2}\right)\right) \right),\\ \nonumber
f_{5n}&=&\frac{e^{-8 i \phi_0 }}{128} \sqrt{\frac{35}{2}}  \left(8 \left(-1+(-1)^n\right) \left(e^{8 i \phi_0 } \cos^8\left(\frac{\theta_0 }{2}\right)+\sin^8\left(\frac{\theta_0
}{2}\right)-e^{2 i \phi_0 } \sin^4\left(\frac{\theta_0 }{2}\right) \sin^2\left(\theta_0 \right)\right)+16e^{4 i \phi_0 }\left(5\sin^2\left(\frac{\theta_0 }{2}\right)\right.\right.\\ \nonumber &&\left.\left.+11~e^{ i n\pi }  \sin^2\left(\frac{\theta_0 }{2}\right) +2 \left(1-e^{ i n\pi }\right) e^{2 i \phi_0 } \cos^2\left(\frac{\theta_0 }{2}\right)\right) \cos^4\left(\frac{\theta_0
}{2}\right)\sin^2\left(\frac{\theta_0 }{2}\right)\right),\\ \nonumber
f_{6n}&=&\frac{e^{-\frac{3}{4} i n \pi -8 i \phi_0 } }{4} \left[\sqrt{2} e^{8 i \phi_0 } \left(1+e^{i n \pi }\right) \cos^8\left(\frac{\theta_0 }{2}\right)+(1-i)
\left(-1+e^{i n \pi }\right)\left( e^{7 i \phi_0 } \cos^7\left(\frac{\theta_0 }{2}\right) \sin\left(\frac{\theta_0 }{2}\right)+ e^{i \phi_0 } \cos\left(\frac{\theta_0 }{2}\right) \sin^7\left(\frac{\theta_0 }{2}\right)\right.\right.\\ \nonumber
&&\left.\left.+7 \left(e^{5 i \phi_0}\cos^5\left(\frac{\theta_0 }{2}\right) \sin^3\left(\frac{\theta_0 }{2}\right)+e^{3 i \phi_0 }  \cos^3\left(\frac{\theta_0 }{2}\right) \sin^5\left(\frac{\theta_0 }{2}\right)\right)\right)-\sqrt{2}
\left(1+e^{i n \pi }\right) \sin^8\left(\frac{\theta_0 }{2}\right)\right],\\ \nonumber
f_{7n}&=&\left(\frac{1}{16}+\frac{i}{16}\right) e^{-\frac{3}{4} i n \pi -8 i \phi_0 } \left[\sqrt{2} \left(-1+e^{i n \pi }\right)\left(-\sin^8\left(\frac{\theta_0}{2}\right)+e^{8 i \phi_0 }\cos^8\left(\frac{\theta_0 }{2}\right)-14 \left(e^{2 i \phi_0 }  \cos^2\left(\frac{\theta_0 }{2}\right) \sin^6\left(\frac{\theta_0 }{2}\right)+
e^{6 i \phi_0 } \cos^6\left(\frac{\theta_0 }{2}\right)\right.\right.\right.\\ \nonumber
&&\left.\left.\left.\sin^2\left(\frac{\theta_0 }{2}\right)\right)\right)+(8-8 i)  \left(1+e^{i n \pi }\right) \left(e^{i \phi_0 }\cos\left(\frac{\theta_0 }{2}\right)\sin^7\left(\frac{\theta_0 }{2}\right)+e^{7 i \phi_0 }  \cos^7\left(\frac{\theta_0 }{2}\right)\sin\left(\frac{\theta_0 }{2}\right)\right)\right], \\ \nonumber
f_{8n}&=&\left(\frac{1}{8}-\frac{i}{8}\right) \sqrt{7} e^{-\frac{3}{4} i n \pi -7 i \phi_0 } \left(\left(1-e^{i
n \pi }\right)\sin^2\left(\frac{\theta_0 }{2}\right)+2 e^{\frac{1}{4} i \pi } e^{i \phi_0 } \left(1+e^{i n \pi }\right) \cos\left(\frac{\theta_0
}{2}\right)\sin\left(\frac{\theta_0 }{2}\right)+e^{2 i \phi_0 } \left(-1+e^{i n \pi }\right) \cos^2\left(\frac{\theta_0 }{2}\right)\right)
 \\ \nonumber && \left(-\sin^4\left(\frac{\theta_0 }{2}\right)+e^{4 i \phi_0 } \cos^4\left(\frac{\theta_0 }{2}\right)\right)\sin\left(\theta_0\right)~~~\mbox{and}\\ \nonumber
f_{9n}&=&\left(\frac{1}{256}+\frac{i}{256}\right) \sqrt{7} e^{-\frac{3}{4} i n \pi -8 i \phi_0 } \left(1-\cos(\theta_0 )+e^{2
i \phi_0 } (1+\cos(\theta_0 )\right) \left[8 \sqrt{2}  \left(-1+e^{i n \pi }\right)\left( e^{6 i \phi_0 }\cos^6\left(\frac{\theta_0 }{2}\right)-
\sin^6\left(\frac{\theta_0 }{2}\right)+\right.\right.\\ \nonumber
&&\left.\left.6 \sqrt{2}~ e^{2 i \phi_0 }  \sin^2\left(\frac{\theta_0 }{2}\right) \sin(\theta_0)\right)-2e^{3 i \phi_0 } \left((-8+8 i) \left(1+e^{i n \pi }\right)\sin\left(\frac{\theta_0 }{2}\right)+3 \sqrt{2} e^{i \phi_0 } \left(-1+e^{i n \pi
}\right) \cos\left(\frac{\theta_0 }{2}\right)\right) \cos\left(\frac{\theta_0 }{2}\right)\sin^2\left(\theta_0 \right)\right].
\end{eqnarray}
\subsection{Expressions for the Coefficients for the case of  $9$ qubit}
The unitary operator $\mathcal{U}$ is block diagonal in two blocks $\mathcal{U}_{+}~(\mathcal{U}_{-})$ having dimension $5\times5~(5\times5)$ in $\ket{\phi}$ basis \cite{sharma2024exactly}. The blocks are given as follows:
\begin{equation}
\mathcal{U}_{\pm}=\frac {e^ {\frac{{\mp}i \pi  }{4}}}{16}\left(
\begin{array}{ccccc}
1 & \mp 3i~ & {6} & \mp{2i} \sqrt{21}  & {3} \sqrt{14}  \\
 \pm 3 i & -{7}  & \pm {10 i}  & -2\sqrt{21}  & \pm {i}\sqrt{14}  \\
 6  & \mp {10 i} & 8  & 0 & -2{\sqrt{14}} \\
\pm {2i} \sqrt{21} & - 2\sqrt{21} & 0 & 8 & \mp {2i}\sqrt{6} \\
 3\sqrt{14}  & \mp i\sqrt{14}  & -2{\sqrt{14}}  & \pm {2i} \sqrt{6}  & {6}  \\
\end{array}
\right).
\end{equation}
The eigenvalues of $\mathcal{U_{+}}\left(\mathcal{U_{-}}\right)$ are  ${e^{\frac{3i \pi }{4}}\left\{1,e^{-\frac{2i \pi }{3}},e^{-\frac{2i \pi }{3}},e^{\frac{2i \pi }{3}},e^{\frac{2i \pi }{3}}\right\}}\left(e^{\frac{i \pi }{4}}\left\{-1,e^{\frac{i \pi }{3}},e^{\frac{i \pi }{3}},e^{-\frac{i \pi }{3}},e^{-\frac{i \pi }{3}}\right\}\right)$
and the eigenvectors are $\left\lbrace\left[\frac{3}{\sqrt{14}},\frac{\pm i}{\sqrt{14}},0,0,1 \right]^T,\left[-\frac{3}{\sqrt{14}},\mp \frac{5 i}{\sqrt{14}},2 i \sqrt{\frac{6}{7}},0,1 \right]^T,\left[\pm\frac{1}{\sqrt{7}},-\frac{3 i}{\sqrt{7}},\pm i \sqrt{\frac{3}{7}},1,0 \right]^T,\left[-\frac{3}{\sqrt{14}},\mp \frac{5 i}{\sqrt{14}},-2 i \sqrt{\frac{6}{7}},0,1 \right]^T
,\left[\mp\frac{1}{\sqrt{7}}, \frac{3 i}{\sqrt{7}},\pm i \sqrt{\frac{3}{7}},1,0\right]^T\right\rbrace$. The $n${th} time evolution of  $\mathcal{U}_{\pm}$ is given as follows:
\begin{equation}
 \mathcal{U}_{\pm}^n= e^{\frac{3 i n\pi}{4}}{\left[
\begin{array}{ccccc}
 \frac{1}{8} \left(3+5 \cos \left(\frac{2 n \pi }{3}\right)\right) & \mp\frac{1}{4} i \sin^2 \left(\frac{n \pi }{3}\right) & \frac{1}{4}\sqrt{3} \sin \left(\frac{2 n \pi }{3}\right) & \mp\frac{1}{4} i \sqrt{7} \sin \left(\frac{2 n \pi }{3}\right) & \frac{1}{2} \sqrt{\frac{7}{2}}\sin ^2\left(\frac{n \pi }{3}\right) \\
 \pm \frac{1}{4} i \sin^2 \left(\frac{n \pi }{3}\right) & \frac{1}{24} \left(1+23 \cos\left(\frac{2 n \pi }{3}\right)\right) & \pm {5 i \sin \left(\frac{2n\pi }{3}\right)}/{4 \sqrt{3}} & -\frac{1}{4} \sqrt{7} \sin \left(\frac{2 n \pi }{3}\right) & \pm\frac{1}{6} i \sqrt{\frac{7}{2}} \sin^2\left(\frac{n\pi }{3}\right) \\
 -\frac{1}{4} \sqrt{3} \sin \left(\frac{2 n \pi }{3}\right) & \pm {5 i \sin \left(\frac{2 n \pi }{3}\right)}/{4 \sqrt{3}} & \cos\left(\frac{2n \pi }{3}\right) & 0 & \frac{1}{2} \sqrt{\frac{7}{6}} \sin \left(\frac{2 n \pi }{3}\right) \\
 \mp \frac{1}{4} i \sqrt{7} \sin \left(\frac{2 n \pi }{3}\right) & \frac{1}{4} \sqrt{7} \sin \left(\frac{2 n \pi }{3}\right) & 0 & \cos \left(\frac{2n \pi }{3}\right) & \pm {i \sin \left(\frac{2 n \pi }{3}\right)}/{2 \sqrt{2}} \\
 \frac{1}{2} \sqrt{\frac{7}{2}} \sin^2 \left(\frac{n \pi }{3}\right) & \mp\frac{1}{6} i \sqrt{\frac{7}{2}} \sin^2 \left(\frac{n \pi }{3}\right)& -\frac{1}{2} \sqrt{\frac{7}{6}} \sin \left(\frac{2 n \pi }{3}\right) & \pm {i \sin \left(\frac{2 n \pi }{3}\right)}/{2 \sqrt{2}} & \frac{1}{12}\left(7+5 \cos \left(\frac{2 n \pi }{3}\right)\right) \\
\end{array}
\right]}.
\end{equation}
The state $\ket{\psi_n}$ is obtained by applying the $n${th} iteration of unitary operator $\mathcal{U}$ to the initial state $\ket{\psi}$ and is expressed as,
\begin{eqnarray}\nonumber
\ket{\psi_n}&=&\mathcal{U}^n\ket{\psi}\\ \nonumber
&=& \bar{c}_{1n}\ket{\phi_1^+}+\bar{c}_{2n}\ket{\phi_2^+}+\bar{c}_{3n}\ket{\phi_3^+}+\bar{c}_{4n}\ket{\phi_4^+}+\bar{c}_{5n}\ket{\phi_5^+} +\bar{c}_{6n}\ket{\phi_1^-}+\bar{c}_{7n}\ket{\phi_2^-}+\bar{c}_{8n}\ket{\phi_3^-}+\bar{c}_{9n}\ket{\phi_4^-}+\bar{c}_{10n}\ket{\phi_5^-},
\end{eqnarray}
where the coefficients  $\bar{c}_{jn}$ are  calculated as follows:
\begin{equation}\label{Eq:qkt535}
 \bar{c}_{jn}=\sum_{q=1}^{\frac{N+1}{2}}\mathcal{U}^n_{j,q}~ a_q+\sum_{q=\frac{N+3}{2}}^{N+1}\mathcal{U}^n_{j,q}~ b_{q-\frac{N+1}{2}},1\leq j\leq N+1.
 \end{equation}
The expressions of the coefficients $\bar{c}_{jn}$ for $9$ qubits can be calculated using   Eqs. (\ref{Eq:arbitaray22}), (\ref{Eq:arbitaray32}) and (\ref{Eq:qkt535}), as follows:
\begin{eqnarray}\nonumber
\bar{c}_{1n}&=&\frac{21e^{-\frac{7}{12} i n \pi }}{8\sqrt{2}}  \left(-1-e^{\frac{2 i n \pi }{3}}+2 e^{\frac{4 i n \pi }{3}}\right)
\left(e^{-4 i \phi_0 } \cos^5\left(\frac{\theta_0 }{2}\right) \sin^4\left(\frac{\theta_0 }{2}\right)-i  e^{-5 i \phi_0 }
\cos^4\left(\frac{\theta_0 }{2}\right) \sin^5\left(\frac{\theta_0 }{2}\right)\right)+\frac{7\sqrt{6}i~ e^{-\frac{1}{4} i n \pi }}{4} \\ \nonumber
&& \sin\left(\frac{n \pi }{3}\right) \left( e^{-3 i \phi_0 } \cos^6\left(\frac{\theta_0
}{2}\right) \sin^3\left(\frac{\theta_0 }{2}\right)+i e^{-6 i \phi_0 } \cos^3\left(\frac{\theta_0 }{2}\right) \sin^6\left(\frac{\theta_0
}{2}\right)\right)-\frac{3\sqrt{6}~ e^{-\frac{1}{4} i n \pi } }{4} \sin\left(\frac{n \pi }{3}\right) \left( e^{-2 i \phi_0 } \cos^7\left(\frac{\theta_0
}{2}\right) \right.\\ \nonumber
&&\left.\sin^2\left(\frac{\theta_0 }{2}\right)- i  e^{-7 i \phi_0 } \cos^2\left(\frac{\theta_0 }{2}\right) \sin^7\left(\frac{\theta_0
}{2}\right)\right)+\frac{3e^{-\frac{1}{4} i n \pi }}{8\sqrt{2}}  \left(i \left(\cos\left(\frac{n \pi }{3}\right)-\cos(n \pi )\right)+\sin(n \pi )\right)
\left( e^{-i \phi_0 } \cos^8\left(\frac{\theta_0 }{2}\right) \sin\left(\frac{\theta_0 }{2}\right)\right.\\ \nonumber
&&\left.+i e^{-8 i \phi_0
} \cos\left(\frac{\theta_0 }{2}\right) \sin^8\left(\frac{\theta_0 }{2}\right)\right)+\frac{e^{-\frac{1}{4} i n \pi }}{8\sqrt{2}} \left(3
e^{i n \pi }+5 \cos\left(\frac{n \pi }{3}\right)\right) \left(\cos^9\left(\frac{\theta_0 }{2}\right)-i e^{-9 i \phi_0
} \sin^9\left(\frac{\theta_0 }{2}\right)\right),\\ \nonumber
\bar{c}_{2n}&=&\frac{7~i~ e^{-\frac{7}{12} i n \pi }}{8\sqrt{2} } \left(-1-e^{\frac{2 i n \pi }{3}}+2 e^{\frac{4 i n \pi }{3}}\right)
\left( e^{-4 i \phi_0 } \cos^5\left(\frac{\theta_0 }{2}\right) \sin^4\left(\frac{\theta_0 }{2}\right)-i  e^{-5 i \phi_0 }
\cos^4\left(\frac{\theta_0 }{2}\right) \sin^5\left(\frac{\theta_0 }{2}\right)\right)+\frac{7\sqrt{6} e^{-\frac{1}{4} i n \pi }}{4}  \sin\left(\frac{n \pi }{3}\right) \\ \nonumber
&&\left( e^{-3 i \phi_0 } \cos^6\left(\frac{\theta_0
}{2}\right) \sin^3\left(\frac{\theta_0 }{2}\right)+i  e^{-6 i \phi_0 } \cos^3\left(\frac{\theta_0 }{2}\right) \sin^6\left(\frac{\theta_0
}{2}\right)\right)+\frac{ i e^{-\frac{1}{4} i n \pi }}{8\sqrt{2}} \left(e^{i n \pi }-\cos\left(\frac{n \pi }{3}\right)\right) \left(\cos^9\left(\frac{\theta_0 }{2}\right)-i e^{-9 i \phi_0
} \sin^9\left(\frac{\theta_0 }{2}\right)\right)\\ \nonumber && -\frac{15 i e^{-\frac{1}{4} i n \pi }} {4}\sqrt{\frac{2}{3}}\sin\left(\frac{n \pi }{3}\right) \left(e^{-2 i \phi_0 } \cos^7\left(\frac{\theta_0
}{2}\right) \sin^2\left(\frac{\theta_0 }{2}\right)- i  e^{-7 i \phi_0 } \cos^2\left(\frac{\theta_0 }{2}\right) \sin^7\left(\frac{\theta_0
}{2}\right)\right)+\frac{e^{-\frac{1}{4} i n \pi }}{8}\left(e^{i n \pi }+23 \cos\left(\frac{n \pi }{3}\right)\right) \\ \nonumber
&&  \left( e^{-i \phi_0 } \cos^8\left(\frac{\theta_0
}{2}\right) \sin\left(\frac{\theta_0 }{2}\right)+i e^{-8 i \phi_0 } \cos\left(\frac{\theta_0 }{2}\right) \sin^8\left(\frac{\theta_0
}{2}\right)\right),\\ \nonumber
\bar{c}_{3n}&=&e^{-\frac{1}{4} i n \pi }\left[-\frac{21}{2\sqrt{6}}   \sin\left(\frac{n \pi }{3}\right) \left( e^{-4
i \phi_0 } \cos^5\left(\frac{\theta_0 }{2}\right) \sin^4\left(\frac{\theta_0 }{2}\right)- i  e^{-5 i \phi_0 } \cos^4\left(\frac{\theta_0
}{2}\right) \sin^5\left(\frac{\theta_0 }{2}\right)\right)+3 \sqrt{2}\cos\left(\frac{n \pi }{3}\right) \left(
e^{-2 i \phi_0 }\cos^7\left(\frac{\theta_0 }{2}\right) \right.\right.\\ \nonumber
&&\left.\left. \sin^2\left(\frac{\theta_0 }{2}\right)-i e^{-7 i \phi_0 } \cos^2\left(\frac{\theta_0
}{2}\right)\sin^7\left(\frac{\theta_0 }{2}\right)\right)+\frac{\sqrt{3}}{4\sqrt{2}}   \sin\left(\frac{n \pi }{3}\right)\left(\cos^9\left(\frac{\theta_0 }{2}\right)-i e^{-9 i \phi_0
} \sin^9\left(\frac{\theta_0 }{2}\right)\right)-\frac{5\sqrt{3} i}{4\sqrt{2}} \sin\left(\frac{n \pi }{3}\right) \right.\\ \nonumber
&&\left.\left( e^{-i \phi_0 } \cos^8\left(\frac{\theta_0 }{2}\right)
\sin\left(\frac{\theta_0 }{2}\right)+i e^{-8 i \phi_0 } \cos\left(\frac{\theta_0 }{2}\right) \sin^8\left(\frac{\theta_0
}{2}\right)\right)\right],\\ \nonumber
\end{eqnarray}
\begin{eqnarray}\nonumber
\bar{c}_{4n}&=&e^{-\frac{1}{4} i n \pi}\left[-\frac{ 3 \sqrt{7}i}{2 \sqrt{2}}  \sin\left(\frac{n \pi }{3}\right) \left( e^{-4 i \phi_0 } \cos^5\left(\frac{\theta_0
}{2}\right) \sin^4\left(\frac{\theta_0 }{2}\right)-i e^{-5 i \phi_0 } \cos^4\left(\frac{\theta_0 }{2}\right) \sin^5\left(\frac{\theta_0
}{2}\right)\right)+\sqrt{42}\cos\left(\frac{n \pi }{3}\right) \left( e^{-3 i \phi_0 } \cos^6\left(\frac{\theta_0
}{2}\right) \right.\right.\\ \nonumber
&&\left.\left.\sin^3\left(\frac{\theta_0 }{2}\right)+i  e^{-6 i \phi_0 } \cos^3\left(\frac{\theta_0 }{2}\right) \sin^6\left(\frac{\theta_0
}{2}\right)\right)-
\frac{3\sqrt{7} }{4\sqrt{2}}  \sin\left(\frac{n \pi }{3}\right) \left(e^{-i \phi_0 } \cos^8\left(\frac{\theta_0
}{2}\right) \sin\left(\frac{\theta_0 }{2}\right)+i e^{-8 i \phi_0 } \cos\left(\frac{\theta_0 }{2}\right) \sin^8\left(\frac{\theta_0
}{2}\right)\right)\right.\\ \nonumber
&&\left.+\frac{ i \sqrt{7}}{4\sqrt{2}} \sin\left(\frac{n \pi }{3}\right) \left(\cos^9\left(\frac{\theta_0 }{2}\right)-i e^{-9 i \phi_0
} \sin^9\left(\frac{\theta_0 }{2}\right)\right)\right],\\ \nonumber
\bar{c}_{5n}&=&\frac{3 \sqrt{7}~e^{-\frac{1}{4} i n \pi }}{12}  \left(7 e^{i n \pi }+5 \cos\left(\frac{n \pi }{3}\right)\right) \left(
e^{-4 i \phi_0 } \cos^5\left(\frac{\theta_0 }{2}\right) \sin^4\left(\frac{\theta_0 }{2}\right)-i  e^{-5 i \phi_0 } \cos^4\left(\frac{\theta_0
}{2}\right) \sin^5\left(\frac{\theta_0 }{2}\right)\right)-\frac{\sqrt{21}~i e^{-\frac{1}{4} i n \pi }}{2 } \sin\left(\frac{n \pi }{3}\right)\\ \nonumber
&& \left(
e^{-3 i \phi_0 } \cos^6\left(\frac{\theta_0 }{2}\right) \sin^3\left(\frac{\theta_0 }{2}\right)+i e^{-6 i \phi_0 } \cos^3\left(\frac{\theta_0
}{2}\right) \sin^6\left(\frac{\theta_0 }{2}\right)\right)+\frac{\sqrt{21}~e^{-\frac{1}{4} i n \pi }}{2}  \sin\left(\frac{n \pi }{3}\right) \left( e^{-2 i \phi_0 } \cos^7\left(\frac{\theta_0
}{2}\right) \sin^2\left(\frac{\theta_0 }{2}\right)-\right.\\ \nonumber
&&\left. i e^{-7 i \phi_0 } \cos^2\left(\frac{\theta_0 }{2}\right) \sin^7\left(\frac{\theta_0
}{2}\right)\right)+\frac{ \sqrt{7}~e^{-\frac{1}{4} i n \pi }}{8}  \left(i \left(\cos\left(\frac{n \pi }{3}\right)-\cos(n \pi )\right)\right) \left(e^{-i \phi_0 } \cos^8\left(\frac{\theta_0 }{2}\right) \sin\left(\frac{\theta_0 }{2}\right)+ i
e^{-8 i \phi_0 } \cos\left(\frac{\theta_0 }{2}\right) \right.\\ \nonumber
&& \left.\sin^8\left(\frac{\theta_0 }{2}\right)\right)+\frac{\sqrt{7}~e^{-\frac{7}{12} i n \pi }}{16}
 \left(-1+e^{\frac{2 i n \pi }{3}}\right) \left(1+2 e^{\frac{2 i n \pi }{3}}\right) \left(\cos^9\left(\frac{\theta_0 }{2}\right)-i e^{-9 i \phi_0
} \sin^9\left(\frac{\theta_0 }{2}\right)\right),\\ \nonumber
\bar{c}_{6n}&=&\frac{21~e^{-\frac{1}{12} i n \pi }}{8\sqrt{2}}   \left(-1-e^{\frac{2 i n \pi }{3}}+2 e^{\frac{4 i n \pi }{3}}\right)
\left(e^{-4 i \phi_0 } \cos^5\left(\frac{\theta_0 }{2}\right) \sin^4\left(\frac{\theta_0 }{2}\right)+ i  e^{-5 i \phi_0 }
\cos^4\left(\frac{\theta_0 }{2}\right) \sin^5\left(\frac{\theta_0 }{2}\right)\right)-\frac{7i \sqrt{6}~e^{\frac{i n \pi }{4}}}{4}   \sin\left(\frac{n \pi }{3}\right) \\ \nonumber
&&\left(e^{-3 i \phi_0 } \cos^6\left(\frac{\theta_0
}{2}\right) \sin^3\left(\frac{\theta_0 }{2}\right)-i  e^{-6 i \phi_0 } \cos^3\left(\frac{\theta_0 }{2}\right) \sin^6\left(\frac{\theta_0
}{2}\right)\right)-\frac{3\sqrt{6}~ e^{\frac{i n \pi }{4}}}{4}  \sin\left(\frac{n \pi }{3}\right) \left(e^{-2 i \phi_0 } \cos^9\left(\frac{\theta_0
}{2}\right) \sin^2\left(\frac{\theta_0 }{2}\right)+\right.\\ \nonumber
&&\left. i e^{-7 i \phi_0 } \cos^2\left(\frac{\theta_0 }{2}\right) \sin^7\left(\frac{\theta_0
}{2}\right)\right)+\frac{3i~e^{\frac{i n \pi }{4}}}{8\sqrt{2}}   \left(e^{i n \pi }-\cos\left(\frac{n \pi }{3}\right)\right) \left(e^{-i \phi_0 } \cos^8\left(\frac{\theta_0
}{2}\right) \sin\left(\frac{\theta_0 }{2}\right)-i e^{-8 i \phi_0 } \cos\left(\frac{\theta_0 }{2}\right) \sin^8\left(\frac{\theta_0
}{2}\right)\right)+\\ \nonumber
&&\frac{e^{\frac{i n \pi }{4}}}{8\sqrt{2}}  \left(3 e^{i n \pi }+5 \cos\left(\frac{n \pi }{3}\right)\right) \left(\cos^9\left(\frac{\theta_0 }{2}\right)-i e^{-9 i \phi_0
} \sin^9\left(\frac{\theta_0 }{2}\right)\right), \\ \nonumber
\bar{c}_{7n}&=&e^{\frac{i n \pi }{4}}\left[\left(i \left(\cos\left(\frac{n \pi }{3}\right)-\cos(n
\pi)\right)\right)\left( \frac{7}{4\sqrt{2}}\left( e^{-4 i \phi_0 } \cos^5\left(\frac{\theta_0 }{2}\right) \sin^4\left(\frac{\theta_0 }{2}\right)+
i e^{-5 i \phi_0 } \cos^4\left(\frac{\theta_0 }{2}\right) \sin^5\left(\frac{\theta_0 }{2}\right)\right)+\frac{1}{8\sqrt{2}}\left(i e^{-9 i \phi_0 } \sin^9\left(\frac{\theta_0 }{2}\right)\right.\right.\right.\\ \nonumber
&&\left.\left.\left.+\cos^9\left(\frac{\theta_0 }{2}\right)\right)\right)+\frac{1 }{8\sqrt{2}} \left(e^{i n \pi }+23 \cos\left(\frac{n \pi }{3}\right)\right)\left(e^{-i \phi_0 } \cos^8\left(\frac{\theta_0
}{2}\right)\sin\left(\frac{\theta_0 }{2}\right)- i e^{-8 i \phi_0 } \cos\left(\frac{\theta_0 }{2}\right) \sin^8\left(\frac{\theta_0
}{2}\right)\right)+\frac{7\sqrt{6}}{4}  \sin\left(\frac{n \pi }{3}\right)\right.\\ \nonumber && \left.\left(e^{-3 i \phi_0 } \cos^6\left(\frac{\theta_0
}{2}\right)\sin^3\left(\frac{\theta_0 }{2}\right)-ie^{-6 i \phi_0 } \cos^3\left(\frac{\theta_0 }{2}\right) \sin^6\left(\frac{\theta_0
}{2}\right)+5 i \left( e^{-2 i \phi_0 } \cos^7\left(\frac{\theta_0
}{2}\right) \sin^2\left(\frac{\theta_0 }{2}\right)+i  e^{-7 i \phi_0 } \cos^2\left(\frac{\theta_0 }{2}\right) \sin^7\left(\frac{\theta_0
}{2}\right)\right)\right)\right],\\ \nonumber
\bar{c}_{8n}&=&e^{\frac{i n \pi }{4}}\left\lbrace \sin\left(\frac{n \pi }{3}\right)\left[-\frac{21}{2\sqrt{6}}  \left(e^{-4 i
\phi_0 } \cos^5\left(\frac{\theta_0 }{2}\right) \sin^4\left(\frac{\theta_0 }{2}\right)+i  e^{-5 i \phi_0 } \cos^4\left(\frac{\theta_0
}{2}\right) \sin^5\left(\frac{\theta_0 }{2}\right)\right)+\frac{\sqrt{3}}{4\sqrt{2}}   \left(\cos^9\left(\frac{\theta_0 }{2}\right)-i e^{-9 i \phi_0
} \sin^9\left(\frac{\theta_0 }{2}\right)\right)\right.\right.\\ \nonumber &&\left.\left.+\frac{5 i\sqrt{3}}{4 \sqrt{2}} \left(e^{-i \phi_0 } \cos^8\left(\frac{\theta_0 }{2}\right)
\sin\left(\frac{\theta_0 }{2}\right)-i e^{-8 i \phi_0 } \cos\left(\frac{\theta_0 }{2}\right) \sin^8\left(\frac{\theta_0
}{2}\right)\right)\right]+3 \sqrt{2} \cos\left(\frac{n \pi }{3}\right) \left( e^{-2
i \phi_0 } \cos^7\left(\frac{\theta_0 }{2}\right) \sin^2\left(\frac{\theta_0 }{2}\right)\right.\right.\\ \nonumber &&\left.\left.+i  e^{-7 i \phi_0 } \cos^2\left(\frac{\theta_0
}{2}\right) \sin^7\left(\frac{\theta_0 }{2}\right)\right)\right\rbrace,\\ \nonumber
\bar{c}_{9n}&=&\frac{3\sqrt{7}e^{\frac{i n \pi }{4}} \sin\left(\frac{n \pi }{3}\right)}{4\sqrt{2}} \left({2}i  \left(  e^{-4 i \phi_0 } \cos^5\left(\frac{\theta_0
}{2}\right) \sin^4\left(\frac{\theta_0 }{2}\right)+ie^{-5 i \phi_0 } \cos^4\left(\frac{\theta_0 }{2}\right) \sin^5\left(\frac{\theta_0
}{2}\right)\right)- \left( e^{-i \phi_0 } \cos^8\left(\frac{\theta_0
}{2}\right) \sin\left(\frac{\theta_0 }{2}\right)-\right.\right.\\  \nonumber&& \left.\left.i e^{-8 i \phi_0 } \cos\left(\frac{\theta_0 }{2}\right) \sin^8\left(\frac{\theta_0
}{2}\right)\right)-{i}  \left(\cos^9\left(\frac{\theta_0 }{2}\right)-i e^{-9 i \phi_0
} \sin^9\left(\frac{\theta_0 }{2}\right)\right)\right) +2\sqrt{21}e^{\frac{i n \pi }{4}} \cos\left(\frac{n \pi }{3}\right) \left( e^{-3 i \phi_0 } \cos^6\left(\frac{\theta_0
}{2}\right) \sin^3\left(\frac{\theta_0 }{2}\right)-\right.\\  \nonumber&& \left.i  e^{-6 i \phi_0 } \cos^3\left(\frac{\theta_0 }{2}\right) \sin^6\left(\frac{\theta_0
}{2}\right)\right)~~~\mbox{and}\\ \nonumber
\end{eqnarray}
\begin{eqnarray}\nonumber
\bar{c}_{10n}&=&\frac{3 \sqrt{7}~e^{\frac{i n \pi }{4}}}{12}  \left(7 e^{i n \pi }+5 \cos\left(\frac{n \pi }{3}\right)\right) \left(
e^{-4 i \phi_0 } \cos^5\left(\frac{\theta_0 }{2}\right) \sin^4\left(\frac{\theta_0 }{2}\right)+ i  e^{-5 i \phi_0 } \cos^4\left(\frac{\theta_0
}{2}\right) \sin^5\left(\frac{\theta_0 }{2}\right)\right)+\frac{3~e^{\frac{i n \pi }{4}}}{2} \sqrt{\frac{7}{3}}  \sin\left(\frac{n \pi }{3}\right)\\ \nonumber
&& \left(e^{-2 i \phi_0 } \cos^7\left(\frac{\theta_0
}{2}\right) \sin^2\left(\frac{\theta_0 }{2}\right)+i e^{-7 i \phi_0 } \cos^2\left(\frac{\theta_0 }{2}\right) \sin^7\left(\frac{\theta_0
}{2}\right)\right)+\frac{i~\sqrt{7}e^{-\frac{1}{12} i n \pi }}{16}  \left(-1-e^{\frac{2 i n \pi }{3}}+2 e^{\frac{4 i n \pi }{3}}\right)
\left(e^{-i \phi_0 } \cos^8\left(\frac{\theta_0 }{2}\right)\right.\\ && \nonumber \left. \sin\left(\frac{\theta_0 }{2}\right)- i e^{-8 i \phi_0
} \cos\left(\frac{\theta_0 }{2}\right) \sin^8\left(\frac{\theta_0 }{2}\right)\right)+\frac{ \sqrt{7}~e^{-\frac{1}{12} i n \pi }}{16}  \left(-1+e^{\frac{2 i n \pi }{3}}\right) \left(1+2 e^{\frac{2 i n \pi }{3}}\right)
\left(\cos^9\left(\frac{\theta_0 }{2}\right)-i e^{-9 i \phi_0
} \sin^9\left(\frac{\theta_0 }{2}\right)\right) \\  \nonumber &&+\frac{i\sqrt{21}~ e^{\frac{i n \pi }{4}}}  {2}\sin\left(\frac{n \pi }{3}\right)\left(
e^{-3 i \phi_0 } \cos^6\left(\frac{\theta_0 }{2}\right) \sin^3\left(\frac{\theta_0 }{2}\right)-i  e^{-6 i \phi_0 } \cos\left(\frac{\theta_0
}{2}\right)^3 \sin^6\left(\frac{\theta_0 }{2}\right)\right).
\end{eqnarray}
\subsection{Expressions for the Coefficients for the case of  $10$ qubit}
The unitary operator $\mathcal{U}$ is block diagonal in two blocks $\mathcal{U}_{+}~(\mathcal{U}_{-})$ having dimension $6\times6~(5\times5)$ in $\ket{\phi}$ basis \cite{sharma2024exactly}. The blocks are given as follows:
\begin{equation}
\mathcal{U}_{+}=\frac{1}{8\sqrt{2}}\left(
\begin{array}{cccccc}
 0 & - \sqrt{{5}}~ e^{\frac{3 i \pi}{4}} & 0 & -2 \sqrt{{15}}~ e^{\frac{3 i \pi}{4}} & 0 & -{3}\sqrt{{7}}~
e^{\frac{3 i \pi}{4}} \\
  \sqrt{{5}} ~e^{\frac{-3 i \pi}{4}} & 0 & 9~ e^{\frac{-3 i \pi}{4}} & 0 &  \sqrt{42} ~e^{\frac{-3 i \pi}{4}} & 0 \\
 0 & -9~ e^{\frac{3 i \pi}{4}} & 0 & -2\sqrt{{3}}~ e^{\frac{3 i \pi}{4}} & 0 &  \sqrt{{35}}
~e^{\frac{3 i \pi}{4}} \\
 2 \sqrt{{15}}~ e^{\frac{-3 i \pi}{4}} & 0 &  2\sqrt{3} e^{\frac{-3 i \pi}{4}} & 0 & - 2\sqrt{14}~ e^{\frac{-3 i \pi}{4}} & 0 \\
 0 & -\sqrt{42}~ e^{\frac{3 i \pi}{4}} & 0 & 2\sqrt{14}~ e^{\frac{3 i \pi}{4}} & 0 & - \sqrt{30}~ e^{\frac{3 i \pi}{4}}
\\
 {3}\sqrt{{7}}~ e^{\frac{-3 i \pi}{4}} & 0 & - \sqrt{{35}}~ e^{\frac{-3 i \pi}{4}} & 0 &  \sqrt{30}~ e^{\frac{-3 i \pi}{4}} & 0 \\
\end{array}
\right)~~~\mbox{and}
\end{equation}
\begin{equation}
\mathcal{U}_{-}={ \frac{1}{16}\left(
\begin{array}{ccccc}
 e^{\frac{3 i \pi}{4}} & 0 & {3} \sqrt{5}~ e^{\frac{3 i \pi}{4}} & 0 &  \sqrt{{210}}~ e^{\frac{3 i \pi}{4}} \\
 0 & -8~ e^{-\frac{3 i \pi}{4}} & 0 & -8 \sqrt{3}~ e^{-\frac{3 i \pi}{4}} & 0 \\
 {3} \sqrt{5}~ e^{\frac{3 i \pi}{4}} & 0 & {13}~ e^{\frac{3 i \pi}{4}} & 0 & - \sqrt{{42}}~ e^{\frac{3 i \pi}{4}}
\\
 0 & -8 \sqrt{3} ~e^{-\frac{3 i \pi}{4}} & 0 & 8~ e^{-\frac{3 i \pi}{4}} & 0 \\
 \sqrt{{210}}~ e^{\frac{3 i \pi}{4}} & 0 & -\sqrt{{42}}~ e^{\frac{3 i \pi}{4}} & 0 & 2~ e^{\frac{3 i \pi}{4}} \\
\end{array}
\right)}.
\end{equation}
 The eigenvalues for $\mathcal{U_{+}}\left(\mathcal{U_{-}}\right)$ are ${\{i,i,i,-i,-i,-i\}}\left(\left\{(-1)^{1/4},-(-1)^{3/4},(-1)^{3/4},(-1)^{3/4},-(-1)^{1/4}\right\}\right)$ and the eigenvectors are $\left\lbrace\left[-\frac{1+i}{\sqrt{7}},-\sqrt{\frac{5}{7}},(1+i) \sqrt{\frac{5}{7}},0,0,1 \right]^T,\left[\sqrt{\frac{5}{42}},(-4+4 i) \sqrt{\frac{2}{21}},3 \sqrt{\frac{3}{14}},0,1,0 \right]^T, \left[\left(-\frac{1}{2}-\frac{i}{2}\right) \sqrt{\frac{5}{3}},\frac{2}{\sqrt{3}},\left(-\frac{1}{2}-\frac{i}{2}\right) \sqrt{3},1,0,0 \right]^T,\right.$\\ $\left.
\left[\frac{1+i}{\sqrt{7}},-\sqrt{\frac{5}{7}},(-1-i) \sqrt{\frac{5}{7}},0,0,1 \right]^T,\left[\sqrt{\frac{5}{42}},(4-4 i) \sqrt{\frac{2}{21}},3 \sqrt{\frac{3}{14}},0,1,0 \right]^T,\left[\left(\frac{1}{2}+\frac{i}{2}\right) \sqrt{\frac{5}{3}},\frac{2}{\sqrt{3}},\left(\frac{1}{2}+\frac{i}{2}\right) \sqrt{3},1,0,0 \right]^T \right\rbrace $\\ $ \left(\left\lbrace\left[0, \sqrt{3},0,1,0\right]^T,\left[-\sqrt{\frac{15}{14}},0,\sqrt{\frac{3}{14}},0,1\right]^T, \left[\sqrt{\frac{14}{15}},0,0,0,1\right]^T,\left[\frac{1}{\sqrt{5}},0,1,0,0\right]^T,\left[0,-\frac{1}{\sqrt{3}},0,1,0\right]^T \right\rbrace\right)$.  The $n$th time evolution  of $\mathcal{U}_{\pm}$ is  given  as follows :
\begin{equation}
{\mathcal{U}_{+}^n=\left[
\begin{array}{cccccc}
 \cos\left(\frac{n \pi }{2}\right) & \bar{a}_1 \left(\frac{1}{16}-\frac{i }{16}\right) \sqrt{5}  & 0 & \bar{a}_1  \left(\frac{1}{8}-\frac{i}{8}\right) \sqrt{15}
& 0 & \bar{a}_1 \left(\frac{3}{16}-\frac{3 i}{16}\right) \sqrt{7} \\
\bar{a}_1  \left(-\frac{1}{16}-\frac{i}{16}\right) \sqrt{5}  & \cos\left(\frac{n \pi }{2}\right) & \bar{a}_1 \left(-\frac{9}{16}-\frac{9 i}{16}\right)  & 0 &
-\frac{\bar{a}_1}{8} (-1)^{1/4} \sqrt{21}  & 0 \\
 0 & \bar{a}_1\left(\frac{9}{16}-\frac{9 i}{16}\right)  & \cos\left(\frac{n \pi }{2}\right) & \bar{a}_1\left(\frac{1}{8}-\frac{i}{8}\right) \sqrt{3} & 0 &
\bar{a}_1 \left(-\frac{1}{16}+\frac{i}{16}\right) \sqrt{35} \\
 \bar{a}_1 \left(-\frac{1}{8}-\frac{i}{8}\right) \sqrt{15}  & 0 & \bar{a}_1 \left(-\frac{1}{8}-\frac{i}{8}\right) \sqrt{3}  & \cos\left(\frac{n \pi }{2}\right)
& \bar{a}_1 \left(\frac{1}{4}+\frac{i}{4}\right) \sqrt{\frac{7}{2}}  & 0 \\
 0 &  \bar{a}_1\left(\frac{1}{8}-\frac{i}{8}\right) \sqrt{\frac{21}{2}} & 0 & \frac{\bar{a}_1 }{4} (-1)^{3/4} \sqrt{7} & \cos \left(\frac{n \pi }{2}\right)&\bar{a}_1  \left(\frac{1}{8}-\frac{i}{8}\right) \sqrt{\frac{15}{2}} \\
\bar{a}_1  \left(-\frac{3}{16}-\frac{3 i}{16}\right) \sqrt{7}  & 0 & \bar{a}_1 \left(\frac{1}{16}+\frac{i}{16}\right) \sqrt{35}  & 0 &  \frac{ -\bar{a}_1}{8} (-1)^{1/4} \sqrt{15}
 & \cos \left(\frac{n \pi }{2}\right) \\
\end{array}
\right]}
\end{equation}
\begin{equation}
\mbox{and}~~\mathcal{U}_{-}^n={ \frac{e^{\frac{ i n \pi }{4}}}{32} \left[
\begin{array}{ccccc}
  e^{\frac{ i n \pi }{2}} \left(17+15 e^{i n \pi }\right) & 0 & -{3\bar{a}_2} \sqrt{5}~ e^{\frac{ i n \pi }{2}}  & 0 & -2{\bar{a}_2}
\sqrt{\frac{105}{2}}~ e^{\frac{i n \pi }{2}}  \\
 0 & 8~ e^{\frac{i n \pi }{4}} \left(3+e^{i n \pi }\right) & 0 & -8{\bar{a}_2} \sqrt{3}   & 0 \\
 -{3\bar{a}_2} \sqrt{5}~ e^{\frac{i n \pi }{2}}  & 0 & e^{\frac{ i n \pi }{2}} \left(29+3 e^{i n \pi }\right) & 0 & 2{\bar{a}_2}
\sqrt{\frac{21}{2}} ~e^{\frac{ i n \pi }{2}}  \\
 0 & -8{\bar{a}_2} \sqrt{3}  & 0 & 8 \left(1+3 e^{i n \pi }\right) & 0 \\
 -2{\bar{a}_2} \sqrt{\frac{105}{2}} e^{\frac{ i n \pi }{2}}  & 0 & 2{\bar{a}_2} \sqrt{\frac{21}{2}} ~e^{\frac{ i n \pi }{2}}  & 0 & 2~
e^{\frac{ i n \pi }{2}} \left(9+7 e^{i n \pi }\right) \\
\end{array}
\right]},
\end{equation}
where \({\bar{a}_1=\sin\left(\frac{n \pi }{2}\right)}\) and $\bar{a}_2= e^{i n \pi }-1$. Applying the unitary operator $\mathcal{U}$ $n$ times on the state $\ket{\psi}$ we get,
\begin{eqnarray}
\ket{\psi_n}&=&\mathcal{U}^n\ket{\psi_0}\\ \nonumber
&=& d_{1n}\ket{\phi_0^+}+d_{2n}\ket{\phi_1^+}+d_{3n}\ket{\phi_2^+}+d_{4n}\ket{\phi_3^+} +d_{5n}\ket{\phi_4^+}+d_{6n}\ket{\phi_5^+}+d_{7n}\ket{\phi_0^-}+d_{8n}\ket{\phi_1^-}\\ \nonumber&&+d_{9n}\ket{\phi_2^-}+d_{10n}\ket{\phi_3^-}+d_{11n}\ket{\phi_4^-},\nonumber
\end{eqnarray}
where the coefficients are given as follows:
\begin{equation}\label{Eq:qkt546}
d_{jn}=\sum_{q=1}^{\frac{N+2}{2}}\mathcal{U}^n_{j,q}~ a_q+\sum_{q=\frac{N+4}{2}}^{N+1}\mathcal{U}^n_{j,q}~ b_{q-\frac{N+2}{2}},1\leq j\leq N+1.
\end{equation}
The expressions of the coefficients $d_{jn}$ for $10$ qubits can be calculated using  Eqs. (\ref{Eq:arbitaray12}), (\ref{Eq:arbitaray15}), (\ref{Eq:arbitaray42}) and (\ref{Eq:qkt546}), as follows:
\begin{eqnarray}\nonumber
d_{1n}&=&\frac{1-i}{8}\left[63 e^{-5 i \phi_0 } \cos^5\left(\frac{\theta_0 }{2}\right) \sin\left(\frac{n
\pi }{2}\right) \sin^5\left(\frac{\theta_0 }{2}\right)+30 \sin\left(\frac{n \pi }{2}\right)
\left( e^{-3 i \phi_0 } \cos^7\left(\frac{\theta_0 }{2}\right) \sin^3\left(\frac{\theta_0 }{2}\right)+ e^{-7 i \phi_0 }
\cos^3\left(\frac{\theta_0 }{2}\right) \sin^7\left(\frac{\theta_0 }{2}\right)\right)\right.\\ \nonumber && \left.+\frac{5}{2}  \sin\left(\frac{n \pi }{2}\right) \left(e^{-i \phi_0 } \cos^9\left(\frac{\theta_0
}{2}\right) \sin\left(\frac{\theta_0 }{2}\right)+ e^{-9 i \phi_0 } \cos\left(\frac{\theta_0 }{2}\right) \sin^9\left(\frac{\theta_0
}{2}\right)\right)\right]+\frac{\cos\left(\frac{n \pi }{2}\right)}{\sqrt{2}} \left(\cos^{10}\left(\frac{\theta_0 }{2}\right)-e^{-10 i
\phi_0 } \sin^{10}\left(\frac{\theta_0 }{2}\right)\right),\\ \nonumber
d_{2n}&=&-\frac{21\sqrt{5}e^{\frac{i \pi }{4} } }{8}\sin\left(\frac{n \pi }{2}\right) \left( e^{-4 i \phi_0
} \cos^6\left(\frac{\theta_0 }{2}\right) \sin^4\left(\frac{\theta_0 }{2}\right)-e^{-6 i \phi_0 } \cos^4\left(\frac{\theta_0 }{2}\right)
\sin^6\left(\frac{\theta_0 }{2}\right)\right)-\frac{27 \sqrt{5}e^{\frac{i \pi }{4}}}{16} \sin\left(\frac{n \pi }{2}\right) \left(
e^{-2 i \phi_0 } \cos^8\left(\frac{\theta_0 }{2}\right) \right.\\ \nonumber && \left.\sin^2\left(\frac{\theta_0 }{2}\right) -e^{-8 i \phi_0 } \cos^2\left(\frac{\theta_0
}{2}\right) \sin^8\left(\frac{\theta_0 }{2}\right)\right)+\cos\left(\frac{n \pi }{2}\right) \left(\sqrt{5} e^{-i \phi_0 } \cos^9\left(\frac{\theta_0 }{2}\right) \sin\left(\frac{\theta_0
}{2}\right)+\sqrt{5} e^{-9 i \phi_0 } \cos\left(\frac{\theta_0 }{2}\right) \sin^9\left(\frac{\theta_0 }{2}\right)\right)
\\ \nonumber&&-\left(\frac{1}{16}+\frac{i}{16}\right)\sqrt{\frac{5}{2}} \sin\left(\frac{n \pi }{2}\right) \left(\cos^{10}\left(\frac{\theta_0 }{2}\right)-e^{-10 i
\phi_0 } \sin^{10}\left(\frac{\theta_0 }{2}\right)\right),\\ \nonumber
d_{3n}&=&\left(-\frac{21}{8}+\frac{21 i}{8}\right) \sqrt{5} e^{-5 i \phi_0 } \cos^5\left(\frac{\theta_0 }{2}\right) \sin\left(\frac{n
\pi }{2}\right) \sin^5\left(\frac{\theta_0 }{2}\right)+\left(\frac{3\sqrt{5}}{4}-\frac{3\sqrt{5}i}{4}\right)  \sin\left(\frac{n \pi }{2}\right)
\left( e^{-3 i \phi_0 } \cos^7\left(\frac{\theta_0 }{2}\right) \sin^3\left(\frac{\theta_0 }{2}\right)+ e^{-7 i \phi_0 }\right.\\ \nonumber && \left.
\cos^3\left(\frac{\theta_0 }{2}\right) \sin^7\left(\frac{\theta_0 }{2}\right)\right)+3 \sqrt{\frac{5}{2}}\cos\left(\frac{n \pi }{2}\right) \left( e^{-2 i \phi_0 } \cos^8\left(\frac{\theta_0 }{2}\right) \sin^2\left(\frac{\theta_0
}{2}\right)- e^{-8 i \phi_0 } \cos^2\left(\frac{\theta_0 }{2}\right) \sin^8\left(\frac{\theta_0 }{2}\right)\right)+\left(\frac{9}{16}-\frac{9
i}{16}\right) \sin\left(\frac{n \pi }{2}\right) \\ \nonumber && \left(\sqrt{5} e^{-i \phi_0 } \cos^9\left(\frac{\theta_0 }{2}\right) \sin\left(\frac{\theta_0
}{2}\right)+\sqrt{5} e^{-9 i \phi_0 } \cos\left(\frac{\theta_0 }{2}\right) \sin^9\left(\frac{\theta_0 }{2}\right)\right),\\ \nonumber
d_{4n}&=&\left(\frac{7}{4}+\frac{7~i}{4}\right) \sqrt{\frac{15}{2}} \sin\left(\frac{n \pi }{2}\right) \left( e^{-4
i \phi_0 } \cos^6\left(\frac{\theta_0 }{2}\right) \sin^4\left(\frac{\theta_0 }{2}\right)- e^{-6 i \phi_0 } \cos^4\left(\frac{\theta_0
}{2}\right) \sin^6\left(\frac{\theta_0 }{2}\right)\right)+2 \sqrt{15}\cos\left(\frac{n \pi }{2}\right) \left( e^{-3 i \phi_0} \cos^7\left(\frac{\theta_0
}{2}\right)\right.\\ \nonumber && \left. \sin^3\left(\frac{\theta_0 }{2}\right)+e^{-7 i \phi_0 } \cos^3\left(\frac{\theta_0 }{2}\right) \sin^{7}\left(\frac{\theta_0
}{2}\right)\right)-\left(\frac{3}{8}+\frac{3~i}{8}\right)\sqrt{\frac{15}{2}} \sin\left(\frac{n \pi }{2}\right) \left( e^{-2 i \phi_0 } \cos^{8}\left(\frac{\theta_0
}{2}\right) \sin^2\left(\frac{\theta_0 }{2}\right)- e^{-8 i \phi_0 } \cos^2\left(\frac{\theta_0 }{2}\right)\right. \\ \nonumber &&\left. \sin^{8}\left(\frac{\theta_0
}{2}\right)\right)-\left(\frac{1}{8}+\frac{i}{8}\right) \sqrt{\frac{15}{2}} \sin\left(\frac{n \pi }{2}\right)\left(\cos^{10}\left(\frac{\theta_0 }{2}\right)-e^{-10 i
\phi_0 } \sin^{10}\left(\frac{\theta_0 }{2}\right)\right),\\ \nonumber
\end{eqnarray}
\begin{eqnarray}\nonumber
d_{5n}&=&\left(\frac{3}{4}-\frac{3 i}{4}\right) \sqrt{\frac{105}{2}} e^{-5 i \phi_0 } \cos^5\left(\frac{\theta_0 }{2}\right) \sin\left(\frac{n
\pi }{2}\right) \sin^5\left(\frac{\theta_0 }{2}\right)+\sqrt{105}\cos\left(\frac{n \pi }{2}\right) \left( e^{-4 i \phi_0 } \cos^6\left(\frac{\theta_0
}{2}\right)\sin^4\left(\frac{\theta_0 }{2}\right)- e^{-6 i \phi_0 } \cos^4\left(\frac{\theta_0 }{2}\right) \right.\\ \nonumber
&&\left.\sin^6\left(\frac{\theta_0
}{2}\right)\right)+\frac{\sqrt{105}~e^{\frac{3i \pi }{4}}}{2}  \sin\left(\frac{n \pi }{2}\right) \left( e^{-3 i \phi_0 } \cos^7\left(\frac{\theta_0
}{2}\right)\sin^3\left(\frac{\theta_0 }{2}\right)+e^{-7 i \phi_0 } \cos^3\left(\frac{\theta_0 }{2}\right) \sin^7\left(\frac{\theta_0
}{2}\right)\right)+\left(\frac{1}{8}-\frac{i}{8}\right) \sqrt{\frac{105}{2}} \sin\left(\frac{n \pi }{2}\right) \\ \nonumber &&  \left(e^{-i \phi_0 }
\cos^9\left(\frac{\theta_0 }{2}\right)\sin\left(\frac{\theta_0 }{2}\right)+ e^{-9 i \phi_0 } \cos\left(\frac{\theta_0 }{2}\right)
\sin^9\left(\frac{\theta_0 }{2}\right)\right),\\ \nonumber
d_{6n}&=&6 \sqrt{7} e^{-5 i \phi_0 } \cos\left(\frac{n \pi }{2}\right) \cos^5\left(\frac{\theta_0 }{2}\right) \sin^5\left(\frac{\theta_0
}{2}\right)-\frac{15\sqrt{15}~e^{\frac{i \pi }{4} }}{8}   \sin\left(\frac{n \pi }{2}\right) \left( e^{-4 i \phi_0 } \cos^6\left(\frac{\theta_0
}{2}\right) \sin^4\left(\frac{\theta_0 }{2}\right)- e^{-6 i \phi_0 } \cos^4\left(\frac{\theta_0 }{2}\right) \right.\\ \nonumber
&& \left.\sin^6\left(\frac{\theta_0
}{2}\right)\right)+\left(\frac{3(1+i)}{16}\right) \sqrt{7} \sin\left(\frac{n \pi }{2}\right)\left(5 \left( e^{-2 i \phi_0 } \cos^8\left(\frac{\theta_0
}{2}\right) \sin^2\left(\frac{\theta_0 }{2}\right)- e^{-8 i \phi_0 } \cos^2\left(\frac{\theta_0 }{2}\right) \sin\left(\frac{\theta_0
}{2}\right)^8\right)- \right. \\ \nonumber && \left.\frac{1}{\sqrt{2}}\left(\cos^{10}\left(\frac{\theta_0 }{2}\right)-e^{-10 i
\phi_0 } \sin^{10}\left(\frac{\theta_0 }{2}\right)\right)\right),\\ \nonumber
d_{7n}&=&\frac{105~ e^{-\frac{1}{4} i n \pi }}{16\sqrt{2}}  \left(-1+e^{i n \pi }\right) \left[\left( e^{-4 i \phi_0
} \cos^6\left(\frac{\theta_0 }{2}\right) \sin^4\left(\frac{\theta_0 }{2}\right)+ e^{-6 i \phi_0 } \cos^4\left(\frac{\theta_0 }{2}\right)
\sin^6\left(\frac{\theta_0 }{2}\right)\right)+\frac{1}{6}\left(e^{-2 i \phi_0 } \cos^8\left(\frac{\theta_0
}{2}\right) \sin^2\left(\frac{\theta_0 }{2}\right) \right.\right.\\ \nonumber && \left. \left.+e^{-8 i \phi_0 } \cos^2\left(\frac{\theta_0 }{2}\right) \sin^8\left(\frac{\theta_0
}{2}\right)\right)\right]+\frac{e^{-\frac{1}{4} i n \pi }}{32\sqrt{2}} \left(15+17 e^{i n \pi }\right) \left(\cos^{10}\left(\frac{\theta_0 }{2}\right)-e^{-10 i
\phi_0 } \sin^{10}\left(\frac{\theta_0 }{2}\right)\right),\\ \nonumber
d_{8n}&=&\frac{3\sqrt{5}~e^{-\frac{3}{4} i n \pi }\left(-1+e^{i n \pi }\right)}{2}    \left( e^{-3 i \phi_0 } \cos^7\left(\frac{\theta_0
}{2}\right) \sin^3\left(\frac{\theta_0 }{2}\right)- e^{-7 i \phi_0 } \cos^3\left(\frac{\theta_0 }{2}\right) \sin^7\left(\frac{\theta_0
}{2}\right)\right)+\frac{\sqrt{5}}{4} \left(3 e^{\frac{1}{4} i n \pi }+e^{-\frac{3}{4} i n \pi }\right)\\ \nonumber &&  \left( e^{-i \phi_0 } \cos^9\left(\frac{\theta_0
}{2}\right) \sin\left(\frac{\theta_0 }{2}\right)- e^{-9 i \phi_0 } \cos\left(\frac{\theta_0 }{2}\right) \sin^9\left(\frac{\theta_0
}{2}\right)\right),\\ \nonumber
d_{9n}&=&\frac{3~e^{-\frac{1}{4} i n \pi }}{32} \sqrt{\frac{5}{2}}\left[ -14 \left(-1+e^{i n \pi }\right) \left( e^{-4 i \phi_0
} \cos^6\left(\frac{\theta_0 }{2}\right) \sin^4\left(\frac{\theta_0 }{2}\right)+ e^{-6 i \phi_0 } \cos^4\left(\frac{\theta_0 }{2}\right)
\sin^6\left(\frac{\theta_0 }{2}\right)\right)+ \left(3+29 e^{i n \pi }\right) \left(
e^{-2 i \phi_0 } \right.\right. \\ \nonumber && \left.\left. \cos^8\left(\frac{\theta_0 }{2}\right) \sin^2\left(\frac{\theta_0 }{2}\right)+ e^{-8 i \phi_0 }\cos^2\left(\frac{\theta_0
}{2}\right)\sin^8\left(\frac{\theta_0 }{2}\right)\right)+ \left(-1+e^{i n \pi }\right) \left(\cos^{10}\left(\frac{\theta_0 }{2}\right)+e^{-10
i \phi_0 } \sin^{10}\left(\frac{\theta_0 }{2}\right)\right)\right],\\ \nonumber
d_{10n}&=&\frac{ \sqrt{15}~e^{-\frac{3}{4} i n \pi } }{2}\left( \left(3+e^{i n \pi }\right) \left( e^{-3 i \phi_0 } \cos^7 \left(\frac{\theta_0
}{2}\right)\sin^3\left(\frac{\theta_0 }{2}\right)-e^{-7 i \phi_0 } \cos^3\left(\frac{\theta_0 }{2}\right) \sin^7 \left(\frac{\theta_0
}{2}\right)\right)-\frac{\left(1-e^{i n \pi }\right)}{2}   \left( e^{-i \phi_0 } \cos^9\left(\frac{\theta_0
}{2}\right)  \right.\right. \\ \nonumber && \left. \left.\sin\left(\frac{\theta_0 }{2}\right)- e^{-9 i \phi_0 } \cos\left(\frac{\theta_0 }{2}\right) \sin^9 \left(\frac{\theta_0
}{2}\right)\right)\right)~~~\mbox{and}\\ \nonumber
d_{11n}&=&\frac{\sqrt{105}~e^{-\frac{1}{4} i n \pi }}{32}\left[\left(14+18 e^{i n \pi }\right) \left( e^{-4 i \phi_0 } \cos^6\left(\frac{\theta_0
}{2}\right) \sin^4\left(\frac{\theta_0 }{2}\right)+ e^{-6 i \phi_0 } \cos^4\left(\frac{\theta_0 }{2}\right) \sin^6\left(\frac{\theta_0
}{2}\right)\right)+ 3\left(1-e^{i n \pi }\right)  \left( e^{-2 i \phi_0
}  \right.\right. \\ \nonumber && \left. \left.\cos^8\left(\frac{\theta_0 }{2}\right)\sin^2\left(\frac{\theta_0 }{2}\right)+e^{-8 i \phi_0 } \cos^2\left(\frac{\theta_0}{2}\right) \sin^8\left(\frac{\theta_0 }{2}\right)\right)+  \left(-1+e^{i n \pi }\right) \left(\cos^{10}\left(\frac{\theta_0 }{2}\right)+e^{-10
i \phi_0 } \sin^{10}\left(\frac{\theta_0 }{2}\right)\right)\right].
\end{eqnarray}
\section{The case for $J=1/2$}
\subsection{Expressions for the Coefficients for the case of $4$ qubit}
Using Eq. (3) from the main text, the unitary operator $\mathcal{U}$ for $4$ qubits in $\ket{\phi}$  basis, for the parameter $J=1/2$ and $\tau=\pi/4$ in $\ket{\phi}$ basis \cite{sharma2024signatures} can be written as follows:
\begin{equation}
\mathcal{U}= \left(
\begin{array}{ccccc}
  -1&  0 & 0 & 0&0 \\
 0 & e^{\frac{-3i\pi}{4}}/2 & \sqrt{3}~ e^{\frac{-3i\pi}{4}}/2 & 0 &0\\
 0 &  \sqrt{3}~ e^{\frac{i\pi}{4}}/2 & - e^{\frac{i\pi}{4}}/2 & 0&0 \\
 0 & 0 & 0  & 0 &1 \\
 0 & 0 & 0  & - e^{\frac{-3i\pi}{4}}&0 \\
\end{array}
\right).
 \end{equation}
 The eigenvalues of the $\mathcal{U}$  are $\left\{\pm e^{-\frac{i  \pi}{8}},e^{-\frac{5i  \pi}{12}},e^{\frac{11i  \pi}{12}},-1\right\}$  and the eigenvectors are  $\left\lbrace\left[0,0,0,-e^{\frac{7i\pi}{8}},1\right]^T,\right.\\ \left.\left[0,0,0,e^{\frac{7i\pi}{8}},1\right]^T,\left[0,-i,1,0,0\right]^T,\left[0,i,1,0,0\right]^T,\left[1,0,0,0,0\right]^T \right\rbrace $. The $n$th time evolution of the  blocks $\mathcal{U_{\pm}}$ is given as follows:
 \begin{equation}
  \left(
\begin{array}{ccccc}
 (-1)^n & 0 & 0 & 0 & 0 \\
 0 & e^{\frac{i n \pi }{4}} \cos\left(\frac{2 n \pi }{3}\right) & -e^{\frac{i n \pi }{4}} \sin\left(\frac{2 n \pi }{3}\right) & 0 & 0
\\
 0 & e^{\frac{i n \pi }{4}} \sin\left(\frac{2 n \pi }{3}\right) & e^{\frac{i n \pi }{4}} \cos\left(\frac{2 n \pi }{3}\right) & 0 & 0
\\
 0 & 0 & 0 & \frac{1}{2} \left((-1)^{n/8}+e^{-\frac{7}{8} i n \pi }\right) & \frac{1}{2} (-1)^{7/8} \left(-(-1)^{n/8}+e^{-\frac{7}{8} i n \pi }\right)
\\
 0 & 0 & 0 & \frac{1}{2} (-1)^{1/8} e^{-\frac{7}{8} i n \pi } \left(-1+e^{i n \pi }\right) & \frac{1}{2} \left((-1)^{n/8}+e^{-\frac{7}{8} i n \pi
}\right) \\
\end{array}
\right).
 \end{equation}
 The analytical calculations of eigenvalues, eigenvectors, and the $n$th time evolution of unitary operator for the parameters $J=1/2$ and $\tau=\pi/4$ are shown in the supplementary material of Ref. \cite{sharma2024signatures}. Here, we write  them again for  better understanding and clarity. Applying the unitary operator $\mathcal{U}$ $n$ times on the state $\ket{\psi}$ we get,
\begin{eqnarray}
\ket{\psi_n}&=&\mathcal{U}^n\ket{\psi}\\ \nonumber
&=& \bar{p}_{1n}\ket{\phi_0^+}+\bar{p}_{2n}\ket{\phi_1^+}+\bar{p}_{3n}\ket{\phi_2^+}+\bar{p}_{4n}\ket{\phi_0^-} +\bar{p}_{5n}\ket{\phi_1^-},\nonumber
\end{eqnarray}
where the coefficients  are given as follows:
\begin{equation}\label{Eq:qkt556}
\bar{p}_{jn}=\sum_{q=1}^{\frac{N+2}{2}}\mathcal{U}^n_{j,q}~ a_q+\sum_{q=\frac{N+4}{2}}^{N+1}\mathcal{U}^n_{j,q}~ b_{q-\frac{N+2}{2}},1\leq j\leq N+1.
\end{equation}
The expressions of the coefficients $\bar{p}_{jn}$ for $4$ qubits can be calculated using  Eqs. (\ref{Eq:arbitaray12}), (\ref{Eq:arbitaray15}), (\ref{Eq:arbitaray42}) and (\ref{Eq:qkt556}), as follows:
\begin{eqnarray}\nonumber
\bar{p}_{1n}&=&e^{i n \pi } \left(\sqrt{2} e^{-i \phi_0 } \cos^3\left(\frac{\theta_0 }{2}\right) \sin\left(\frac{\theta_0 }{2}\right)-\sqrt{2}
e^{-3 i \phi_0 } \cos\left(\frac{\theta_0 }{2}\right) \sin^3\left(\frac{\theta_0 }{2}\right)\right),\\ \nonumber
\bar{p}_{2n}&=&-\sqrt{6} e^{\frac{i n \pi }{4}-2 i \phi_0 } \cos^2\left(\frac{\theta_0 }{2}\right) \sin\left(\frac{2 n \pi }{3}\right)
\sin^2\left(\frac{\theta_0 }{2}\right)+e^{\frac{i n \pi }{4}} \cos\left(\frac{2 n \pi }{3}\right) \left(\frac{\cos^4\left(\frac{\theta_0
}{2}\right)}{\sqrt{2}}+\frac{e^{-4 i \phi_0 } \sin^4\left(\frac{\theta_0 }{2}\right)}{\sqrt{2}}\right),\\ \nonumber
\bar{p}_{3n}&=&\sqrt{6} e^{\frac{i n \pi }{4}-2 i \phi_0 } \cos\left(\frac{2 n \pi }{3}\right) \cos^2\left(\frac{\theta_0 }{2}\right)
\sin^2\left(\frac{\theta_0 }{2}\right)+e^{\frac{i n \pi }{4}} \sin\left(\frac{2 n \pi }{3}\right) \left(\frac{\cos^4\left(\frac{\theta_0
}{2}\right)}{\sqrt{2}}+\frac{e^{-4 i \phi_0 } \sin^4\left(\frac{\theta_0 }{2}\right)}{\sqrt{2}}\right),\\ \nonumber
\bar{p}_{4n}&=&e^{-\frac{3}{8} i n \pi } \left[\sqrt{2}\cos\left(\frac{n \pi }{2}\right) \left(e^{-i \phi_0 } \cos^3\left(\frac{\theta_0
}{2}\right) \sin\left(\frac{\theta_0 }{2}\right)+ e^{-3 i \phi_0 } \cos\left(\frac{\theta_0 }{2}\right) \sin^3\left(\frac{\theta_0
}{2}\right)\right)+\frac{e^{\frac{3 i \pi }{8}} \sin\left(\frac{n \pi }{2}\right)}{\sqrt{2}} \right.\\ \nonumber &&\left.\left(\cos^4\left(\frac{\theta_0
}{2}\right)-e^{-4 i \phi_0 } \sin^4\left(\frac{\theta_0 }{2}\right)\right)\right]~~~\mbox{and}\\ \nonumber
\bar{p}_{5n}&=&e^{-\frac{3}{8} i n \pi }\left[-\sqrt{2}~e^{-\frac{3 i \pi }{8}} \sin\left(\frac{n \pi }{2}\right) \left( e^{-i \phi_0 } \cos^3\left(\frac{\theta_0
}{2}\right) \sin\left(\frac{\theta_0 }{2}\right)+ e^{-3 i \phi_0 } \cos\left(\frac{\theta_0 }{2}\right) \sin^3\left(\frac{\theta_0
}{2}\right)\right)+ \frac{\cos\left(\frac{n \pi }{2}\right)}{\sqrt{2}} \right.\\ \nonumber &&\left.\left(\cos^4\left(\frac{\theta_0
}{2}\right)-e^{-4 i \phi_0 } \sin^4\left(\frac{\theta_0 }{2}\right)\right)\right].
\end{eqnarray}
\subsection{Expressions for the Coefficients for the case of  $6$ qubit}
In $\ket{\phi}$ basis, the unitary operator $\mathcal{U}$ is block diagonal in two blocks $\mathcal{U}_{+}~(\mathcal{U}_{-})$ having dimension $4\times4~(3\times3)$  \cite{sharma2024signatures}. The blocks are given as follows:
\begin{equation}
\mathcal{U_+}= \frac{-e^{\frac{i \pi }{8}}}{2\sqrt{2}} \left(
\begin{array}{cccc}
 0 &  \sqrt{3} & 0 & \sqrt{5} \\
 \sqrt{3}~e^{\frac{i \pi }{4}} & 0 & \sqrt{5}~e^{\frac{i \pi }{4}} & 0 \\
 0 &  \sqrt{5} & 0 & - \sqrt{3} \\
 -\sqrt{5}~e^{\frac{i \pi }{4}} & 0 & \sqrt{3}~e^{\frac{i \pi }{4}}  & 0 \\
\end{array}
\right)~~\mbox{and}
 \end{equation}
 \begin{equation}
\mathcal{U_-}= \frac{ e^{\frac{i \pi }{8}}}{4} \left(
\begin{array}{ccc}
 1 & 0 &  \sqrt{15} \\
 0 & 4~e^{\frac{i \pi }{4}} & 0 \\
  \sqrt{15} & 0 & -1 \\
\end{array}
\right).
 \end{equation}
The eigenvalues of the $\mathcal{U_+} \left( \mathcal{U_-}\right)$  are $\left\{(-1)^{1/4}, -(-1)^{1/4}, (-1)^{3/4}, -(-1)^{3/4}\right\}\left(\left\{(-1)^{3/8}, (-1)^{1/8}, -(-1)^{1/8}\right\}\right)$  and the eigenvectors are  $\left\lbrace\left[\sqrt{\frac{3}{5}},\sqrt{\frac{3}{5}},\frac{(-1)^{3/8}}{2}  \sqrt{\frac{5}{2}},-\frac{(-1)^{3/8}}{2}  \sqrt{\frac{5}{2}}\right]^T, \left[-2(-1)^{1/8} \sqrt{\frac{2}{5}},2 (-1)^{1/8} \sqrt{\frac{2}{5}},0,0\right]^T, \left[1,1,-\frac{(-1)^{3/8}}{2}  \sqrt{\frac{3}{2}},\frac{(-1)^{3/8}}{2}
\sqrt{\frac{3}{2}}\right]^T,\left[0,0,1,1\right]^T  \right\rbrace $\\$  \left(\left\lbrace\left[-\sqrt{\frac{3}{5}},\sqrt{\frac{5}{3}},0\right]^T, \left[0,0,1\right]^T, \left[1,1,0\right]^T\right\rbrace\right)$. The $n$th time evolution of the  blocks $\mathcal{U_{\pm}}$ is given as follows:
 \begin{equation}
\mathcal{U}_{+}^n= {\frac{e^{\frac{i n \pi }{4}}}{16}\left[
\begin{array}{cccc}
 {a_n}  \left(3+5 e^{\frac{i n \pi }{2}}\right)  & 2\sqrt{{6}}~b_n{e^{\frac{7i  \pi }{8}}}
  & 4{i \sqrt{15} e^{\frac{3i n \pi }{4}}}  \sin\left(\frac{n \pi }{4}\right)\cos\left(\frac{n \pi }{2}\right) & 2\sqrt{{10}}~b_n~{e^{(\frac{ i n \pi }{2}+\frac{3i  \pi }{8})}}   \\
 -2\sqrt{{6}}~b_n~e^{\frac{i  \pi }{8}}   & 8a_n   &-2\sqrt{{10}}~b_n~ {e^{\frac{i  \pi }{8}}}    & 0 \\
 {4i \sqrt{15} e^{\frac{3i n \pi }{4}}}  \sin\left(\frac{n \pi }{4}\right)\cos\left(\frac{n \pi }{2}\right) &2\sqrt{{10}}~b_n{e^{\frac{7i  \pi }{8}}}
 & a_n \left(5+3 e^{\frac{i n \pi
}{2}}\right)  & -2\sqrt{{6}}~b_n~{e^{(\frac{ i n \pi }{2}+\frac{3i  \pi }{8})}}    \\
 -2\sqrt{{10}}~b_n~{e^{(\frac{ i n \pi }{2}+\frac{5i  \pi }{8})}}     & 0 & 2\sqrt{{6}}~b_n~{e^{(\frac{ i n \pi }{2}+\frac{i  \pi }{8})}}
  & {8a_n~e^{\frac{i n \pi }{2}}}  \\
\end{array}
\right]}
 \end{equation}
 \begin{equation}
 \mbox{and}~~~~
 \mathcal{U}_{-}^n =\frac{e^{\frac{i n \pi }{8}}}{8}{\left[
\begin{array}{ccc}
 5 + 3~e^{i n \pi } & 0 &  \sqrt{15}\left( 1-e^{{ i n \pi }}\right) \\
 0 & 8 ~ e^{\frac{ i n \pi }{4}} & 0 \\
 \sqrt{15}\left( 1-e^{{ i n \pi }}\right) & 0 & 3 + 5~ e^{i n \pi } \\
\end{array}
\right]},
 \end{equation}
where $a_n$=$1+e^{i n \pi }$ and $b_n$=$1-e^{i n \pi }$. Applying the unitary operator $\mathcal{U}$ $n$ times on the state $\ket{\psi}$ we get,
\begin{eqnarray}
\ket{\psi_n}&=&\mathcal{U}^n\ket{\psi}\\ \nonumber
&=& \bar{g}_{1n}\ket{\phi_0^+}+\bar{g}_{2n}\ket{\phi_1^+}+\bar{g}_{3n}\ket{\phi_2^+}+\bar{g}_{4n}\ket{\phi_3^+} +\bar{g}_{5n}\ket{\phi_0^-}+\bar{g}_{6n}\ket{\phi_1^-}+\bar{g}_{7n}\ket{\phi_2^-},\nonumber
\end{eqnarray}
where the coefficients are given as follows:
\begin{equation}\label{Eq:qkt568}
\bar{g}_{jn}=\sum_{q=1}^{\frac{N+2}{2}}\mathcal{U}^n_{j,q}~ a_q+\sum_{q=\frac{N+4}{2}}^{N+1}\mathcal{U}^n_{j,q}~ b_{q-\frac{N+2}{2}},1\leq j\leq N+1.
\end{equation}
The expressions of the coefficients $\bar{g}_{jn}$ for $6$ qubits can be calculated using Eqs. (\ref{Eq:arbitaray12}), (\ref{Eq:arbitaray15}), (\ref{Eq:arbitaray42}) and (\ref{Eq:qkt568}), as follows:
\begin{eqnarray}\nonumber
\bar{g}_{1n}&=&\frac{5~ e^{\frac{3 i \pi }{8}-\frac{i n \pi }{4}-3 i \phi_0 }}  {2 \sqrt{2}}\left(-1+e^{i n \pi }\right)\cos^3\left(\frac{\theta_0 }{2}\right)
\sin^3\left(\frac{\theta_0 }{2}\right)+\frac{ {15}i}{8\sqrt{2}}  \left(\sin\left(\frac{n \pi }{4}\right)-\sin\left(\frac{3
n \pi }{4}\right)\right) \left( e^{-2 i \phi_0 } \cos^4\left(\frac{\theta_0 }{2}\right) \sin^2\left(\frac{\theta_0 }{2}\right)-
e^{-4 i \phi_0 } \right.\\ \nonumber &&\left.\cos^2\left(\frac{\theta_0 }{2}\right) \sin^4\left(\frac{\theta_0 }{2}\right)\right)-\frac{3~e^{\frac{7 i \pi }{8}-\frac{3 i n \pi }{4}} \left(1-e^{i n \pi }\right)}{4\sqrt{{2}}}   \left( e^{-i \phi_0 } \cos^5\left(\frac{\theta_0
}{2}\right) \sin\left(\frac{\theta_0 }{2}\right)+ e^{-5 i \phi_0 } \cos\left(\frac{\theta_0 }{2}\right) \sin^5\left(\frac{\theta_0
}{2}\right)\right)+\frac{1}{4\sqrt{2}} \cos\left(\frac{n \pi }{2}\right) \\ \nonumber && \left(4 \cos\left(\frac{n \pi }{4}\right)+i \sin\left(\frac{n
\pi }{4}\right)\right) \left(\cos^6\left(\frac{\theta_0 }{2}\right)-e^{-6 i \phi_0 } \sin^6\left(\frac{\theta_0 }{2}\right)\right),\\ \nonumber
\bar{g}_{2n}&=&-\frac{5\sqrt{3}e^{\frac{i \pi }{8}-\frac{3 i n \pi }{4}}}{8}   \left(-1+e^{i n \pi }\right) \left(
e^{-2 i \phi_0 } \cos^4\left(\frac{\theta_0 }{2}\right) \sin^2\left(\frac{\theta_0 }{2}\right)- e^{-4 i \phi_0 } \cos^2\left(\frac{\theta_0
}{2}\right) \sin^4\left(\frac{\theta_0 }{2}\right)\right)+\frac{\sqrt{3}~e^{-\frac{3}{4} i n \pi }}{2}  \left(1+e^{i n \pi }\right)  \\ \nonumber &&\left(
e^{-i \phi_0 }\cos^5\left(\frac{\theta_0 }{2}\right) \sin\left(\frac{\theta_0 }{2}\right)+e^{-5 i \phi_0 } \cos\left(\frac{\theta_0
}{2}\right) \sin^5\left(\frac{\theta_0 }{2}\right)\right)-\frac{1}{4} \sqrt{\frac{3}{2}} e^{\frac{i \pi }{8}-\frac{3 i n \pi }{4}} \left(-1+e^{i n \pi }\right) \left(\frac{\cos^6\left(\frac{\theta_0
}{2}\right)}{\sqrt{2}}-\frac{e^{-6 i \phi_0 } \sin^6\left(\frac{\theta_0 }{2}\right)}{\sqrt{2}}\right),\\ \nonumber
\end{eqnarray}
\begin{eqnarray}\nonumber
\bar{g}_{3n}&=&-\frac{1}{2} \sqrt{\frac{15}{2}} e^{\frac{3 i \pi }{8}-\frac{i n \pi }{4}-3 i \phi_0 } \left(-1+e^{i n \pi }\right) \cos^3\left(\frac{\theta_0
}{2}\right) \sin^3\left(\frac{\theta_0 }{2}\right)+\frac{1}{4}\sqrt{\frac{15}{2}} \cos\left(\frac{n \pi }{2}\right) \left(4 \cos\left(\frac{n \pi
}{4}\right)-i \sin\left(\frac{n \pi }{4}\right)\right) \left( e^{-2 i \phi_0 } \cos^4\left(\frac{\theta_0 }{2}\right)
\right.\\ \nonumber && \left.\sin^2\left(\frac{\theta_0 }{2}\right)-e^{-4 i \phi_0 } \cos^2\left(\frac{\theta_0 }{2}\right) \sin^4\left(\frac{\theta_0
}{2}\right)\right)+\frac{ e^{\frac{7 i \pi }{8}-\frac{3 i n \pi }{4}}}{4} \sqrt{\frac{15}{2}} \left(-1+e^{i n \pi }\right) \left( e^{-i \phi_0 } \cos^5\left(\frac{\theta_0
}{2}\right) \sin\left(\frac{\theta_0 }{2}\right)+ e^{-5 i \phi_0 } \cos\left(\frac{\theta_0 }{2}\right) \right.\\ \nonumber && \left.\sin^5\left(\frac{\theta_0
}{2}\right)\right)+\frac{i \sqrt{15}}{8}  \left(\sin\left(\frac{n \pi }{4}\right)-\sin\left(\frac{3 n \pi }{4}\right)\right) \left(\frac{\cos^6\left(\frac{\theta_0
}{2}\right)}{\sqrt{2}}-\frac{e^{-6 i \phi_0 } \sin^6\left(\frac{\theta_0 }{2}\right)}{\sqrt{2}}\right),\\ \nonumber
\bar{g}_{4n}&=&\sqrt{5} e^{-3 i \phi_0 } \left(e^{-\frac{1}{4} i n \pi }+e^{\frac{3 i n \pi }{4}}\right) \cos^3\left(\frac{\theta_0 }{2}\right)
\sin^3\left(\frac{\theta_0 }{2}\right)+\frac{3\sqrt{5}~e^{\frac{5 i \pi }{8}-\frac{i n \pi }{4}}}{8} \left(-1+e^{i n \pi }\right)
\left( e^{-2 i \phi_0 } \cos^4\left(\frac{\theta_0 }{2}\right) \sin^2\left(\frac{\theta_0 }{2}\right)-
e^{-4 i \phi_0 } \right.\\ \nonumber && \left.\cos^2\left(\frac{\theta_0 }{2}\right) \sin^4\left(\frac{\theta_0 }{2}\right)\right)-\frac{1}{4} \sqrt{\frac{5}{2}} e^{\frac{5 i \pi }{8}-\frac{i n \pi }{4}} \left(-1+e^{i n \pi }\right) \left(\frac{\cos^6\left(\frac{\theta_0
}{2}\right)}{\sqrt{2}}-\frac{e^{-6 i \phi_0 } \sin^6\left(\frac{\theta_0 }{2}\right)}{\sqrt{2}}\right),\\ \nonumber
\bar{g}_{5n}&=&\frac{ e^{-\frac{7}{8} i n \pi }}{8}  \left[\frac{15}{\sqrt{2}}\left(-1+e^{i n \pi }\right) \left( e^{-2 i \phi_0 } \cos^4\left(\frac{\theta_0
}{2}\right) \sin^2\left(\frac{\theta_0 }{2}\right)+ e^{-4 i \phi_0 } \cos^2\left(\frac{\theta_0 }{2}\right) \sin^4\left(\frac{\theta_0
}{2}\right)\right)+ \frac{\left(3+5 e^{i n \pi }\right)}{\sqrt{2}} \right. \\ \nonumber  &&\left.\left(\cos^6\left(\frac{\theta_0 }{2}\right)+e^{-6
i \phi_0 } \sin^6\left(\frac{\theta_0 }{2}\right)\right)\right],
~~~\bar{g}_{6n}=e^{\frac{3 i n \pi }{8}} \left(\sqrt{3} e^{-i \phi_0 } \cos^5\left(\frac{\theta_0 }{2}\right) \sin\left(\frac{\theta_0
}{2}\right)-\sqrt{3} e^{-5 i \phi_0 } \cos\left(\frac{\theta_0 }{2}\right) \sin^5\left(\frac{\theta_0 }{2}\right)\right)~~\mbox{and}\\ \nonumber
\bar{g}_{7n}&=&\frac{\sqrt{15} e^{-\frac{7}{8} i n \pi }}{8} \left[ \sqrt{\frac{15}{2}}\left(5+3 e^{i n \pi }\right) \left( e^{-2 i \phi_0 } \cos^4\left(\frac{\theta_0
}{2}\right)\sin^2\left(\frac{\theta_0 }{2}\right)+ e^{-4 i \phi_0 } \cos^2\left(\frac{\theta_0 }{2}\right) \sin^4\left(\frac{\theta_0
}{2}\right)\right) -\frac{\left(1-e^{i n \pi }\right)} {\sqrt{2}} \right. \\ \nonumber  &&\left.\left(\cos^6\left(\frac{\theta_0 }{2}\right)+e^{-6
i \phi_0 } \sin^6\left(\frac{\theta_0 }{2}\right)\right)\right].
\end{eqnarray}
\subsection{Expressions for the Coefficients for the case of  $8$ qubit}
In $\ket{\phi}$ basis, the unitary operator $\mathcal{U}$ is block diagonal in two blocks $\mathcal{U}_{+}~(\mathcal{U}_{-})$ having dimension $5\times5~(4\times4)$  \cite{sharma2024signatures}. The blocks are given as follows:
\begin{equation}
\mathcal{U}_{+}=\frac{1}{8}\left(
\begin{array}{ccccc}
 i & 0 & 2~i \sqrt{7} & 0 & i~ \sqrt{35} \\
 0 & -6 ~e^{\frac{i \pi }{4}} & 0 & -2 \sqrt{7}~ e^{\frac{i \pi }{4}} & 0 \\
 -2~i \sqrt{7} & 0 & -4~i & 0 & 2~i \sqrt{5} \\
 0 & -2 \sqrt{7}~ e^{\frac{i \pi }{4}} & 0 & 6~ e^{\frac{i \pi }{4}} & 0 \\
 i~ \sqrt{35} & 0 & -2~i \sqrt{5} & 0 & 3 ~i \\
\end{array}
\right)~ \mbox{and}
\end{equation}
\begin{equation}
\mathcal{U}_{-}=\frac{-1}{2\sqrt{2}}\left(
\begin{array}{cccc}
 0 & i & 0 &  i ~\sqrt{7} \\
 -e^{\frac{i \pi }{4}} & 0 & -\sqrt{7}~ e^{\frac{i \pi }{4}} & 0 \\
 0 & - i ~\sqrt{7} & 0 & i \\
 - \sqrt{7}~ e^{\frac{i \pi }{4}} & 0 & e^{\frac{i \pi }{4}} & 0 \\
\end{array}
\right).
\end{equation}
The eigenvalues of $\mathcal{U_{+}}$ $\left(\mathcal{U_{-}}\right)$ are $\left\lbrace i,e^{\frac{7i \pi }{6}},e^{\frac{- i \pi }{6}},e^{\frac{5 i \pi }{4}},e^{\frac{ i \pi }{4}}\right\rbrace$ $\left(\left\lbrace e^{\frac{7i \pi }{8}},e^{\frac{15i \pi }{8}},e^{\frac{11i \pi }{8}},e^{\frac{3i \pi }{8}}\right\rbrace\right)$ and the eigenvectors are\\
 $\left\lbrace\left[ \sqrt{\frac{5}{7}},-\sqrt{\frac{7}{5}},-\sqrt{\frac{7}{5}},0,0\right]^T, \left[0,0,0,\sqrt{7},-\frac{1}{\sqrt{7}}\right]^T,\left[0,-2
i \sqrt{\frac{3}{5}},2 i \sqrt{\frac{3}{5}},0,0 \right]^T,\left[0,0,0,1,1 \right]^T ,\left[1,1,1,0,0\right]^T\right\rbrace $\\ $\left(\left\lbrace\left[2 (-1)^{5/8} \sqrt{\frac{2}{7}},-2 (-1)^{5/8} \sqrt{\frac{2}{7}},0,0 \right]^T,\left[\frac{1}{\sqrt{7}},\frac{1}{\sqrt{7}},-\sqrt{7},-\sqrt{7} \right]^T,\left[0,0,2
(-1)^{1/8} \sqrt{2},-2 (-1)^{1/8} \sqrt{2} \right]^T,\left[1,1,1,1 \right]^T\right\rbrace \right)$.
The $n$th time evolution of $\mathcal{U}_{+}$ and $\mathcal{U}_{-}$ is given as follows:
\begin{equation}
 \mathcal{U}_{+}^n=\frac{ e^{\frac{i n \pi }{2}}}{24}{\left[
\begin{array}{ccccc}
  \left(10+14 \cos\left(\frac{2 n \pi }{3}\right)\right) & 0 & 4 \sqrt{{21}}  \sin\left(\frac{2 n \pi }{3}\right) & 0 & -{4\sqrt{35}\sin^2\left(\frac{ n \pi }{3}\right)}
\\
 0 & 3{e^{\frac{-i n \pi }{4}}}  \left(1+7 e^{i n \pi }\right) & 0 & {-3\sqrt{7} e^{\frac{-i n \pi }{4}}} \left(1-e^{i n \pi }\right)
& 0 \\
- 4 \sqrt{{21}}  \sin\left(\frac{2 n \pi }{3}\right) & 0 & 24\cos\left(\frac{2 n \pi }{3}\right) & 0 & 4\sqrt{15}  \sin\left(\frac{2 n \pi }{3}\right) \\
 0 & {-3\sqrt{7} e^{\frac{-i n \pi }{4}}} \left(1-e^{i n \pi }\right) & 0 & 3{e^{\frac{-i n \pi }{4}} } \left(7+e^{i n \pi }\right)
& 0 \\
 -{4\sqrt{35}\sin^2\left(\frac{ n \pi }{3}\right)}  & 0 & -4\sqrt{15} \sin\left(\frac{2n\pi}{3}\right) & 0 & \left(14+10 \cos\left(\frac{2 n \pi }{3}\right)\right)
\\
\end{array}
\right]}
\end{equation}
\begin{equation}
 \mbox{and}~~\mathcal{U}_{-}^n=\frac{e^{\frac{7 i n \pi }{8}}}{8}{\left[
\begin{array}{cccc}
   4a_n & -\sqrt{2}b_n e^{\frac{5 i  \pi }{8}}
& 0 &  \sqrt{{14}}b_n e^{\frac{5i  \pi }{8}}  \\
 \sqrt{2}b_n e^{\frac{3 i  \pi }{8}} & \frac{a_n}{2}  \left(7~e^{\frac{-i n
\pi }{2}}+1\right) &  \sqrt{14}b_n e^{(\frac{- i n \pi }{2}+\frac{7 i  \pi }{8})}   &
-i \sqrt{7}\bar{c_n} e^{\frac{ i n \pi }{4}} \\
 0 & -\sqrt{14}b_n e^{(\frac{- i n \pi }{2}+\frac{ i  \pi }{8})} & 4a_n e^{\frac{- i n \pi }{2}}  & \sqrt{2}b_n e^{(\frac{- i n \pi }{2}+\frac{ i  \pi }{8})} \\
 \sqrt{{14}}b_n e^{\frac{3i  \pi }{8}}    & -i \sqrt{7}\bar{c_n} e^{\frac{ i n \pi }{4}} & -\sqrt{2}b_n e^{(\frac{- i n \pi }{2}+\frac{ 7i  \pi }{8})}  & \frac{ a_n}{2} \left(e^{\frac{-i n \pi }{2}}+7 \right)  \\
\end{array}
\right]},
\end{equation}
where $a_n$=$1+e^{i n \pi }$, $b_n=e^{i n \pi }-1$ and $\bar{c}_n=\sin\left(\frac{n\pi}{4}\right)-\sin\left(\frac{3n\pi}{4}\right)$. The state $\ket{\psi_n}$ can be calculated by  applying  unitary operator $\mathcal{U}$ $n$ times  on the state $\psi$ we get,
\begin{eqnarray}
\ket{\psi_n}&=&\mathcal{U}^n\ket{\psi_0}\\ \nonumber
&=& \bar{f}_{1n}\ket{\phi_0^+}+\bar{f}_{2n}\ket{\phi_1^+}+\bar{f}_{3n}\ket{\phi_2^+}+\bar{f}_{4n}\ket{\phi_3^+} +\bar{f}_{5n}\ket{\phi_4^+}+\bar{f}_{6n}\ket{\phi_0^-}+\bar{f}_{7n}\ket{\phi_1^-}+\bar{f}_{8n}\ket{\phi_2^-}+\bar{f}_{8n}\ket{\phi_3^-},\nonumber
\end{eqnarray}
where the coefficients are  given as follows:
\begin{equation}\label{Eq:qkt5799}
\bar{f}_{jn}=\sum_{q=1}^{\frac{N+2}{2}}\mathcal{U}^n_{j,q}~ a_q+\sum_{q=\frac{N+4}{2}}^{N+1}\mathcal{U}^n_{j,q}~ b_{q-\frac{N+2}{2}},1\leq j\leq N+1.
\end{equation}
The expressions of the coefficients $\bar{f}_{jn}$ for $8$ qubits can be calculated using Eqs. (\ref{Eq:arbitaray12}), (\ref{Eq:arbitaray15}), (\ref{Eq:arbitaray42}) and (\ref{Eq:qkt5799}), as follows:
\begin{eqnarray}\nonumber
\bar{f}_{1n}&=&\frac{e^{-\frac{5}{6} i n \pi -8 i \phi_0 } }{24 \sqrt{2}}\left( \left(7+7 e^{\frac{2 i n \pi }{3}}+10 e^{\frac{4 i n \pi }{3}}\right)\left(e^{8 i \phi_0 }
\cos^8\left(\frac{\theta_0 }{2}\right)+ \sin^8\left(\frac{\theta_0 }{2}\right)\right)+7
i \sqrt{3} e^{2 i \phi_0 } \left(-1+e^{\frac{2 i n \pi }{3}}\right) \sin^4\left(\frac{\theta_0 }{2}\right)\sin^2\left(\theta_0 \right)\right.\\ \nonumber
&&\left. +\frac{7~e^{4 i \phi_0 }}{2}  \left(-1+e^{\frac{2 i n \pi }{3}}\right) \left(5\sin^2\left(\frac{\theta_0 }{2}\right)+10 e^{\frac{2 i
n \pi }{3}}\sin^2\left(\frac{\theta_0 }{2}\right)+2 i \sqrt{3} e^{2 i \phi_0 } \cos^2\left(\frac{\theta_0 }{2}\right)\right) \cos^2\left(\frac{\theta_0
}{2}\right)\sin^2\left(\theta_0 \right)\right),\\ \nonumber
\bar{f}_{2n}&=&\frac{ e^{-\frac{3}{4} i n \pi -7 i \phi_0 }}{8} \left(-\sin^2\left(\frac{\theta_0 }{2}\right)+e^{2 i \phi_0 } \cos^2\left(\frac{\theta_0
}{2}\right)\right) \left[7\sin^4\left(\frac{\theta_0 }{2}\right)+e^{i n \pi }\sin^4\left(\frac{\theta_0 }{2}\right)+e^{2 i \phi_0 } \cos^2\left(\frac{\theta_0
}{2}\right) \left(14\sin^2\left(\frac{\theta_0 }{2}\right)\right.\right.\\ \nonumber
&&\left.\left.-6 e^{i n \pi }\sin^2\left(\frac{\theta_0 }{2}\right)+~e^{2 i \phi_0 } \left(7+e^{i
n \pi }\right) \cos^2\left(\frac{\theta_0 }{2}\right)\right)\right] \sin(\theta_0),\\ \nonumber
\bar{f}_{3n}&=&\frac{1}{96} \sqrt{\frac{7}{2}} e^{-\frac{5}{6} i n \pi -8 i \phi_0 } \left[-8 i \sqrt{3}  \left(-1+e^{\frac{2 i n \pi }{3}}\right)\left( e^{8 i \phi_0 }\cos^8\left(\frac{\theta_0 }{2}\right)- \sin^8\left(\frac{\theta_0 }{2}\right)\right)+24 e^{2 i \phi_0 } \left(1+e^{\frac{2 i n \pi }{3}}\right) \sin^4\left(\frac{\theta_0
}{2}\right) \sin^2\left(\theta_0 \right)\right.\\ \nonumber
&&\left.+4~e^{4 i \phi_0 } \left(5 i \sqrt{3} \left(-1+e^{\frac{2 i n \pi }{3}}\right)\sin^2\left(\frac{\theta_0 }{2}\right)+6 e^{2 i \phi_0 } \left(1+e^{\frac{2
i n \pi }{3}}\right) \cos^2\left(\frac{\theta_0 }{2}\right)\right) \cos^2\left(\frac{\theta_0 }{2}\right)\sin^2\left(\theta_0 \right)\right],\\ \nonumber
\bar{f}_{4n}&=&-\frac{\sqrt{7} e^{-\frac{3}{4} i n \pi -7 i \phi_0 }}{8}  \left(-\sin^2\left(\frac{\theta_0 }{2}\right) +e^{2 i \phi_0
} \cos^2\left(\frac{\theta_0 }{2}\right)\right) \left[-\sin^4\left(\frac{\theta_0 }{2}\right)+e^{i n \pi }\sin^4\left(\frac{\theta_0 }{2}\right)-2
e^{2 i \phi_0 } \left(1+3 e^{i n \pi }\right) \cos^2\left(\frac{\theta_0 }{2}\right)\right.\\ \nonumber
&&\left.\sin^2\left(\frac{\theta_0 }{2}\right)+~e^{4 i \phi_0 } \left(-1+e^{i
n \pi }\right) \cos^4\left(\frac{\theta_0 }{2}\right)\right] \sin(\theta_0),\\ \nonumber
\end{eqnarray}
\begin{eqnarray}\nonumber
\bar{f}_{5n}&=&\frac{e^{-\frac{5}{6} i n \pi -8 i \phi_0 }}{24} \sqrt{\frac{35}{2}}  \left(e^{8 i \phi_0 } \left(-1-e^{\frac{2 i n \pi }{3}}+2 e^{\frac{4 i n
\pi }{3}}\right)\left( \cos^8\left(\frac{\theta_0 }{2}\right)+ \sin^8\left(\frac{\theta_0
}{2}\right)\right)-i \sqrt{3} e^{2 i \phi_0 } \left(-1+e^{\frac{2 i n \pi }{3}}\right) \sin^4\left(\frac{\theta_0 }{2}\right) \sin^2\left(\theta_0 \right)+\right.\\ \nonumber
&&\left.\frac{e^{4 i \phi_0 }}{2}  \left(5 \sin^2\left(\frac{\theta_0 }{2}\right)+5 e^{\frac{2 i n \pi }{3}} \sin^2\left(\frac{\theta_0 }{2}\right)+14
e^{\frac{4 i n \pi }{3}} \sin^2\left(\frac{\theta_0 }{2}\right)-2 i \sqrt{3} e^{2 i \phi_0 } \left(-1+e^{\frac{2 i n \pi }{3}}\right) \cos^2\left(\frac{\theta_0
}{2}\right)\right)\cos^2\left(\frac{\theta_0 }{2}\right)\sin^2\left(\theta_0 \right)\right),\\ \nonumber
\bar{f}_{6n}&=&\frac{e^{-\frac{1}{8} i (n \pi +64 \phi_0 )}}{2 \sqrt{2}}\left(e^{8 i \phi_0 } \left(1+e^{i n \pi }\right) \cos^8\left(\frac{\theta_0 }{2}\right)+e^{\frac{5 i \pi }{8}+7 i \phi_0} \left(-1+e^{i
n \pi }\right) \cos^7\left(\frac{\theta_0 }{2}\right) \sin\left(\frac{\theta_0 }{2}\right)+\left(-\sin^5\left(\frac{\theta_0 }{2}\right)-e^{i
n \pi }\sin^5\left(\frac{\theta_0 }{2}\right)\right.\right.\\ \nonumber
&&\left.\left.+e^{\frac{5 i \pi }{8}+ i \phi_0}  \left(-1+e^{i n \pi }\right) \cos\left(\frac{\theta_0 }{2}\right)
\left(\sin^4\left(\frac{\theta_0 }{2}\right)+7 e^{2 i \phi_0 } \cos^2\left(\frac{\theta_0 }{2}\right)\sin^2\left(\frac{\theta_0 }{2}\right)+7
e^{4 i \phi_0 } \cos^4\left(\frac{\theta_0 }{2}\right)\right)\right) \sin^3\left(\frac{\theta_0 }{2}\right)\right),\\ \nonumber
\bar{f}_{7n}&=&\frac{e^{-\frac{5}{8} i n \pi }}{16}  \left[2~  e^{\frac{i n \pi }{2}-\frac{ i \pi }{8}} \left(-1+e^{i n \pi }\right) \left(\cos^8\left(\frac{\theta_0
}{2}\right)-e^{-8 i \phi_0 } \sin^8\left(\frac{\theta_0 }{2}\right)\right)+e^{-7 i \phi_0 } \left(7+e^{\frac{i n \pi }{2}}\right) \left(1+e^{i
n \pi }\right) \left(e^{6 i \phi_0 } \cos^6\left(\frac{\theta_0 }{2}\right)\right.\right.\\ \nonumber
&&+\left.\left.\sin^6\left(\frac{\theta_0 }{2}\right)\right) \sin[\theta_0
]-7 e^{\frac{7 i \pi }{8}} e^{-6 i \phi_0 } \left(-1+e^{i n \pi }\right) \left(-\sin^4\left(\frac{\theta_0 }{2}\right)+e^{4 i \phi_0 } \cos^4\left(\frac{\theta_0
}{2}\right)\right)\sin^2\left(\theta_0 \right)+7 i e^{\frac{3 i n \pi }{4}-4 i \phi_0 } \cos\left(\frac{n \pi }{2}\right)  \right.\\ \nonumber
&&\left.\sin\left(\frac{n
\pi }{4}\right)\sin^3(\theta_0 ) (\cos(\phi_0 )+i \cos(\theta_0) \sin(\phi_0 ))\right],\\ \nonumber
\bar{f}_{8n}&=&\frac{1}{4} \sqrt{\frac{7}{2}} e^{-\frac{5}{8} i n \pi -7 i \phi_0 } \left[-e^{\frac{i \pi }{8}} \left(-1+e^{i n \pi }\right)\sin^2\left(\frac{\theta_0
}{2}\right)+2 e^{i \phi_0 } \left(1+e^{i n \pi }\right) \sin\left(\frac{\theta_0 }{2}\right)\cos\left(\frac{\theta_0 }{2}\right)+e^{\frac{i
\pi }{8}} e^{2 i \phi_0 } \left(-1+e^{i n \pi }\right) \cos^2\left(\frac{\theta_0 }{2}\right)\right] \\ \nonumber
&&\left(-\sin^4\left(\frac{\theta_0 }{2}\right)+e^{4
i \phi_0 } \cos^4\left(\frac{\theta_0 }{2}\right)\right) \sin(\theta_0 )~~~\mbox{and}\\ \nonumber
\bar{f}_{9n}&=&\frac{\sqrt{7} e^{-\frac{5}{8} i n \pi -8 i \phi_0 }}{64}  \left[4  \left(1-e^{i n \pi }\right)\left(
2~e^{\frac{3i \pi }{8}+\frac{i n \pi }{2}} \left(e^{8 i \phi_0 } \cos^8\left(\frac{\theta_0 }{2}\right)-\sin^8\left(\frac{\theta_0 }{2}\right)\right)- e^{\frac{7i \pi }{8}+2i\phi_0}   \left(e^{4 i \phi_0 } \cos^4\left(\frac{\theta_0 }{2}\right)-\sin^4\left(\frac{\theta_0
}{2}\right)\right) \right.\right.\\ \nonumber&&\left.\left.\sin^2\left(\theta_0 \right)\right)+ 4e^{i \phi_0 }  \left(1+e^{i n \pi }\right) \left(-1+e^{\frac{in \pi }{2}}\right)\left(e^{6 i \phi_0 } \cos^6\left(\frac{\theta_0 }{2}\right)+ \sin^6\left(\frac{\theta_0 }{2}\right)\right)
\sin(\theta_0 )+e^{2i \phi_0 } \left(1+7 e^{\frac{i n \pi }{2}}\right)  \left(\sin^2\left(\frac{\theta_0}{2}\right) \right.\right.\\ \nonumber&&\left.\left.+e^{2 i \phi_0 } \cos^2\left(\frac{\theta_0 }{2}\right)\right) \sin^3(\theta_0 )\right].\nonumber
\end{eqnarray}
\subsection{Expressions for the Coefficients for the case of  $10$ qubit}
The unitary operator $\mathcal{U}$ is block diagonalized in two blocks $\mathcal{U}_{+}~(\mathcal{U}_{-})$ having dimension $6\times6~(5\times5)$ in $\ket{\phi}$ basis \cite{sharma2024signatures}. The blocks are given as follows:
\begin{equation}
\mathcal{U}_{+}=-\frac{e^{\frac{3i \pi }{8}}}{8\sqrt{2}} \left(
\begin{array}{cccccc}
 0 &  \sqrt{5} & 0 &  2\sqrt{15} & 0 & 3\text{  }\sqrt{7} \\
 -e^{\frac{i \pi }{4}} \sqrt{5} & 0 & -9 ~e^{\frac{i \pi }{4}} & 0 & -e^{\frac{i \pi }{4}} \sqrt{42} & 0 \\
 0 & 9  & 0 & 2\text{  }\sqrt{3} & 0 & - \sqrt{35} \\
 2~ e^{\frac{i \pi }{4}}\sqrt{15} & 0 & 2~ e^{\frac{i \pi }{4}}\sqrt{3} & 0 & 2~ e^{\frac{i \pi }{4}} \sqrt{14} & 0 \\
 0 &  \sqrt{42} & 0 & - 2\sqrt{14} & 0 &  \sqrt{30} \\
 -3~ e^{\frac{i \pi }{4}} \sqrt{7} & 0 & e^{\frac{i \pi }{4}} \sqrt{35} & 0 & -e^{\frac{i \pi }{4}} \sqrt{30} & 0 \\
\end{array}
\right),
\end{equation}
\begin{equation}
\mbox{and}\mathcal{U}_{-}=\frac{e^{\frac{3i \pi }{8}}}{16} \left(
\begin{array}{ccccc}
 1 & 0 & 3\text{  }\sqrt{5} & 0 & \sqrt{210} \\
 0 & -8~ e^{\frac{i \pi }{4}} & 0 & -8 \sqrt{3}~ e^{\frac{i \pi }{4}}  & 0 \\
 3\text{  }\sqrt{5} & 0 & 13  & 0 & - \sqrt{42} \\
 0 & 8\sqrt{3}~ e^{\frac{i \pi }{4}}  & 0 & -8~ e^{\frac{i \pi }{4}} & 0 \\
  \sqrt{210} & 0 & -\sqrt{42} & 0 & 2 \\
\end{array}
\right).
\end{equation}
The eigenvalues for $\mathcal{U_{+}}\left(\mathcal{U_{-}}\right)$ are ${\{-1,-1,1,1,i,-i\}}\left(\left\{-(-1)^{3/8},(-1)^{3/8},-(-1)^{23/24},(-1)^{3/8},-(-1)^{7/24}\right\}\right)$\\ and the eigenvectors are $\left\lbrace\left[(-1)^{3/8} \sqrt{\frac{2}{7}},\sqrt{\frac{5}{42}},-(-1)^{3/8} \sqrt{\frac{2}{7}},\sqrt{\frac{5}{42}},-\sqrt{\frac{15}{14}},-\sqrt{\frac{15}{14}} \right]^T,\left[-\sqrt{\frac{5}{7}},-\frac{8(-1)^{5/8}}{\sqrt{21}},-\sqrt{\frac{5}{7}},\frac{8 (-1)^{5/8}}{\sqrt{21}},0,0\right]^T,\right.$\\ $\left. \left[-(-1)^{3/8} \sqrt{\frac{10}{7}},3 \sqrt{\frac{3}{14}},(1-i) (-1)^{5/8} \sqrt{\frac{5}{7}},3 \sqrt{\frac{3}{14}},-\sqrt{\frac{3}{14}},-\sqrt{\frac{3}{14}}\right]^T,
\left[0,0,0,0,-\frac{4
(-1)^{1/8}}{\sqrt{7}},\frac{4 (-1)^{1/8}}{\sqrt{7}} \right]^T,\left[0,1,0,1,1,1 \right]^T,\right.$\\ $\left.\left[1,0,1,0,0,0 \right]^T \right\rbrace \left(\left\lbrace\left[0.684653,-0.0809597,-0.0616442~ i,0,0.728869,0\right]^T,\left[0,0,\frac{1}{\sqrt{2}},0,\frac{1}{\sqrt{2}}\right]^T, \left[-0.306186,0.866322,0,\right.\right.\right.$\\$\left.\left.\left. 0.287612,0\right]^T,\left[0,0,0,\frac{i}{\sqrt{2}} ,0,0.\, \frac{-i}{\sqrt{2}}\right]^T,\left[-0.661438,-0.48483-0.0638078~ i,0,0.621313,0\right]^T \right\rbrace\right)$.  The $n$th time evolution  of $\mathcal{U}_{\pm}$ is  given  as follows:
\begin{equation}
\mathcal{U}_{+}^n=\left(
\begin{array}{cccccc}
 \frac{a_n}{64} \left(17+15 e^{\frac{i n \pi }{2}}\right)  & \frac{(-1)^{3/8}b_n}{16}  \sqrt{\frac{5}{2}}  & \frac{3\sqrt{5}a_n}{64}  \left(-1+e^{\frac{i n \pi }{2}}\right)  & \frac{(-1)^{7/8}}{8}  \sqrt{\frac{15}{2}}
e^{\frac{i n \pi }{2}}  & \frac{a_n}{32} \sqrt{\frac{105}{2}} \left(1-e^{\frac{i n \pi }{2}}\right)  & \frac{3(-1)^{3/8}\sqrt{{7}}b_n}{16\sqrt{2}}    \\
 -\frac{(-1)^{5/8}b_n}{16}  \sqrt{\frac{5}{2}}  & \frac{a_n}{2}  & -\frac{9 (-1)^{5/8}~b_n }{16 \sqrt{2}} & 0 & -\frac{(-1)^{5/8} \sqrt{21} ~b_n }{16} & 0 \\
 \frac{-3\sqrt{5}~a_n}{64}  \left(1-e^{\frac{i n \pi }{2}}\right)  & \frac{9 (-1)^{3/8} b_n}{16 \sqrt{2}}
& \frac{a_n}{64} \left(29+3 e^{\frac{i n \pi }{2}}\right)  & \frac{(-1)^{7/8}~b_n}{8}  \sqrt{\frac{3}{2}} e^{\frac{i n \pi
}{2}}  & \frac{a_n}{32} \sqrt{\frac{21}{2}} \left(1-e^{\frac{i n \pi }{2}}\right)  & -\frac{(-1)^{3/8} ~\sqrt{{35}}b_n}{16\sqrt{2}}
  \\
 \frac{-(-1)^{1/8}~\sqrt{{15}}b_n}{8\sqrt{2}}   e^{\frac{i n \pi }{2}} & 0 & -\frac{(-1)^{1/8}~b_n}{8}  \sqrt{\frac{3}{2}}
e^{\frac{i n \pi }{2}}  & e^{i n \pi } \cos\left(\frac{n \pi }{2}\right) & \frac{(-1)^{1/8} \sqrt{7}~b_n}{8}  e^{\frac{i
n \pi }{2}}  & 0 \\
 \frac{a_n}{32} \sqrt{\frac{105}{2}} \left(1-e^{\frac{i n \pi }{2}}\right)  & \frac{(-1)^{3/8} \sqrt{21}~b_n}{16}   & \frac{a_n}{32} \sqrt{\frac{21}{2}} \left(1-e^{\frac{i n \pi }{2}}\right)  & -\frac{(-1)^{7/8} \sqrt{7}~b_n}{8}
e^{\frac{i n \pi }{2}} & \frac{a_n}{32} \left(9+7 e^{\frac{i n \pi }{2}}\right)  & \frac{(-1)^{3/8} \sqrt{15}~b_n}{16}
  \\
 -\frac{3(-1)^{5/8} ~b_n}{16} \sqrt{\frac{7}{2}}  & 0 & \frac{(-1)^{5/8}~b_n}{16}  \sqrt{\frac{35}{2}}
& 0 & -\frac{(-1)^{5/8} \sqrt{15}~b_n}{16}   & \frac{a_n}{2}  \\
\end{array}
\right)
\end{equation}
\begin{equation}
\mbox{and}~~\mathcal{U}_{-}^n=\left(
\begin{array}{ccccc}
 f_n & 0 & -0.209632~b_n(-1)^\frac{3n}{8} . & 0 & -0.452856~b_n (-1)^\frac{3n}{8}
 \\
 0 & 0.5 c_n (-1)^\frac{31n}{24}  & 0 & (-0.5 i)~d_n(-1)^\frac{31n}{24}  & 0 \\
 -0.209632~ b_n(-1)^\frac{3n}{8}  & 0 & e_n & 0 & 0.2025232~b_n(-1)^\frac{3n}{8}
\\
 0 & (0.5 i)~d_n(-1)^\frac{31n}{24}  & 0 & 0.5~c_n(-1)^{31 n/24}  & 0 \\
 -0.452856~ b_n (-1)^\frac{3n}{8}  & 0 & 0.2025232~b_n(-1)^\frac{3n}{8} & 0 & g_n
\\
\end{array}
\right),
\end{equation}
where $a_n$=$1+e^{i n \pi }$,~$f_n$=$(-1)^\frac{3n}{8}  \left(0.53125 +0.46875 (-1)^n\right)$,~$g_n$=$(-1)^\frac{3n}{8} \left(0.5625+0.4375 (-1)^n\right)$, $b_n=e^{i n \pi }-1$,~ $c_n$=$1+(-1)^{2 n/3}$, $d_n$=$(-1)^{2 n/3}-1$ and $e_n$=$(-1)^\frac{3n}{8} \left(0.90625+0.09375(-1)^n\right)$. The initial state $\ket{\psi}$ after the $n${th} implementations of the unitary operator $\mathcal{U}$ can be expressed as follows:
\begin{eqnarray}
\ket{\psi_n}&=&\mathcal{U}^n\ket{\psi}\\ \nonumber
&=& \bar{d}_{1n}\ket{\phi_0^+}+\bar{d}_{2n}\ket{\phi_1^+}+\bar{d}_{3n}\ket{\phi_2^+}+\bar{d}_{4n}\ket{\phi_3^+} +\bar{d}_{5n}\ket{\phi_4^+}+\bar{d}_{6n}\ket{\phi_5^+}+\bar{d}_{7n}\ket{\phi_0^-}+\bar{d}_{8n}\ket{\phi_1^-}+\\ \nonumber&&\bar{d}_{9n}\ket{\phi_2^-}+\bar{d}_{10n}\ket{\phi_3^-}+\bar{d}_{11n}\ket{\phi_4^-},\nonumber
\end{eqnarray}
where the coefficients are given as follows:
\begin{equation}\label{Eq:qkt588}
\bar{d}_{jn}=\sum_{q=1}^{\frac{N+2}{2}}\mathcal{U}^n_{j,q}~ a_q+\sum_{q=\frac{N+4}{2}}^{N+1}\mathcal{U}^n_{j,q}~ b_{q-\frac{N+2}{2}},1\leq j\leq N+1.
\end{equation}
The expressions of the coefficients $\bar{d}_{jn}$ for $10$ qubits can be calculated using  Eqs. (\ref{Eq:arbitaray12}), (\ref{Eq:arbitaray15}), (\ref{Eq:arbitaray42}) and (\ref{Eq:qkt588}), as follows:
\begin{eqnarray}\nonumber
\bar{d}_{1n}&=&\frac{63 \left(-1+(-1)^n\right) e^{\frac{3 i \pi }{8}-5 i \phi_0 } \cos^5\left(\frac{\theta_0 }{2}\right) \sin^5\left(\frac{\theta_0
}{2}\right)}{8 \sqrt{2}}-\frac{15 e^{\frac{3 i \pi }{8}}}{2\sqrt{2}}  \sin\left(\frac{n \pi }{2}\right) \left(e^{-3 i \phi_0 } \cos^7\left(\frac{\theta_0
}{2}\right) \sin^3\left(\frac{\theta_0 }{2}\right)+ e^{-7 i \phi_0 } \cos^3\left(\frac{\theta_0 }{2}\right) \sin^7\left(\frac{\theta_0
}{2}\right)\right)\\ \nonumber &&-\frac{45 \left(-1-(-1)^n+(-i)^n+i^n\right)}{64\sqrt{2}} \left[6 \left( e^{-4 i \phi_0 } \cos^6\left(\frac{\theta_0
}{2}\right) \sin^4\left(\frac{\theta_0 }{2}\right)- e^{-6 i \phi_0 } \cos^4\left(\frac{\theta_0 }{2}\right) \sin^6\left(\frac{\theta_0
}{2}\right)\right)+\left(e^{-2 i \phi_0 } \cos^8\left(\frac{\theta_0
}{2}\right)\sin^2\left(\frac{\theta_0 }{2}\right)\right.\right.\\ \nonumber&& \left. \left. - e^{-8 i \phi_0 } \cos^2\left(\frac{\theta_0 }{2}\right) \sin^8\left(\frac{\theta_0
}{2}\right)\right)\right]+\frac{1}{64\sqrt{2}}
\left(17+17 (-1)^n+15 (-i)^n+15 i^n\right)\left(\cos^{10}\left(\frac{\theta_0 }{2}\right)-e^{-10 i
\phi_0 } \sin^{10}\left(\frac{\theta_0 }{2}\right)\right)+\\ \nonumber
&&\frac{5 e^{\frac{3 i \pi }{8}}}{16\sqrt{2}} \left(-1+(-1)^n\right) \left(e^{-i \phi_0 } \cos^9\left(\frac{\theta_0 }{2}\right)
\sin\left(\frac{\theta_0 }{2}\right)+ e^{-9 i \phi_0 } \cos\left(\frac{\theta_0 }{2}\right) \sin^9\left(\frac{\theta_0 }{2}\right)\right),\\ \nonumber
\end{eqnarray}
\begin{eqnarray}\nonumber
\bar{d}_{2n}&=&-\frac{\sqrt{5}\left(-1+(-1)^n\right) e^{\frac{5 i \pi }{8}}}{16}   \left[\left(e^{-4 i \phi_0 } \cos^6\left(\frac{\theta_0
}{2}\right) \sin^4\left(\frac{\theta_0 }{2}\right)- e^{-6 i \phi_0 } \cos^4\left(\frac{\theta_0 }{2}\right) \sin^6\left(\frac{\theta_0
}{2}\right)\right)- \left(\frac{\cos^{10}\left(\frac{\theta_0 }{2}\right)}{2}-\frac{e^{-10 i \phi_0 } \sin^{10}\left(\frac{\theta_0 }{2}\right)}{2}\right)\right.\\  \nonumber&&\left.-\frac{27 }{2}  \left( e^{-2 i \phi_0 } \cos^8\left(\frac{\theta_0
}{2}\right) \sin^2\left(\frac{\theta_0 }{2}\right)- e^{-8 i \phi_0 } \cos^2\left(\frac{\theta_0 }{2}\right) \sin^8\left(\frac{\theta_0
}{2}\right)\right)\right]+\frac{\sqrt{5}\left(1+(-1)^n\right)}{2}  \left( e^{-i \phi_0 } \cos^9\left(\frac{\theta_0 }{2}\right) \sin\left(\frac{\theta_0 }{2}\right)\right.\\  \nonumber&&\left.+
e^{-9 i \phi_0 } \cos\left(\frac{\theta_0 }{2}\right) \sin^9\left(\frac{\theta_0 }{2}\right)\right),\\ \nonumber
\bar{d}_{3n}&=&-\frac{21}{8} \sqrt{\frac{5}{2}} \left(-1+(-1)^n\right) e^{\frac{3 i \pi }{8}-5 i \phi_0 } \cos^5\left(\frac{\theta_0 }{2}\right)
\sin^5\left(\frac{\theta_0 }{2}\right)-\frac{3}{64} \sqrt{\frac{5}{2}} \left(-1-(-1)^n+(-i)^n+i^n\right)\left[7 \left( e^{-4 i \phi_0 } \cos^6\left(\frac{\theta_0
}{2}\right) \sin^4\left(\frac{\theta_0 }{2}\right)\right.\right.\\  \nonumber&&\left.\left.- e^{-6 i \phi_0 } \cos^4\left(\frac{\theta_0 }{2}\right) \sin\left(\frac{\theta_0
}{2}\right)^6\right)+\left(\cos^{10}\left(\frac{\theta_0 }{2}\right)-e^{-10 i
\phi_0 } \sin^{10}\left(\frac{\theta_0 }{2}\right)\right)\right]-\frac{3}{4} \sqrt{\frac{5}{2}} e^{\frac{7 i \pi }{8}+\frac{i n \pi }{2}} \left(-1+e^{i n \pi }\right) \left( e^{-3 i \phi_0 } \cos^7\left(\frac{\theta_0
}{2}\right) \right.\\ \nonumber&&\left.+\sin^3\left(\frac{\theta_0 }{2}\right)+ e^{-7 i \phi_0 } \cos^3\left(\frac{\theta_0 }{2}\right) \sin^7\left(\frac{\theta_0
}{2}\right)\right)+
\frac{3}{64} \sqrt{\frac{5}{2}}\left(29+29 (-1)^n+3 (-i)^n+3 i^n\right) \left( e^{-2 i \phi_0 } \cos^8\left(\frac{\theta_0 }{2}\right) \sin^2\left(\frac{\theta_0
}{2}\right)+\right.\\ \nonumber
&&\left.- e^{-8 i \phi_0 } \cos^2\left(\frac{\theta_0 }{2}\right) \sin^8\left(\frac{\theta_0 }{2}\right)\right)+\frac{9\sqrt{5}
\left(-1+(-1)^n\right) e^{\frac{3 i \pi }{8}}}{16 \sqrt{2}} \left( e^{-i \phi_0 } \cos^9\left(\frac{\theta_0 }{2}\right) \sin\left(\frac{\theta_0
}{2}\right)+ e^{-9 i \phi_0 } \cos\left(\frac{\theta_0 }{2}\right) \sin^9\left(\frac{\theta_0 }{2}\right)\right), \\ \nonumber
\bar{d}_{4n}&=&\frac{\sqrt{15}e^{\frac{5 i \pi }{8}} \sin\left(\frac{n \pi }{2}\right) }{8}  \left[14\left( e^{-4 i \phi_0 } \cos^6\left(\frac{\theta_0
}{2}\right) \sin^4\left(\frac{\theta_0 }{2}\right)- e^{-6 i \phi_0 } \cos^4\left(\frac{\theta_0 }{2}\right) \sin^6\left(\frac{\theta_0
}{2}\right)\right)-3   \left(\left( e^{-2 i \phi_0 } \cos^8\left(\frac{\theta_0
}{2}\right)\sin^2\left(\frac{\theta_0 }{2}\right)-\right.\right.\right. \\ \nonumber && \left.\left.\left. e^{-8 i \phi_0 } \cos^2\left(\frac{\theta_0 }{2}\right) \sin^8\left(\frac{\theta_0
}{2}\right)\right)-\frac{1}{3}  \left({\cos^{10}\left(\frac{\theta_0
}{2}\right)}-{e^{-10 i \phi_0 } \sin^{10}\left(\frac{\theta_0 }{2}\right)}\right)\right)\right] +2 \sqrt{15}\cos\left(\frac{n \pi }{2}\right) \left( e^{-3 i \phi_0 } \cos^7\left(\frac{\theta_0 }{2}\right) \sin^3\left(\frac{\theta_0
}{2}\right)\right. \\ \nonumber && \left.+ e^{-7 i \phi_0 } \cos^3\left(\frac{\theta_0 }{2}\right) \sin^7\left(\frac{\theta_0 }{2}\right)\right),\\ \nonumber
\bar{d}_{5n}&=&\frac{3\sqrt{105}\left(-1+(-1)^n\right) e^{\frac{3 i \pi }{8}-5 i \phi_0 }}{8}   \cos^5\left(\frac{\theta_0 }{2}\right)
\sin^5\left(\frac{\theta_0 }{2}\right)+\frac{\sqrt{105}\left(9+9 (-1)^n+7 (-i)^n+7 i^n\right)}{32}  \left( e^{-4 i \phi_0 } \cos^6\left(\frac{\theta_0
}{2}\right) \sin^4\left(\frac{\theta_0 }{2}\right)\right. \\ \nonumber && \left. -e^{-6 i \phi_0 } \cos^4\left(\frac{\theta_0 }{2}\right) \sin^6\left(\frac{\theta_0
}{2}\right)\right)+\frac{\sqrt{105}}{2}  e^{\frac{3 i \pi }{8}} \sin\left(\frac{n \pi }{2}\right) \left( e^{-3 i \phi_0 } \cos^7\left(\frac{\theta_0
}{2}\right) \sin^3\left(\frac{\theta_0 }{2}\right)+ e^{-7 i \phi_0 } \cos^3\left(\frac{\theta_0 }{2}\right) \sin^7\left(\frac{\theta_0
}{2}\right)\right) \\ \nonumber && +\frac{\sqrt{105}}{16}  \left(-1+(-1)^n\right) e^{\frac{3 i \pi }{8}} \left( e^{-i \phi_0 } \cos^9\left(\frac{\theta_0 }{2}\right)
\sin\left(\frac{\theta_0 }{2}\right)+  e^{-9 i \phi_0 } \cos\left(\frac{\theta_0 }{2}\right) \sin^9\left(\frac{\theta_0 }{2}\right)\right)-\frac{\sqrt{105}}{64}
 \left(-1-(-1)^n+(-i)^n+i^n\right)\\  \nonumber && \left(\left(\cos^{10}\left(\frac{\theta_0 }{2}\right)-e^{-10 i \phi_0 }
\sin^{10}\left(\frac{\theta_0 }{2}\right)\right)-3 \left( e^{-2 i \phi_0 } \cos^8\left(\frac{\theta_0
}{2}\right) \sin^2\left(\frac{\theta_0 }{2}\right)-e^{-8 i \phi_0 } \cos^2\left(\frac{\theta_0 }{2}\right) \sin^8\left(\frac{\theta_0
}{2}\right)\right)\right),\\ \nonumber
\bar{d}_{6n}&=&3 \sqrt{7} \left(1+(-1)^n\right) e^{-5 i \phi_0 } \cos^5\left(\frac{\theta_0 }{2}\right) \sin^5\left(\frac{\theta_0
}{2}\right)-\frac{15\sqrt{7} \left(-1+(-1)^n\right) e^{\frac{5 i \pi }{8}}}{16} \left[ \left( e^{-4 i \phi_0 } \cos^6\left(\frac{\theta_0 }{2}\right)
\sin^4\left(\frac{\theta_0 }{2}\right)- e^{-6 i \phi_0 } \cos^4\left(\frac{\theta_0 }{2}\right)\right. \right.\\ \nonumber
&&\left.\left.\sin^6\left(\frac{\theta_0 }{2}\right)\right)+\frac{1 }{2}\left( \left( e^{-2 i \phi_0 } \cos^8\left(\frac{\theta_0
}{2}\right)\sin^2\left(\frac{\theta_0 }{2}\right)- e^{-8 i \phi_0 } \cos^2\left(\frac{\theta_0 }{2}\right) \sin\left(\frac{\theta_0
}{2}\right)^8\right)\right]-\frac{1}{5}  \left({\cos^{10}\left(\frac{\theta_0 }{2}\right)}-{e^{-10
i \phi_0 } \sin^{10}\left(\frac{\theta_0 }{2}\right)}\right)\right),\\ \nonumber
\bar{d}_{7n}&=&\frac{15 e^{-\frac{5}{8} i n \pi } \left(-1+e^{i n \pi }\right)}{32\sqrt{2}} \left( 14\left( e^{-4 i \phi_0
} \cos^6\left(\frac{\theta_0 }{2}\right) \sin^4\left(\frac{\theta_0 }{2}\right)+ e^{-6 i \phi_0 } \cos^4\left(\frac{\theta_0 }{2}\right)
\sin^6\left(\frac{\theta_0 }{2}\right)\right)+3 \left(e^{-2 i \phi_0 } \cos^8\left(\frac{\theta_0}{2}\right) \sin^2\left(\frac{\theta_0 }{2}\right)\right.\right.\\ \nonumber && \left.\left.+ e^{-8 i \phi_0 } \cos^2\left(\frac{\theta_0 }{2}\right) \sin^8\left(\frac{\theta_0}{2}\right)\right)\right)+\frac{1}{32} e^{-\frac{5}{8} i n \pi } \left(15+17 e^{i n \pi }\right) \left(\frac{\cos^{10}\left(\frac{\theta_0 }{2}\right)}{\sqrt{2}}+\frac{e^{-10
i \phi_0 } \sin^{10}\left(\frac{\theta_0 }{2}\right)}{\sqrt{2}}\right),\\ \nonumber
\bar{d}_{8n}&=&e^{-\frac{3}{8} i n \pi }\left(2 \sqrt{15} \sin\left(\frac{n \pi }{3}\right) \left(e^{-3 i \phi_0 } \cos^7\left(\frac{\theta_0
}{2}\right) \sin^3\left(\frac{\theta_0 }{2}\right)- e^{-7 i \phi_0 } \cos^3\left(\frac{\theta_0 }{2}\right) \sin^7\left(\frac{\theta_0
}{2}\right)\right)+\sqrt{5} \cos\left(\frac{n \pi }{3}\right) \left( e^{-i \phi_0 } \cos^9\left(\frac{\theta_0 }{2}\right)
\right.\right.\\ \nonumber && \left.\left.\sin\left(\frac{\theta_0 }{2}\right)- e^{-9 i \phi_0 } \cos\left(\frac{\theta_0 }{2}\right) \sin^9\left(\frac{\theta_0 }{2}\right)\right)\right),\\ \nonumber
\end{eqnarray}
\begin{eqnarray}\nonumber
\bar{d}_{9n}&=&\frac{21}{16} \sqrt{\frac{5}{2}} \left(-e^{\frac{3 i n \pi }{8}}+e^{-\frac{5}{8} i n \pi }\right) \left( e^{-4
i \phi_0 } \cos^6\left(\frac{\theta_0 }{2}\right) \sin^4\left(\frac{\theta_0 }{2}\right)+ e^{-6 i \phi_0 } \cos^4\left(\frac{\theta_0
}{2}\right) \sin^6\left(\frac{\theta_0 }{2}\right)\right)+\frac{3~e^{-\frac{5}{8} i n \pi }}{32}  \sqrt{\frac{5}{2}} \left(3+29 e^{i n \pi }\right) \\ \nonumber && \left(
e^{-2 i \phi_0 }\cos^8\left(\frac{\theta_0 }{2}\right) \sin^2\left(\frac{\theta_0 }{2}\right)+e^{-8 i \phi_0 } \cos^2\left(\frac{\theta_0
}{2}\right) \sin^8\left(\frac{\theta_0 }{2}\right)\right)+\frac{3\sqrt{5} ~e^{-\frac{5}{8} i n \pi }\left(-1+e^{i n \pi }\right)}{32}   \left(\frac{\cos^{10}\left(\frac{\theta_0 }{2}\right)}{\sqrt{2}}+\frac{e^{-10
i \phi_0 } \sin^{10}\left(\frac{\theta_0 }{2}\right)}{\sqrt{2}}\right),\\ \nonumber
\bar{d}_{10n}&=&e^{-\frac{3}{8} i n \pi } \cos\left(\frac{n \pi }{3}\right) \left(2 \sqrt{15} e^{-3 i \phi_0 } \cos^7\left(\frac{\theta_0
}{2}\right) \sin^3\left(\frac{\theta_0 }{2}\right)-2 \sqrt{15} e^{-7 i \phi_0 } \cos^3\left(\frac{\theta_0 }{2}\right) \sin^7\left(\frac{\theta_0
}{2}\right)\right)-e^{-\frac{3}{8} i n \pi } \sin\left(\frac{n \pi }{3}\right) \left(\sqrt{5} e^{-i \phi_0 }\right.\\ \nonumber && \left. \cos^9\left(\frac{\theta_0 }{2}\right)
\sin\left(\frac{\theta_0 }{2}\right)-\sqrt{5} e^{-9 i \phi_0 } \cos\left(\frac{\theta_0 }{2}\right) \sin^9\left(\frac{\theta_0 }{2}\right)\right) \mbox{and}\\ \nonumber
\bar{d}_{11n}&=&\frac{\sqrt{105}~e^{-\frac{5}{8} i n \pi } }{16}  \left(7+9 e^{i n \pi }\right) \left( e^{-4 i \phi_0 } \cos^6\left(\frac{\theta_0
}{2}\right) \sin^4\left(\frac{\theta_0 }{2}\right)+ e^{-6 i \phi_0 } \cos^4\left(\frac{\theta_0 }{2}\right) \sin^6\left(\frac{\theta_0
}{2}\right)\right)+\frac{3\sqrt{105}}{32}  \left(-e^{\frac{3 i n \pi }{8}}+e^{-\frac{5}{8} i n \pi }\right) \\ \nonumber &&\left(
e^{-2 i \phi_0 } \cos^8\left(\frac{\theta_0 }{2}\right) \sin^2\left(\frac{\theta_0 }{2}\right)+ e^{-8 i \phi_0 } \cos^2\left(\frac{\theta_0
}{2}\right) \sin^8\left(\frac{\theta_0 }{2}\right)\right)+\frac{e^{-\frac{5}{8} i n \pi } }{16} \sqrt{\frac{105}{2}} \left(-1+e^{i n \pi }\right) \left(\frac{\cos^{10}\left(\frac{\theta_0 }{2}\right)}{\sqrt{2}}+\frac{e^{-10
i \phi_0 } \sin^{10}\left(\frac{\theta_0 }{2}\right)}{\sqrt{2}}\right).
\end{eqnarray}
\end{document}